\documentclass[vecphys]{svmult}
\usepackage{amsmath}

\usepackage{makeidx}         
\usepackage{graphicx}        
\usepackage{multicol}        
\usepackage[bottom]{footmisc}

\def\lya{Ly$\alpha$~}
\def\ljeans{\lambda_{\rm J}}
\def\centering{ }
\def\ion{ }
\def\citep{\cite}
\def\citet{\cite}
\def\HII{H II~}
\def\sun{\odot}

\def\ga{>}
\def\gtrsim{>}
\def\la{<}
\def\del{\delta}
\def\lesssim{<}
\def\kjeans{k_{\rm J}}
\def\mjeans{M_{\rm J}}
\def\beq{\begin{equation}}
\def\eeq{\end{equation}}
\def\ba{\begin{eqnarray}}
\def\ee{\end{equation}}

\def\Msun{$M_\odot$}

\def\beqa{\begin{eqnarray}}
\def\eeqa{\end{eqnarray}}
\def\xb{{\bf x}}
\def\rb{{\bf r}}
\def\vb{{\bf v}}
\def\ub{{\bf u}}
\def\kb{{\bf k}}
\def\Omm{{\Omega_m}}
\def\Ommz{{\Omega_m^{\,z}}}
\def\Omr{{\Omega_r}}
\def\Omk{{\Omega_k}}
\def\Oml{{\Omega_{\Lambda}}}
\def\nb{\bar{n}}

\def\Ng{N_\gamma}

\def\HI{\rm H\,I~}
\def\cN{c_{\rm N}}
\def\kB{k}
\def\Ni{N_{\rm ion}}
\def\fg{f_{\rm gas}}
\def\fe{f_{\rm eject}}
\def\fin{f_{\rm int}}
\def\fw{f_{\rm wind}}
\def\NGST{{\it JWST}\,}
\def\dh{{\delta_{H}}}
\def\n{{\bf n}}
\def\k{{\bf k}}

\newcommand{\bx}{{\bf x}}

\newcommand{\br}{{\bf r}}
\newcommand{\bk}{{\bf k}}
\newcommand{\Lya}{Ly$\alpha$~}

\newcommand{\td}{{\tilde{\delta}}}

\makeindex             

\begin{document}

\title*{First Light}
\author{Abraham Loeb\inst{1}}
\institute{Department of Astronomy, Harvard University, 60 Garden St.,
Cambridge, MA 02138 \texttt{aloeb@cfa.harvard.edu}}
%
%
\maketitle

\begin{abstract}

The first dwarf galaxies, which constitute the building blocks of the
collapsed objects we find today in the Universe, had formed hundreds of
millions of years after the big bang. This pedagogical review describes the
early growth of their small-amplitude seed fluctuations from the epoch of
inflation through dark matter decoupling and matter-radiation equality, to
the final collapse and fragmentation of the dark matter on all mass scales
above $\sim 10^{-4}M_\odot$. The condensation of baryons into halos in the
mass range of $\sim 10^5$--$10^{10}M_\odot$ led to the formation of the
first stars and the re-ionization of the cold hydrogen gas, left over from
the big bang.  The production of heavy elements by the first stars started
the metal enrichment process that eventually led to the formation of rocky
planets and life.

A wide variety of instruments currently under design [including
large-aperture infrared telescopes on the ground or in space ({\it JWST}),
and low-frequency arrays for the detection of redshifted 21cm radiation],
will establish better understanding of the first sources of light during an
epoch in cosmic history that was largely unexplored so far. Numerical
simulations of reionization are computationally challenging, as they
require radiative transfer across large cosmological volumes as well as
sufficently high resolution to identify the sources of the ionizing
radiation. The technological challenges for observations and the
computational challenges for numerical simulations, will motivate intense
work in this field over the coming decade.

\noindent
{\bf Disclaimer:} {\it This review was written as an introductory text for
a series of lectures at the SAAS-FEE 2006 winter school, and so it includes
a limited sample of references on each subject. It does not intend to
provide a comprehensive list of all up-to-date references on the topics
under discussion, but rather to raise the interest of beginning graduate
students in the related literature.}

\end{abstract}

%


\section{\bf Opening Remarks}
\label{sec1}

When I open the daily newspaper as part of my morning routine, I often see
lengthy descriptions of conflicts between people on borders, properties, or
liberties. Today's news is often forgotten a few days later.  But when one
opens ancient texts that have appealed to a broad audience over a longer
period of time, such as the Bible, what does one often find in the opening
chapter?...  a discussion of how the constituents of the Universe
(including light, stars and life) were created.  Although humans are often
occupied with mundane problems, they are curious about the big picture. As
citizens of the Universe, we cannot help but wonder how the first sources
of light formed, how life came to existence, and whether we are alone as
intelligent beings in this vast space.  As astronomers in the twenty first
century, we are uniquely positioned to answer these big questions with
scientific instruments and a quantitative methodology. In this pedagogical
review, intended for students preparing to specialize in cosmology, I will
describe current ideas about one of these topics: {\it the appearance of
the first sources of light and their influence on the surrounding
Universe}. This topic is one of the most active frontiers in present-day
cosmology. As such it is an excellent area for a PhD thesis of a graduate
student interested in cosmology.  I will therefore highlight the unsolved
questions in this field as much as the bits we understand.

\section{\bf Excavating the Universe for Clues About Its History}

When we look at our image reflected off a mirror at a distance of 1 meter,
we see the way we looked 6 nano-seconds ago, the light travel time to the
mirror and back. If the mirror is spaced $10^{19}~{\rm cm}=3$pc away, we
will see the way we looked twenty one years ago. Light propagates at a
finite speed, and so by observing distant regions, we are able to see how
the Universe looked like in the past, a light travel time ago. The
statistical homogeneity of the Universe on large scales guarantees that
what we see far away is a fair statistical representation of the conditions
that were present in in our region of the Universe a long time ago.

\begin{figure}
\centering
\includegraphics[height=6cm]{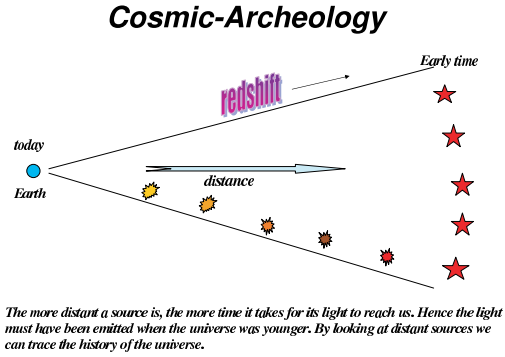}
\caption{Cosmology is like archeology. The deeper one looks, the older is
the layer that one is revealing, owing to the finite propagation speed of
light.}
\label{fig:1}       
\end{figure}

This fortunate situation makes cosmology an empirical science. We do not
need to guess how the Universe evolved. Using telescopes we can simply see
the way it appeared at earlier cosmic times. Since a greater distance means
a fainter flux from a source of a fixed luminosity, the observation of the
earliest sources of light requires the development of sensitive instruments
and poses challenges to observers.

We can in principle image the Universe only if it is transparent. Earlier
than 0.4 million years after the big bang, the cosmic plasma was ionized
and the Universe was opaque to Thomson scattering by the dense gas of free
electrons that filled it. Thus, telescopes cannot be used to image the
infant Universe at earlier times (or redshifts $\ga 10^3$). The earliest
possible image of the Universe was recorded by COBE and WMAP (see Fig. 2).

\begin{figure}
\centering
\includegraphics[height=6cm]{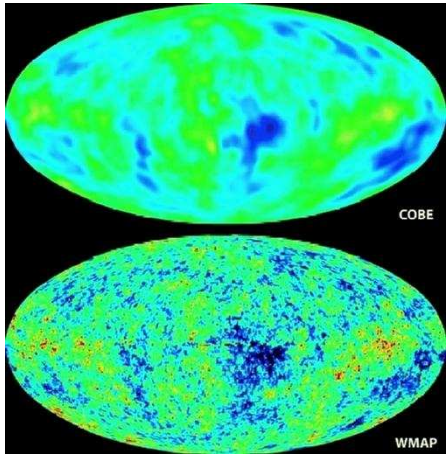}
\caption{Images of the Universe shortly after it became transparent, taken
by the {\it COBE} and {\it WMAP} satellites (see http://map.gsfc.nasa.gov/
for details).  The slight density inhomogeneties in the otherwise uniform
Universe, imprinted hot and cold brightness map of the cosmic microwave
background.  The existence of these anisotropies was predicted three
decades before the technology for taking this image 
became available in a number of theoretical papers, 
including \cite{zel,sachs,sciama,silk,peeb}.}
\end{figure}

\begin{figure}
\centering
\includegraphics[height=6cm]{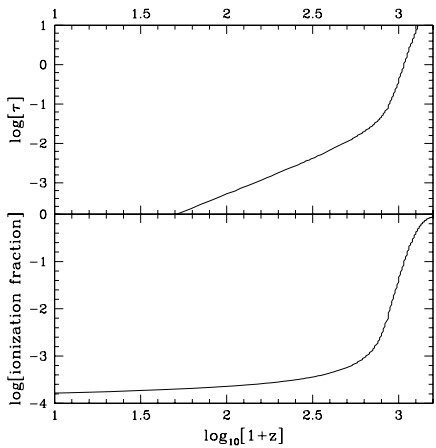}
\caption{The optical depth of the Universe to electron scattering (upper
panel) and the ionization fraction (lower panel) as a function of redshift
before reionization.  Observatories of electromagnetic radiation cannot
image the opaque Universe beyond a redshift of $z\sim 1100$.}
\end{figure}

\section{Bakground Cosmological Model}

\subsection{The Expanding Universe}
\label{sec2.1}

The modern physical description of the Universe as a whole can be traced
back to Einstein, who argued theoretically for the so-called ``cosmological
principle'': that the distribution of matter and energy must be homogeneous
and isotropic on the largest scales. Today isotropy is well established
(see the review by Wu, Lahav, \& Rees 1999 \cite{Wu99}) for the distribution of faint
radio sources, optically-selected galaxies, the X-ray background, and most
importantly the cosmic microwave background (hereafter, CMB; see, e.g.,
Bennett et al.\ 1996 \cite{Be96}). The constraints on homogeneity are less strict, but
a cosmological model in which the Universe is isotropic but significantly
inhomogeneous in spherical shells around our special location, is also
excluded \cite{Goodman95}.

In General Relativity, the metric for a space which is spatially
homogeneous and isotropic is the Friedman-Robertson-Walker metric, which
can be written in the form \beq \label{RW}
ds^2=dt^2-a^2(t)\left[\frac{dR^2}{1-k\,R^2}+R^2
\left(d\theta^2+\sin^2\theta\,d\phi^2\right)\right]\ , \eeq where $a(t)$ is
the cosmic scale factor which describes expansion in time, and
$(R,\theta,\phi)$ are spherical comoving coordinates. The constant $k$
determines the geometry of the metric; it is positive in a closed Universe,
zero in a flat Universe, and negative in an open Universe. Observers at
rest remain at rest, at fixed $(R,\theta,\phi)$, with their physical
separation increasing with time in proportion to $a(t)$. A given observer
sees a nearby observer at physical distance $D$ receding at the Hubble
velocity $H(t)D$, where the Hubble constant at time $t$ is
$H(t)=d\,a(t)/dt$. Light emitted by a source at time $t$ is observed at
$t=0$ with a redshift $z=1/a(t)-1$, where we set $a(t=0) \equiv 1$
for convenience (but note that old textbooks may use a different 
convention).

The Einstein field equations of General Relativity yield the Friedmann
equation (e.g., Weinberg 1972 \cite{We72}; Kolb \& Turner 1990
\cite{Kolb90}) \beq H^2(t)=\frac{8 \pi G}{3}\rho-\frac{k}{a^2}\ ,\eeq which
relates the expansion of the Universe to its matter-energy content. For
each component of the energy density $\rho$, with an equation of state
$p=p(\rho)$, the density $\rho$ varies with $a(t)$ according to the
equation of energy conservation \beq d (\rho R^3)=-p d(R^3)\ . \eeq With
the critical density \beq \rho_C(t) \equiv \frac{3 H^2(t)}{8 \pi G} \eeq
defined as the density needed for $k=0$, we define the ratio of the total
density to the critical density as \beq \Omega \equiv \frac{\rho}{\rho_C}\
. \eeq With $\Omm$, $\Oml$, and $\Omr$ denoting the present contributions
to $\Omega$ from matter (including cold dark matter as well as a
contribution $\Omega_b$ from baryons), vacuum density (cosmological
constant), and radiation, respectively, the Friedmann equation becomes \beq
\frac{H(t)}{H_0}= \left[ \frac{\Omm} {a^3}+ \Oml+ \frac{\Omr}{a^4}+
\frac{\Omk}{a^2}\right]\ , \eeq where we define $H_0$ and
$\Omega_0=\Omm+\Oml+\Omr$ to be the present values of $H$ and $\Omega$,
respectively, and we let \beq \Omk \equiv -\frac{k}{H_0^2}=1-\Omega_m. \eeq
In the particularly simple Einstein-de Sitter model ($\Omm=1$,
$\Oml=\Omr=\Omk=0$), the scale factor varies as $a(t) \propto
t^{2/3}$. Even models with non-zero $\Oml$ or $\Omk$ approach the
Einstein-de Sitter behavior at high redshift, i.e.\ when $(1+z) \gg
|\Omm^{-1}-1|$ (as long as $\Omr$ can be neglected). In this high-$z$
regime the age of the Universe is,
\begin{equation}
t\approx {2\over 3 H_0 \sqrt{\Omega_m}} \left(1+z\right)^{-3/2}~.
\end{equation}
The Friedmann equation implies that models with $\Omk=0$ converge to the
Einstein-de Sitter limit faster than do open models.

In the standard hot Big Bang model, the Universe is initially hot and the
energy density is dominated by radiation. The transition to matter
domination occurs at $z \sim 3500$, but the Universe remains hot enough
that the gas is ionized, and electron-photon scattering effectively couples
the matter and radiation. At $z \sim 1100$ the temperature drops below
$\sim 3000$K and protons and electrons recombine to form neutral
hydrogen. The photons then decouple and travel freely until the present,
when they are observed as the CMB \cite{WMAP}.

\subsection{Composition of the Universe}

According to the standard cosmological model, the Universe started at the
big bang about 14 billion years ago.  During an early epoch of accelerated
superluminal expansion, called inflation, a region of microscopic size was
stretched to a scale much bigger than the visible Universe and our local
geometry became flat. At the same time, primordial density fluctuations
were generated out of quantum mechanical fluctuations of the vacuum.  These
inhomogeneities seeded the formation of present-day structure through the
process of gravitational instability. The mass density of ordinary
(baryonic) matter makes up only a fifth of the matter that led to the
emergence of structure and the rest is the form of an unknown dark matter
component.  Recently, the Universe entered a new phase of accelerated
expansion due to the dominance of some dark vacuum energy density over the
ever rarefying matter density.

\noindent
The basic question that cosmology attempts to answer is: 

\noindent
{\bf What are the ingredients (composition and initial conditions) of the
Universe and what processes generated the observed structures in it?}

\noindent
In detail, we would like to know: 

\noindent
{\it (a)} Did inflation occur and when? If so, what drove it and how did it
end?

\noindent
{\it (b)} What is the nature of of the dark energy and how does it change
over time and space?

\noindent
{\it (c)} What is the nature of the dark matter and how did it regulate the
evolution of structure in the Universe?

Before hydrogen recombined, the Universe was opaque to electromagnetic
radiation, precluding any possibility for direct imaging of its evolution.
The only way to probe inflation is through the fossil record that it left
behind in the form of density perturbations and gravitational waves.
Following inflation, the Universe went through several other milestones
which left a detectable record. These include: baryogenesis (which resulted
in the observed asymmetry between matter and anti-matter), the electroweak
phase transition (during which the symmetry between electromagnetic and
weak interactions was broken), the QCD phase transition (during which
protons and neutrons were assembled out of quarks and gluons), the dark
matter freeze-out epoch (during which the dark matter decoupled from the
cosmic plasma), neutrino decoupling, electron-positron annihilation, and
light-element nucleosynthesis (during which helium, deuterium and lithium
were synthesized). The signatures that these processes left in the Universe
can be used to constrain its parameters and answer the above questions.

Half a million years after the big bang, hydrogen recombined and the
Universe became transparent. The ultimate goal of observational cosmology
is to image the entire history of the Universe since then. Currently, we
have a snapshot of the Universe at recombination from the CMB, and detailed
images of its evolution starting from an age of a billion years until the
present time. The evolution between a million and a billion years has not
been imaged as of yet.

Within the next decade, NASA plans to launch an infrared space telescope
(JWST) that will image the very first sources of light (stars and black
holes) in the Universe, which are predicted theoretically to have formed in
the first hundreds of millions of years. In parallel, there are several
initiatives to construct large-aperture infrared telescopes on the ground
with the same goal in
mind\footnote{http://www.eso.org/projects/owl/}$^{,}$
\footnote{http://celt.ucolick.org/}$^{,}$\footnote{http://www.gmto.org/}.
The neutral hydrogen, relic from cosmological recombination, can be mapped
in three-dimensions through its 21cm line even before the first galaxies
formed \cite{Loeb04}. Several groups are currently constructing
low-frequency radio arrays in an attempt to map the initial inhomogeneities
as well as the process by which the hydrogen was re-ionized by the first
galaxies.

\begin{figure}
\centering
\includegraphics[height=6cm]{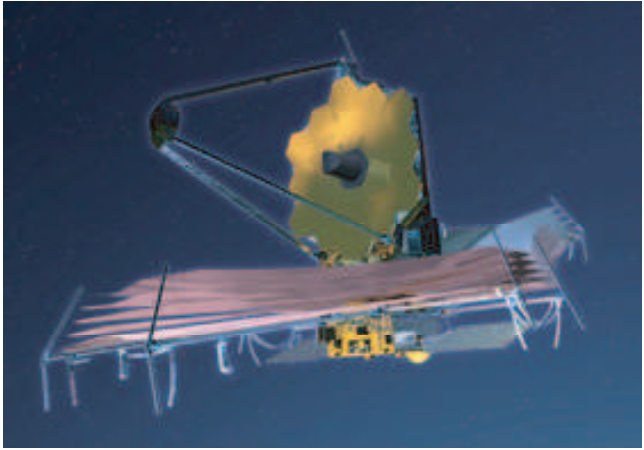}
\caption{A sketch of the current design for the {\it James Webb Space
Telescope}, the successor to the {\it Hubble Space Telescope} to be
launched in 2011 (http://www.jwst.nasa.gov/). The current design includes a
primary mirror made of beryllium which is 6.5 meter in diameter as well as
an instrument sensitivity that spans the full range of infrared wavelengths
of 0.6--28$\mu$m and will allow detection of the first galaxies in the
infant Universe.
The telescope will orbit 1.5 million km from Earth at the Lagrange L2
point.}
\end{figure}

The next generation of ground-based telescopes will have a diameter of
twenty to thirty meter. Together with JWST (that will not be affected by
the atmospheric backgound) they will be able to image the first
galaxies. Given that these galaxies also created the ionized bubbles around
them, the same galaxy locations should correlate with bubbles in the
neutral hydrogen (created by their UV emission). Within a decade it would
be possible to explore the environmental influence of individual galaxies
by using the two sets of instruments in concert \cite{WyBar}.

\begin{figure}
\centering
\includegraphics[height=6cm]{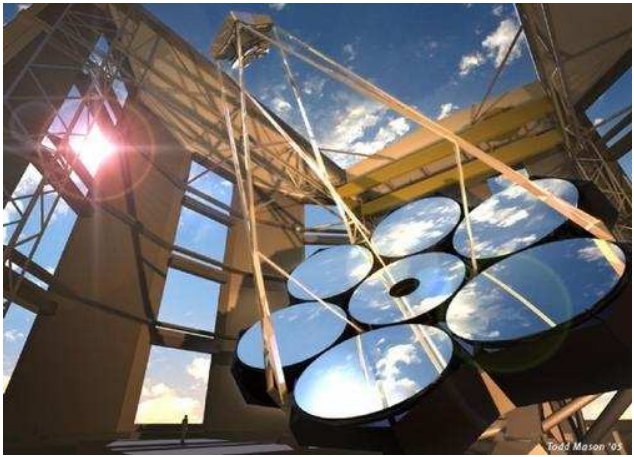}
\caption{Artist conception of the design for one of the future giant
telescopes that could probe the first generation of galaxies from the
ground. The {\it Giant Magellan Telescope} (GMT) will contain seven mirrors
(each 8.4 meter in diameter) and will have the resolving power equivalent
to a 24.5 meter (80 foot) primary mirror. For more details see
http://www.gmto.org/}
\label{gmt}
\end{figure}

The dark ingredients of the Universe can only be probed indirectly through
a variety of luminous tracers.  The distribution and nature of the dark
matter are constrained by detailed X-ray and optical observations of
galaxies and galaxy clusters. The evolution of the dark energy with cosmic
time will be constrained over the coming decade by surveys of Type Ia
supernovae, as well as surveys of X-ray clusters, up to a redshift of two.

On large scales ($\ga 10$Mpc) the power-spectrum of primordial density
perturbations is already known from the measured microwave background
anisotropies, galaxy surveys, weak lensing, and the Ly$\alpha$
forest. Future programs will refine current knowledge, and will search for
additional trademarks of inflation, such as gravitational waves (through
CMB polarization), small-scale structure (through high-redshift galaxy
surveys and 21cm studies), or the Gaussian statistics of the initial
perturbations.


%
%
%
%
The big bang is the only known event where particles with energies
approaching the Planck scale [$(\hbar c^5/G)^{1/2}\sim 10^{19}~{\rm GeV}$]
interacted.  It therefore offers prospects for probing the unification
physics between quantum mechanics and general relativity (to which string
theory is the most-popular candidate).  Unfortunately, the exponential
expansion of the Universe during inflation erases memory of earlier cosmic
epochs, such as the Planck time.

%
%
%

\subsection{Linear Gravitational Growth}
\label{sec2.2}

Observations of the CMB (e.g., Bennett et al.\ 1996 \cite{Be96}) show that the
Universe at recombination was extremely uniform, but with spatial
fluctuations in the energy density and gravitational potential of
roughly one part in $10^5$. Such small fluctuations, generated in the
early Universe, grow over time due to gravitational instability, and
eventually lead to the formation of galaxies and the large-scale
structure observed in the present Universe.

As before, we distinguish between fixed and comoving
coordinates. Using vector notation, the fixed coordinate ${\bf r}$
corresponds to a comoving position $\xb=\rb/a$. In a homogeneous
Universe with density $\rho$, we describe the cosmological expansion
in terms of an ideal pressureless fluid of particles each of which is
at fixed $\xb$, expanding with the Hubble flow $\vb=H(t) \rb$ where
$\vb=d\rb/dt$. Onto this uniform expansion we impose small
perturbations, given by a relative density perturbation \beq
\delta(\xb)=\frac{\rho(\rb)}{\bar{\rho}}-1\ , \eeq where the mean
fluid density is $\bar{\rho}$, with a corresponding peculiar velocity
$\ub \equiv \vb - H \rb$. Then the fluid is described by the
continuity and Euler equations in comoving coordinates \cite{p80,Peebles}:
\beqa \frac{\partial \delta}{\partial t}+\frac{1}{a}{\bf
\nabla} \cdot \left[(1+\delta) \ub\right] &=& 0 \\ \frac{\partial
\ub}{\partial t}+H\ub+\frac{1}{a}(\ub \cdot {\bf \nabla})
\ub&=&-\frac{1}{a}{\bf \nabla}\phi\ .  \eeqa The potential $\phi$ is
given by the Poisson equation, in terms of the density perturbation:
\beq \nabla^2\phi=4 \pi G \bar{\rho} a^2 \delta\ . \eeq This fluid
description is valid for describing the evolution of collisionless
cold dark matter particles until different particle streams cross.
This
``shell-crossing'' typically occurs only after perturbations have
grown to become non-linear, and at that point the individual particle
trajectories must in general be followed. Similarly, baryons can be
described as a pressureless fluid as long as their temperature is
negligibly small, but non-linear collapse leads to the formation of
shocks in the gas.

For small perturbations $\delta \ll 1$, the fluid equations can be
linearized and combined to yield \beq \frac{\partial^2\delta}{\partial
t^2}+2 H \frac{\partial\delta}{\partial t}=4 \pi G \bar{\rho} \delta\
. \eeq This linear equation has in general two independent solutions, only
one of which grows with time. Starting with random initial conditions, this
``growing mode'' comes to dominate the density evolution. Thus, until it
becomes non-linear, the density perturbation maintains its shape in
comoving coordinates and grows in proportion to a growth factor $D(t)$. The
growth factor in the matter-dominated era is given by \cite{p80} \beq
D(t) \propto \frac{\left(\Oml a^3+\Omk a+\Omm\right)^{1/2}}{a^{3/2}}\int_0^a
\frac{a'^{3/2}\, da'}{\left(\Oml a'^3+\Omk a'+\Omm\right)^{3/2}}\ , \eeq where
we neglect $\Omr$ when considering halos forming in the matter-dominated
regime at $z \ll 10^4$. In the Einstein-de Sitter model (or, at high redshift,
in other models as well) the growth factor is simply proportional to
$a(t)$.

The spatial form of the initial density fluctuations can be described in
Fourier space, in terms of Fourier components \beq \delta_\kb = \int d^3x\,
\delta(x) e^{-i \kb \cdot \xb}\ .\eeq Here we use the comoving wavevector
$\kb$, whose magnitude $k$ is the comoving wavenumber which is equal to
$2\pi$ divided by the wavelength. The Fourier description is particularly
simple for fluctuations generated by inflation (e.g., Kolb \& Turner 1990
\cite{Kolb90}). Inflation generates perturbations given by a Gaussian
random field, in which different $\kb$-modes are statistically independent,
each with a random phase. The statistical properties of the fluctuations
are determined by the variance of the different $\kb$-modes, and the
variance is described in terms of the power spectrum $P(k)$ as follows:
\beq \left<\delta_{\kb} \delta_{{\bf k'}}^{*}\right>=\left(2 \pi\right)^3
P(k)\, \delta^{(3)} \left(\kb-{\bf k'}\right)\ , \eeq where $\delta^{(3)}$
is the three-dimensional Dirac delta function.  The gravitational potential
fluctuations are sourced by the density fluctuations through Poisson's
equation.

In standard models, inflation produces a primordial power-law spectrum
$P(k) \propto k^n$ with $n \sim 1$. Perturbation growth in the
radiation-dominated and then matter-dominated Universe results in a
modified final power spectrum, characterized by a turnover at a scale
of order the horizon $cH^{-1}$ at matter-radiation equality, and a
small-scale asymptotic shape of $P(k) \propto k^{n-4}$. The overall
amplitude of the power spectrum is not specified by current models of
inflation, and it is usually set by comparing to the observed CMB
temperature fluctuations or to local measures of large-scale
structure.

Since density fluctuations may exist on all scales, in order to determine
the formation of objects of a given size or mass it is useful to consider
the statistical distribution of the smoothed density field.  Using a window
function $W(\rb)$ normalized so that $\int d^3r\, W(\rb)=1$, the smoothed
density perturbation field, $\int d^3r \delta(\xb) W(\rb)$, itself follows
a Gaussian distribution with zero mean. For the particular choice of a
spherical top-hat, in which $W=1$ in a sphere of radius $R$ and is zero
outside, the smoothed perturbation field measures the fluctuations in the
mass in spheres of radius $R$. The normalization of the present power
spectrum is often specified by the value of $\sigma_8 \equiv \sigma(R=8
h^{-1} {\rm Mpc})$. For the top-hat, the smoothed perturbation field is
denoted $\delta_R$ or $\delta_M$, where the mass $M$ is related to the
comoving radius $R$ by $M=4 \pi \rho_m R^3/3$, in terms of the current mean
density of matter $\rho_m$. The variance $\langle \delta_M \rangle^2$ is
\beq \sigma^2(M)= \sigma^2(R)= \int_0^{\infty}\frac{dk}{2 \pi^2} \,k^2 P(k)
\left[\frac{3 j_1(kR)}{kR} \right]^2\ ,\label{eqsigM}\eeq where
$j_1(x)=(\sin x-x \cos x)/x^2$. The function $\sigma(M)$ plays a crucial
role in estimates of the abundance of collapsed objects, as we describe
later.

Species that decouple from the cosmic plasma (like the dark matter or the
baryons) would show fossil evidence for acoustic oscillations in their
power spectrum of inhomogeneities due to sound waves in the radiation fluid
to which they were coupled at early times.  This phenomenon can be
understood as follows. Imagine a localized point-like perturbation from
inflation at $t=0$. The small perturbation in density or pressure will send
out a sound wave that will reach the sound horizon $c_s t$ at any later
time $t$.  The perturbation will therefore correlate with its surroundings
up to the sound horizon and all $k$-modes with wavelengths equal to this
scale or its harmonics will be correlated.
The scales of the perturbations that grow to become the first collapsed
objects at $z< 100$ cross the horizon in the radiation dominated era
after the dark matter decouples from the cosmic plasma. Next we consider
the imprint of this decoupling on the smallest-scale structure of
the dark matter.

\subsection{The Smallest-Scale Power Spectrum of Cold Dark Matter}

%
A broad range of observational data involving the dynamics of galaxies, the
growth of large-scale structure, and the dynamics and nucleosynthesis of
the Universe as a whole, indicate the existence of dark matter with a mean
cosmic mass density that is $\sim 5$ times larger than the density of the
baryonic matter \cite{Jungman,WMAP}.  The data is consistent with a dark
matter composed of weakly-interacting, massive particles, that decoupled
early and adiabatically cooled to an extremely low temperature by the
present time \cite{Jungman}.  The Cold Dark Matter (CDM) has not been
observed directly as of yet, although laboratory searches for particles
from the dark halo of our own Milky-Way galaxy have been able to restrict
the allowed parameter space for these particles. Since an alternative
more-radical interpretation of the dark matter phenomenology involves a
modification of gravity \cite{Beken}, it is of prime importance to find
direct fingerprints of the CDM particles. One such fingerprint involves the
small-scale structure in the Universe \cite{Green}, on which we focus in
this section.

The most popular candidate for the CDM particle is a Weakly Interacting
Massive Particle (WIMP).  The lightest supersymmetric particle (LSP) could
be a WIMP (for a review see \cite{Jungman}).  The CDM particle mass depends
on free parameters in the particle physics model but typical values cover a
range around $M\sim 100~{\rm GeV}$ (up to values close to a TeV). In many
cases the LSP hypothesis will be tested at the Large Hadron Collider
(e.g. \cite{battaglia}) or in direct detection experiments
(e.g. \cite{baltz}).

The properties of the CDM particles affect their response to the
small-scale primordial inhomogeneities produced during cosmic
inflation. The particle cross-section for scattering off standard model
fermions sets the epoch of their thermal and kinematic decoupling from the
cosmic plasma (which is significantly later than the time when their
abundance freezes-out at a temperature $T\sim M$). Thermal decoupling is
defined as the time when the temperature of the CDM stops following that of
the cosmic plasma while kinematic decoupling is defined as the time when
the bulk motion of the two species start to differ.  For CDM the epochs of
thermal and kinetic decoupling coincide.  They occur when the time it takes
for collisions to change the momentum of the CDM particles equals the
Hubble time.  The particle mass determines the thermal spread in the speeds
of CDM particles, which tends to smooth-out fluctuations on very small
scales due to the free-streaming of particles after kinematic decoupling
\cite{Green,Green2}.  Viscosity has a similar effect before the CDM fluid
decouples from the cosmic radiation fluid \cite{Hofmann}.  An important
effect involves the memory the CDM fluid has of the acoustic oscillations
of the cosmic radiation fluid out of which it decoupled.  Here we consider
the imprint of these acoustic oscillations on the small-scale power
spectrum of density fluctuations in the Universe. Analogous imprints of
acoustic oscillations of the baryons were identified recently in maps of
the CMB \cite{WMAP}, and the distribution of nearby galaxies
\cite{Eisenste}; these signatures appear on much larger scales, since the
baryons decouple much later when the scale of the horizon is larger.  The
discussion in this section follows Loeb \& Zaldarriaga (2005) \cite{Loe05}.

\noindent{\bf Formalism}

Kinematic decoupling of CDM occurs during the radiation-dominated era.  For
example, if the CDM is made of neutralinos with a particle mass of $\sim
100~{\rm GeV}$, then kinematic decoupling occurs at a cosmic temperature of
$T_{\rm d}\sim 10~{\rm MeV}$ \cite{Hofmann,Chen}.  As long as $T_d \ll
100~{\rm MeV}$, we may ignore the imprint of the QCD phase transition
(which transformed the cosmic quark-gluon soup into protons and neutrons)
on the CDM power spectrum \cite{Schmid}.  Over a short period of time
during this transition, the pressure does not depend on density and the
sound speed of the plasma vanishes, resulting in a significant growth for
perturbations with periods shorter than the length of time over which the
sound speed vanishes. The transition occurs when the temperature of the
cosmic plasma is $\sim 100-200~{\rm MeV}$ and lasts for a small fraction of
the Hubble time. As a result, the induced modifications are on scales
smaller than those we are considering here and the imprint of the QCD phase
transition is washed-out by the effects we calculate.

At early times the contribution of the dark matter to the energy density is
negligible. Only at relatively late times when the cosmic temperature drops
to values as low as $\sim 1$ eV, matter and radiation have comparable
energy densities.  As a result, the dynamics of the plasma at earlier times
is virtually unaffected by the presence of the dark matter particles.  In
this limit, the dynamics of the radiation determines the gravitational
potential and the dark matter just responds to that potential.  We will use
this simplification to obtain analytic estimates for the behavior of the
dark matter transfer function.

The primordial inflationary fluctuations lead to acoustic modes in the
radiation fluid during this era. The interaction rate of the particles in
the plasma is so high that we can consider the plasma as a perfect fluid
down to a comoving scale,
\begin{equation}
\lambda_f \sim \eta_{d}/\sqrt{N} \ \ \ \ ;  \ \ \ N \sim n \sigma t_d  ,
\end{equation}
where $\eta_d=\int_{0}^{t_d} dt/a(t)$ is the conformal time (i.e.  the
comoving size of the horizon) at the time of CDM decoupling, $t_d$;
$\sigma$ is the scattering cross section and $n$ is the relevant particle
density.  (Throughout this section we set the speed of light and Planck's
constant to unity for brevity.) The damping scale depends on the species
being considered and its contribution to the energy density, and is the
largest for neutrinos which are only coupled through weak interactions. In
that case $N\sim (T/T_{d}^\nu)^3$ where $T_{d}^\nu\sim 1\ {\rm MeV}$ is the
temperature of neutrino decoupling. At the time of CDM decoupling $N\sim
M/T_d \sim 10^4$ for the rest of the plasma, where $M$ is the mass of the
CDM particle. Here we will consider modes of wavelength larger than
$\lambda_f$, and so we neglect the effect of radiation diffusion damping
and treat the plasma (without the CDM) as a perfect fluid.

The equations of motion for a perfect fluid during the radiation era can be
solved analytically. We will use that solution here, following the notation
of Dodelson \cite{Dodelson}.  As usual we Fourier decompose fluctuations and
study the behavior of each Fourier component separately. For a mode of
comoving wavenumber $k$ in Newtonian gauge, the gravitational potential
fluctuations are given by:
\begin{equation}
\Phi= 3\Phi_p\left[{\sin(\omega\eta) -\omega\eta\cos(\omega\eta)
\over (\omega\eta)^3}\right],
\label{eq:phi}
\end{equation}
where $\omega=k/\sqrt{3}$ is the frequency of a mode and $\Phi_p$ is its
primordial amplitude in the limit $\eta \rightarrow 0$. In this section we
use conformal time $\eta=\int dt/a(t)$ with $a(t)\propto t^{1/2}$ during
the radiation-dominated era.  Expanding the temperature anisotropy in
multipole moments and using the Boltzmann equation to
describe their evolution, the monopole $\Theta_0$ and dipole $\Theta_1$ of
the photon distribution can be written in terms of the gravitational
potential as \cite{Dodelson}:
\begin{eqnarray}
\Theta_0&=&\Phi\left({x^2\over 6}+{1\over 2}\right)+{x\over 2}\Phi^\prime
\nonumber \\ \Theta_1&=&-{x^2\over 6}\left(\Phi^\prime +{1\over x}
\Phi\right)
\label{eq:theta01}
\end{eqnarray}
where $x\equiv k\eta$ and a prime denotes a derivative with respect to $x$.

The solutions in equations (\ref{eq:phi}) and (\ref{eq:theta01}) assume
that both the sound speed and the number of relativistic degrees of freedom
are constant over time. As a result of the QCD phase transition and of
various particles becoming non-relativistic, both of these assumptions are
not strictly correct. The vanishing sound speed during the QCD phase
transition provides the most dramatic effect, but its imprint is on scales
smaller than the ones we consider here because the transition occurs at a
significantly higher temperature and only lasts for a fraction of the
Hubble time \cite{Schmid}.

Before the dark matter decouples kinematically, we will treat it as a fluid
which can exchange momentum with the plasma through particle collisions. At
early times, the CDM fluid follows the motion of the plasma and is involved
in its acoustic oscillations. The continuity and momentum equations for the
CDM can be written as:
\begin{eqnarray}
\dot \delta_c+\theta_c &=& 3 \dot \Phi \nonumber \\ \dot \theta_c + {\dot a
\over a} \theta_c &=& k^2 c_s^2 \delta_c - k^2 \sigma_c - k^2 \Phi +
\tau_c^{-1} (\Theta_1 - \theta_c)
\label{eq:dmcontmom}
\end{eqnarray}
where a dot denotes an $\eta$-derivative, $\delta_c$ is the dark matter density
perturbation, $\theta_c$ is the divergence of the dark matter velocity
field and $\sigma_c$ denotes the anisotropic stress. In writing these
equations we have followed Ref. \cite{Ma}. The term $\tau_c^{-1} (\Theta_1 -
\theta_c)$ encodes the transfer of momentun between the radiation and CDM
fluids and $\tau_c^{-1}$ provides the collisional rate of momentum transfer,
\begin{equation}
\tau_c^{-1}= n \sigma {T \over M} a, 
\end{equation}  
with $n$ being the number density of particles with which the dark matter
is interacting, $\sigma(T)$ the average cross section for interaction and
$M$ the mass of the dark matter particle. The relevant scattering partners
are the standard model leptons which have thermal abundances.  For detailed
expressions of the cross section in the case of supersymmetric (SUSY) dark
matter, see Refs. \cite{Chen,Green2}. For our purpose, it is sufficient to
specify the typical size of the cross section and its scaling with cosmic
time,
\begin{equation}
\sigma\approx {T^2 \over M_\sigma^4}  ,
\end{equation}
where the coupling mass $M_\sigma$ is of the order of the weak-interaction
scale ($\sim 100$ GeV) for SUSY dark matter. This equation should be taken
as the definition of $M_\sigma$, as it encodes all the uncertainties in the
details of the particle physics model into a single parameter. The
temperature dependance of the averaged cross section is a result of the
available phase space. Our results are quite insensitive to the details
other than through the decoupling time. Equating $\tau_c^{-1}/a$ to the
Hubble expansion rate gives the temperature of kinematic decoupling:
\begin{equation}
T_d= \left({M_\sigma^4 M \over M_{pl}}\right)^{1/4}\approx 10 \ {\rm MeV}
\left({M_\sigma \over 100 \ {\rm GeV}}\right) \left( M \over 100\ {\rm GeV}
\right)^{1/4}  .
\end{equation}

The term $k^2 c_s^2 \delta_c$ in Eq. (\ref{eq:dmcontmom}) results from the
pressure gradient force and $c_s$ is the dark matter sound speed.  In the
tight coupling limit, $\tau_c \ll H^{-1}$ we find that $c_s^2\approx f_c
T/M$ and that the shear term is $k^2 \sigma_c \approx f_v c_s^2 \tau_c
\theta_c$.  Here $f_{v}$ and $f_{c}$ are constant factors of order unity.
We will find that both these terms make a small difference on the scales of
interest, so their precise value is unimportant.

By combining both equations in (\ref{eq:dmcontmom}) into a single equation
for $\delta_c$ we get
\begin{eqnarray}
\delta_c^{\prime\prime}&+& {1\over x}\left[1+ F_{\rm
v}(x)\right]\delta_c^\prime +c_s^2(x) \delta_c \nonumber \\ = &S(x)&
-3F_{\rm v}(x)\Phi^\prime+{x_d^4 \over x^5} \left(3\theta_0^\prime
-\delta_c^\prime\right),
\label{eq:delta}
\end{eqnarray}
where $x_d=k\eta_d$ and $\eta_d$ denotes the time of kinematic decoupling
which can be expressed in terms of the decoupling temperature as,
\begin{eqnarray}
\eta_d= {2 t_d (1+z_d) }\approx {M_{pl} \over T_0 T_d} &\approx& 10 \ {\rm
pc} \ \left({T_d \over 10\ {\rm MeV}}\right)^{-1} \nonumber \\ &\propto&
M_\sigma^{-1} M^{-1/4} ,
\end{eqnarray}
with $T_0=2.7$K being the present-day CMB temperature and $z_d$ being the
redshift at kinematic decoupling.  We have also introduced the source
function,
\begin{equation}
S(x)\equiv -3\Phi^{\prime\prime}+\Phi-{3\over x}\Phi^\prime.
\label{eq:s}
\end{equation}
For $x\ll x_d$, the dark matter sound speed is given by 
\begin{equation}
c_s^2(x)=c_s^2(x_d) {x_d\over x}, 
\label{eq:sound}
\end{equation}
where $c_s^2(x_d)$ is the dark matter sound speed at kinematic decoupling
(in units of the speed of light),
\begin{equation}
c_s(x_d) \approx 10^{-2} f_c^{1/2} \left({T_d \over 10\ {\rm MeV}}
\right)^{1/2} \left({M \over 100\ {\rm GeV}} \right)^{-1/2}.
\end{equation}
In writing (\ref{eq:sound}) we have assumed that prior to decoupling the
temperature of the dark matter follows that of the plasma. For the
viscosity term we have,
\begin{equation}
F_{v}(x)= f_{v} c_s^2(x_d) x_d^2  \left({x_d\over x}\right)^5.
\label{eq:F_v_before}
\end{equation}

\noindent{\bf Free streaming after kinematic decoupling}

In the limit of the collision rate being much slower than the Hubble
expansion, the CDM is decoupled and the evolution of its perturbations is
obtained by solving a Boltzman equation: \begin{equation} {\partial f \over
\partial \eta} +{d r_i \over d\eta} {\partial f \over \partial r_i} + {d
q_i \over d\eta} {\partial f \over \partial q_i} =0,
\end{equation}
where $f(\vec r,\vec q,\eta)$ is the distribution function which depends on
position, comoving momentum $\vec q$, and time. The comoving momentum
3-components are ${d x_i / d\eta} = q_i/a$. We use the Boltzman equation to
find the evolution of modes that are well inside the horizon with $x\gg
1$. In the radiation era, the gravitational potential decays after horizon
crossing (see Eq. \ref{eq:phi}). In this limit the comoving momentum
remains constant, ${d q_i /d\eta} =0$ and the Boltzman equation becomes,
\begin{equation}
{\partial f \over \partial \eta} +{q_i \over a} {\partial f \over \partial
r_i} =0.
\end{equation}
We consider a single Fourier mode and write $f$ as, 
\begin{equation}
f(\vec r,\vec q,\eta)=f_0(q) [1+\delta_F(\vec q,\eta) e^{i \vec k \cdot
\vec r}],
\end{equation}
where $f_0(q)$ is the unperturbed distribution,
\begin{equation}
f_0(q) = n_{\rm CDM} \left( {M \over 2\pi T_{\rm CDM} } \right)^{3/2}
\exp\left[-{1\over 2} {M q^2 \over T_{\rm CDM}}\right]
\end{equation} 
where $n_{\rm CDM}$ and $T_{\rm CDM}$ are the present-day density and
temperature of the dark matter.

Our approach is to solve the Boltzman equation with initial conditions
given by the fluid solution at a time $\eta_*$ (which will depend on
$k$). The simplified Boltzman equation can be easily solved to give
$\delta_F(\vec q,\eta)$ as a function of the initial conditions
$\delta_F(\vec q,\eta_*)$,
\begin{equation}
\delta_F(\vec q,\eta)=\delta_F(\vec q,\eta_*) \exp[-i \vec q \cdot \vec k
{\eta_* \over a(\eta_*)} \ln (\eta/\eta_*)] .
\label{solbol}
\end{equation}   

The CDM overdensity $\delta_c$ can then be expressed in terms of the
perturbation in the distribution function as,
\begin{equation}
\delta_c(\eta)={1\over n_{\rm CDM}} \int d^3q \ f_0(q) \ \delta_F(\vec q,\eta).
\end{equation} 
We can use (\ref{solbol}) to obtain the evolution of $\delta_c$ in terms of
its value at $\eta_*$,
\begin{equation}
\delta_c(\eta)= \exp\left[-{1\over 2} {k^2 \over k_f^2} \ln^2({\eta\over
\eta_*})\right] \ \left[\delta|_{\eta_*} + {d \delta \over d
\eta}|_{\eta_*} \eta_* \ln({\eta\over \eta_*})\right] ,
\label{delcbol}
\end{equation}
where $k_f^{-2}= \sqrt{(T_d/ M)} \eta_d$. The exponential term is
responsible for the damping of perturbations as a result of free streaming and
the dispersion of the CDM particles after they decouple from the plasma. The
above expression is only valid during the radiation era. The free streaming
scale is simply given by $\int dt (v/a) \propto \int dt a^{-2}$ which grows
logarithmically during the radiation era as in equation (\ref{delcbol}) but
stops growing in the matter era when $a\propto t^{2/3}$.

Equation (\ref{delcbol}) can be used to show that even during the free
streaming epoch, $\delta_c$ satisfies equation (\ref{eq:delta}) but with a
modified sound speed and viscous term. For $x \gg x_d$ one should use,
\begin{eqnarray}
c_s^2(x) &=& c_s^2(x_d) \left({x_d\over x}\right)^2 \left[1+ x_d^2
c_s^2(x_d) \ln^2({x\over x_d})\right] \nonumber \\ F_{v}(x)&=&2 c_s^2(x_d)
x_d^2 \ln\left({x_d\over x}\right)
\end{eqnarray}
The differences between the above scalings and those during the tight
coupling regime are a result of the fact that the dark matter temperature
stops following the plasma temperature but rather scales as $a^{-2}$ after
thermal decoupling, which coincides with the kinematic decoupling.
We ignore the effects of heat transfer during the
fluid stage of the CDM because its temperature is controlled by the much
larger heat reservoir of the radiation-dominated plasma at that stage.

To obtain the transfer function we solve the dark matter fluid equation
until decoupling and then evolve the overdensity using equation
(\ref{delcbol}) up to the time of matter--radiation equality. In practice,
we use the fluid equations up to $x_*=10\  {\rm max}(x_d, 10)$ so as to switch
into the free streaming solution well after the gravitational potential has
decayed. In the fluid equations, we smoothly match the sound speed and
viscosity terms at $x = x_d$. As mentioned earlier, because $c_s(x_d)$ is
so small and we are interested in modes that are comparable to the size of
the horizon at decoupling, i.e. $x_d \sim {\rm few}$, both the dark
matter sound speed and the associated viscosity play only a minor
role, and our simplified treatment is adequate.

\begin{figure}
\centering
\includegraphics[height=6cm]{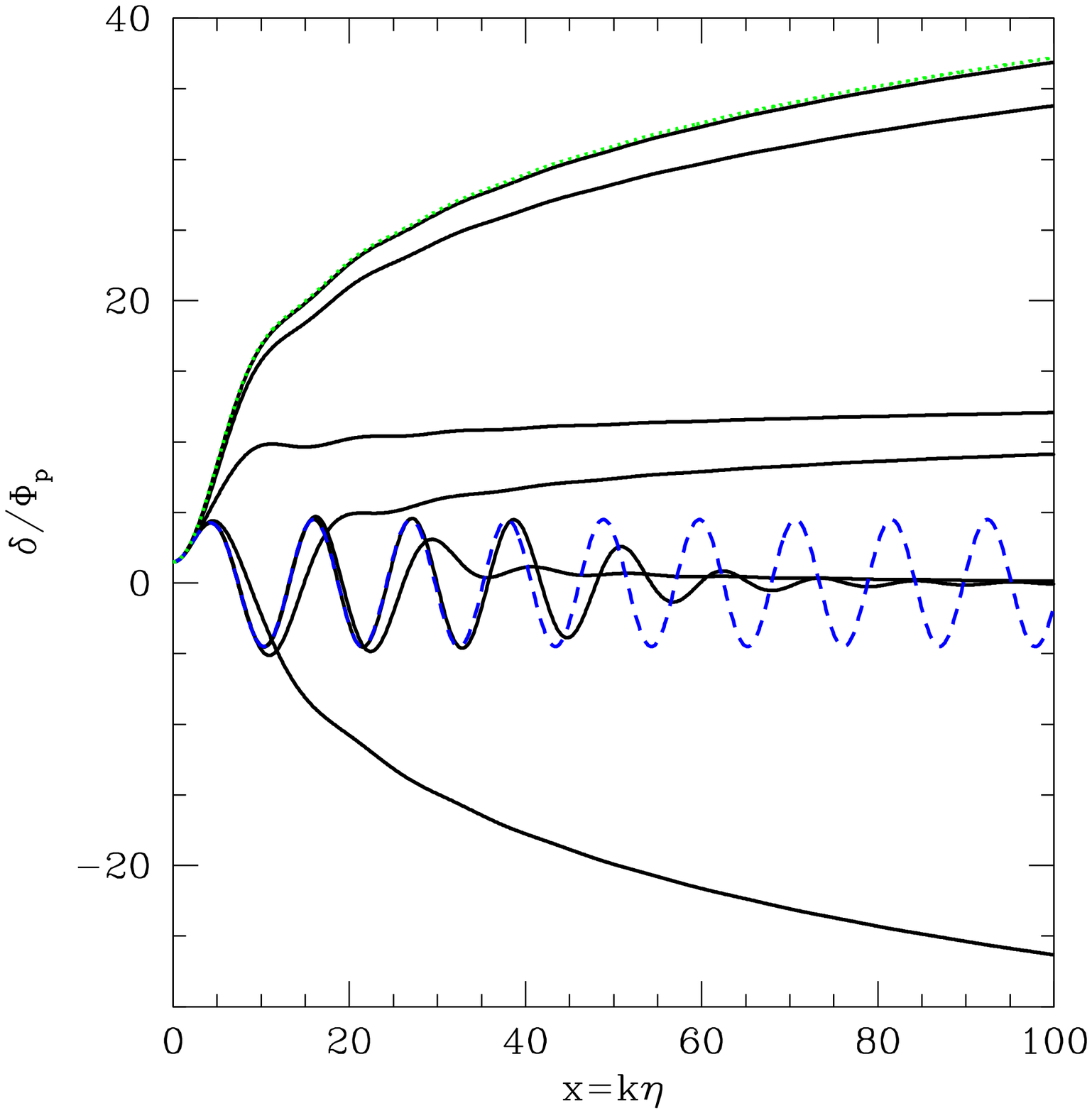}
\caption{The normalized amplitude of CDM fluctuations $\delta/\Phi_P$ for a
variety of modes with comoving wavenumbers
$\log(k\eta_d)=(0,1/3,2/3,1,4/3,5/3,2)$ as a function of $x\equiv k\eta$,
where $\eta=\int_0^t dt/a(t)$ is the conformal time coordinate. The dashed
line shows the temperature monopole $3\theta_0$ and the uppermost (dotted)
curve shows the evolution of a mode that is uncoupled to the cosmic
plasma.}
\label{figtime}
\end{figure}

In Figure \ref{figtime} we illustrate the time evolution of modes during
decoupling for a variety of $k$ values. The situation is clear. Modes that
enter the horizon before kinematic decoupling oscillate with the
radiation fluid. This behavior has two important effects. In the absence of
the coupling, modes receive a ``kick" by the source term $S(x)$ as they
cross the horizon. After that they grow logarithmically. In our case, modes
that entered the horizon before kinematic decoupling follow the plasma
oscillations and thus miss out on both the horizon ``kick" and the
beginning of the logarithmic growth. Second, the decoupling from the
radiation fluid is not instantaneous and this acts to further damp the
amplitude of modes with $x_d \gg 1$. This effect can be understood as
follows.  Once the oscillation frequency of the mode becomes high compared
to the scattering rate, the coupling to the plasma effectively damps the
mode. In that limit one can replace the forcing term $\Theta_0^\prime$ by
its average value, which is close to zero. Thus in this regime, the
scattering is forcing the amplitude of the dark matter oscillations to
zero.  After kinematic decoupling the modes again grow logarithmically but
from a very reduced amplitude. {\it The coupling with the plasma induces
both oscillations and damping of modes that entered the horizon before
kinematic decoupling. This damping is different from the free streaming
damping that occurs after kinematic decoupling}.

\begin{figure}
\centering
\includegraphics[height=6cm]{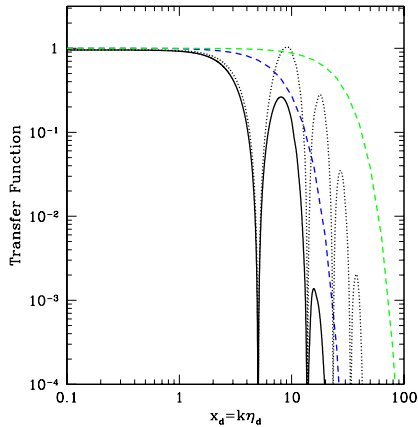}
\caption{Transfer function of the CDM density perturbation amplitude
(normalized by the primordial amplitude from inflation).  We show two
cases: {\it (i)} $T_d/M=10^{-4}$ and $T_d/T_{\rm eq}=10^7$; {\it (ii)}
$T_d/M=10^{-5}$ and $T_d/T_{\rm eq}=10^7$. In each case the oscillatory
curve is our result and the other curve is the free-streaming only result
that was derived previously in the literature [4,7,8].  }
\label{figtransfer}
\end{figure}

In Figure \ref{figtransfer} we show the resulting transfer function of the
CDM overdensity. The transfer function is defined as the ratio between the
CDM density perturbation amplitude $\delta_c$ when the effect of the
coupling to the plasma is included and the same quantity in a model where
the CDM is a perfect fluid down to arbitrarily small scales (thus, the
power spectrum is obtained by multiplying the standard result by the square
of the transfer function).  This function shows both the oscillations and
the damping signature mentioned above. The peaks occur at multipoles of the
horizon scale at decoupling,
\begin{equation} k_{peak}= (8,15.7,24.7,..) \eta_d^{-1} \propto {M_{pl} \over
T_0 T_d}.
\end{equation} This same scale determines the ``oscillation" damping. The
free streaming damping scale is,
\begin{equation}
\eta_d c_d(\eta_d) \ln(\eta_{eq}/\eta_d) \propto {M_{pl} M^{1/2} \over T_0
T_d^{3/2}} \ln(T_d/T_{\rm eq}),
\end{equation}
where $T_{\rm eq}$ is the temperature at matter radiation equality,
$T_{\rm eq}\approx 1\ {\rm eV}$. The free streaming scale is parametrically
different from the ``oscillation" damping scale. However for our fiducial
choice of parameters for the CDM particle they roughly coincide.

The CDM damping scale is significantly smaller than the scales observed
directly in the Cosmic Microwave Background or through large scale
structure surveys. For example, the ratio of the damping scale to the scale
that entered the horizon at Matter-radiation equality is
$\eta_{d}/\eta_{eq}\sim T_{eq}/T_d \sim 10^{-7}$ and to our present horizon
$\eta_{d}/\eta_{0}\sim (T_{eq} T_0)^{1/2}/T_{d} \sim 10^{-9}$. In the
context of inflation, these scales were created 16 and 20 {\it e}--folds
apart. Given the large extrapolation, one could certainly imagine that a
change in the spectrum could alter the shape of the power spectrum around
the damping scale.  However, for smooth inflaton potentials with small
departures from scale invariance this is not likely to be the case. On
scales much smaller than the horizon at matter radiation equality, the
spectrum of perturbations density before the effects of the damping are
included is approximately,
\begin{eqnarray}
\Delta^2(k)&\propto& \exp\left[(n-1)\ln(k\eta_{eq})+ {1\over 2} \alpha^2
\ln(k\eta_{eq})^2 + \cdots\right]\nonumber \\ && \times \ln^2(k\eta_{eq}/8)
\end{eqnarray}
where the first term encodes the shape of the primordial spectrum and the
second the transfer function. Primordial departures from scale invariance
are encoded in the slope $n$ and its running $\alpha$. The effective slope
at scale $k$ is then,
\begin{equation}
{\partial \ln \Delta^2 \over \partial \ln k}= (n-1)+\alpha \ln(k\eta_{eq})
+ {2 \over \ln(k\eta_{eq}/8)}.
\end{equation}
For typical values of $(n-1) \sim 1/60$ and $\alpha\sim 1/60^2$ the slope
is still positive at $k\sim \eta_d^{-1}$, so the cut-off in the power will
come from the effects we calculate rather than from the shape of the
primordial spectrum. However given the large extrapolation in scale, one
should keep in mind the possibility of significant effects resulting from
the mechanisms that generates the density perturbations.

\noindent{\bf Implications}
We have found that acoustic oscillations, a relic from the epoch when the
dark matter coupled to the cosmic radiation fluid, truncate the CDM power
spectrum on a comoving scale larger than effects considered before, such as
free-streaming and viscosity \cite{Green,Green2,Hofmann}. For SUSY dark
matter, the minimum mass of dark matter clumps that form in the Universe is
therefore increased by more than an order of magnitude to a value of
\footnote{Our definition of the cut-off mass follows the convention of the
Jeans mass, which is defined as the mass enclosed within a sphere of radius
$\lambda_{\rm J}/2$ where $\lambda_{\rm J}\equiv 2\pi/k_{\rm J}$ is the
Jeans wavelength \cite{Haiman}.}
\begin{eqnarray}
M_{\rm cut}&=& {4\pi\over 3} \left({\pi \over k_{\rm cut}}\right)^{3}
\Omega_M \rho_{\rm crit} \nonumber \\ &\simeq& 10^{-4} \left({T_d\over
10~{\rm MeV}}\right)^{-3} M_\odot,
\label{mcut}
\end{eqnarray}
where $\rho_{\rm crit}=(H_0^2/8\pi G)=9\times 10^{-30}~{\rm g~cm^{-3}}$ is
the critical density today, and $\Omega_M$ is the matter density for the
concordance cosmological model \cite{WMAP}. We define the cut-off
wavenumber $k_{\rm cut}$ as the point where the transfer function first
drops to a fraction $1/e$ of its value at $k\rightarrow 0$. This
corresponds to $k_{\rm cut}\approx 3.3 \ \eta_d^{-1}$.

\begin{figure}
\centering
\includegraphics[height=6cm]{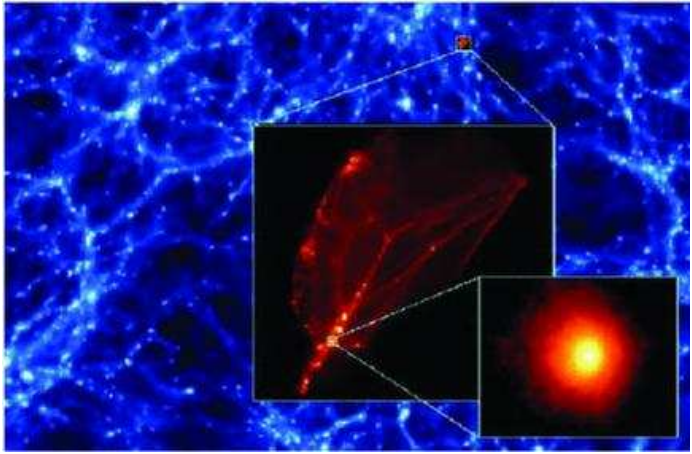}
\label{figMoore}
\caption{A slice through a numerical simulation of the first 
dark matter condensations to form in the Universe. Colors represent
the dark matter density at $z=26$. The simulated volume is 60 comoving pc
on a side, simulated with 64 million particles each weighing
$1.2\times 10^{-10}M_\odot$ (!). (from Diemand, Moore, \& Stadel 2005 \cite{Moore}). }
\end{figure} 

Recent numerical simulations \cite{Moore,Gao} of the earliest and smallest
objects to have formed in the Universe (see Fig. \ref{figMoore}), need to
be redone for the modified power spectrum that we calculated in this
section.  Although it is difficult to forecast the effects of the acoustic
oscillations through the standard Press-Schechter formalism \cite{Press},
it is likely that the results of such simulations will be qualitatively the
same as before except that the smallest clumps would have a mass larger
than before (as given by Eq. \ref{mcut}).

Potentially, there are several observational signatures of the smallest CDM
clumps.  As pointed out in the literature \cite{Moore,Stoehr}, the smallest
CDM clumps could produce $\gamma$-rays through dark-matter annihilation in
their inner density cusps, with a flux in excess of that from nearby dwarf
galaxies.  If a substantial fraction of the Milky Way halo is composed of
CDM clumps with a mass $\sim 10^{-4} M_\odot$, the nearest clump is
expected to be at a distance of $\sim 4\times 10^{17}$ cm.  Given that the
characteristic speed of such clumps is a few hundred ${\rm km~s^{-1}}$, the
$\gamma$-ray flux would therefore show temporal variations on the
relatively long timescale of a thousand years.  Passage of clumps through
the solar system should also induce fluctuations in the detection rate of
CDM particles in direct search experiments.
Other observational effects have rather limited prospects for
detectability.  Because of their relatively low-mass and large size ($\sim
10^{17}~{\rm cm}$), the CDM clumps are too diffuse to produce any
gravitational lensing signatures (including {\it femto-}lensing
\cite{Gould}), even at cosmological distances.

The smallest CDM clumps should not affect the intergalactic baryons which
have a much larger Jeans mass. However, once objects above
$\sim 10^6M_\odot$ start to collapse at redshifts $z<30$, the baryons would
be able to cool inside of them via molecular hydrogen transitions and the
interior baryonic Jeans mass would drop.  The existence of dark matter
clumps could then seed the formation of the first stars inside these
objects \cite{Bromm}.


\subsection{Structure of the Baryons}

\noindent{\bf Early Evolution of Baryonic Perturbations on Large Scales}

The baryons are coupled through Thomson scattering to the radiation fluid
until they become neutral and decouple. After cosmic recombination, they
start to fall into the potential wells of the dark matter and their early
evolution was derived by Barkana \& Loeb (2005) \cite{BLinf}.

On large scales, the dark matter (dm) and the baryons (b) are affected only
by their combined gravity and gas pressure can be ignored. The evolution of
sub-horizon linear perturbations is described in the matter-dominated
regime by two coupled second-order differential equations \cite{Peebles}:
\beqa \ddot{\delta}_{\rm dm} + 2H \dot {\delta}_{\rm dm} & = & 4 \pi G
\bar{\rho}_m \left(f_{\rm b} \delta_{\rm b} + f_{\rm dm} \delta_{\rm
dm}\right)\ , \nonumber \\ \ddot{\delta}_{\rm b}+ 2H \dot {\delta}_{\rm b}
& = & 4 \pi G \bar{\rho}_m \left(f_{\rm b} \delta_{\rm b} + f_{\rm dm}
\delta_{\rm dm}\right)\ , \label{coupled}\eeqa where $\delta_{\rm dm}(t)$
and $\delta_{\rm b}(t)$ are the perturbations in the dark matter and
baryons, respectively, the derivatives are with respect to cosmic time $t$,
$H(t)=\dot{a}/a$ is the Hubble constant with $a=(1+z)^{-1}$, and we assume
that the mean mass density $\bar{\rho}_m(t)$ is made up of respective mass
fractions $f_{\rm dm}$ and $f_{\rm b}=1-f_{\rm dm}$. Since these linear
equations contain no spatial gradients, they can be solved spatially for
$\delta_{\rm dm}(\bx,t)$ and $\delta_{\rm b}(\bx,t)$ or in Fourier space
for $\td_{\rm dm}(\bk,t)$ and $\td_{\rm b}(\bk,t)$.

Defining $\delta_{\rm tot}
\equiv f_{\rm b} \delta_{\rm b} + f_{\rm dm} \delta_{\rm dm}$ and
$\delta_{\rm b-} \equiv \delta_{\rm b} - \delta_{\rm tot}$ , we find
\beqa \ddot{\delta}_{\rm tot} + 2H \dot {\delta}_{\rm tot} & = & 4 \pi
G \bar{\rho}_m \delta_{\rm tot}\ , \nonumber \\ \ddot{\delta}_{\rm
b-}+ 2H \dot{\delta}_{\rm b-} & = & 0\ .  \eeqa Each of these
equations has two independent solutions. The equation for $\delta_{\rm
tot}$ has the usual growing and decaying solutions, which we denote
$D_1(t)$ and $D_4(t)$, respectively, while the $\delta_{\rm b-}$
equation has solutions $D_2(t)$ and $D_3(t)$; we number the solutions
in order of declining growth rate (or increasing decay rate). We
assume an Einstein-de Sitter, matter-dominated Universe in the
redshift range $z=20$--150, since the radiation contributes less than
a few percent at $z < 150$, while the cosmological constant and the
curvature contribute to the energy density less than a few percent at
$z > 3$. In this regime $a \propto t^{2/3}$ and the solutions are
$D_1(t)=a/a_i$ and $D_4(t)=(a/a_i)^{-3/2}$ for $\delta_{\rm tot}$, and
$D_2(t)=1$ and $D_3(t)=(a/a_i)^{-1/2}$ for $\delta_{\rm b-}$, where we
have normalized each solution to unity at the starting scale factor
$a_i$, which we set at a redshift $z_i=150$. The observable baryon
perturbation can then be written as \beq \td_{\rm b}(\bk,t) =
\td_{\rm b-} + \td_{\rm tot} = \sum_{m=1}^4 \td_{m}(\bk)\, D_m(t)\
, \label{eq:delb} \eeq and similarly for the dark matter perturbation,
\beq \td_{\rm dm}={1\over f_{\rm dm}}\left(\td_{\rm
tot}-f_b\td_b\right)= \sum_{m=1}^4 \td_{m}(\bk)\, C_m(t)\ ,
\label{eq:deldm} 
\eeq where $C_i=D_i$ for $i=1,4$ and $C_i=-(f_{\rm b}/f_{\rm dm})D_i$ for
$i=2,3$. We may establish the values of $\td_{m}(\bk)$ by inverting the
$4\times 4$ matrix ${\bf A}$ that relates the 4-vector
$(\td_1,\td_2,\td_3,\td_4)$ to the 4-vector that represents the initial
conditions $(\td_{\rm b},\td_{\rm dm},\dot{\td}_{\rm b},\dot{\td}_{\rm
dm})$ at the initial time.

\begin{figure}
\includegraphics[width=84mm]{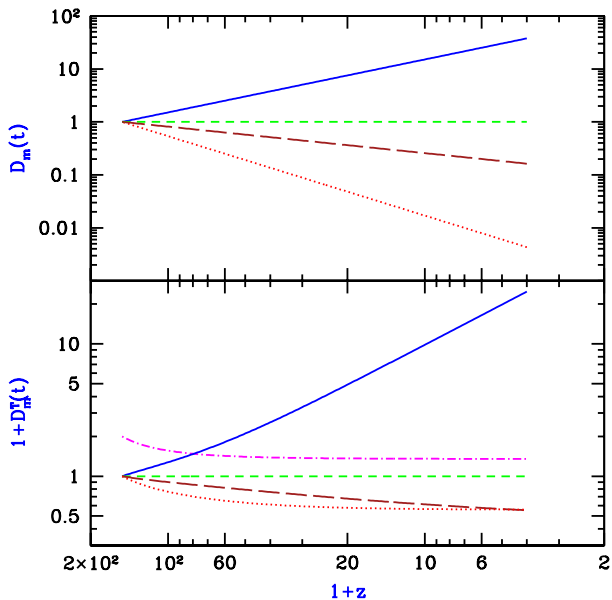}
\caption{Redshift evolution of the amplitudes of the independent modes of
the density perturbations (upper panel) and the temperature perturbations
(lower panel), starting at redshift 150 (from Barkana \& Loeb 2005
\cite{BLinf}). We show $m=1$ (solid curves), $m=2$ (short-dashed curves),
$m=3$ (long-dashed curves), $m=4$ (dotted curves), and $m=0$ (dot-dashed
curve). Note that the lower panel shows one plus the mode amplitude.}
\label{fig:Tevol}
\end{figure}

\begin{figure}
\includegraphics[width=84mm]{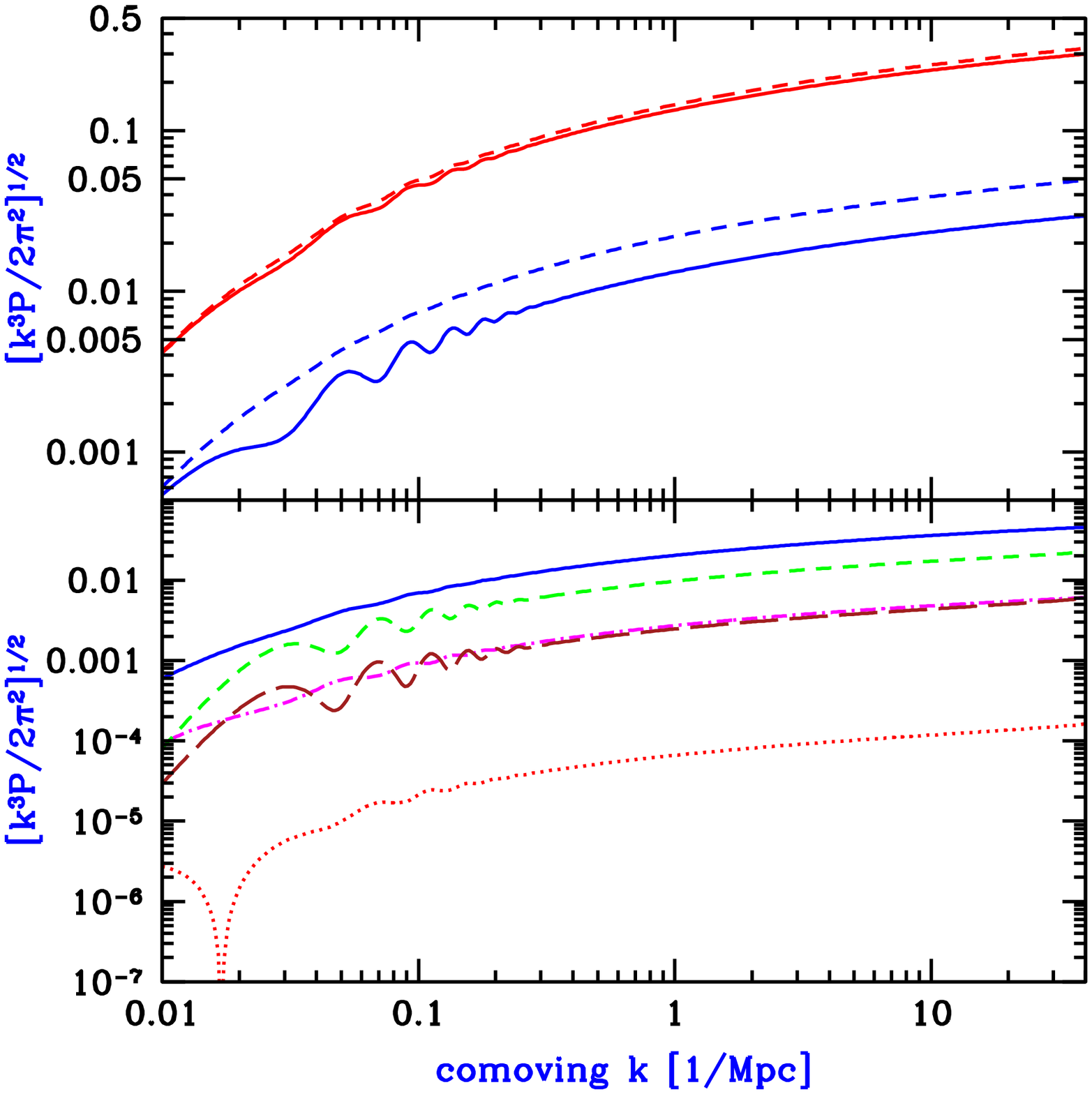}
\caption{Power spectra and initial perturbation amplitudes versus
wavenumber (from \cite{BLinf}). The upper panel shows $\td_{\rm b}$ (solid
curves) and $\td_{\rm dm}$ (dashed curves) at $z=150$ and 20 (from bottom
to top).  The lower panel shows the initial ($z=150$) amplitudes of $\td_1$
(solid curve), $\td_2$ (short-dashed curve), $\td_3$ (long-dashed curve),
$\td_4$ (dotted curve), and $\td_T(t_i)$ (dot-dashed curve). Note that if
$\td_1$ is positive then so are $\td_3$ and $\td_T(t_i)$, while $\td_2$ is
negative at all $k$, and $\td_4$ is negative at the lowest $k$ but is
positive at $k > 0.017$ Mpc$^{-1}$.}
\label{fig:del}
\end{figure}

Next we describe the fluctuations in the sound speed of the cosmic gas
caused by Compton heating of the gas, which is due to scattering of the
residual electrons with the CMB photons. The evolution of the temperature
$T$ of a gas element of density $\rho_b$ is given by the first law of
thermodynamics: \beq dQ = \frac{3} {2} \kB dT - \kB T d \log \rho_{\rm b}\
, \eeq where $dQ$ is the heating rate per particle. Before the first
galaxies formed, \beq \frac{dQ} {d t} = 4 \frac{\sigma_{\rm T}\, c} {m_e}
\, \kB (T_\gamma - T) \rho_\gamma x_e(t) \ , \eeq where $\sigma_T$ is the
Thomson cross-section, $x_e(t)$ is the electron fraction out of the total
number density of gas particles, and $\rho_\gamma$ is the CMB energy
density at a temperature $T_\gamma$. In the redshift range of interest, we
assume that the photon temperature ($T_\gamma = T_\gamma^0/a$) is spatially
uniform, since the high sound speed of the photons (i.e., $c/\sqrt{3}$)
suppresses fluctuations on the sub-horizon scales that we consider, and the
horizon-scale $\sim 10^{-5}$ fluctuations imprinted at cosmic recombination
are also negligible compared to the smallWe establish the values of
$\td_{m}(\bk)$ by inverting the $4\times 4$ matrix ${\bf A}$ that relates
the 4-vector $(\td_1,\td_2,\td_3,\td_4)$ to the 4-vector that represents
the initial conditions $(\td_{\rm b},\td_{\rm dm},\dot{\td}_{\rm
b},\dot{\td}_{\rm dm})$ at the initial time.  er-scale fluctuations in the
gas density and temperature. Fluctuations in the residual electron fraction
$x_e(t)$ are even smaller.  Thus, \beq \frac{dT} {dt} = \frac{2} {3} T
\frac{d \log \rho_{\rm b}} {dt} + \frac{x_e(t)}{t_\gamma}\, (T_\gamma -
T)\, a^{-4}\ , \eeq where $t_\gamma^{-1} \equiv \bar{\rho}_\gamma^0
({8\sigma_{\rm T}\, c}/{3 m_e}) = 8.55 \times 10^{-13} {\rm yr}^{-1}$.
After cosmic recombination, $x_e(t)$ changes due to the slow recombination
rate of the residual ions: \beq {d x_e(t)\over dt} = - \alpha_B(T) x_e^2(t)
\bar{n}_H (1+y)\ , \eeq where $\alpha_B(T)$ is the case-B recombination
coefficient of hydrogen, $\bar{n}_H$ is the mean number density of hydrogen
at time $t$, and $y=0.079$ is the helium to hydrogen number density
ratio. This yields the evolution of the mean temperature, ${d \bar{T}}/{dt}
= - 2 H \bar{T} + {x_e(t)}{t_\gamma^{-1}}\, (T_\gamma - \bar{T})\,
a^{-4}$. In prior analyses \cite{Peebles, Ma} a spatially uniform
speed of sound was assumed for the gas at each redshift. Note that we refer
to $\delta p/ \delta \rho$ as the square of the sound speed of the fluid,
where $\delta p$ is the pressure perturbation, although we are analyzing
perturbations driven by gravity rather than sound waves driven by pressure
gradients.

Instead of assuming a uniform sound speed, we find the first-order
perturbation equation, \beq \frac{d \delta_T} {d t} = \frac{2}{3}
\frac{d \delta_b} {dt} - \frac{x_e(t)} {t_\gamma} \frac{T_\gamma}
{\bar{T}} a^{-4} \delta_T\ , \label{eq:order1} \eeq
where we defined the fractional temperature perturbation $\delta_T$.  Like
the density perturbation equations, this equation can be solved separately
at each $\bx$ or at each $\bk$. Furthermore, the solution $\delta_T (t)$ is
a linear functional of $\delta_b(t)$ [for a fixed function $x_e(t)$].
Thus, if we choose an initial time $t_i$ then using Eq.~(\ref{eq:delb}) we
can write the solution in Fourier space as \beq \td_T (\bk,t) =
\sum_{m=1}^4 \td_{m}(\bk)\, D^T_m(t) + \td_T (\bk,t_i)\, D^T_0(t)\ ,
\label{eq:delT} \eeq where $D^T_m(t)$ is the solution of
Eq.~(\ref{eq:order1}) with $\delta_T = 0$ at $t_i$ and with the
perturbation mode $D_m(t)$ substituted for $\delta_b(t)$, while $D^T_0(t)$
is the solution with no perturbation $\delta_b(t)$ and with $\delta_T = 1$
at $t_i$. By modifying the CMBFAST code (http://www.cmbfast.org/), we can
numerically solve Eq.~(\ref{eq:order1}) along with the density perturbation
equations for each $\bk$ down to $z_i=150$, and then match the solution to
the form of Eq.~(\ref{eq:delT}).

\begin{figure}
\includegraphics[width=84mm]{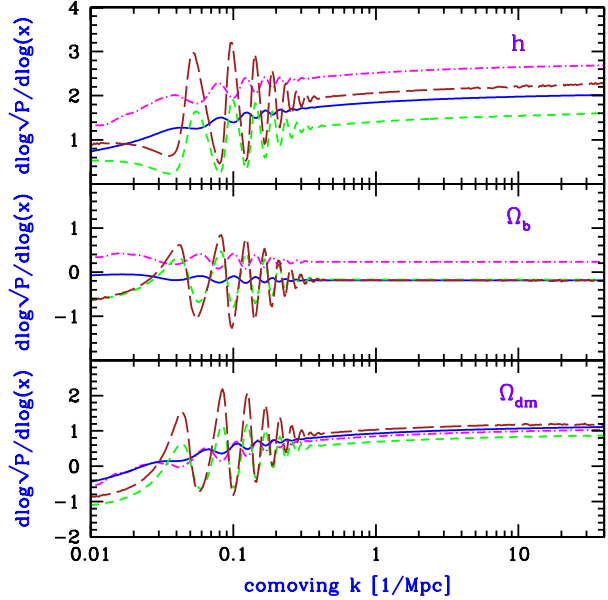}
\caption{Relative sensitivity of perturbation amplitudes at $z=150$ to
cosmological parameters (from \cite{BLinf}). For variations in a parameter
$x$, we show $d{\rm log}\sqrt{P(k)}/d{\rm log}(x)$. We consider variations
in $\Omega_{\rm dm} h^2$ (upper panel), in $\Omega_b h^2$ (middle panel),
and in the Hubble constant $h$ (lower panel). When we vary each parameter
we fix the other two, and the variations are all carried out in a flat
$\Omega_{\rm total}=1$ universe. We show the sensitivity of $\td_1$ (solid
curves), $\td_2$ (short-dashed curves), $\td_3$ (long-dashed curves), and
$\td_T(t_i)$ (dot-dashed curves).}
\label{fig:sens}
\end{figure}

Figure~\ref{fig:Tevol} shows the time evolution of the various independent
modes that make up the perturbations of density and temperature, starting
at the time $t_i$ corresponding to $z_i=150$. $D^T_2(t)$ is identically
zero since $D_2(t)=1$ is constant, while $D^T_3(t)$ and $D^T_4(t)$ are
negative. Figure~\ref{fig:del} shows the amplitudes of the various
components of the initial perturbations. We consider comoving wavevectors
$k$ in the range 0.01 -- 40 Mpc$^{-1}$, where the lower limit is set by
considering sub-horizon scales at $z=150$ for which photon perturbations
are negligible compared to $\delta_{\rm dm}$ and $\delta_{\rm b}$, and the
upper limit is set by requiring baryonic pressure to be negligible compared
to gravity. $\td_2$ and $\td_3$ clearly show a strong signature of the
large-scale baryonic oscillations, left over from the era of the
photon-baryon fluid before recombination, while $\td_1$, $\td_4$, and
$\td_T$ carry only a weak sign of the oscillations. For each quantity, the
plot shows $[k^3 P(k)/(2 \pi^2)]^{1/2}$, where $P(k)$ is the corresponding
power spectrum of fluctuations.  $\td_4$ is already a very small correction
at $z=150$ and declines quickly at lower redshift, but the other three
modes all contribute significantly to $\td_{\rm b}$, and the $\td_T(t_i)$
term remains significant in $\td_T(t)$ even at $z \la 100$. Note that at
$z=150$ the temperature perturbation $\td_T$ has a different shape with
respect to $k$ than the baryon perturbation $\td_{\rm b}$, showing that
their ratio cannot be described by a scale-independent speed of sound.

The power spectra of the various perturbation modes and of
$\td_T(t_i)$ depend on the initial power spectrum of density
fluctuations from inflation and on the values of the fundamental
cosmological parameters ($\Omega_{\rm dm}$, $\Omega_b$,
$\Omega_{\Lambda}$, and $h$). If these independent power spectra can
be measured through 21cm fluctuations, this will probe the basic
cosmological parameters through multiple combinations, allowing
consistency checks that can be used to verify the adiabatic nature and
the expected history of the perturbations. Figure~\ref{fig:sens}
illustrates the relative sensitivity of $\sqrt{P(k)}$ to variations in
$\Omega_{\rm dm} h^2$, $\Omega_b h^2$, and $h$, for the quantities
$\td_1$, $\td_2$, $\td_3$, and $\td_T(t_i)$. Not shown is $\td_4$,
which although it is more sensitive (changing by order unity due to
$10\%$ variations in the parameters), its magnitude always remains
much smaller than the other modes, making it much harder to
detect. Note that although the angular scale of the baryon
oscillations constrains also the history of dark energy through the
angular diameter distance, we have focused here on other cosmological
parameters, since the contribution of dark energy relative to matter
becomes negligible at high redshift.

\noindent
{\bf Cosmological Jeans Mass}

The Jeans length $\ljeans$ was originally defined (Jeans 1928 \cite{Jeans28}) in Newtonian
gravity as the critical wavelength that separates oscillatory and
exponentially-growing density perturbations in an infinite, uniform, and
stationary distribution of gas. On scales $\ell$ smaller than $\ljeans$,
the sound crossing time, $\ell/c_s$ is shorter than the gravitational
free-fall time, $(G\rho)^{-1/2}$, allowing the build-up of a pressure force
that counteracts gravity. On larger scales, the pressure gradient force is
too slow to react to a build-up of the attractive gravitational force.  The
Jeans mass is defined as the mass within a sphere of radius $\ljeans/2$,
$\mjeans=(4\pi/3)\rho(\ljeans/2)^3$.  In a perturbation with a mass greater
than $\mjeans$, the self-gravity cannot be supported by the pressure
gradient, and so the gas is unstable to gravitational collapse. The
Newtonian derivation of the Jeans instability suffers from a conceptual
inconsistency, as the unperturbed gravitational force of the uniform
background must induce bulk motions (compare to Binney \& Tremaine 1987
\cite{Bi87}).  However, this inconsistency is remedied when the analysis is
done in an expanding Universe.

The perturbative derivation of the Jeans instability criterion can be
carried out in a cosmological setting by considering a sinusoidal
perturbation superposed on a uniformly expanding background.  Here, as
in the Newtonian limit, there is a critical wavelength $\ljeans$ that
separates oscillatory and growing modes.  Although the expansion of
the background slows down the exponential growth of the amplitude to a
power-law growth, the fundamental concept of a minimum mass that can
collapse at any given time remains the same (see, e.g. Kolb \& Turner
1990 \cite{Kolb90}; Peebles 1993 \cite{Peebles}).

We consider a mixture of dark matter and baryons with density parameters
$\Omega_{\rm dm}^{\,z}=\bar{\rho}_{\rm dm}/\rho_{\rm c}$ and $\Omega_{\rm
b}^{\,z}=\bar{\rho}_{\rm b}/\rho_{\rm c}$, where $\bar{\rho}_{\rm dm}$ is
the average dark matter density, $\bar{\rho}_{\rm b}$ is the average
baryonic density, $\rho_{\rm c}$ is the critical density, and $\Omega_{\rm
dm}^{\,z} + \Omega_{\rm b}^{\,z} = \Ommz$ is given by
equation(\ref{Ommz}). We also assume spatial fluctuations in the gas and
dark matter densities with the form of a single spherical Fourier mode
on a scale much smaller than the horizon,
\begin{eqnarray}
\frac{\rho_{\rm dm}(r,t)-\bar{\rho}_{\rm dm}(t)}
{\bar{\rho}_{\rm dm}(t)} & = &
\delta_{\rm dm}(t) \frac{\sin(kr)}{kr}\ , \\
\frac{\rho_{\rm b}(r,t)-\bar{\rho}_{\rm b}(t)}
{\bar{\rho}_{\rm b}(t)} & = &
\delta_{\rm b}(t) \frac{\sin(kr)}{kr}\ ,
\end{eqnarray}
where $\bar{\rho}_{\rm dm}(t)$ and $\bar{\rho}_{\rm b}(t)$ are the
background densities of the dark matter and baryons, $\delta_{\rm
dm}(t)$ and $\delta_{\rm b}(t)$ are the dark matter and baryon
overdensity amplitudes, $r$ is the comoving radial coordinate, and $k$
is the comoving perturbation wavenumber.  We adopt an ideal gas
equation-of-state for the baryons with a specific heat ratio
$\gamma$=$5/3$.  Initially, at time $ t=t_{\rm i}$, the gas
temperature is uniform $ T_{\rm b}(r,t_{\rm i})$=$ T_{\rm i}$, and the
perturbation amplitudes are small $\delta_{\rm dm,i},\delta_{\rm
b,i}\ll 1$.  We define the region inside the first zero of
$\sin(kr)/(kr)$, namely $0<kr<\pi$, as the collapsing ``object''.

The evolution of the temperature of the baryons $T_{\rm b}(r,t)$ in
the linear regime is determined by the coupling of their free
electrons to the CMB through Compton
scattering, and by the adiabatic expansion of the gas.  Hence, $
T_{\rm b}(r,t)$ is generally somewhere between the CMB temperature,
$T_{\gamma}\propto (1+z)^{-1}$ and the adiabatically-scaled
temperature $T_{\rm ad}\propto (1+z)^{-2}$.  In the limit of tight
coupling to $T_{\gamma}$, the gas temperature remains uniform.  On the
other hand, in the adiabatic limit, the temperature develops a
gradient according to the relation \beq T_{\rm b} \propto
\rho_b^{(\gamma-1)}.  \eeq

The evolution of a cold dark matter
overdensity, $\delta_{\rm dm}(t)$, in the linear regime is described by the
equation (\ref{coupled}),
\begin{equation}
       {\ddot{\delta}}_{\rm dm}
 + 2H{\dot \delta}_{\rm dm}
       ={3\over 2}H^2 \left(\Omega_{\rm b} \delta_{\rm b} +
\Omega_{\rm dm}
        \delta_{\rm dm}\right)\label{dm}
\end{equation}
whereas the evolution of the overdensity of the baryons, $\delta_{\rm
b}(t)$, with the inclusion of their pressure force is described by
(see \S 9.3.2 of \cite{Kolb90}),
\begin{equation}
        {\ddot{\delta}}_{\rm b}+
       2H{\dot \delta}_{\rm b} ={3\over 2}H^2
\left(\Omega_{\rm b} \delta_{\rm b} +
       \Omega_{\rm dm} \delta_{\rm dm}\right) -\frac{\kB T_{\rm
       i}}{\mu m_p} \left({k\over a}\right)^2
       \left(\frac{a_{\rm i}}{a} \right)^{(1+\beta)} \left(\delta_{\rm
       b}+{2\over 3}\beta [\delta_{\rm b}-\delta_{\rm
       b,i}]\right)\label{b}.
\end{equation}
Here, $H(t)={\dot a}/a$ is the Hubble parameter at a cosmological time
$t$, and $\mu=1.22$ is the mean molecular weight of the neutral
primordial gas in atomic units. The parameter $\beta$ distinguishes
between the two limits for the evolution of the gas temperature. In
the adiabatic limit $\beta=1$, and when the baryon temperature is
uniform and locked to the background radiation, $\beta=0$. The last
term on the right hand side (in square brackets) takes into account
the extra pressure gradient force in $\nabla(\rho_{\rm b} T)=(T \nabla
\rho_{\rm b}+\rho_{\rm b}\nabla T)$, arising from the temperature
gradient which develops in the adiabatic limit. The Jeans wavelength
$\ljeans=2\pi/\kjeans$ is obtained by setting the right-hand side of
equation~(\ref{b}) to zero, and solving for the critical wavenumber
$\kjeans$.  As can be seen from equation~(\ref{b}), the critical
wavelength $\ljeans$ (and therefore the mass $\mjeans$) is in general
time-dependent.  We infer from equation~(\ref{b}) that as time
proceeds, perturbations with increasingly smaller initial wavelengths
stop oscillating and start to grow.

To estimate the Jeans wavelength, we equate the right-hand-side of
equation~(\ref{b}) to zero. We further approximate $\delta_{\rm
b}\sim \delta_{\rm dm}$, and consider sufficiently high redshifts at
which the Universe is matter dominated and flat, $(1+z)\gg {\rm
max}[(1-\Omega_m-\Omega_\Lambda)/\Omega_m,
(\Omega_\Lambda/\Omega_m)^{1/3}]$. In this regime, $\Omega_{\rm
b}\ll\Omm \approx 1$, $H\approx 2/(3t)$, and
$a=(1+z)^{-1}\approx (3H_0{\sqrt{\Omega_m}}/2)^{2/3}t^{2/3}$, where
$\Omega_m=\Omega_{\rm dm}+\Omega_b$ is the total matter
density parameter.  Following cosmological recombination at $z\approx
10^3$, the residual ionization of the cosmic gas keeps its temperature
locked to the CMB temperature (via Compton scattering) down to a
redshift of \cite{Peebles}
\begin{equation}
1+z_t\approx 160 (\Omega_b h^2/0.022)^{2/5}\ .
\end{equation}
In the redshift range between recombination and $z_t$, $\beta=0$ and
\beq
\kjeans\equiv (2\pi/\ljeans)=[2\kB T_{\gamma}(0)/3\mu
m_p]^{-1/2} {\sqrt{\Omega_m}} H_0\ ,
\eeq
so that the Jeans mass is therefore redshift independent and obtains the
value (for the total mass of baryons and dark matter)
\begin{equation}
M_{\rm J}\equiv {4\pi\over 3} \left({\ljeans\over 2}\right)^3
{\bar\rho}(0)
= 1.35\times 10^5 \left({\Omega_mh^2\over 0.15}\right)^{-1/2} M_\odot\ .
\end{equation}

Based on the similarity of $\mjeans$ to the mass of a globular
cluster, Peebles \& Dicke (1968) \cite{Pe68} suggested that globular clusters form
as the first generation of baryonic objects shortly after cosmological
recombination. Peebles \& Dicke assumed a baryonic Universe, with a
nonlinear fluctuation amplitude on small scales at $z\sim 10^3$, a
model which has by now been ruled out. The lack of a dominant mass of dark
matter inside globular clusters makes it unlikely that they formed
through direct cosmological collapse, and more likely that they
resulted from fragmentation during the process of galaxy formation.

\begin{figure}
\centering
\includegraphics[height=6cm]{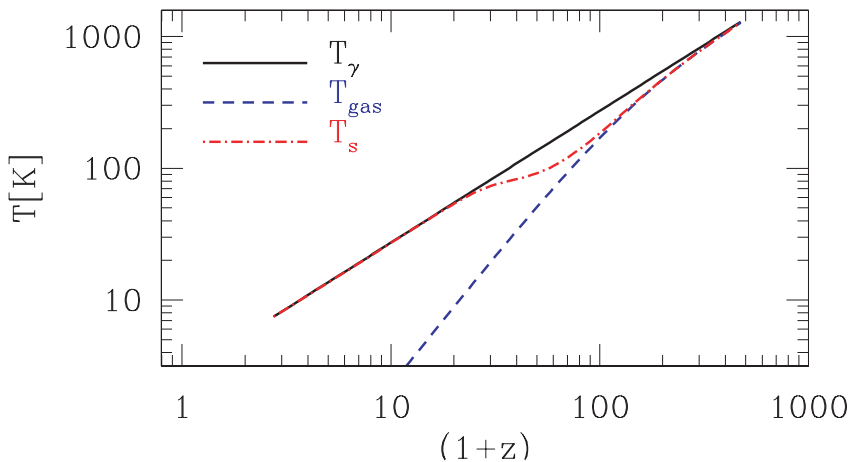}
\caption{Thermal history of the baryons, left over from the big bang,
before the first galaxies formed.  The
residual fraction of free electrons couple the gas temperture $T_{\rm gas}$
to the cosmic microwave background temperature [$T_\gamma\propto (1+z)$]
until a redshift $z\sim 200$. Subsequently the gas temperature cools
adiabatically at a faster rate [$T_{\rm gas}\propto (1+z)^2$].  Also
shown is the spin temperature of the 21cm transition of hydrogen $T_{\rm
s}$ which interpolates between the gas and radiation temperature
and will be discussed in detail later in this review.}
\end{figure}

At $z\la z_t$, the gas temperature declines adiabatically as
$[(1+z)/(1+z_t)]^2$ (i.e., $\beta=1$) and the total Jeans mass obtains
the value,
\begin{equation}
\mjeans= 4.54\times 10^3\left({\Omega_mh^2\over 0.15}\right)^{-1/2}
\left({\Omega_b h^2\over 0.022}\right)^{-3/5}
\left({1+z\over 10}\right)^{3/2}~M_\odot.
\label{eq:m_j}
\end{equation}

It is not clear how the value of the Jeans mass derived above relates
to the mass of collapsed, bound objects. The above analysis is
perturbative (Eqs.~\ref{dm} and \ref{b} are valid only as long as
$\delta_{\rm b}$ and $\delta_{\rm dm}$ are much smaller than unity),
and thus can only describe the initial phase of the collapse.  As
$\delta_{\rm b}$ and $\delta_{\rm dm}$ grow and become larger than
unity, the density profiles start to evolve and dark matter shells may
cross baryonic shells \cite{Haiman94} due to their
different dynamics. Hence the amount of mass enclosed within a given
baryonic shell may increase with time, until eventually the dark
matter pulls the baryons with it and causes their collapse even
for objects below the Jeans mass.


Even within linear theory, the Jeans mass is related only to the evolution
of perturbations at a given time. When the Jeans mass itself varies with
time, the overall suppression of the growth of perturbations depends on a
time-weighted Jeans mass. Gnedin \& Hui (1998) \cite{Gnedin98} showed that
the correct time-weighted mass is the filtering mass $M_F=(4 \pi/3)\,
\bar{\rho}\, (2 \pi a/k_F)^3$, in terms of the comoving wavenumber $k_F$
associated with the ``filtering scale'' (note the change in convention from
$\pi/k_J$ to $2\pi/k_F$). The wavenumber $k_F$ is related to the Jeans
wavenumber $\kjeans$ by \beq \frac{1}{k^2_F (t)}=\frac{1}{D(t)} \int_0^t
dt' \, a^2(t') \frac{\ddot{D} (t')+ 2 H(t') \dot{D}(t')} {k_J^2 (t')} \,
\int_{t'}^t \frac{dt''} {a^2(t'')}\ , \eeq where $D(t)$ is the linear
growth factor.  At high redshift (where $\Ommz \rightarrow 1$), this
relation simplifies to \cite{Gnedin2000b} \beq \frac{1}{k^2_F (t)}=
\frac{3}{a} \int_0^a \frac{d a'}{k_J^2(a')} \left( 1-\sqrt{ \frac{a'} {a}}~
\right)\ . \eeq Then the relationship between the linear overdensity of the
dark matter $\delta_{\rm dm}$ and the linear overdensity of the baryons
$\delta_b$, in the limit of small $k$, can be written as \cite{Gnedin98}
\beq \frac{\delta_b} {\delta_{\rm dm}} = 1-\frac{k^2}{k_F^2}+O (k^4)\
. \eeq

Linear theory specifies whether an initial perturbation, characterized
by the parameters $ k$, $\delta_{\rm dm,i}$, $\delta_{\rm b,i}$ and
$t_{\rm i}$, begins to grow.  To determine the minimum mass of
nonlinear baryonic objects resulting from the shell-crossing and
virialization of the dark matter, we must use a different model which
examines the response of the gas to the gravitational potential of a
virialized dark matter halo.

\subsection{Formation of Nonlinear Objects}
\label{sec2.3}

\subsection{Spherical Collapse}

Let us consider a spherically symmetric density or velocity perturbation of
the smooth cosmological background, and examine the dynamics of a test
particle at a radius $r$ relative to the center of symmetry.  Birkhoff's
(1923) \cite{Birk23} theorem implies that we may ignore the mass outside this radius in
computing the motion of our particle.  We further find that the
relativistic equations of motion describing the system reduce to the usual
Friedmann equation for the evolution of the scale factor of a homogeneous
Universe, but with a density parameter $\Omega$ that now takes account of
the additional mass or peculiar velocity.  In particular, despite the
arbitrary density and velocity profiles given to the perturbation, only the
total mass interior to the particle's radius and the peculiar velocity at
the particle's radius contribute to the effective value of $\Omega$.  We
thus find a solution to the particle's motion which describes its departure
from the background Hubble flow and its subsequent collapse or expansion.
This solution holds until our particle crosses paths with one from a
different radius, which happens rather late for most initial profiles.

As with the Friedmann equation for a smooth Universe, it is possible to
reinterpret the problem into a Newtonian form.  Here we work in an inertial
(i.e.  non-comoving) coordinate system and consider the force on the
particle as that resulting from a point mass at the origin (ignoring
the possible presence of a vacuum energy density):
\begin{equation}\label{eq:1a}
\frac {d^2r} {dt^2} = - \frac {GM}{r^2}, 
\end{equation}
where $G$ is Newton's constant, $r$ is the distance of the particle from
the center of the spherical perturbation, and $M$ is the total mass within
that radius.  As long as the radial shells do not cross each other, the
mass $M$ is constant in time.  The initial density profile determines $M$,
while the initial velocity profile determines $dr/dt$ at the initial time.
As is well-known, there are three branches of solutions: one in which the
particle turns around and collapses, another in which it reaches an
infinite radius with some asymptotically positive velocity, and a third
intermediate case in which it reachs an infinite radius but with a velocity
that approaches zero.  These cases may be written as \cite{Gu72}:
\begin{align}
\label{eq:2} 
\left.
\begin{array}{r}
r = A(\cos \eta -1)  \\  
t = B(\eta - \sin \: \eta) 
\end{array} 
\right\}
&&{\rm Closed} &&(0 \leq \eta \leq 2 \pi )
\end{align}
\begin{align}\label{eq:3}
\left.
\begin{array}{r}
r = A{\eta ^2} /2 \\ 
t = {B\eta^3} / 6 
\end{array}
\right\}
&&{\rm Flat} &&(0 \leq \eta \leq \infty ) 
\end{align}
\begin{align}\label{eq:4}
\left.
\begin{array}{r}
r = A({\rm cosh} \: \eta -1)  \\  
t = B({\rm sinh} \: \eta - \eta) 
\end{array} 
\right\}
&&{\rm Open} &&(0 \leq \eta \leq \infty )
\end{align}
where ${A^3} = GM{B^2}$ applies in all cases.  All three solutions have
$r^3 = 9GM{t^2}/2$ as $t$ goes to zero, which matches the linear theory
expectation that the perturbation amplitude get smaller as one goes back
in time.  In the closed case, the shell turns around at time $\pi B$ and
radius $2A$ and collapses to zero radius at time $2\pi B$.

We are now faced with the problem of relating the spherical collapse
parameters $A, B,$ and $M$ to the linear theory density perturbation
$\delta$ \cite{p80}.  We do this by returning to the equation of
motion.  Consider that at an early epoch (i.e. scale factor $a_i \ll$ 1),
we are given a spherical patch of uniform overdensity $\delta_i$ (the
so-called `top-hat' perturbation).  If $\Omega$ is essentially unity at
this time and if the perturbation is pure growing mode, then the initial
velocity is radially inward with magnitude $\delta_i H( t_i )r/3$, where
$H(t_i)$ is the Hubble constant at the initial time and $r$ is the radius
from the center of the sphere. This can be easily seen from the continuity
equation in spherical coordinates.  The equation of motion (in noncomoving
coordinates) for a particle beginning at radius $r_i$ is simply
\begin{equation}\label{eq:5} 
\frac {d^2r}{dt^2} = - \frac{GM}{r^2} + \frac {\Lambda r}{3}, 
\end{equation}
where $M = (4 \pi /3) r^3_i \rho_i (1 + \delta_i) $ and $\rho_i$
is the background density of the Universe at time $t_i$.  We next define
the dimensionless radius $x = ra_i/r_i$ and rewrite equation (\ref{eq:5})
as
\begin{equation}\label{eq:6}
\frac {l}{H^2_0} \frac {d^2x}{dt^2}=- \frac {\Omega_m}{2x^2}(1+ \delta_i
)+\Omega_\Lambda x.
\end{equation}
Our initial conditions for the integration of this orbit are 
\begin{equation} \label{eq:7}
x(t_i) = a_i
\end{equation}
\begin{equation}\label{eq:8}
\frac {dx}{dt}(t_i) = H(t_1)x \left( 1- \frac {\delta_i}{3} \right) =H_0 a_i \left( 1- \frac 
{\delta_i}{3} 
\right) 
\sqrt { \frac {\Omega_m} {a^3_i} + \frac {\Omega_k} {a^2_i} + \Omega_\Lambda },
\end{equation}
where $H(t_1)=H_0[\Omega_m/a^3(t_1)+(1-\Omega_m)]^{1/2}$ is the Hubble
parameter for a flat Universe at a a cosmic time $t_1$.  Integrating
equation (\ref{eq:6}) yields
\begin{equation}\label{eq:9}
\frac {1}{H^2_0} {\left( \frac {dx}{dt} \right)}^2 =\frac {\Omega_m}{x}(1+ \delta_i )+\Omega_\Lambda x^2 
+ 
K,
\end{equation}
where $K$ is a constant of integration.  Evaluating this at the initial
time and dropping terms of $O(a_i)$ (but $\delta_i \sim a_i$, so we keep
ratios of order unity), we find
\begin{equation}\label{eq:10}
K = - \frac {5 \delta_i }{3a_i} \Omega_m + \Omega_k.	
\end{equation}
If $K$ is sufficiently negative, the particle will turn-around and the
sphere will collapse at a time
\begin{equation}\label{eq:11}
H_0 t_{coll} = 2 \int_0^{a_{\rm max}} da \left( \Omega_m/a + K + \Omega_\Lambda a^2 \right)^{-1/2},
\end{equation}
where $a_{\rm max}$ is the value of $a$ which sets the denominator of the
integral to zero.

For the case of $\Lambda$ = 0, we can determine the spherical collapse
parameters $A$ and $B.$ $K > 0 \: (K < 0)$ produces an open (closed) model.
Comparing coefficients in the energy equations [eq. (\ref{eq:9}) and the
integration of (\ref{eq:1a})], one finds
\begin{equation}\label{eq:12}
A = \frac {\Omega_mr_i}{2a_i} \left| \frac {5 \delta_i }{3a_i} \Omega_m - \Omega_k\right|^{-1}	
\end{equation}
\begin{equation}\label{eq:13} 
B = \frac {\Omega_m}{2H_0} \left| \frac {5 \delta_i }{3a_i} \Omega_m - \Omega_k \right|^{-3/2},		
\end{equation}
where $\Omega_k = 1 - \Omega_m$.  In particular, in an $\Omega = 1$
Universe, where $1 + z = (3H_0t/2)^{-2/3}$, we find that a shell collapses
at redshift $1 + z_c = 0.5929\delta_i/a_i$, or in other words a shell
collapsing at redshift $z_c$ had a linear overdensity extrapolated to the
present day of $\delta_0 = 1.686(1 + z_c).$

While this derivation has been for spheres of constant density, we may treat 
a general spherical density profile $\delta_i(r)$ up until shell crossing \cite{Gu72}.  
A particular radial shell evolves according to the mass interior to it; therefore, we
define the average overdensity $\overline{\delta_i}$
\begin{equation}\label{eq:14}
\overline{\delta_i}(R) = \frac {3}{4 \pi R^3} \int_0^R d^3r\delta_i(r), 
\end{equation}
so that we may use $\overline{\delta_i}$ in place of ${\delta_i}$ in the
above formulae.  If $\overline{\delta_i}$ is not monotonically decreasing
with $R$, then the spherical top-hat evolution of two different radii will
predict that they cross each other at some late time; this is known as
shell crossing and signals the breakdown of the solution.  Even
well-behaved $\overline{\delta_i}$ profiles will produce shell crossing if
shells are allowed to collapse to $r = 0$ and then reexpand, since these
expanding shells will cross infalling shells.  In such a case, first-time
infalling shells will never be affected prior to their turn-around; the
more complicated behavior after turn-around is a manifestation of
virialization.  While the end state for general initial conditions cannot
be predicted, various results are known for a self-similar collapse, in which
$\delta (r)$ is a power-law \cite{Fi84,Be85}, as well as for the case of 
secondary infall models \cite{Go75,Gu77,Ho85}.

\subsection{Halo Properties}

The small density fluctuations evidenced in the CMB grow over time as
described in the previous subsection, until the perturbation $\delta$
becomes of order unity, and the full non-linear gravitational problem
must be considered. The dynamical collapse of a dark matter halo can
be solved analytically only in cases of particular symmetry. If we
consider a region which is much smaller than the horizon $cH^{-1}$,
then the formation of a halo can be formulated as a problem in
Newtonian gravity, in some cases with minor corrections coming from
General Relativity. The simplest case is that of spherical symmetry,
with an initial ($t=t_i\ll t_0$) top-hat of uniform overdensity
$\delta_i$ inside a sphere of radius $R$. Although this model is
restricted in its direct applicability, the results of spherical
collapse have turned out to be surprisingly useful in understanding
the properties and distribution of halos in models based on cold dark
matter.

The collapse of a spherical top-hat perturbation is described by the
Newtonian equation (with a correction for the cosmological constant) \beq
\frac{d^2r}{dt^2}=H_0^2 \Oml\, r-\frac{GM}{r^2}\ , \eeq where $r$ is the
radius in a fixed (not comoving) coordinate frame, $H_0$ is the present-day
Hubble constant, $M$ is the total mass enclosed within radius $r$, and the
initial velocity field is given by the Hubble flow $dr/dt=H(t) r$. The
enclosed $\delta$ grows initially as $\delta_L=\delta_i D(t)/D(t_i)$, in
accordance with linear theory, but eventually $\delta$ grows above
$\delta_L$. If the mass shell at radius $r$ is bound (i.e., if its total
Newtonian energy is negative) then it reaches a radius of maximum expansion
and subsequently collapses. As demonstrated in the previous section, at the
moment when the top-hat collapses to a point, the overdensity predicted by
linear theory is $\delta_L\, = 1.686$ in the Einstein-de Sitter model, with
only a weak dependence on $\Omm$ and $\Oml$. Thus a tophat collapses at
redshift $z$ if its linear overdensity extrapolated to the present day
(also termed the critical density of collapse) is \beq \delta_{\rm
crit}(z)=\frac{1.686}{D(z)}\ ,
\label{deltac} \eeq where we set $D(z=0)=1$.

Even a slight violation of the exact symmetry of the initial perturbation
can prevent the tophat from collapsing to a point. Instead, the halo
reaches a state of virial equilibrium by violent relaxation (phase
mixing). Using the virial theorem $U=-2K$ to relate the potential energy
$U$ to the kinetic energy $K$ in the final state (implying that the virial
radius is half the turnaround radius - where the kinetic energy vanishes),
the final overdensity relative to the critical density at the collapse
redshift is $\Delta_c=18\pi^2 \simeq 178$ in the Einstein-de Sitter model,
modified in a Universe with $\Omm+\Oml=1$ to the fitting formula (Bryan \&
Norman 1998 \cite{BM98}) \beq \Delta_c=18\pi^2+82 d-39 d^2\ , \eeq where $d\equiv
\Ommz-1$ is evaluated at the collapse redshift, so that \beq
\Ommz=\frac{\Omm (1+z)^3}{\Omm (1+z)^3+\Oml+\Omk (1+z)^2}\ .
\label{Ommz} \eeq

A halo of mass $M$ collapsing at redshift $z$ thus has a virial radius
\beq r_{\rm vir}=0.784 \left(\frac{M}{10^8\ h^{-1} \ M_{\sun}
}\right)^{1/3} \left[\frac{\Omm} {\Ommz}\ \frac{\Delta_c}
{18\pi^2}\right]^{-1/3} \left (\frac{1+z}{10} \right)^{-1}\ h^{-1}\
{\rm kpc}\ , \label{rvir}\eeq and a corresponding circular velocity,
\beq V_c=\left(\frac{G M}{r_{\rm vir}}\right)^{1/2}= 23.4 \left(
\frac{M}{10^8\ h^{-1} \ M_{\sun} }\right)^{1/3} \left[\frac {\Omm}
{\Ommz}\ \frac{\Delta_c} {18\pi^2}\right]^{1/6} \left( \frac{1+z} {10}
\right)^{1/2}\ {\rm km\ s}^{-1}\ . \label{Vceqn} \eeq In these
expressions we have assumed a present Hubble constant written in the
form $H_0=100\, h\mbox{ km s}^{-1}\mbox{Mpc}^{-1}$. We may also define
a virial temperature \beq \label{tvir} T_{\rm vir}=\frac{\mu m_p
V_c^2}{2 \kB}=1.98\times 10^4\ \left(\frac{\mu}{0.6}\right)
\left(\frac{M}{10^8\ h^{-1} \ M_{\sun} }\right)^{2/3} \left[ \frac
{\Omm} {\Ommz}\ \frac{\Delta_c} {18\pi^2}\right]^{1/3}
\left(\frac{1+z}{10}\right)\ {\rm K} \ , \eeq where $\mu$ is the mean
molecular weight and $m_p$ is the proton mass. Note that the value of
$\mu$ depends on the ionization fraction of the gas; for a fully
ionized primordial gas $\mu=0.59$, while a gas with ionized hydrogen
but only singly-ionized helium has $\mu=0.61$. The binding energy of
the halo is approximately\footnote{The coefficient of $1/2$ in
equation~(\ref{Ebind}) would be exact for a singular isothermal
sphere, $\rho(r)\propto 1/r^2$.} \beq \label{Ebind} E_b= {1\over 2}
\frac{GM^2}{r_{\rm vir}} = 5.45\times 10^{53} \left(\frac{M}{10^8\
h^{-1} \ M_{\sun} }\right)^{5/3} \left[ \frac {\Omm} {\Ommz}\
\frac{\Delta_c} {18\pi^2}\right]^{1/3} \left(\frac{1+z}{10}\right)
h^{-1}\ {\rm erg}\ . \eeq Note that the binding energy of the baryons
is smaller by a factor equal to the baryon fraction $\Omega_b/\Omm$.

Although spherical collapse captures some of the physics governing the
formation of halos, structure formation in cold dark matter models procedes
hierarchically. At early times, most of the dark matter is in low-mass
halos, and these halos continuously accrete and merge to form high-mass
halos. Numerical simulations of hierarchical halo formation indicate a
roughly universal spherically-averaged density profile for the resulting
halos (Navarro, Frenk, \& White 1997, hereafter NFW \cite{Na97}), though
with considerable scatter among different halos (e.g., \cite{Bu00}). The NFW profile has the form \beq \rho(r)=\frac{3 H_0^2} {8 \pi G}
(1+z)^3 \frac{\Omm}{\Ommz} \frac{\delta_c} {\cN x (1+\cN x)^2}\ ,
\label{NFW} \eeq where $x=r/r_{\rm vir}$, and the characteristic density
$\delta_c$ is related to the concentration parameter $\cN$ by \beq
\delta_c=\frac{\Delta_c}{3} \frac{\cN^3} {\ln(1+\cN)-\cN/(1+\cN)} \ . \eeq
The concentration parameter itself depends on the halo mass $M$, at a given
redshift $z$ \cite{WBPKD02}.

More recent N-body simulations indicate deviations from the original NFW
profile; for details and refined fitting formula see \cite{nav04}.


\section{Nonlinear Growth}


\subsection{The Abundance of Dark Matter Halos}
\label{sec2.4}

In addition to characterizing the properties of individual halos, a
critical prediction of any theory of structure formation is the
abundance of halos, i.e.\ the number density of halos as a function of
mass, at any redshift. This prediction is an important step toward
inferring the abundances of galaxies and galaxy clusters. While the
number density of halos can be measured for particular cosmologies in
numerical simulations, an analytic model helps us gain physical
understanding and can be used to explore the dependence of abundances
on all the cosmological parameters. 

A simple analytic model which successfully matches most of the numerical
simulations was developed by Press \& Schechter (1974) \cite{Press}. The model is based
on the ideas of a Gaussian random field of density perturbations, linear
gravitational growth, and spherical collapse. To determine the abundance of
halos at a redshift $z$, we use $\delta_M$, the density field smoothed on a
mass scale $M$, as defined in \S \ref{sec2.2}. Since $\delta_M$ is
distributed as a Gaussian variable with zero mean and standard deviation
$\sigma(M)$ [which depends only on the present linear power spectrum, see
equation~(\ref{eqsigM})], the probability that $\delta_M$ is greater than
some $\delta$ equals \beq \int_{\delta}^{\infty}d\delta_M \frac{1}{\sqrt{2
\pi}\, \sigma(M)} \exp \left[- \frac{\delta_M^2} {2
\,\sigma^2(M)}\right]={1\over 2} {\rm erfc}\left(\frac{\delta} {\sqrt{2}
\,\sigma(M) } \right)\ . \label{PS1} \eeq The fundamental ansatz is to
identify this probability with the fraction of dark matter particles which
are part of collapsed halos of mass greater than $M$, at redshift
$z$. There are two additional ingredients: First, the value used for
$\delta$ is $\delta_{\rm crit}(z)$ given in equation~(\ref{deltac}), which
is the critical density of collapse found for a spherical top-hat
(extrapolated to the present since $\sigma(M)$ is calculated using the
present power spectrum); and second, the fraction of dark matter in halos
above $M$ is multiplied by an additional factor of 2 in order to ensure
that every particle ends up as part of some halo with $M>0$. Thus, the
final formula for the mass fraction in halos above $M$ at redshift $z$ is
\beq
\label{PSerfc} F(>M | z)={\rm erfc}\left(\frac{\delta_{\rm crit}(z)} 
{\sqrt{2}\,\sigma(M) } \right)\ . \eeq

This ad-hoc factor of 2 is necessary, since otherwise only positive
fluctuations of $\delta_M$ would be included. Bond et al.\ (1991) \cite{bond91}
found an alternate derivation of this correction factor, using a
different ansatz. In their derivation, the factor of 2 has a more
satisfactory origin, namely the so-called ``cloud-in-cloud'' problem:
For a given mass $M$, even if $\delta_M$ is smaller than $\delta_{\rm
crit}(z)$, it is possible that the corresponding region lies inside a
region of some larger mass $M_L>M$, with $\delta_{M_L}>\delta_{\rm
crit}(z)$. In this case the original region should be counted as
belonging to a halo of mass $M_L$. Thus, the fraction of particles
which are part of collapsed halos of mass greater than $M$ is larger
than the expression given in equation~(\ref{PS1}). Bond et al.\ showed
that, under certain assumptions, the additional contribution results
precisely in a factor of 2 correction.

\begin{figure}
\centering
\includegraphics[height=6cm]{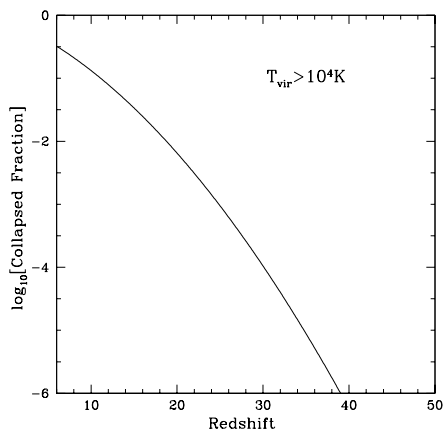}
\caption{Fraction of baryons that assembled into dark matter halos with a
virial temperature of $T_{\rm vir}>10^4$K as a function of redshift. These
baryons are above the temperature threshold for gas cooling and
fragmentation via atomic transitions.  After reionization the temperature
barrier for star formation in galaxies is raised because the photo-ionized
intergalactic medium is already heated to $\sim 10^{4}$K and it can
condense only into halos with $T_{\rm vir}>10^5$K.}
\label{collapsed}
\end{figure}

Differentiating the fraction of dark matter in halos above $M$ yields
the mass distribution. Letting $dn$ be the comoving number density of
halos of mass between $M$ and $M+dM$, we have \beq\frac{dn}{dM}=
\sqrt{\frac{2}{\pi}}\, \frac{\rho_m}{M}\, \frac{-d(\ln \sigma)}{dM}
\,\nu_c\, e^{-\nu_c^2/2}\ , \eeq where $\nu_c=\delta_{\rm crit}(z)/
\sigma(M)$ is the number of standard deviations which the critical
collapse overdensity represents on mass scale $M$. Thus, the abundance
of halos depends on the two functions $\sigma(M)$ and $\delta_{\rm
crit} (z)$, each of which depends on the energy content of the
Universe and the values of the other cosmological parameters. Since
recent observations confine the standard set of parameters to a
relatively narrow range, we illustrate the abundance of halos and
other results for a single set of parameters: $\Omm=0.3$, $\Oml=0.7$,
$\Omega_b=0.045$, $\sigma_8=0.9$, a primordial power spectrum index
$n=1$ and a Hubble constant $h=0.7$. 

\begin{figure}%
\centering
\includegraphics[height=6cm]{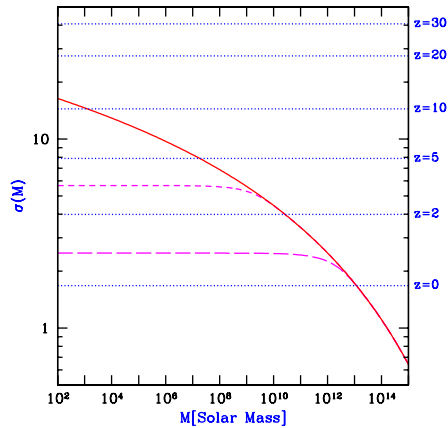}
\caption{Mass fluctuations and collapse thresholds in cold dark matter
models. The horizontal dotted lines show the value of the extrapolated
collapse overdensity $\delta_{\rm crit}(z)$ at the indicated
redshifts. Also shown is the value of $\sigma(M)$ for the cosmological
parameters given in the text (solid curve), as well as $\sigma(M)$ for
a power spectrum with a cutoff below a mass $M=1.7\times 10^8
M_{\sun}$ (short-dashed curve), or $M=1.7\times 10^{11} M_{\sun}$
(long-dashed curve). The intersection of the horizontal lines with the
other curves indicate, at each redshift $z$, the mass scale (for each
model) at which a $1-\sigma$ fluctuation is just collapsing at $z$
(see the discussion in the text).}
\label{fig2a}
\end{figure}

Figure \ref{fig2a} shows $\sigma(M)$ and $\delta_{\rm crit}(z)$, with
the input power spectrum computed from Eisenstein \& Hu (1999) \cite{Eis99}. The
solid line is $\sigma(M)$ for the cold dark matter model with the
parameters specified above. The horizontal dotted lines show the value
of $\delta_{\rm crit}(z)$ at $z=0, 2, 5, 10, 20$ and 30, as indicated
in the figure. From the intersection of these horizontal lines with
the solid line we infer, e.g., that at $z=5$ a $1-\sigma$ fluctuation
on a mass scale of $2\times 10^7 M_{\sun}$ will collapse. On the other
hand, at $z=5$ collapsing halos require a $2-\sigma$ fluctuation on a
mass scale of $3\times 10^{10} M_{\sun}$, since $\sigma(M)$ on this
mass scale equals about half of $\delta_{\rm crit}(z=5)$. Since at
each redshift a fixed fraction ($31.7\%$) of the total dark matter
mass lies in halos above the $1-\sigma$ mass, Figure \ref{fig2a} shows
that most of the mass is in small halos at high redshift, but it
continuously shifts toward higher characteristic halo masses at lower
redshift. Note also that $\sigma(M)$ flattens at low masses because of
the changing shape of the power spectrum. Since $\sigma \rightarrow
\infty$ as $M \rightarrow 0$, in the cold dark matter model all the
dark matter is tied up in halos at all redshifts, if sufficiently
low-mass halos are considered.
  
Also shown in Figure \ref{fig2a} is the effect of cutting off the
power spectrum on small scales. The short-dashed curve corresponds to
the case where the power spectrum is set to zero above a comoving
wavenumber $k=10\, {\rm Mpc}^{-1}$, which corresponds to a mass
$M=1.7\times 10^8 M_{\sun}$. The long-dashed curve corresponds to a
more radical cutoff above $k=1\, {\rm Mpc}^{-1}$, or below
$M=1.7\times 10^{11} M_{\sun}$. A cutoff severely reduces the
abundance of low-mass halos, and the finite value of $\sigma(M=0)$
implies that at all redshifts some fraction of the dark matter does
not fall into halos. At high redshifts where $\delta_{\rm crit}(z) \gg
\sigma(M=0)$, all halos are rare and only a small fraction of the dark
matter lies in halos. In particular, this can affect the abundance of
halos at the time of reionization, and thus the observed limits on
reionization constrain scenarios which include a small-scale cutoff in
the power spectrum \cite{BHO00}.

\begin{figure}
\centering
\includegraphics[height=6cm]{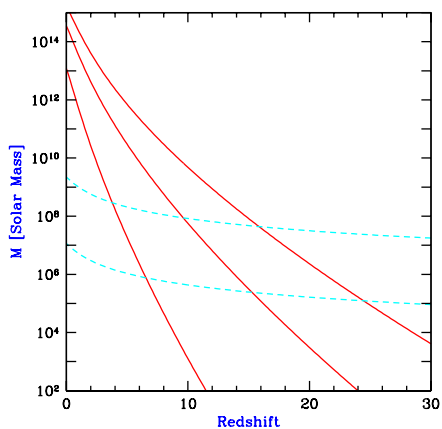}
\caption{Characteristic properties of collapsing halos: Halo mass.
The solid curves show the mass of collapsing halos which correspond to
$1-\sigma$, $2-\sigma$, and $3-\sigma$ fluctuations (in order from
bottom to top). The dotted curves show the mass corresponding to the
minimum temperature required for efficient cooling with primordial
atomic species only (upper curve) or with the addition of molecular
hydrogen (lower curve).}
\label{fig2b}
\end{figure}

\begin{figure}
\centering
\includegraphics[height=6cm]{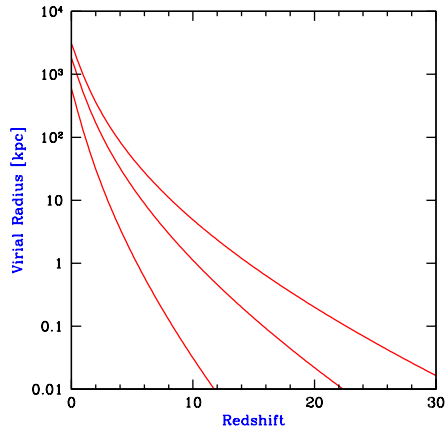}
\caption{Characteristic properties of collapsing halos: Halo virial
radius. The curves show the virial radius of collapsing halos which
correspond to $1-\sigma$, $2-\sigma$, and $3-\sigma$ fluctuations (in
order from bottom to top).}
\label{fig2c}
\end{figure}

\begin{figure}
\centering
\includegraphics[height=6cm]{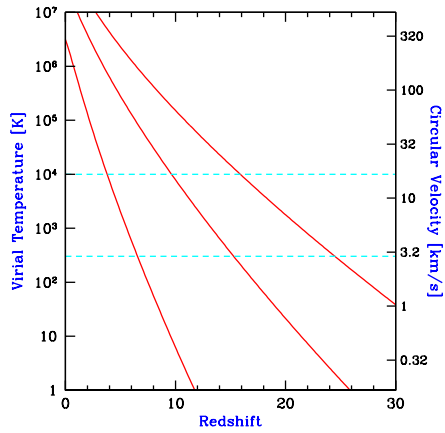}
\caption{Characteristic properties of collapsing halos: Halo virial
temperature and circular velocity. The solid curves show the virial
temperature (or, equivalently, the circular velocity) of collapsing
halos which correspond to $1-\sigma$, $2-\sigma$, and $3-\sigma$
fluctuations (in order from bottom to top). The dotted curves show the
minimum temperature required for efficient cooling with primordial
atomic species only (upper curve) or with the addition of molecular
hydrogen (lower curve).}
\label{fig2d}
\end{figure}

\begin{figure}
\centering
\includegraphics[height=6cm]{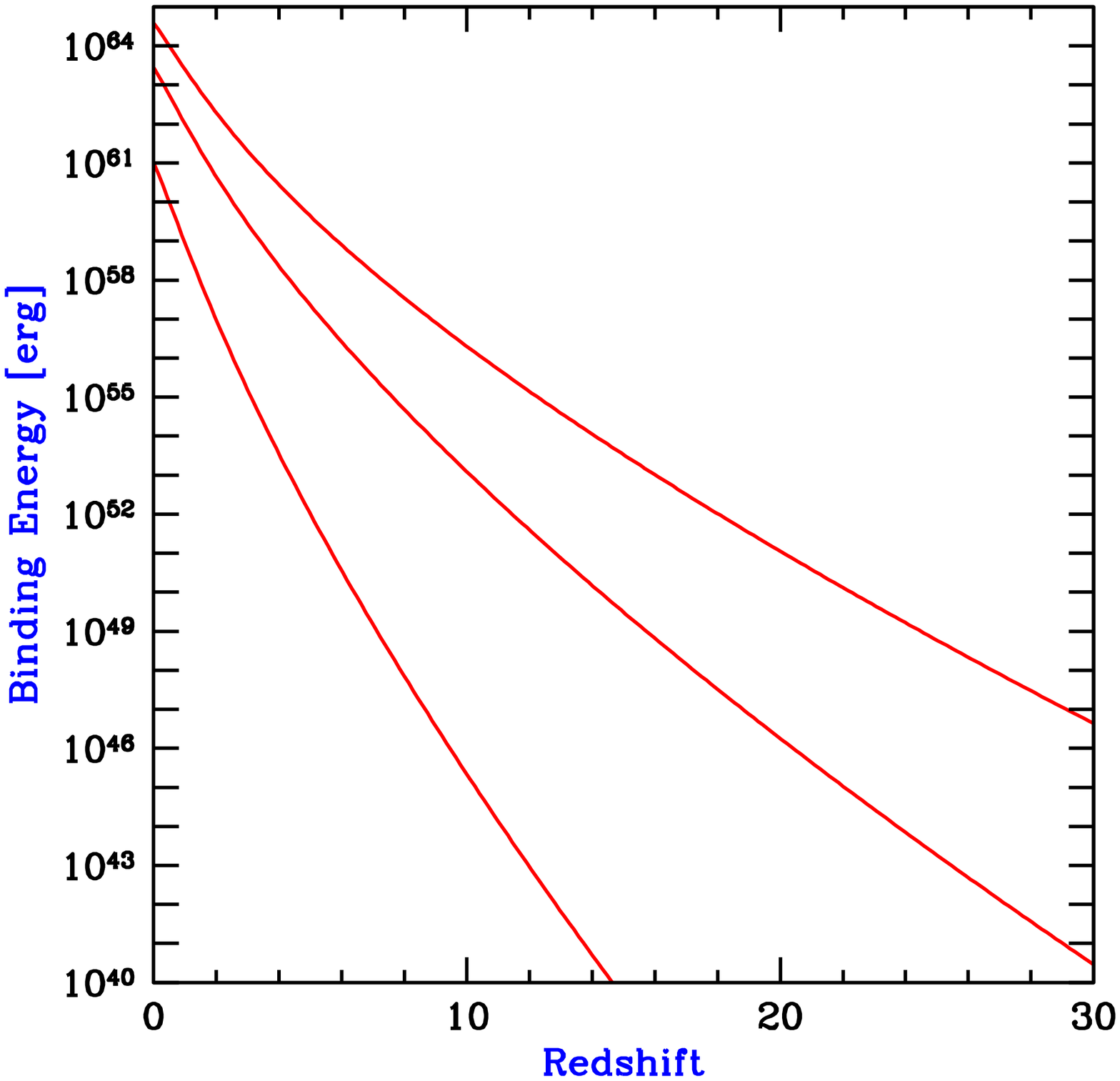}
\caption{Characteristic properties of collapsing halos: Halo binding
energy. The curves show the total binding energy of collapsing halos
which correspond to $1-\sigma$, $2-\sigma$, and $3-\sigma$
fluctuations (in order from bottom to top).}
\label{fig2e}
\end{figure}

In figures \ref{fig2b} -- \ref{fig2e} we show explicitly the properties of
collapsing halos which represent $1-\sigma$, $2-\sigma$, and $3-\sigma$
fluctuations (corresponding in all cases to the curves in order from bottom
to top), as a function of redshift. No cutoff is applied to the power
spectrum. Figure \ref{fig2b} shows the halo mass, Figure \ref{fig2c} the
virial radius, Figure \ref{fig2d} the virial temperature (with $\mu$ in
equation~(\ref{tvir}) set equal to $0.6$, although low temperature halos
contain neutral gas) as well as circular velocity, and Figure \ref{fig2e}
shows the total binding energy of these halos. In figures \ref{fig2b} and
\ref{fig2d}, the dotted curves indicate the minimum virial temperature
required for efficient cooling with primordial atomic species only (upper
curve) or with the addition of molecular hydrogen (lower curve). Figure
\ref{fig2e} shows the binding energy of dark matter halos. The binding
energy of the baryons is a factor $\sim \Omega_b/\Omega_m\sim 15\%$
smaller, if they follow the dark matter. Except for this constant factor,
the figure shows the minimum amount of energy that needs to be deposited
into the gas in order to unbind it from the potential well of the dark
matter. For example, the hydrodynamic energy released by a single
supernovae, $\sim 10^{51}~{\rm erg}$, is sufficient to unbind the gas in
all $1-\sigma$ halos at $z\ga 5$ and in all $2-\sigma$ halos at $z\ga 12$.

At $z=5$, the halo masses which correspond to $1-\sigma$, $2-\sigma$,
and $3-\sigma$ fluctuations are $1.8\times 10^7 M_{\sun}$, $3.0\times
10^{10} M_{\sun}$, and $7.0\times 10^{11} M_{\sun}$, respectively. The
corresponding virial temperatures are $2.0 \times 10^3$ K, $2.8 \times
10^5$ K, and $2.3 \times 10^6$ K. The equivalent circular velocities
are 7.5 ${\rm km\ s}^{-1}$, 88 ${\rm km\ s}^{-1}$, and 250 ${\rm km\
s}^{-1}$. At $z=10$, the $1-\sigma$, $2-\sigma$, and $3-\sigma$
fluctuations correspond to halo masses of $1.3\times 10^3 M_{\sun}$,
$5.7\times 10^7 M_{\sun}$, and $4.8\times 10^9 M_{\sun}$,
respectively. The corresponding virial temperatures are 6.2 K, $7.9
\times 10^3$ K, and $1.5 \times 10^5$ K. The equivalent circular
velocities are 0.41 ${\rm km\ s}^{-1}$, 15 ${\rm km\ s}^{-1}$, and 65
${\rm km\ s}^{-1}$. Atomic cooling is efficient at $T_{\rm vir} \ga
10^4$ K, or a circular velocity $V_c \ga 17\ {\rm km\ s}^{-1}$. This
corresponds to a $1.2-\sigma$ fluctuation and a halo mass of
$2.1\times 10^8 M_{\sun}$ at $z=5$, and a $2.1-\sigma$ fluctuation and
a halo mass of $8.3\times 10^7 M_{\sun}$ at $z=10$. Molecular hydrogen
provides efficient cooling down to $T_{\rm vir} \sim 300$ K, or a
circular velocity $V_c \sim 2.9\ {\rm km\ s}^{-1}$. This corresponds
to a $0.81-\sigma$ fluctuation and a halo mass of $1.1\times 10^6
M_{\sun}$ at $z=5$, and a $1.4-\sigma$ fluctuation and a halo mass of
$4.3\times 10^5 M_{\sun}$ at $z=10$.

\begin{figure}
\centering
\includegraphics[height=6cm]{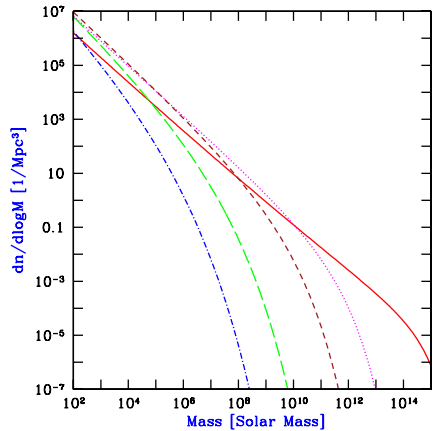}
\caption{Halo mass function at several redshifts: $z=0$ (solid curve),
$z=5$ (dotted curve), $z=10$ (short-dashed curve), $z=20$ (long-dashed
curve), and $z=30$ (dot-dashed curve).}
\label{fig2f}
\end{figure}

In Figure  \ref{fig2f} we show the halo mass function $dn/d\ln(M)$ at
several different redshifts: $z=0$ (solid curve), $z=5$ (dotted
curve), $z=10$ (short-dashed curve), $z=20$ (long-dashed curve), and
$z=30$ (dot-dashed curve). Note that the mass function does not
decrease monotonically with redshift at all masses. At the lowest
masses, the abundance of halos is higher at $z>0$ than at $z=0$.

\subsection {The Excursion-Set (Extended Press-Schechter) Formalism}

The usual Press-Schechter formalism makes no attempt to deal with the
correlations between halos or between different mass scales.  In
particular, this means that while it can generate a distribution of halos
at two different epochs, it says nothing about how particular halos in one
epoch are related to those in the second.  We therefore would like some
method to predict, at least statistically, the growth of individual halos
via accretion and mergers.  Even restricting ourselves to spherical
collapse, such a model must utilize the full spherically-averaged density
profile around a particular point.  The potential correlations between the
mean overdensities at different radii make the statistical description
substantially more difficult.

The excursion set formalism (Bond et al. 1991 \cite{bond91}) seeks to
describe the statistics of halos by considering the statistical properties
of $\overline{\delta}(R)$, the average overdensity within some spherical
window of characteristic radius $R$, as a function of $R$.  While the
Press-Schechter model depends only on the Gaussian distribution of
$\overline{\delta}$ for one particular $R$, the excursion set considers all
$R$.  Again the connection between a value of the linear regime $\delta$
and the final state is made via the spherical collapse solution, so that
there is a critical value $\delta_c (z)$ of $\overline{\delta}$ which is
required for collapse at a redshift $z$.

For most choices of window function, the functions $\overline{\delta}(R)$
are correlated from one $R$ to another such that it is prohibitively
difficult to calculate the desired statistics directly [although Monte
Carlo realizations are possible \cite{bond91}].
However, for one particular choice of a window function, the correlations
between different $R$ greatly simplify and many interesting quantities may
be calculated \cite{bond91,LC93}.  The key is to use a $k$-space top-hat
window function, namely $W_k = 1$ for all $k$ less than some critical $k_c$
and $W_k = 0$ for $k > k_c$.  This filter has a spatial form of $W(r) \:
\propto \: j_1 (k_c r) / k_c r$, which implies a volume $6 \pi^2 /
k^{3}_{c}$ or mass $6\pi^2 \rho_b /k^{3}_{c}$.  The characteristic radius
of the filter is $\sim k^{-1}_{c}$, as expected.  Note that in real space,
this window function converges very slowly, due only to a sinusoidal
oscillation, so the region under study is rather poorly localized.

The great advantage of the sharp $k$-space filter is that the difference at
a given point between $\overline{\delta}$ on one mass scale and that on
another mass scale is statistically independent from the value on the
larger mass scale.  With a Gaussian random field, each
$\delta_k$ is Gaussian distributed independently from the others.  For this
filter,
\begin{equation}\label{eq:15}
\overline{\delta}(M) = \int_{k<k_c(M)} \frac {d^3k}{(2 \pi )^3} \delta_k, 
\end{equation}
meaning that the overdensity on a particular scale is simply the sum of the
random variables $\delta_k$ interior to the chosen $k_c$.  Consequently,
the difference between the $\overline{\delta}(M)$ on two mass scales is
just the sum of the $\delta_k$ in the spherical shell between the two
$k_c$, which is independent from the sum of the $\delta_k$ interior to the
smaller $k_c$.  Meanwhile, the distribution of $\overline{\delta}(M)$ given
no prior information is still a Gaussian of mean zero and variance
\begin{equation}\label{eq:16}
\sigma^2(M) = \frac {1}{2 \pi^2} \int_{k<k_c(M)} dk \: k^2P(k). 
\end{equation}

If we now consider $\overline{\delta}$ as a function of scale $k_c$, we see
that we begin from $\overline{\delta} = 0$ at $k_c = 0 \: (M = \infty )$
and then add independently random pieces as $k_c$ increases.  This
generates a random walk, albeit one whose stepsize varies with $k_c$.  We
then assume that, at redshift $z$, a given function $\overline{\delta}(k_c
)$ represents a collapsed mass $M$ corresponding to the $k_c$ where the
function first crosses the critical value $\delta_c (z)$.  With this
assumption, we may use the properties of random walks to calculate the
evolution of the mass as a function of redshift.

It is now easy to rederive the Press-Schechter mass function, including the
previously unexplained factor of 2 \cite{bond91,LC93,Whi94}.  The fraction
of mass elements included in halos of mass less than $M$ is just the
probability that a random walk remains below $\delta_c (z)$ for all $k_c$
less than $K_c$, the filter cutoff appropriate to $M$.  This probability
must be the complement of the sum of the probabilities that: {\it (a)}
$\overline{\delta} (K_c) > \delta_c(z)$; or that {\it (b)}
$\overline{\delta}(K_c) < \delta_c(z)$ but $\overline{\delta}(k'_c) >
\delta_c(z)$ for some $k'_c < K_c$.  But these two cases in fact have equal
probability; any random walk belonging to class {\it (a)} may be reflected
around its first upcrossing of $\delta_c (z)$ to produce a walk of class
{\it (b)}, and vice versa.  Since the distribution of
$\overline{\delta}(K_c)$ is simply Gaussian with variance $\sigma^2(M)$,
the fraction of random walks falling into class {\it (a)} is simply $(1/
\sqrt{2\pi\sigma^2}) \int_{\delta_c (z)}^{\infty} d \delta \; \exp \{ -
\delta^2 / 2 \sigma^2 (M) \}$.  Hence, the fraction of mass elements
included in halos of mass less than $M$ at redshift $z$ is simply
\begin{equation}\label{eq:17} 
F(<M)=1-2 \times \frac {1} { \sqrt{2\pi\sigma^2}} \int_{\delta_c
(z)}^{\infty} d \delta \: \exp \{ - \delta^2 /2 \sigma^2 (M) \}
\end{equation} 
which may be differentiated to yield the Press-Schechter mass function. We
may now go further and consider how halos at one redshift are related to
those at another redshift.  If we are given that a halo of mass $M_2$
exists at redshift $z_2$, then we know that the random function
$\overline{\delta}(k_c)$ for each mass element within the halo first
crosses $\delta(z_2)$ at $k_{c2}$ corresponding to $M_2$.  Given this
constraint, we may study the distribution of $k_c$ where the function
$\overline{\delta}(k_c)$ crosses other thresholds.  It is particularly easy
to construct the probability distribution for when trajectories first cross
some $\delta_c (z_1) \: > \: \delta_c (z_2)$ (implying $z_1 \: > \: z_2$);
clearly this occurs at some $k_{c1} \: > k_{c2}$.  This problem reduces to
the previous one if we translate the origin of the random walks from
$(k_c,\overline{\delta}) = (0,0) \: {\rm to} \: (k_{c2},\delta_c(z_2))$.
We therefore find the distribution of halo masses $M_1$ that a mass element
finds itself in at redshift $z_1$ given that it is part of a larger halo of
mass $M_2$ at a later redshift $z_2$ is \cite{bond91,Bow91})
\begin{equation}\label{eq:18}\begin{split} 
& \frac { dP } { dM_1 } (M_1,z_1 | M_2,z_2) = \\ & \sqrt {\frac {2}{\pi}}
\frac {\delta_c (z_1) - \delta_c (z_2)} { [ \sigma^2 (M_1) - \sigma^2
(M_2)]^{3/2} } \left | \frac { d \sigma (M_1) } { d M_1} \right | \exp \left
\{ - \frac { [ \delta_c (z_1) - \delta_c (z_2) ] ^2 } { 2 [ \sigma^2 (M_1)
- \sigma^2 (M_2) ] } \right \}.
\end{split}\end{equation} 
This may be rewritten as saying that the quantity 
\begin{equation}\label{eq:19}
\tilde {v} = \frac {\delta_c (z_1) - \delta_c (z_2)} { \sqrt {\sigma^2
(M_1) - \sigma^2 (M_2)} }
\end{equation} 
is distributed as the positive half of a Gaussian with unit variance;
equation (\ref{eq:19}) may be inverted to find $M_1(\tilde {v})$.

We seek to interpret the statistics of these random walks as those of
merging and accreting halos.  For a single halo, we may imagine that as we
look back in time, the object breaks into ever smaller pieces, similar to
the branching of a tree.  Equation (\ref{eq:18}) is the distribution of the
sizes of these branches at some given earlier time.  However, using this
description of the ensemble distribution to generate Monte Carlo
realizations of single merger trees has proven to be difficult.  In all
cases, one recursively steps back in time, at each step breaking the final
object into two or more pieces.  An elaborate scheme (Kauffmann \& White
1993 \cite{KW93}) picks a large number of progenitors from the ensemble
distribution and then randomly groups them into sets with the correct total
mass.  This generates many (hundreds) possible branching schemes of equal
likelihood.  A simpler scheme (Lacey \& Cole 1993 \cite{LC93}) assumes that
at each time step, the object breaks into two pieces.  One value from the
distribution (\ref{eq:18}) then determines the mass ratio of the two
branchs.

One may also use the distribution of the ensemble to derive some additional
analytic results.  A useful example is the distribution of the epoch at
which an object that has mass $M_2$ at redshift $z_2$ has accumulated half
of its mass \cite{LC93}.  The probability that the formation time is
earlier than $z_1$ is equal to the probability that at redshift $z_1$ a
progenitor whose mass exceeds $M_2/2$ exists:
\begin{equation}\label{eq:20}  
P(z_f > z_1) = \int_{M_2/2}^{M_2} \frac{M_2}{M} \frac {dP}{dM}
(M,z_1|M_2,z_2)dM,
\end{equation} 
where $dP/dM$ is given in equation (\ref{eq:18}).  The factor of $M_2/M$
corrects the counting from mass weighted to number weighted; each halo of
mass $M_2$ can have only one progenitor of mass greater than $M_2/2$.
Differentiating equation (\ref{eq:20}) with respect to time gives the
distribution of formation times.  This analytic form is an excellent match
to scale-free N-body simulations \cite{LC94}.  On the
other hand, simple Monte Carlo implementations of equation (\ref{eq:18})
produce formation redshifts about 40\% higher \cite{LC93}.  As there may be
correlations between the various branches, there is no unique Monte Carlo
scheme.

Numerical tests of the excursion set formalism are quite encouraging.  Its
predictions for merger rates are in very good agreement with those measured
in scale-free N-body simulations for mass scales down to around 10\% of the
nonlinear mass scale (that scale at which $\sigma_M= 1$ ), and
distributions of formation times closely match the analytic predictions
\cite{LC94}.  The model appears to be a promising method for tracking the
merging of halos, with many applications to cluster and galaxy formation
modeling.  In particular, one may use the formalism as the foundation of
semi-analytic galaxy formation models \cite{KWG93}.  The excursion set
formalism may also be used to derive the correlations of halos in the
nonlinear regime \cite{mw96}.

\subsection{Response of Baryons to Nonlinear Dark Matter Potentials}

The dark matter is assumed to be cold and to dominate gravity, and so
its collapse and virialization proceeds unimpeded by pressure
effects. In order to estimate the minimum mass of baryonic objects, we
must go beyond linear perturbation theory and examine the baryonic
mass that can accrete into the final gravitational potential well
of the dark matter.

For this purpose, we assume that the dark matter had already
virialized and produced a gravitational potential $\phi({\bf r})$ at a
redshift $z_{\rm vir}$ (with $\phi\rightarrow0$ at large distances,
and $\phi<0$ inside the object) and calculate the resulting
overdensity in the gas distribution, ignoring cooling (an assumption
justified by spherical collapse simulations which indicate that
cooling becomes important only after virialization; see Haiman et al.\
1996 \cite{Haiman}).

After the gas settles into the dark matter potential well, it
satisfies the hydrostatic equilibrium equation,
\begin{equation}
\nabla p_{\rm b} = -\rho_{\rm b} \nabla \phi
\label{eq:hyd}
\end{equation}
where $p_{\rm b}$ and $\rho_{\rm b}$ are the pressure and mass density
of the gas.  At $z \la 100$ the gas temperature is decoupled from the
CMB, and its pressure evolves adiabatically (ignoring atomic or
molecular cooling),
\begin{equation}
{p_{\rm b}\over {\bar p}_{\rm b}} = \left({\rho_{\rm b}\over {\bar
\rho}_{\rm b}}\right)^{5/3}
\label{eq:adi}
\end{equation}
where a bar denotes the background conditions. We substitute
equation~(\ref{eq:adi}) into~(\ref{eq:hyd}) and get the solution,
\begin{equation}
{\rho_{\rm b}\over {\bar \rho}_{\rm b}}= \left(1- {2\over 5}{\mu
m_p\phi\over {\kB \bar T}}\right)^{3/2}
\end{equation}
where ${\bar T}= {\bar p}_{\rm b} \mu m_p/(\kB {\bar \rho}_{\rm b})$
is the background gas temperature.  If we define $T_{\rm vir}=
-{1\over 3}m_p\phi/\kB$ as the virial temperature for a potential
depth $-\phi$, then the overdensity of the baryons at the
virialization redshift is
\begin{equation}
\delta_{\rm b} = {\rho_{\rm b}\over {\bar \rho}_{\rm b}} - 1 =
\left(1+
{6\over 5}{T_{\rm vir}\over {\bar T}}\right)^{3/2} - 1 .
\label{eq:del}
\end{equation}
This solution is approximate for two reasons: (i) we assumed that the
gas is stationary throughout the entire region and ignored the
transitions to infall and the Hubble expansion at the interface
between the collapsed object and the background intergalactic medium
(henceforth IGM), and (ii) we ignored entropy production at the
virialization shock surrounding the object.  Nevertheless, the result
should provide a better estimate for the minimum mass of collapsed
baryonic objects than the Jeans mass does, since it incorporates the
nonlinear potential of the dark matter.

We may define the threshold for the collapse of baryons by the
criterion that their mean overdensity, $\delta_{\rm b}$, exceeds a
value of 100, amounting to $\ga 50\%$ of the baryons that would
assemble in the absence of gas pressure, according to the spherical
top-hat collapse model. Equation~(\ref{eq:del}) then
implies that $T_{\rm vir} > 17.2\, {\bar T}$.

As mentioned before, the gas temperature evolves at $z\la 160$ according to
the relation ${\bar T}\approx 170 [(1+z) /100]^2\ {\rm K}$. This implies
that baryons are overdense by $\delta_{\rm b} > 100$ only inside halos with
a virial temperature $T_{\rm vir}\ga 2.9\times 10^3~[(1+z)/100]^2\ {\rm
K}$. Based on the top-hat model, this implies a minimum halo mass for
baryonic objects of
\begin{equation}
M_{\rm min}= 5.0 \times 10^3  \left({\Omega_m h^2\over
0.15}\right)^{-1/2} \left({\Omega_b h^2\over 0.022}\right)^{-3/5}
\left({1+z\over 10}\right)^{3/2}~M_\odot,
\label{eq:M_min}
\end{equation}
where we consider sufficiently high redshifts so that $\Ommz \approx
1$. This minimum mass is coincidentally almost identical to the naive
Jeans mass calculation of linear theory in equation~(\ref{eq:m_j})
despite the fact that it incorporates shell crossing by the dark
matter, which is not accounted for by linear theory. Unlike the Jeans
mass, the minimum mass depends on the choice for an overdensity
threshold [taken arbitrarily as $\delta_{\rm b}>100$ in
equation~(\ref{eq:M_min})]. To estimate the minimum halo mass which
produces any significant accretion we set, e.g., $\delta_{\rm b}=5$,
and get a mass which is lower than $M_{\rm min}$ by a factor of 27.

Of course, once the first stars and quasars form they heat the
surrounding IGM by either outflows or radiation. As a result, the
Jeans mass which is relevant for the formation of new objects changes
\cite{G097,g00}). The most dramatic change
occurs when the IGM is photo-ionized and is consequently heated to a
temperature of $\sim(1$--$2)\times 10^4$ K. 

\section{\bf Fragmentation of the First Gaseous Objects to Stars}
\label{sec4}

\subsection{Star Formation}
\label{sec4.1}

As mentioned in the preface, the fragmentation of the first gaseous
objects is a well-posed physics problem with well specified initial
conditions, for a given power-spectrum of primordial density
fluctuations.
This problem is ideally suited for three-dimensional computer
simulations, since it cannot be reliably addressed in idealized 1D or
2D geometries.

Recently, two groups have attempted detailed 3D simulations of the
formation process of the first stars in a halo of $\sim 10^6 M_\odot$ by
following the dynamics of both the dark matter and the gas components,
including H$_2$ chemistry and cooling. Bromm, Coppi, \& Larson (1999)
\cite{BCL99} have used a Smooth Particle Hydrodynamics (SPH) code to
simulate the collapse of a top-hat overdensity with a prescribed solid-body
rotation (corresponding to a spin parameter $\lambda=5\%$) and additional
small perturbations with $P(k)\propto k^{-3}$ added to the top-hat
profile. Abel et al.\ (2002) \cite{ABN02} isolated a high-density filament
out of a larger simulated cosmological volume and followed the evolution of
its density maximum with exceedingly high resolution using an Adaptive Mesh
Refinement (AMR) algorithm.

\begin{figure} 
\centering
\includegraphics[height=6cm]{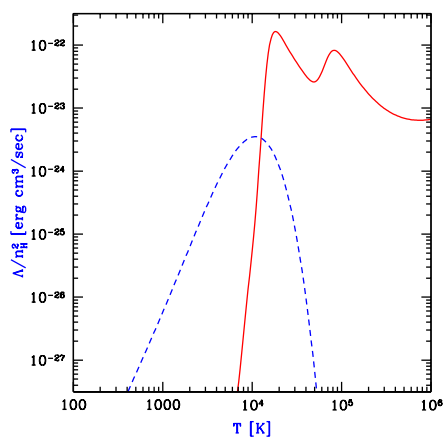}
\caption{Cooling rates as a function of temperature for a primordial
gas composed of atomic hydrogen and helium, as well as molecular
hydrogen, in the absence of any external radiation. We assume a
hydrogen number density $n_H=0.045\ {\rm cm}^{-3}$, corresponding to
the mean density of virialized halos at $z=10$. The plotted quantity
$\Lambda/n_H^2$ is roughly independent of density (unless $n_H \ga 10\
{\rm cm}^{-3}$), where $\Lambda$ is the volume cooling rate (in
erg/sec/cm$^3$). The solid line shows the cooling curve for an atomic
gas, with the characteristic peaks due to collisional excitation of
\ion{H}{1} and \ion{He}{2}. The dashed line shows the additional
contribution of molecular cooling, assuming a molecular abundance
equal to $1\%$ of $n_H$.}
\label{cooling}
\end{figure}

\begin{figure} 
\centering
\includegraphics[height=6cm]{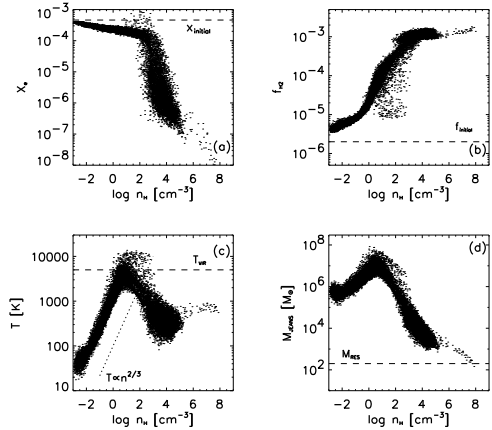}
\caption{Numerical results from Bromm et al.\ (1999) \cite{BCL99}, showing gas
properties at $z=31.2$ for a collapsing slightly inhomogeneous top-hat
region with a prescribed solid-body rotation.  {\bf (a)} Free electron
fraction (by number) vs.\ hydrogen number density (in cm$^{-3}$). At
densities exceeding $n\sim 10^{3}$ cm$^{-3}$, recombination is very
efficient, and the gas becomes almost completely neutral.\ {\bf (b)}
Molecular hydrogen fraction vs.\ number density. After a quick initial
rise, the H$_{2}$ fraction approaches the asymptotic value of $f\sim
10^{-3}$, due to the H$^{-}$ channel.  {\bf (c)} Gas temperature vs.\ 
number density. At densities below $\sim 1$ cm$^ {-3}$, the gas
temperature rises because of adiabatic compression until it reaches
the virial value of $T_{vir}\simeq 5000$ K.  At higher densities,
cooling due to H$_{2}$ drives the temperature down again, until the
gas settles into a quasi-hydrostatic state at $T\sim 500$ K and $n\sim
10^{4}$ cm$^{-3}$.  Upon further compression due to accretion and the
onset of gravitational collapse, the gas shows a further modest rise
in temperature. {\bf (d)} Jeans mass (in $M_{\odot}$) vs.\ number
density. The Jeans mass reaches a value of $M_{J}\sim 10^{3}M_{\odot}$
for the quasi-hydrostatic gas in the center of the potential well, and
reaches the resolution limit of the simulation, $M_{\rm res}\simeq 200
M_{\odot}$, for densities close to $n=10^{8}$ cm$^{-3}$.  }
\label{fig4a}
\end{figure}

\begin{figure} 
\centering
\includegraphics[height=6cm]{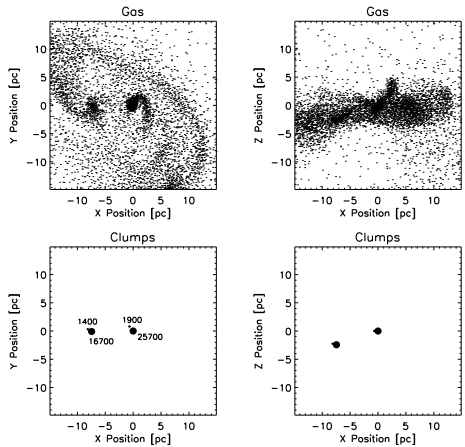}
\caption{Gas and clump morphology at $z=28.9$ in
the simulation of Bromm et al.\ (1999) \cite{BCL99}.
{\it Top row:} The remaining gas in the diffuse phase.
{\it Bottom row:} Distribution of clumps. The numbers next to the dots
denote clump mass in units of $M_{\odot}$. 
{\it Left panels:} Face-on view.
{\it Right panels:} Edge-on view.
The length of the box is 30 pc.
The gas has settled into a flattened
configuration with two dominant clumps of mass close to 
$20,000 M_{\odot}$. During the subsequent evolution, the clumps
survive
without merging, and grow in mass only slightly by accretion of
surrounding
gas. }
\label{fig4b}
\end{figure}

The generic results of Bromm et al.\ (1999 \cite{BCL99}; see also Bromm 2000 \cite{Bro00}) are
illustrated in Figure \ref{fig4a}. The collapsing region forms a disk which
fragments into many clumps. The clumps have a typical mass $\sim
10^2$--$10^3M_\odot$. This mass scale corresponds to the Jeans mass for a
temperature of $\sim 500$K and the density $\sim 10^4~{\rm cm^{-3}}$ where
the gas lingers because its cooling time is longer than its collapse time
at that point (see Fig. \ref{fig4b}). Each clump accretes mass slowly
until it exceeds the Jeans mass and collapses at a roughly constant
temperature (isothermally) due to H$_2$ cooling that brings the gas to a
fixed temperature floor.  The clump formation efficiency is high in this
simulation due to the synchronized collapse of the overall top-hat
perturbation.

\noindent
 
Bromm (2000) \cite{Bro00} has simulated the collapse of one of the
above-mentioned clumps with $\sim 1000 M_\odot$ and demonstrated that it
does not tend to fragment into sub-components. Rather, the clump core of
$\sim 100M_\odot$ free-falls towards the center leaving an extended
envelope behind with a roughly isothermal density profile.  At very high
gas densities, three-body reactions become important in the chemistry of
H$_2$.  Omukai \& Nishi (1998) \cite{ON1998} have included these reactions
as well as radiative transfer and followed the collapse in spherical
symmetry up to stellar densities. Radiation pressure from nuclear burning
at the center is unlikely to reverse the infall as the stellar mass builds
up.  These calculations indicate that each clump may end as a single
massive star; however, it is conceivable that angular momentum may
eventually halt the collapsing cloud and lead to the formation of a binary
stellar system instead.

The Jeans mass, which is defined based on small fluctuations in a
background of {\it uniform}\/ density, does not strictly apply in the
context of collapsing gas cores. We can instead use a slightly modified
critical mass known as the Bonnor-Ebert mass \cite{Bo56,Eb55}. For baryons
in a background of uniform density $\rho_b$, perturbations are unstable to
gravitational collapse in a region more massive than the Jeans
mass. Instead of a uniform background, we consider a spherical,
non-singular, isothermal, self-gravitating gas in hydrostatic equilibrium,
i.e., a centrally-concentrated object which more closely resembles the gas
cores found in the above-mentioned simulations. In this case, small
fluctuations are unstable and lead to collapse if the sphere is more
massive than the Bonnor-Ebert mass $M_{\rm BE}$, given by the same
expression the Jeans Mass but with a different coefficient (1.2
instead of 2.9) and with $\rho_b$ denoting in this case the gas (volume)
density at the surface of the sphere,
 \beq M_{\rm BE}=1.2\,
\frac{1}{\sqrt{\rho_b}}\,\left( \frac{k T} {G \mu m_p} \right)^{3/2}\
. \label{MJb} \eeq

\begin{figure} 
\centering
\includegraphics[height=6cm]{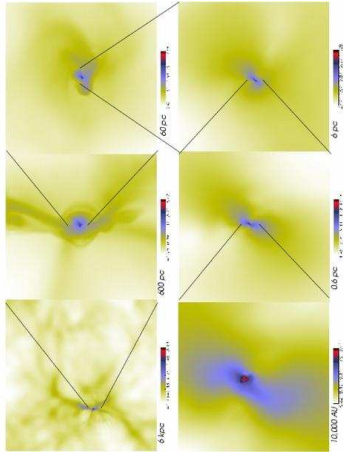}
\caption{Zooming in on the core of a star forming region with the {\it
Adaptive Mesh Refinement} simulation of Abel et al.\ (2000)
\cite{Abel}. The panels show different length scales, decreasing clockwise
by an order of magnitude between adjacent panels. Note the large dynamic
range of scales which are being resolved, from 6 kpc (top left panel) down
to 10,000 AU (bottom left panel).}
\label{fig4c}
\end{figure}

\begin{figure} 
\centering
\includegraphics[height=6cm]{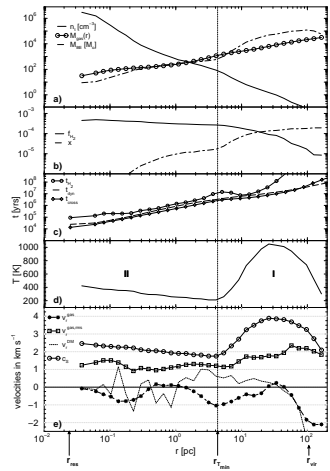}
\caption{Gas profiles from the simulation of Abel et al.\
(2000)\cite{Abel}.  The cell size on the finest grid corresponds to $0.024$
pc, while the simulation box size corresponds to 6.4 kpc.  Shown are
spherically-averaged mass-weighted profiles around the baryon density peak
shortly before a well defined fragment forms ($z=19.1$). Panel (a) shows
the baryonic number density, enclosed gas mass in solar mass, and the local
Bonnor-Ebert mass $M_{\rm BE}$ (see text).
Panel (b) plots the molecular hydrogen fraction (by number) $f_{H_2}$
and the free electron fraction $x$.  The H$_2$ cooling time,
$t_{H_2}$, the time it takes a sound wave to travel to the center,
$t_{\rm cross}$, and the free--fall time $t_{\rm ff}=[3\pi/(32G\rho
)]^{1/2}$ are given in panel (c). Panel (d) gives the temperature in K
as a function of radius.  The bottom panel gives the local sound
speed, $c_s$ (solid line with circles), the rms radial velocities of
the dark matter (dashed line) and the gas (dashed line with asterisks)
as well as the rms gas velocity (solid line with square symbols). The
vertical dotted line indicates the radius ($\sim 5$ pc) at which the
gas has reached its minimum temperature allowed by H$_2$ cooling. The
virial radius of the $5.6\times 10^{6}M_\odot$ halo is 106 pc.  }
\label{fig4d}
\end{figure}

In their simulation, Abel et al.\ (2000)\cite{Abel} adopted the actual
cosmological density perturbations as initial conditions. The simulation
focused on the density peak of a filament within the IGM, and evolved it to
very high densities (Fig. \ref{fig4c}). Following the initial collapse of
the filament, a clump core formed with $\sim 200M_\odot$, amounting to only
$\sim 1\%$ of the virialized mass.  Subsequently due to slow cooling, the
clump collapsed subsonically in a state close to hydrostatic equilibrium
(see Fig. \ref{fig4d}).  Unlike the idealized top-hat simulation of Bromm
et al.\ (2001) \cite{BKL2001}, the collapse of the different clumps within
the filament is not synchronized.  Once the first star forms at the center
of the first collapsing clump, it is likely to affect the formation of
other stars in its vicinity.

As soon as nuclear burning sets in the core of the proto-star, the
radiation emitted by the star starts to affect the infall of the
surrounding gas towards it.  The radiative feedback involves
photo-dissociation of H$_2$, Ly$\alpha$ radiation pressure, and
photo-evaporation of the accretion disk. Tan \& McKee \cite{Tan2003}
studied these effects by extrapolating analytically the infall of gas from
the final snapshot of the above resolution-limited simulations to the scale
of a proto-star; they concluded that nuclear burning (and hence the
feedback) starts when the proton-star accretes $\sim 30M_\odot$ and
accretion is likely to be terminated when the star reaches $\sim
200M_\odot$.

\noindent
  
\noindent
  
If the clumps in the above simulations end up forming individual very
massive stars, then these stars will likely radiate copious amounts of
ionizing radiation \cite{CBA84,TS00,BKL2001} and expel strong
winds.  Hence, the stars will have a large effect on their interstellar
environment, and feedback is likely to control the overall star formation
efficiency.  This efficiency is likely to be small in galactic potential
wells which have a virial temperature lower than the temperature of
photoionized gas, $\sim 10^4$K. In such potential wells, the gas may go
through only a single generation of star formation, leading to a
``suicidal'' population of massive stars.

The final state in the evolution of these stars is uncertain; but if their
mass loss is not too extensive, then they are likely to end up as black
holes \cite{CBA84,FWH00}. The remnants may provide the seeds of quasar black
holes \cite{Lar99}.  Some of the massive stars may end their lives by
producing gamma-ray bursts. If so then the broad-band afterglows of these
bursts could provide a powerful tool for probing the epoch of reionization
\cite{LR00,CL00}).
There is no better way to end the dark ages than with $\gamma$-ray burst
fireworks.

{\it Where are the first stars or their remnants located today?} The very
first stars formed in rare high-$\sigma$ peaks and hence are likely to
populate the cores of present-day galaxies \cite{WS99}. However, the bulk
of the stars which formed in low-mass systems at later times are expected
to behave similarly to the collisionless dark matter particles and populate
galaxy halos \cite{Loe98}.

\subsection{The Mass Function of Stars}

Currently, we do not have direct observational constraints on how
the first stars, the so-called Population~III stars, formed at
the end of the cosmic dark ages. It is, therefore, instructive to
briefly summarize what we have learned about star formation in the
present-day Universe, where theoretical reasoning is guided by a
wealth of observational data (see \cite{Pud2002} for a recent review).

\begin{figure}
  \includegraphics[height=.3\textheight]{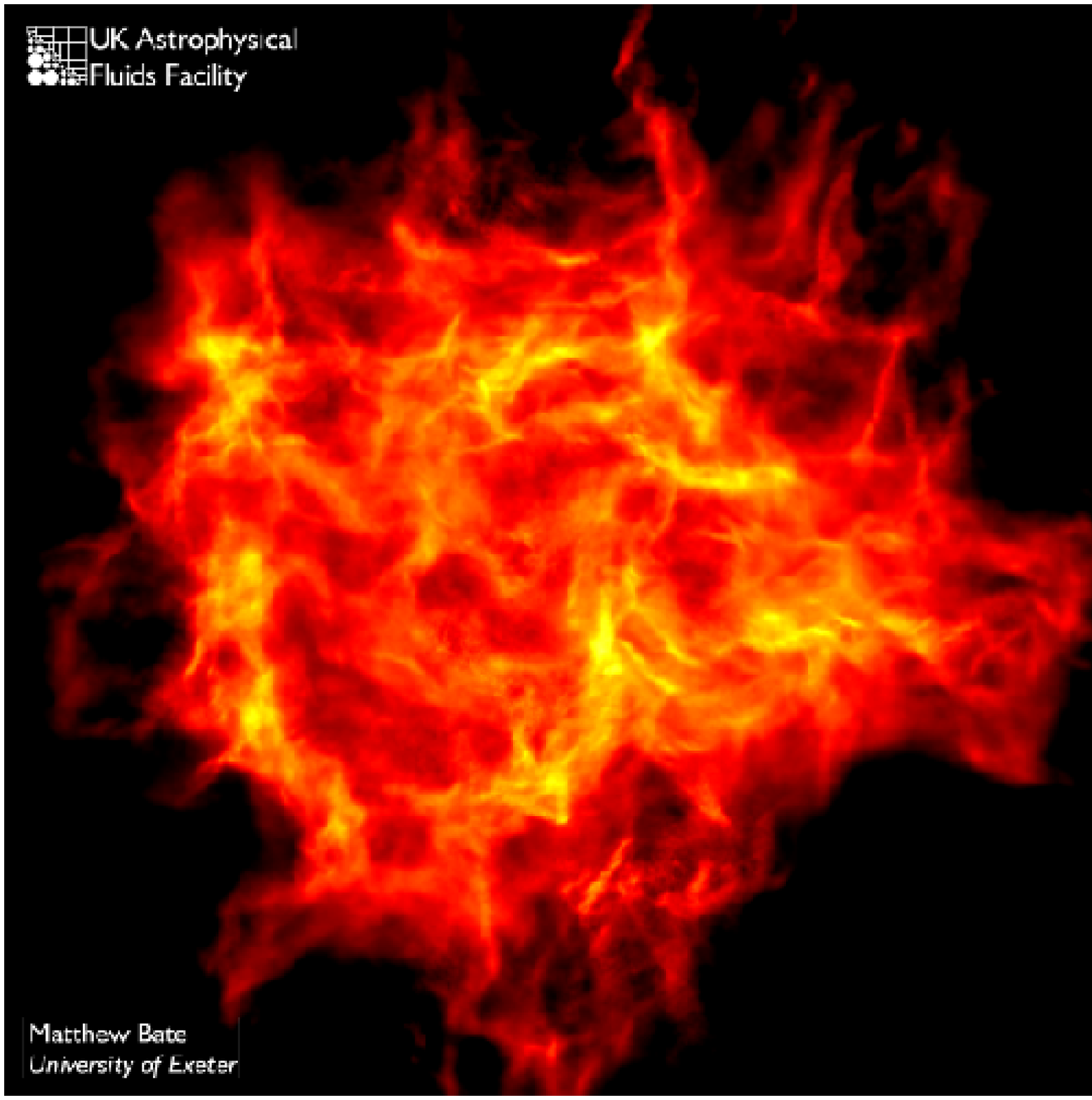}
  \includegraphics[height=.3\textheight]{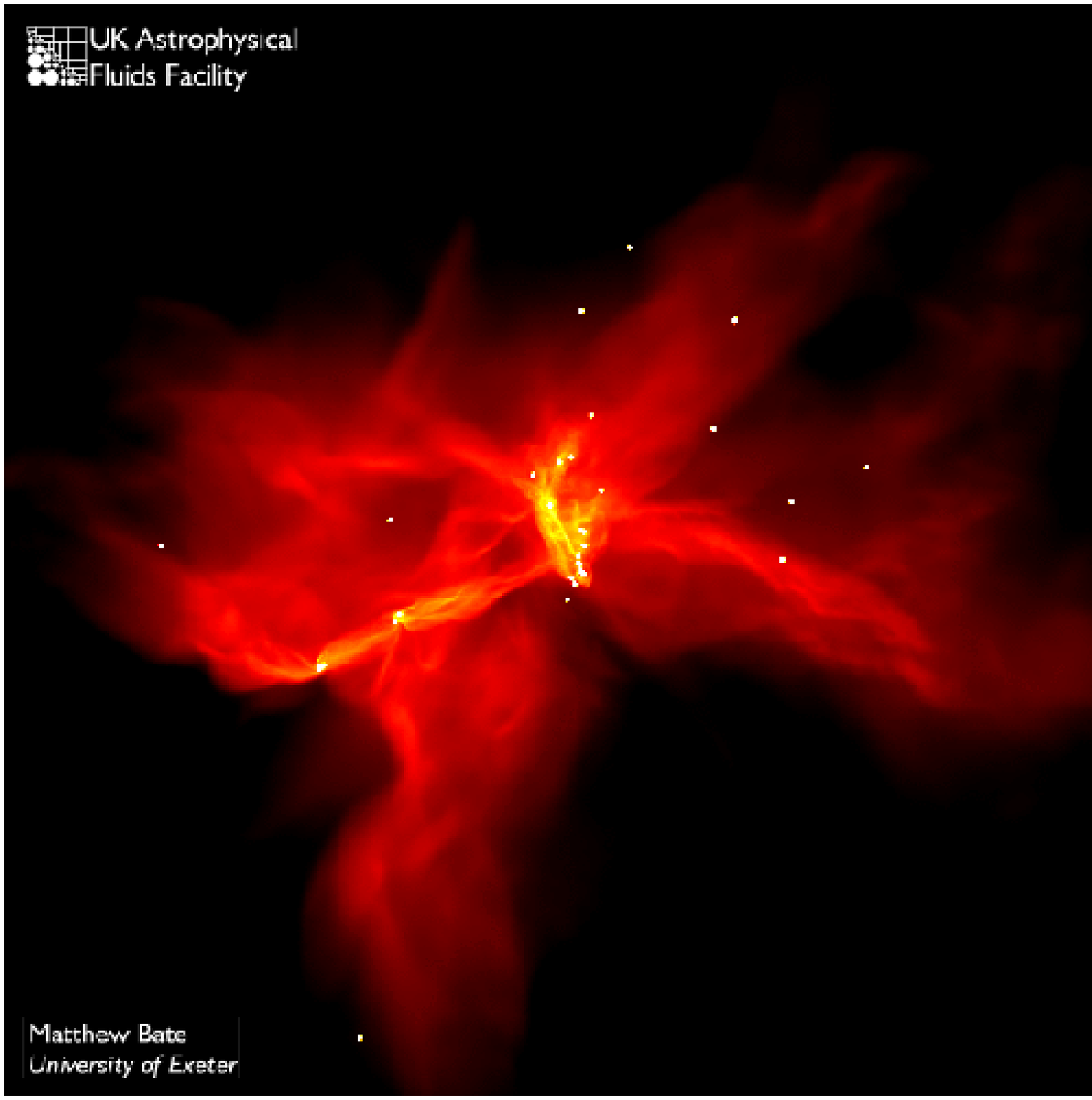}
\caption{A hydrodynamic simulation of the collapse and fragmentation
of a turbulent molecular cloud in the present-day Universe (from
\cite{BBB2003}).  The cloud has a mass of $50 M_{\odot}$.  The panels
show the column density through the cloud, and span a scale of 0.4 pc
across.  {\it Left:} The initial phase of the collapse. The turbulence
organizes the gas into a network of filaments, and decays thereafter
through shocks.  {\it Right:} A snapshot taken near the end of the
simulation, after 1.4 initial free-fall times of $2\times 10^{5}$yr.
Fragmentation has resulted in $\sim 50$ stars and brown dwarfs.
The star formation efficiency is $\sim 10$\% on the scale of the
overall cloud, but can be much larger in the dense sub-condensations.
This result is in good agreement with what is observed in local star-forming
regions.}
\label{firstfig}
\end{figure}

Population~I stars form out of cold, dense molecular gas that is
structured in a complex, highly inhomogeneous way. The molecular
clouds are supported against gravity by turbulent velocity fields and
pervaded on large scales by magnetic fields.  Stars tend to form in
clusters, ranging from a few hundred up to $\sim 10^{6}$ stars. It
appears likely that the clustered nature of star formation leads to
complicated dynamics and tidal interactions that transport angular
momentum, thus allowing the collapsing gas to overcome the classical
centrifugal barrier \cite{Lar2002}.  The initial 
mass function (IMF) of Pop~I stars is
observed to have the approximate Salpeter form (e.g., \cite{Kr02})
\begin{equation}
\frac{{\rm d}N}{{\rm d log}M}\propto M^{x} \mbox{\ ,}
\end{equation}
where
\begin{equation}
x\simeq \left\{
\begin{array}{rl}
-1.35 & \mbox{for \ }M\ge 0.5 M_{\odot}\\
0.0 & \mbox{for \ }0.007 \le M\le 0.5 M_{\odot}\\
\end{array}
\right. \mbox{\ .}
\end{equation}
The lower cutoff in mass corresponds roughly to the opacity limit for
fragmentation. This limit reflects the minimum fragment mass, set when the
rate at which gravitational energy is released during the collapse exceeds
the rate at which the gas can cool (e.g., \cite{MJR1976}).  The most
important feature of the observed IMF is that $\sim 1 M_{\odot}$ is the
characteristic mass scale of Pop~I star formation, in the sense that most
of the mass goes into stars with masses close to this value. In Figure
\ref{firstfig}, we show the result from a recent hydrodynamical simulation of
the collapse and fragmentation of a molecular cloud core
\cite{BBB2002,BBB2003}.  This simulation illustrates the highly dynamic and
chaotic nature of the star formation process\footnote{See http://
www.ukaff.ac.uk/starcluster for an animation.}.

The metal-rich chemistry, magnetohydrodynamics, and radiative
transfer involved in present-day star formation is complex, and we
still lack a comprehensive theoretical framework that predicts the IMF
from first principles. Star formation in the high redshift Universe,
on the other hand, poses a theoretically more tractable problem due to
a number of simplifying features, such as: (i) the initial absence of
heavy metals and therefore of dust; and (ii) the absence of
dynamically-significant magnetic fields, in the pristine gas left over
from the big bang. The cooling of the primordial gas does then only
depend on hydrogen in its atomic and molecular form.  Whereas in the
present-day interstellar medium, the initial state of the star forming
cloud is poorly constrained, the corresponding initial conditions for
primordial star formation are simple, given by the popular
$\Lambda$CDM model of cosmological structure formation. We now turn to
a discussion of this theoretically attractive and important problem.

{\it How did the first stars form?} A complete answer to this question
would entail a theoretical prediction for the Population~III IMF, which is
rather challenging. Let us start by addressing the simpler problem of
estimating the characteristic mass scale of the first stars. As mentioned
before, this mass scale is observed to be $\sim 1 M_{\odot}$ in the
present-day Universe.  

Bromm \& Loeb (2004) \cite{BL04} carried out idealized simulations of the
protostellar accretion problem and estimated the final mass of a
Population~III star.  Using the smoothed particle hydrodynamics (SPH)
method, they included the chemistry and cooling physics relevant for the
evolution of metal-free gas (see \cite{BCL02} for details). Improving on
earlier work \cite{BCL99,BCL02} by initializing the simulations according
to the $\Lambda$CDM model, they focused on an isolated overdense region
that corresponds to a 3$\sigma-$peak \cite{BL04}: a halo containing a total
mass of $10^{6}M_{\odot}$, and collapsing at a redshift $z_{\rm vir}\simeq
20$. In these runs, one high-density clump has formed at the center of the
minihalo, possessing a gas mass of a few hundred solar masses.  Soon after
its formation, the clump becomes gravitationally unstable and undergoes
runaway collapse. Once the gas clump has exceeded a threshold density of
$10^{7}$ cm$^{-3}$, it is replaced by a sink particle which is a
collisionless point-like particle that is inserted into the simulation.
This choice for the density threshold ensures that the local Jeans mass is
resolved throughout the simulation.  The clump (i.e., sink particle) has an
initial mass of $M_{\rm Cl}\simeq 200M_{\odot}$, and grows subsequently by
ongoing accretion of surrounding gas.  High-density clumps with such masses
result from the chemistry and cooling rate of molecular hydrogen, H$_{2}$,
which imprint characteristic values of temperature, $T\sim 200$~K, and
density, $n\sim 10^{4}$ cm$^{-3}$, into the metal-free gas \cite{BCL02}.
Evaluating the Jeans mass for these characteristic values results in
$M_{J}\sim \mbox{\ a few \ }\times 10^{2}M_{\odot}$, which is close to the
initial clump masses found in the simulations.

\begin{figure}
  \includegraphics[height=.3\textheight]{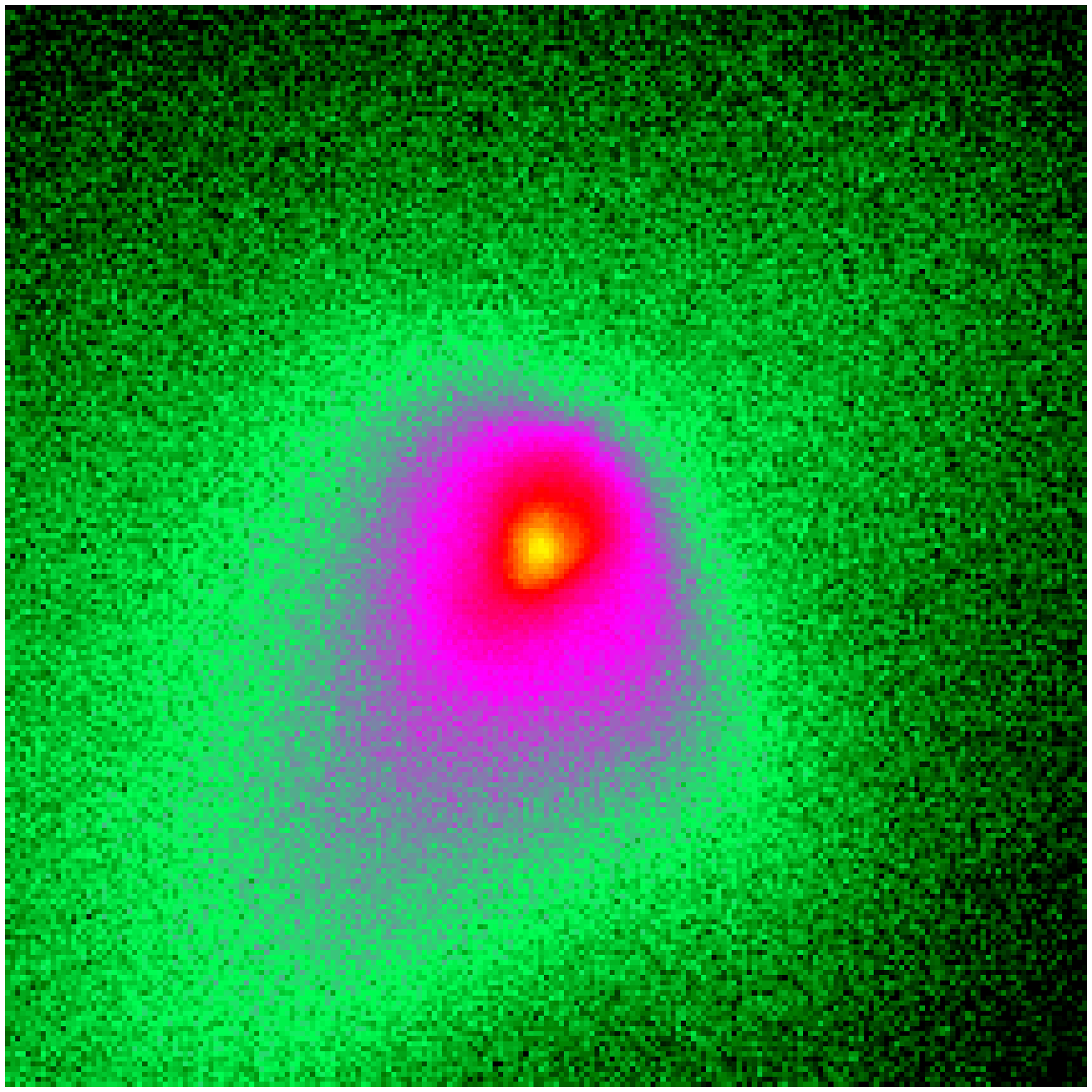}
  \includegraphics[height=.3\textheight]{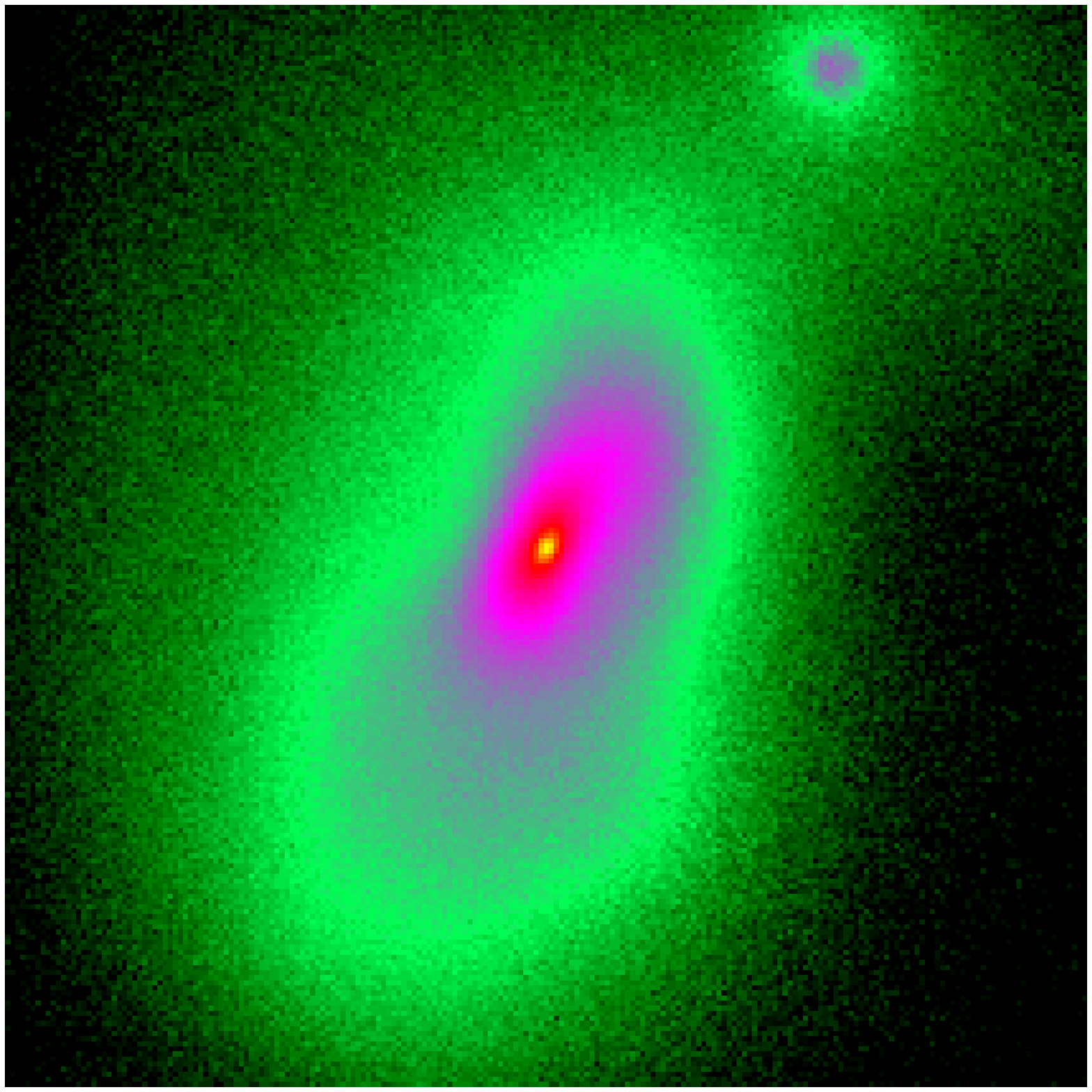}
\caption{Collapse and fragmentation of a primordial cloud (from
\cite{BL04}).  Shown is the projected gas density at a redshift $z\simeq
21.5$, briefly after gravitational runaway collapse has commenced in the
center of the cloud.  {\it Left:} The coarse-grained morphology in a box
with linear physical size of 23.5~pc.  At this time in the unrefined
simulation, a high-density clump (sink particle) has formed with an initial
mass of $\sim 10^{3}M_{\odot}$.  {\it Right:} The refined morphology in
a box with linear physical size of 0.5~pc.  The central density peak,
vigorously gaining mass by accretion, is accompanied by a secondary clump.}
\label{2ab}
\end{figure}

\begin{figure}[ht]
  \includegraphics[height=.3\textheight]{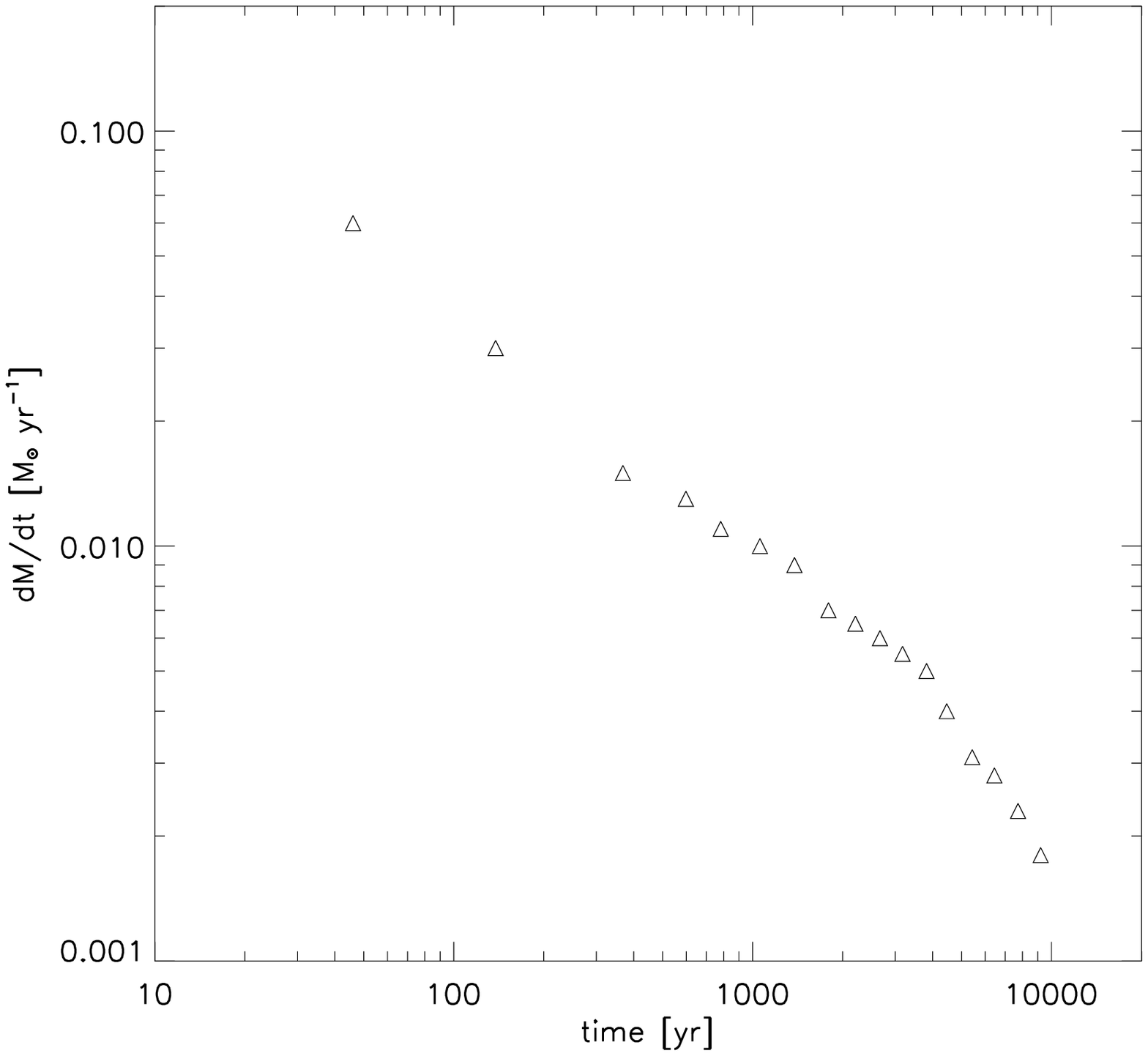}
  \includegraphics[height=.3\textheight]{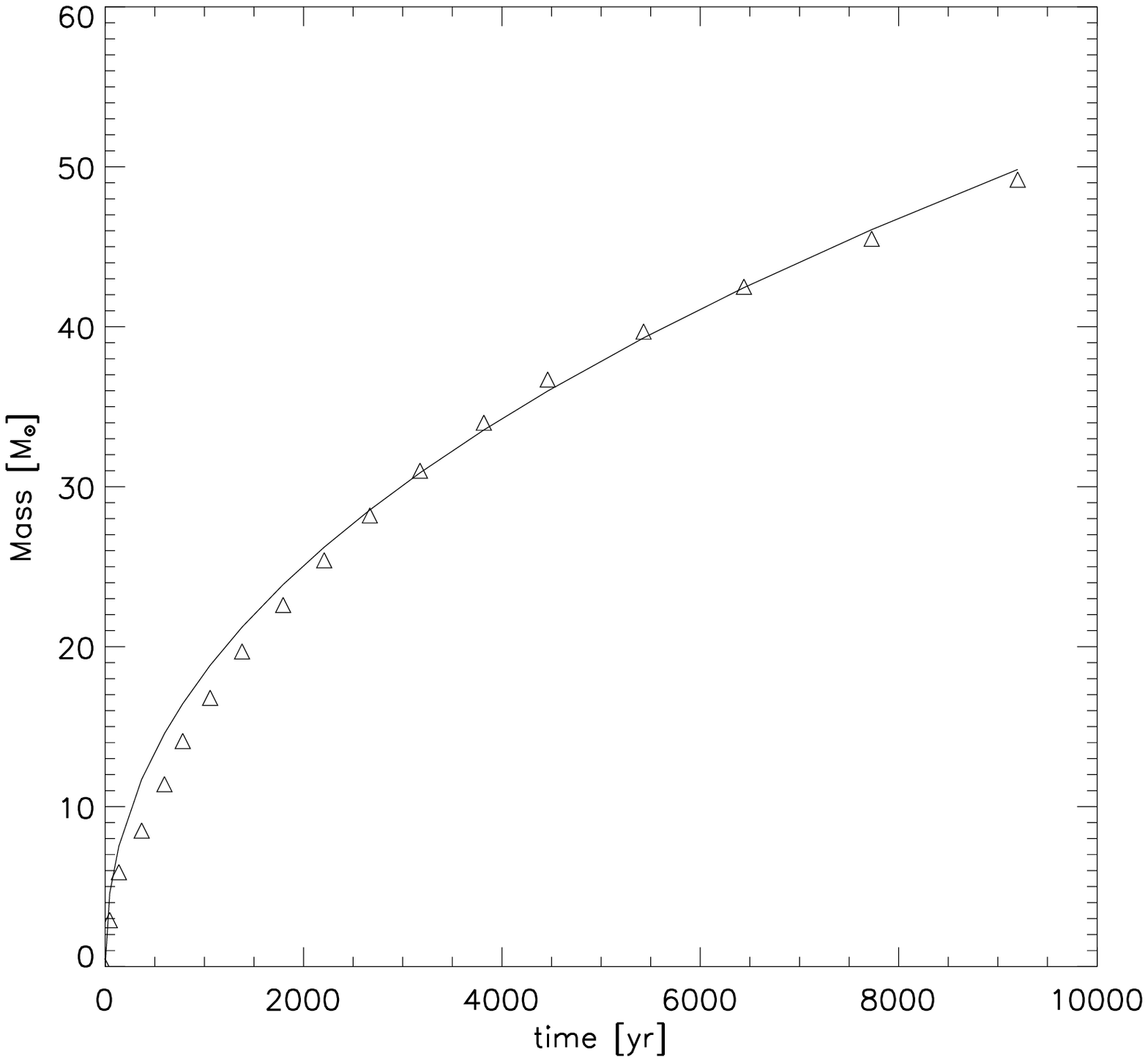}
\caption{Accretion onto a primordial protostar (from \cite{BL04}).  The
morphology of this accretion flow is shown in Fig.~\ref{2ab}.  {\it Left:}
Accretion rate (in $M_{\odot}$~yr$^{-1}$) vs. time (in yr) since molecular
core formation.  {\it Right:} Mass of the central core (in $M_{\odot}$)
vs. time.  {\it Solid line:} Accretion history approximated as:
$M_{\ast}\propto t^{0.45}$.  Using this analytical approximation, we
extrapolate that the protostellar mass has grown to $\sim 150 M_{\odot}$
after $\sim 10^{5}$~yr, and to $\sim 700 M_{\odot}$ after $\sim 3\times
10^{6}$~yr, the total lifetime of a very massive star.  }
\label{fig3ab}
\end{figure}

The high-density clumps are clearly not stars yet. To probe the subsequent
fate of a clump, Bromm \& Loeb (2004) \cite{BL04} have re-simulated the
evolution of the central clump with sufficient resolution to follow the
collapse to higher densities. Figure~\ref{2ab} ({\it right panel})
shows the gas density on a scale of 0.5~pc, which is two orders of magnitude
smaller than before. Several features are evident in this plot. First, the
central clump does not undergo further sub-fragmentation, and is likely to
form a single Population~III star. Second, a companion clump is visible at
a distance of $\sim 0.25$~pc. If negative feedback from the first-forming
star is ignored, this companion clump would undergo runaway collapse on its
own approximately $\sim 3$~Myr later.  This timescale is comparable to the
lifetime of a very massive star (VMS)\cite{BKL2001}.  If the second clump
was able to survive the intense radiative heating from its neighbor, it
could become a star before the first one explodes as a supernova
(SN). Whether more than one star can form in a low-mass halo thus crucially
depends on the degree of synchronization of clump formation.  Finally, the
non-axisymmetric disturbance induced by the companion clump, as well as the
angular momentum stored in the orbital motion of the binary system, allow
the system to overcome the angular momentum barrier for the collapse of the
central clump (see \cite{Lar2002}).

The recent discovery of stars like HE0107-5240 with a mass of $0.8
M_{\odot}$ and an iron abundance of ${\rm [Fe/H]} = -5.3$ \cite{Cr02} shows
that at least some low mass stars could have formed out of extremely
low-metallicity gas.  The above simulations show that although the majority
of clumps are very massive, a few of them, like the secondary clump in
Fig.~\ref{2ab}, are significantly less massive. Alternatively, low-mass
fragments could form in the dense, shock-compressed shells that surround
the first hypernovae \cite{MBH03}.

{\it How massive were the first stars?} Star formation typically
proceeds from the `inside-out', through the accretion of gas onto a
central hydrostatic core.  Whereas the initial mass of the hydrostatic
core is very similar for primordial and present-day star formation
\cite{ON1998}, the accretion process -- ultimately responsible for
setting the final stellar mass, is expected to be rather different. On
dimensional grounds, the accretion rate is simply related to the sound
speed cubed over Newton's constant (or equivalently given by the ratio
of the Jeans mass and the free-fall time): $\dot{M}_{\rm acc}\sim
c_s^3/G \propto T^{3/2}$. A simple comparison of the temperatures in
present-day star forming regions ($T\sim 10$~K) with those in
primordial ones ($T\sim 200-300$~K) already indicates a difference in
the accretion rate of more than two orders of magnitude.

The above refined simulation enables one to study the three-dimensional
accretion flow around the protostar (see also
\cite{OP2001,Rip2002,Tan2003}).  The gas may now reach densities of
$10^{12}$ cm$^{-3}$ before being incorporated into a central sink
particle. At these high densities, three-body reactions \cite{PSS1983} 
convert the gas into a fully molecular form.  Figure~\ref{fig3ab}
shows how the molecular core grows in mass over the first $\sim 10^{4}$~yr
after its formation. The accretion rate ({\it left panel}) is initially
very high, $\dot{M}_{\rm acc}\sim 0.1 M_{\odot}$~yr$^{-1}$, and
subsequently declines according to a power law, with a possible break at
$\sim 5000$~yr. The mass of the molecular core ({\it right panel}), taken
as an estimator of the proto-stellar mass, grows approximately as:
$M_{\ast}\sim \int \dot{M}_{\rm acc}{\rm d}t \propto t^{0.45}$. A rough
upper limit for the final mass of the star is then: $M_{\ast}(t=3\times
10^{6}{\rm yr})\sim 700 M_{\odot}$. In deriving this upper bound, we have
conservatively assumed that accretion cannot go on for longer than the
total lifetime of a massive star.

{\it Can a Population~III star ever reach this asymptotic mass limit?}  The
answer to this question is not yet known with any certainty, and it depends
on whether the accretion from a dust-free envelope is eventually terminated
by feedback from the star (e.g., \cite{OP2001,Rip2002,Tan2003,OI2002}). The
standard mechanism by which accretion may be terminated in metal-rich gas,
namely radiation pressure on dust grains \cite{WC1987}, is evidently not
effective for gas with a primordial composition. Recently, it has been
speculated that accretion could instead be turned off through the formation
of an H~II region \cite{OI2002}, or through the photo-evaporation of the
accretion disk \cite{Tan2003}. The termination of the accretion process
defines the current unsolved frontier in studies of Population~III star
formation. Current simulations indicate that the first stars were
predominantly very massive ($\ga 30M_\odot$), and consequently rather
different from present-day stellar populations. The crucial question then
arises: {\it How and when did the transition take place from the early
formation of massive stars to that of low-mass stars at later times?}  We
address this problem next.

The very first stars, marking the cosmic Renaissance of structure
formation, formed under conditions that were much simpler than the
highly complex environment in present-day molecular clouds.
Subsequently, however, the situation rapidly became more complicated
again due to the feedback from the first stars on the IGM.  Supernova
explosions dispersed the nucleosynthetic products from the first
generation of stars into the surrounding gas (e.g., \cite{MFR01,MFM02,
TSD02}), including also dust grains produced in the explosion itself
\cite{LH97,TodF01}.  Atomic and molecular cooling became much more
efficient after the addition of these metals. Moreover, the
presence of ionizing cosmic rays, as well as of UV and X-ray
background photons, modified the thermal and chemical behavior of
the gas in important ways (e.g., \cite{MBA01,MBA03}).

Early metal enrichment was likely the dominant effect that brought
about the transition from Population~III to Population~II star
formation.  Recent numerical simulations of collapsing primordial
objects with overall masses of $\sim 10^{6}M_{\odot}$, have shown that
the gas has to be enriched with heavy elements to a minimum level of
$Z_{\rm crit}\simeq 10^{-3.5}Z_{\odot}$, in order to have any effect
on the dynamics and fragmentation properties of the system
\cite{Om00,BFCL01,BrL03}.  Normal, low-mass (Population~II) stars are
hypothesized to only form out of gas with metallicity $Z\ge Z_{\rm
crit}$.  Thus, the characteristic mass scale for star formation is
expected to be a function of metallicity, with a discontinuity at
$Z_{\rm crit}$ where the mass scale changes by $\sim$ two orders of
magnitude. The redshift where this transition occurs has important
implications for the early growth of cosmic structure, and the
resulting observational signature (e.g.,
\cite{Wyithe03,FL03,MBH03,Sch02}) 
include the extended 
nature of reionization \cite{FL05}.


For additional detailes about the properties of the first stars, see the
comprehensive review by Bromm \& Larson (2004) \cite{Bromm}.

\subsection{Gamma-ray Bursts: Probing the First Stars One 
Star at a Time}

Gamma-Ray Bursts (GRBs) are believed to originate in compact remnants
(neutron stars or black holes) of massive stars. Their high luminosities
make them detectable out to the edge of the visible Universe
\cite{CL00,LR00}.  GRBs offer the opportunity to detect the most distant
(and hence earliest) population of massive stars, the so-called
Population~III (or Pop~III), one star at a time.  In the hierarchical
assembly process of halos which are dominated by cold dark matter (CDM),
the first galaxies should have had lower masses (and lower stellar
luminosities) than their low-redshift counterparts.  Consequently, the
characteristic luminosity of galaxies or quasars is expected to decline
with increasing redshift. GRB afterglows, which already produce a peak flux
comparable to that of quasars or starburst galaxies at $z\sim 1-2$, are
therefore expected to outshine any competing source at the highest
redshifts, when the first dwarf galaxies have formed in the Universe.

\begin{figure}
\centering
\includegraphics[height=6cm]{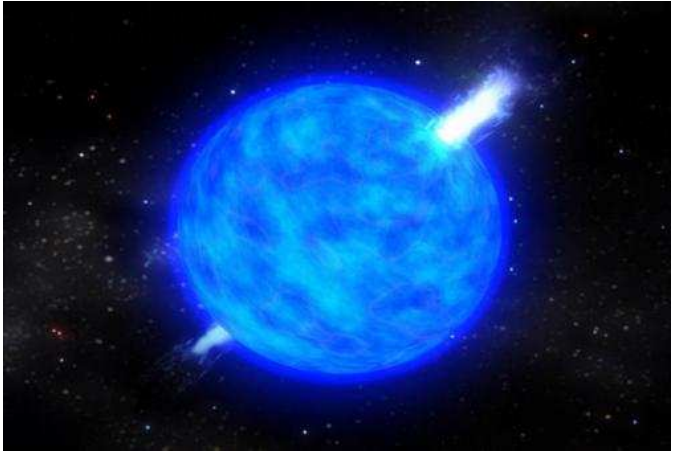}
\caption{Illustration of a long-duration gamma-ray burst in the popular
``collapsar'' model.
The collapse of the core of a massive star (which lost
its hydrogen envelope) to a black hole generates two opposite jets moving
out at a speed close to the speed of light. The jets drill a hole in the
star and shine brightly towards an observer who happened to be located
within with the collimation cones of the jets. The jets emenating from a
single massive star are so bright that they can be seen across the Universe
out to the epoch when the first stars have formed.  Upcoming observations
by the {\it Swift} satellite will have the sensitivity to reveal whether
the first stars served as progenitors of gamma-ray bursts
(for updates see http://swift.gsfc.nasa.gov/).}
\label{grb}
\end{figure}

The first-year polarization data from the {\it Wilkinson Microwave
Anisotropy Probe} ({\it WMAP}) indicates an optical depth to electron
scattering of $\sim 17\pm 4$\% after cosmological recombination
\cite{Kog03,WMAP}.  This implies that the first stars must have formed at
a redshift $z\sim $10--20, and reionized a substantial fraction of the
intergalactic hydrogen around that time \cite{Cen03,CFW03,SL03,WL03,YBH04}.
Early reionization can be achieved with plausible star formation parameters
in the standard $\Lambda$CDM cosmology; in fact, the required optical depth
can be achieved in a variety of very different ionization histories since
{\it WMAP} places only an integral constraint on these histories
\cite{HH03}. One would like to probe the full history of reionization in
order to disentangle the properties and formation history of the stars that
are responsible for it. GRB afterglows offer the opportunity to detect
stars as well as to probe the metal
enrichment level \cite{FL03} of the intervening IGM.

GRBs, the electromagnetically-brightest explosions in the Universe, should
be detectable out to redshifts $z>10$ \cite{CL00,LR00}.
High-redshift GRBs can be identified through infrared
photometry, based on the Ly$\alpha$ break induced by absorption of their
spectrum at wavelengths below $1.216\, \mu {\rm m}\, [(1+z)/10]$. Follow-up
spectroscopy of high-redshift candidates can then be performed on a
10-meter-class telescope. Recently, the ongoing {\it Swift} mission
\cite{Geh04} has detected a GRB originating at $z\simeq 6.3$
(e.g., \cite{Hai05}), thus demonstrating the viability of
GRBs as probes of the early Universe.

There are four main advantages of GRBs relative to traditional cosmic
sources such as quasars:

\noindent {\it (i)} The GRB afterglow flux at a given observed time lag
after the $\gamma$-ray trigger is not expected to fade significantly with
increasing redshift, since higher redshifts translate to earlier times in
the source frame, during which the afterglow is intrinsically brighter
\cite{CL00}. For standard afterglow lightcurves and spectra, the
increase in the luminosity distance with redshift is compensated by this
{\it cosmological time-stretching} effect.

\begin{figure}
\centering
\includegraphics[height=10cm]{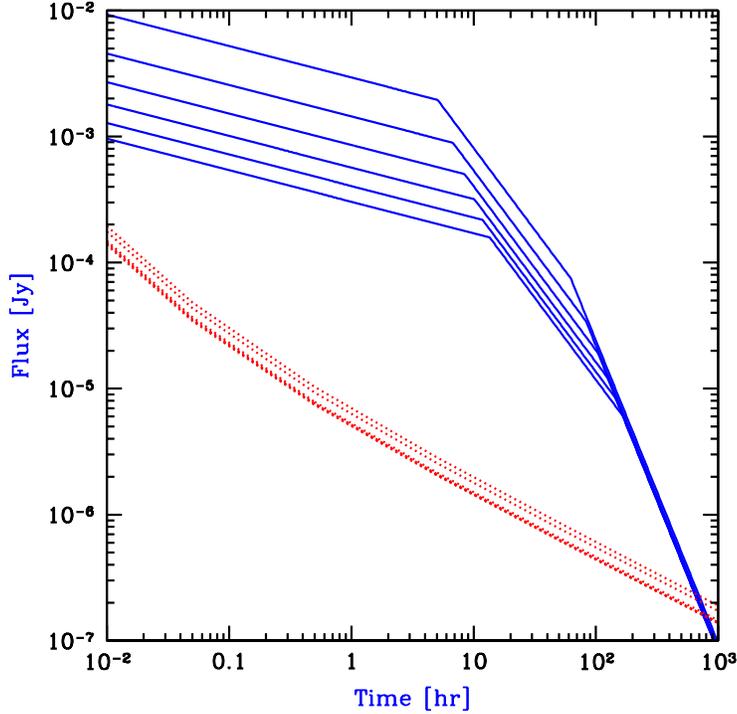}
\caption{GRB afterglow flux as a function of time since the $\gamma$-ray
trigger in the observer frame (taken from \cite{BL04}). The flux
(solid curves) is calculated at the redshifted Ly$\alpha$ wavelength. The
dotted curves show the planned detection threshold for the {\it James Webb
Space Telescope} ({\it JWST}), assuming a spectral resolution $R=5000$
with the near infrared spectrometer, a signal to noise ratio of 5 per
spectral resolution element, and an exposure time equal to $20\%$ of the
time since the GRB explosion (see  http://www.ngst.stsci.edu/nms/main/~).
Each set of curves shows a sequence of redshifts, namely $z=5$, 7, 9, 11,
13, and 15, respectively, from top to bottom.}
\label{fig1val}
\end{figure}

\noindent {\it (ii)} As already mentioned, in the standard $\Lambda$CDM
cosmology, galaxies form hierarchically, starting from small masses and
increasing their average mass with cosmic time. Hence, the characteristic
mass of quasar black holes and the total stellar mass of a galaxy were
smaller at higher redshifts, making these sources intrinsically fainter
\cite{WL02}.  However, GRBs are believed to originate from a stellar mass
progenitor and so the intrinsic luminosity of their engine should not
depend on the mass of their host galaxy. GRB afterglows are therefore
expected to outshine their host galaxies by a factor that gets larger with
increasing redshift.

\noindent {\it (iii)} Since the progenitors of GRBs are believed to be
stellar, they likely originate in the most common star-forming galaxies at
a given redshift rather than in the most massive host galaxies, as is the
case for bright quasars \cite{GRBquasar}. Low-mass host galaxies
induce only a weak ionization effect on the surrounding IGM and do not
greatly perturb the Hubble flow around them. Hence, the Ly$\alpha$ damping
wing should be closer to the idealized unperturbed IGM case
and its detailed spectral shape should be easier
to interpret. Note also that unlike the case of a quasar, a GRB afterglow
can itself ionize at most $\sim 4\times 10^4 E_{51} M_\odot$ of hydrogen if
its UV energy is $E_{51}$ in units of $10^{51}$ ergs (based on the
available number of ionizing photons), and so it should have a negligible
cosmic effect on the surrounding IGM. 

\noindent
{\it (iv)} GRB afterglows have smooth (broken power-law) continuum spectra
unlike quasars which show strong spectral features (such as broad emission
lines or the so-called ``blue bump'') that complicate the extraction of IGM
absorption features. In particular, the continuum extrapolation into the
Ly$\alpha$ damping wing (the so-called Gunn-Peterson absorption trough)
during the epoch of reionization is much more straightforward for the
smooth UV spectra of GRB afterglows than for quasars with an underlying
broad Ly$\alpha$ emission line \cite{GRBquasar}.

The optical depth of the uniform IGM to Ly$\alpha$ absorption is
given by (Gunn \& Peterson 1965 \cite{GP}),
\begin{equation}
\tau_{s}={\pi e^2 f_\alpha \lambda_\alpha n_{\HI}(z_s) \over m_e
cH(z_s)} \approx 6.45\times 10^5 x_{\HI} \left({\Omega_bh\over
0.03}\right)\left({\Omega_m\over 0.3}\right)^{-1/2} \left({1+z_s\over
10}\right)^{3/2} \label{G-P}
\end{equation}
where $H\approx 100h~{\rm km~s^{-1}~Mpc^{-1}}\Omega_m^{1/2}(1+z_s)^{3/2}$
is the Hubble parameter at the source redshift $z_s>>1$, $f_\alpha=0.4162$
and $\lambda_\alpha=1216$\AA~ are the oscillator strength and the
wavelength of the Ly$\alpha$ transition; $n_{\HI}(z_s)$ is the neutral
hydrogen density at the source redshift (assuming primordial abundances);
$\Omega_m$ and $\Omega_b$ are the present-day density parameters of all
matter and of baryons, respectively; and $x_{\HI}$ is the average fraction
of neutral hydrogen. In the second equality we have implicitly considered
high-redshifts, $(1+z)\gg {\rm
max}\left[(1-\Omega_m-\Omega_\Lambda)/\Omega_m,
(\Omega_\Lambda/\Omega_m)^{1/3}\right]$, at which the vacuum energy density
is negligible relative to matter ($\Omega_\Lambda\ll \Omega_m$) and the
Universe is nearly flat; for $\Omega_m=0.3, \Omega_\Lambda=0.7$ this
corresponds to the condition $z\gg 1.3$ which is well satisfied by the
reionization redshift.

At wavelengths longer than \lya at the source, the optical depth
obtains a small value; these photons redshift away from the line
center along its red wing and never resonate with the line core on
their way to the observer. The red damping wing of the Gunn-Peterson
trough (Miralda-Escud\'e 1998 \cite{Mir98})
\begin{equation}
\tau(\lambda_{\rm obs})=\tau_s \left(\Lambda\over
4\pi^2\nu_\alpha\right) {\tilde \lambda}_{\rm obs}^{3/2}\left[
I({\tilde\lambda}_{\rm obs}^{-1}) -
I([(1+z_{i})/(1+z_s)]{\tilde\lambda}_{\rm obs}^{-1})\right]~~{\rm
for}~~{\tilde\lambda}_{\rm obs}\geq 1~,
\label{eq:shift}
\end{equation}
where $\tau_s$ is given in equation~(\ref{G-P}), also we define
\begin{equation}
{\tilde \lambda}_{\rm obs}\equiv {\lambda_{\rm obs}\over
(1+z_s)\lambda_\alpha}
\end{equation}
and
\begin{equation}
I(x)\equiv {x^{9/2}\over 1-x}+{9\over 7}x^{7/2}+{9\over 5}x^{5/2}+ 3
x^{3/2}+9 x^{1/2}-{9\over 2} \ln\left[ {1+x^{1/2}\over 1-x^{1/2}}
\right]\ .
\end{equation}

\begin{figure}
\centering
\includegraphics[height=10cm]{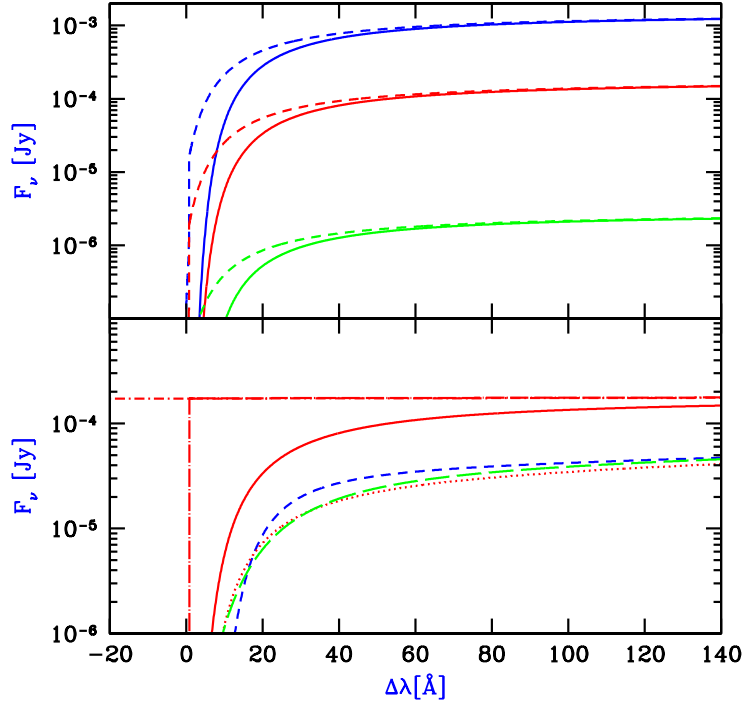}
\caption{Expected spectral shape of the Ly$\alpha$ absorption trough due to
intergalactic absorption in GRB afterglows (taken from
\cite{BL04}). The spectrum is presented in terms of the flux density
$F_{\nu}$ versus relative observed wavelength $\Delta \lambda$, for a
source redshift $z=7$ (assumed to be prior to the final reionization phase)
and the typical halo mass $M=4 \times 10^8 M_{\odot}$ expected for GRB host
galaxies that cool via atomic transitions. {\it Top panel:} Two examples
for the predicted spectrum including IGM HI absorption (both resonant and
damping wing), for host galaxies with (i) an age $t_S=10^7$ yr, a UV escape
fraction $f_{\rm esc}=10\%$ and a Scalo initial mass function (IMF) in
solid curves, or (ii) $t_S=10^8$ yr, $f_{\rm esc}=90\%$ and massive
($>100M_\odot$) Pop III stars in dashed curves. The observed time after the
$\gamma$-ray trigger is one hour, one day, and ten days, from top to
bottom, respectively. {\it Bottom panel:} Predicted spectra one day after a
GRB for a host galaxy with $t_S=10^7$ yr, $f_{\rm esc}=10\%$ and a Scalo
IMF. Shown is the unabsorbed GRB afterglow (dot-short dashed curve), the
afterglow with resonant IGM absorption only (dot-long dashed curve), and
the afterglow with full (resonant and damping wing) IGM absorption (solid
curve). Also shown, with 1.7 magnitudes of extinction, are the afterglow
with full IGM absorption (dotted curve), and attempts to reproduce this
profile with a damped Ly$\alpha$ absorption system in the host galaxy
(dashed curves).  (Note, however, that damped absorption of this type could
be suppressed by the ionizing effect of the afterglow UV radiation on the
surrounding interstellar medium of its host galaxy\cite{PL98}.) Most
importantly, the overall spectral shape of the Ly$\alpha$ trough carries
precious information about the neutral fraction of the IGM at the source
redshift; averaging over an ensemble of sources with similar redshifts can
reduce ambiguities in the interpretation of each case due to particular
local effects.}
\label{fig2val}
\end{figure}

Although the nature of the central engine that powers the relativistic jets
of GRBs is still unknown, recent evidence indicates that long-duration GRBs
trace the formation of massive stars (e.g.,
\cite{T97,Wij98,BN00,Kul00,BKD02,Nat05}) and in particular that long-duration
GRBs are associated with Type Ib/c supernovae \cite{Sta03}. Since the first
stars in the Universe are predicted to be predominantly massive
\cite{ABN02,BCL02,Bromm}, their death might give rise to large numbers of
GRBs at high redshifts.  In contrast to quasars of comparable brightness,
GRB afterglows are short-lived and release $\sim 10$ orders of magnitude
less energy into the surrounding IGM. Beyond the scale of their host
galaxy, they have a negligible effect on their cosmological
environment\footnote{Note, however, that feedback from a single GRB or
supernova on the gas confined within early dwarf galaxies could be
dramatic, since the binding energy of most galaxies at $z>10$ is lower than
$10^{51}~{\rm ergs}$ \cite{BL01}.}. Consequently, they are ideal probes
of the IGM during the reionization epoch.  Their rest-frame UV spectra can
be used to probe the ionization state of the IGM through the spectral shape
of the Gunn-Peterson (Ly$\alpha$) absorption trough, or its metal
enrichment history through the intersection of enriched bubbles of
supernova (SN) ejecta from early galaxies \cite{FL03}.  Afterglows that are
unusually bright ($>10$mJy) at radio frequencies should also show a
detectable forest of 21~cm absorption lines due to enhanced HI column
densities in sheets, filaments, and collapsed minihalos within the IGM
\cite{Carilli04,FL02}.

Another advantage of GRB afterglows is that once they fade away, one may
search for their host galaxies. Hence, GRBs may serve as signposts of the
earliest dwarf galaxies that are otherwise too faint or rare on their own
for a dedicated search to find them. Detection of metal absorption lines
from the host galaxy in the afterglow spectrum, offers an unusual
opportunity to study the physical conditions (temperature, metallicity,
ionization state, and kinematics) in the interstellar medium of these
high-redshift galaxies.  As Figure \ref{fig2val} indicates, damped
Ly$\alpha$ absorption within the host galaxy may mask the clear signature
of the Gunn-Peterson trough in some galaxies \cite{BL04}.  A small
fraction ($\sim 10$) of the GRB afterglows are expected to originate at
redshifts $z>5$ \cite{BL02,BL06}.  This subset of afterglows can be
selected photometrically using a small telescope, based on the Ly$\alpha$
break at a wavelength of $1.216\, \mu {\rm m}\, [(1+z)/10]$, caused by
intergalactic HI absorption.  The challenge in the upcoming years will be
to follow-up on these candidates spectroscopically, using a large (10-meter
class) telescope.  GRB afterglows are likely to revolutionize observational
cosmology and replace traditional sources like quasars, as probes of the
IGM at $z>5$.  The near future promises to be exciting for GRB astronomy as
well as for studies of the high-redshift Universe.

It is of great importance to constrain the Pop~III star formation mode, and
in particular to determine down to which redshift it continues to be
prominent. The extent of the Pop~III star formation will affect models of
the initial stages of reionization (e.g.,
\cite{WL03,CFW03,Sok04,YBH04,ABS06}) and metal enrichment (e.g.,
\cite{MBH03,FL03,FL05,Sch03,SSR04}), and will determine whether planned
surveys will be able to effectively probe Pop~III stars (e.g.,
\cite{Sca05}).  The constraints on Pop~III star formation will also
determine whether the first stars could have contributed a significant
fraction to the cosmic near-IR background (e.g.,
\cite{SBK02,SF03,Kas05,MS05,DAK05}).  To constrain high-redshift star
formation from GRB observations, one has to address two
major questions:

\noindent {\it (1)} {\it What is the signature of GRBs that originate in
metal-free, Pop~III progenitors?} Simply knowing that a given GRB came from
a high redshift is not sufficient to reach a definite conclusion as to the
nature of the progenitor. Pregalactic metal enrichment was likely
inhomogeneous, and we expect normal Pop~I and II stars to exist in galaxies
that were already metal-enriched at these high redshifts
\cite{BL06}. Pop~III and Pop~I/II star formation is thus predicted to have
occurred concurrently at $z > 5$. How is the predicted high mass-scale for
Pop~III stars reflected in the observational signature of the resulting
GRBs? Preliminary results from numerical simulations of Pop III star
formation indicate that circumburst densities are systematically higher in
Pop~III environments. GRB afterglows will then be much brighter than for
conventional GRBs. In addition, due to the systematically increased
progenitor masses, the Pop~III distribution may be biased toward
long-duration events.

\noindent {\it (2)}
The modelling of Pop~III cosmic star formation histories has a number of
free parameters, such as the star formation efficiency and the strength of
the chemical feedback. The latter refers to the timescale for, and spatial
extent of, the distribution of the first heavy elements that were produced
inside of Pop~III stars, and subsequently dispersed into the IGM by
supernova blast waves. Comparing with theoretical GRB redshift
distributions one can use the GRB redshift
distribution observed by {\it Swift} to calibrate the free model
parameters. In particular, one can use this strategy to measure the
redshift where Pop~III star formation terminates.

\begin{figure}[tp!]
\includegraphics[height=.4\textheight]{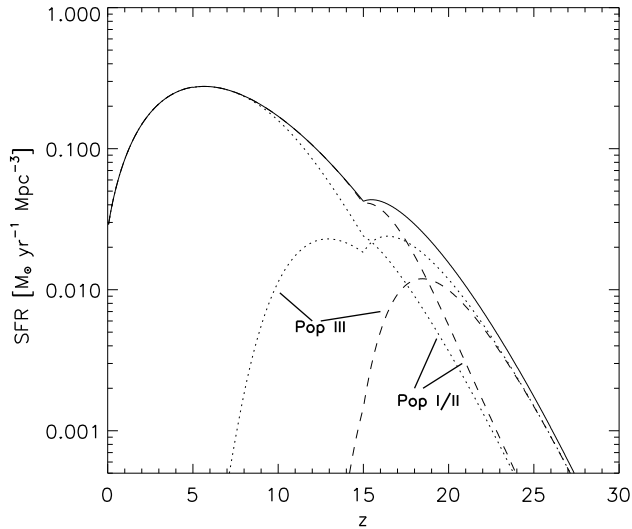}
\caption{Theoretical prediction for the comoving star formation rate
(SFR) in units of $M_{\odot}$~yr$^{-1}$~Mpc$^{-3}$, as a function of
redshift (from \cite{BL06}).  We assume that cooling in primordial gas is
due to atomic hydrogen only, a star formation efficiency of
$\eta_\ast=10\%$, and reionization beginning at $z_{\rm reion}\approx 17$.
{\it Solid line:} Total comoving SFR.  {\it Dotted lines:} Contribution to
the total SFR from Pop~I/II and Pop~III for the case of weak chemical
feedback.  {\it Dashed lines:} Contribution to the total SFR from Pop~I/II
and Pop~III for the case of strong chemical feedback.  Pop~III star
formation is restricted to high redshifts, but extends over a significant
range, $\Delta z\sim 10-15$.
\label{fig1bl}}
\end{figure}

\begin{figure}[t]
\includegraphics[height=.4\textheight]{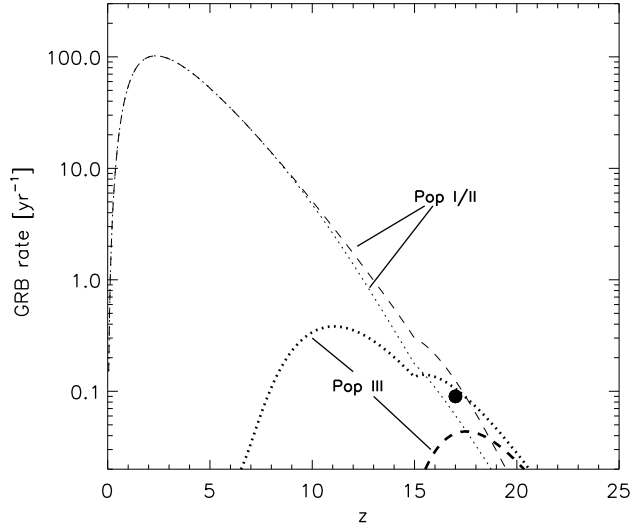}
\caption{Predicted GRB rate to be observed by {\it Swift} (from \cite{BL06}).
Shown is the observed
number of bursts per year, $dN_{\rm GRB}^{\rm obs}/d\ln (1+z)$, as a function
of redshift.  All rates are calculated with a constant GRB efficiency,
$\eta_{\rm GRB}\simeq 2\times 10^{-9}$~bursts $M_{\odot}^{-1}$, using the
cosmic SFRs from Fig.~\ref{fig1bl}.
{\it Dotted lines:} Contribution to the observed GRB
rate from Pop~I/II and Pop~III for the case of weak chemical feedback.
{\it Dashed lines:} Contribution to the GRB rate from Pop~I/II and Pop~III
for the case of strong chemical feedback. 
{\it Filled circle:} GRB rate from Pop~III stars if these were
responsible for reionizing the Universe at $z\sim 17$.  
\label{fig2bl}
}
\end{figure}

Figures~\ref{fig1bl} and \ref{fig2bl} illustrate these issues (based on
\cite{BL06}).  Figure \ref{fig2bl} leads to the robust expectation that
$\sim 10$\% of all {\it Swift} bursts should originate at $z > 5$. This
prediction is based on the contribution from Population~I/II stars which
are known to exist even at these high redshifts. Additional GRBs could be
triggered by Pop~III stars, with a highly uncertain efficiency. Assuming
that long-duration GRBs are produced by the collapsar mechanism, a Pop~III
star with a close binary companion provides a plausible GRB progenitor.
The Pop~III GRB efficiency, reflecting the probability of forming
sufficiently close and massive binary systems, to lie between zero (if
tight Pop~III binaries do not exist) and $\sim 10$ times the empirically
inferred value for Population~I/II (due to the increased fraction of black
hole forming progenitors among the massive Pop~III stars).

A key ingredient in determining the underlying star formation history from
the observed GRB redshift distribution is the GRB luminosity function,
which is only poorly constrained at present.  The improved statistics
provided by {\it Swift} will enable the construction of an empirical
luminosity function. With an improved luminosity function it would be
possible to re-calibrate the theoretical prediction in Figure~2 more
reliably.

In order to predict the observational signature of high-redshift GRBs, it
is important to know the properties of the GRB host systems.  Within
variants of the popular CDM model for structure formation, where small
objects form first and subsequently merge to build up more massive ones,
the first stars are predicted to form at $z\sim 20$--$30$ in minihalos of
total mass (dark matter plus gas) $\sim 10^6 M_{\odot}$
\cite{Teg97,BL01,YBH04}.  These objects are the sites for the formation
of the first stars, and thus are the potential hosts of the
highest-redshift GRBs.  {\it What is the environment in which the earliest
GRBs and their afterglows did occur?}  This problem breaks down into two
related questions: (i) what type of stars (in terms of mass, metallicity,
and clustering properties) will form in each minihalo?, and (ii) how will
the ionizing radiation from each star modify the density structure of the
surrounding gas? These two questions are fundamentally intertwined. The
ionizing photon production strongly depends on the stellar mass, which in
turn is determined by how the accretion flow onto the growing protostar
proceeds under the influence of this radiation field. In other words, the
assembly of the Population~III stars and the development of an HII region
around them proceed simultaneously, and affect each other.  As a
preliminary illustration, Figure~\ref{fig3ab} describes the photo-evaporation
as a self-similar champagne flow \cite{Shu02} with parameters appropriate
for the Pop~III case.

\begin{figure}[t]
\includegraphics[height=.4\textheight]{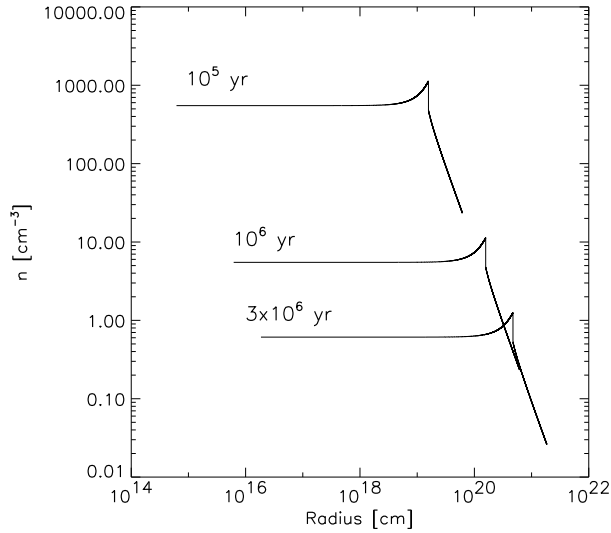}
\caption{Effect of photoheating from a Population~III star on the density
profile in a high-redshift minihalo (from \cite{BrL06}).  The curves,
labeled by the time after the onset of the central point source, are
calculated according to a self-similar model for the expansion of an HII
region. Numerical simulations closely conform to this analytical
behavior. Notice that the central density is significantly reduced by the
end of the life of a massive star, and that a central core has developed
where the density is constant.}
\label{fig3b}
\end{figure}

Notice that the central density is significantly reduced by the end of the
life of a massive star, and that a central core has developed where the
density is nearly constant. Such a flat density profile is markedly
different from that created by stellar winds ($\rho \propto
r^{-2}$). Winds, and consequently mass-loss, may not be important for
massive Population~III stars \cite{BHW01,K02},
and such a flat density profile may be
characteristic of GRBs that originate from metal-free Population~III
progenitors.

\begin{figure}[t]
\includegraphics[height=.3\textheight]{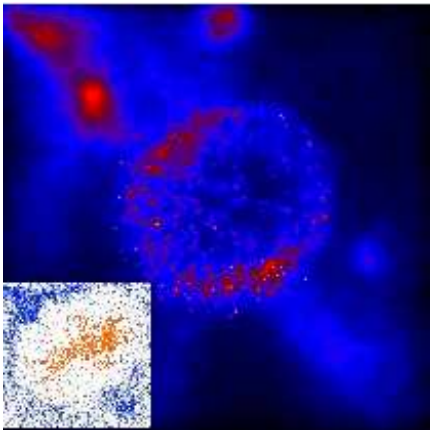}
\caption{Supernova explosion in the high-redshift Universe (from \cite{BYH03}).
The snapshot is taken $\sim 10^6$~yr after the explosion with
total energy $E_{\rm SN}\simeq 10^{53}$~ergs. We show the projected gas
density within a box of linear size 1~kpc. The SN bubble has expanded to a
radius of $\sim 200$~pc, having evacuated most of the gas in the
minihalo. {\it Inset:} Distribution of metals. The stellar ejecta ({\it gray
dots}) trace the metals and are embedded in pristine metal-poor gas ({\it
black dots}).  }
\label{ffig4b}
\end{figure}

The first galaxies may be surrounded by a shell of highly enriched material
that was carried out in a SN-driven wind (see Fig.~\ref{ffig4b}). A GRB in
that galaxy may show strong absorption lines at a velocity separation
associated with the wind velocity.  Simulating these winds and calculating
the absorption profile in the featureless spectrum of a GRB afterglow, will
allow us to use the observed spectra of high-$z$ GRBs and directly probe the
degree of metal enrichment in the vicinity of the first star forming
regions (see \cite{FL03} for a semi-analytic treatment).

As the early afterglow radiation propagates through the interstellar
environment of the GRB, it will likely modify the gas properties close to
the source; these changes could in turn be noticed as time-dependent
spectral features in the spectrum of the afterglow and used to derive the
properties of the gas cloud (density, metal abundance, and size).  The UV
afterglow radiation can induce detectable changes to the interstellar
absorption features of the host galaxy \cite{PL98}; dust destruction could
have occurred due to the GRB X-rays \cite{WD00,FKR01}, and molecules could
have been destroyed near the GRB source \cite{DH02}.  Quantitatively, all
of the effects mentioned above strongly depend on the exact properties of
the gas in the host system.

Most studies to date have assumed a constant efficiency of forming GRBs per
unit mass of stars. This simplifying assumption could lead, under different
circumstances, to an overestimation or an underestimation of the frequency
of GRBs. Metal-free stars are thought to be massive \cite{ABN02,BCL02} and
their extended envelopes may suppress the emergence of relativistic jets
out of their surface (even if such jets are produced through the collapse
of their core to a spinning black hole). On the other hand, low-metallicity
stars are expected to have weak winds with little angular momentum loss
during their evolution, and so they may preferentially yield rotating
central configurations that make GRB jets after core collapse.

{\it What kind of metal-free, Pop~III progenitor stars may lead to GRBs?}
Long-duration GRBs appear to be associated with Type Ib/c supernovae
\cite{Sta03}, namely progenitor massive stars that have lost their hydrogen
envelope. This requirement is explained theoretically in the collapsar
model, in which the relativistic jets produced by core collapse to a black
hole are unable to emerge relativistically out of the stellar surface if
the hydrogen envelope is retained \cite{MWH01}.  The question then arises
as to whether the lack of metal line-opacity that is essential for
radiation-driven winds in metal-rich stars, would make a Pop~III star
retain its hydrogen envelope, thus quenching any relativistic jets and
GRBs.

Aside from mass transfer in a binary system, individual Pop~III stars could
lose their hydrogen envelope due to either: (i) violent pulsations,
particularly in the mass range $100$--$140 M_{\odot}$, or (ii) a wind
driven by helium lines.  The outer stellar layers are in a state where
gravity only marginally exceeds radiation pressure due to
electron-scattering (Thomson) opacity. Adding the small, but still
non-negligible contribution from the bound-free opacity provided by
singly-ionized helium, may be able to unbind the atmospheric
gas. Therefore, mass-loss might occur even in the absence of dust or any
heavy elements.

\subsection{Emission Spectrum of Metal-Free Stars}
\label{sec4.1.2}

The evolution of metal-free (Population III) stars is qualitatively
different from that of enriched (Population I and II) stars. In the
absence of the catalysts necessary for the operation of the CNO cycle,
nuclear burning does not proceed in the standard way. At first,
hydrogen burning can only occur via the inefficient PP chain. To
provide the necessary luminosity, the star has to reach very high
central temperatures ($T_{c}\simeq 10^{8.1}$ K). These temperatures
are high enough for the spontaneous turn-on of helium burning via the
triple-$\alpha$ process. After a brief initial period of
triple-$\alpha$ burning, a trace amount of heavy elements
forms. Subsequently, the star follows the CNO cycle. In constructing
main-sequence models, it is customary to assume that a trace mass
fraction of metals ($Z\sim 10^{-9}$) is already present in the star
(El Eid 1983 \cite{EFO83}; Castellani et al.\ 1983 \cite{CCT83}).

\begin{figure} 
\centering
\includegraphics[height=6cm]{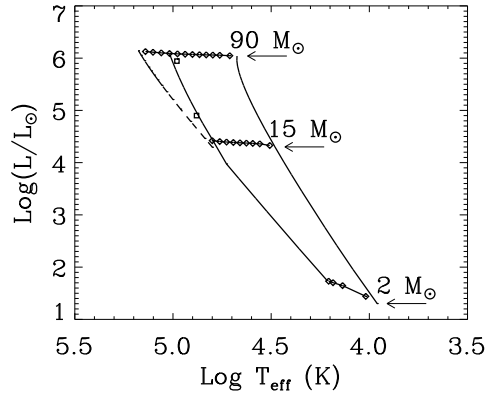}
\caption{Luminosity vs.\ effective temperature for zero-age main
sequences stars in the mass range of $2$--$90M_\odot$ (from Tumlinson
\& Shull 2000 \cite{TS00}). The curves show Pop I ($Z_{\odot}$ = 0.02) and Pop III
stars of mass 2, 5, 8, 10, 15, 20, 25, 30, 35, 40, 50, 60, 70, 80, and
90 \Msun. The diamonds mark decades in metallicity in the approach to
$Z = 0$ from 10$^{-2}$ down to 10$^{-5}$ at 2 \Msun, down to
10$^{-10}$ at 15 \Msun, and down to 10$^{-13}$ at 90 \Msun. The
dashed line along the Pop III ZAMS assumes pure H-He composition,
while the solid line (on the left) marks the upper MS with $Z_{\rm C}
= 10^{-10}$ for the $M \geq 15$ \Msun\ models. Squares mark the points
corresponding to pre-enriched evolutionary models from El Eid et al.\
(1983) \cite{EFO83} at 80 \Msun\ and from Castellani et al.\ (1983) \cite{CCT83} for 25 \Msun.  }
\label{fig4e}
\end{figure}

\begin{figure}
\centering
\includegraphics[height=6cm]{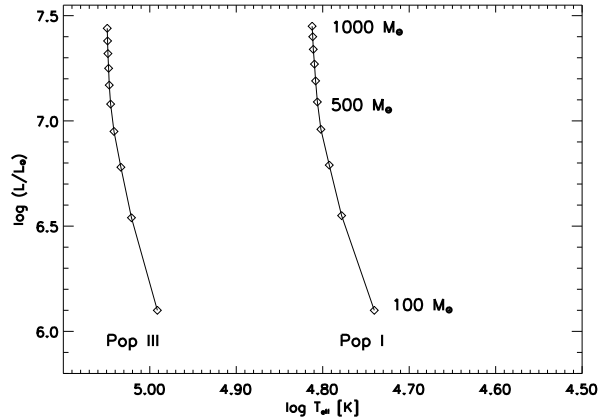}
\caption{Same as Figure \ref{fig4e} but for very massive stars above
$100M_\odot$ (from Bromm, Kudritzki, \& Loeb 2001 \cite{BKL2001}).  {\it Left solid
line:} Pop III zero-age main sequence (ZAMS).  {\it Right solid line:}
Pop I ZAMS. In each case, stellar luminosity (in $L_{\odot}$) is
plotted vs.\ effective temperature (in K).  {\it Diamond-shaped
symbols:} Stellar masses along the sequence, from $100 M_{\odot}$
(bottom) to $1000 M_{\odot}$ (top) in increments of $100 M_{\odot}$.
The Pop III ZAMS is systematically shifted to higher effective
temperature, with a value of $\sim 10^{5}$ K which is approximately
independent of mass. The luminosities, on the other hand, are almost
identical in the two cases.  }
\label{fig4f}
\end{figure}

Figures \ref{fig4e} and \ref{fig4f} show the luminosity $L$ vs.\
effective temperature $T$ for zero-age main sequence stars in the mass
ranges of $2$--$90M_\odot$ (Fig. \ref{fig4e}) and $100$--$1000
M_\odot$ (Fig. \ref{fig4f}). Note that above $\sim 100M_\odot$ the
effective temperature is roughly constant, $T_{\rm eff}\sim 10^5$K,
implying that the spectrum is independent of the mass distribution of
the stars in this regime (Bromm, Kudritzky, \& Loeb 2001 \cite{BKL2001}). As is
evident from these figures (see also Tumlinson \& Shull 2000 \cite{TS00}), both
the effective temperature and the ionizing power of metal-free (Pop
III) stars are substantially larger than those of metal-rich (Pop I)
stars.  Metal-free stars with masses $\ga 20M_\odot$ emit between
$10^{47}$ and $10^{48}$ H I and He I ionizing photons
per second per solar mass of stars, where the lower value applies to
stars of $\sim 20M_\odot$ and the upper value applies to stars of $\ga
100 M_{\odot}$ (see Tumlinson \& Shull 2000 \cite{TS00} and Bromm et al.\ 2001 \cite{BKL2001} for
more details). Over a lifetime of $\sim 3\times 10^6$ years these
massive stars produce $10^4$--$10^5$ ionizing photons per stellar
baryon. However, this powerful UV emission is suppressed as soon as
the interstellar medium out of which new stars form is enriched by
trace amounts of metals. Even though the collapsed fraction of baryons
is small at the epoch of reionization, it is likely that most of the
stars responsible for the reionization of the Universe formed out of
enriched gas.

\noindent
  
\noindent
  
\begin{figure} 
\centering
\includegraphics[height=6cm]{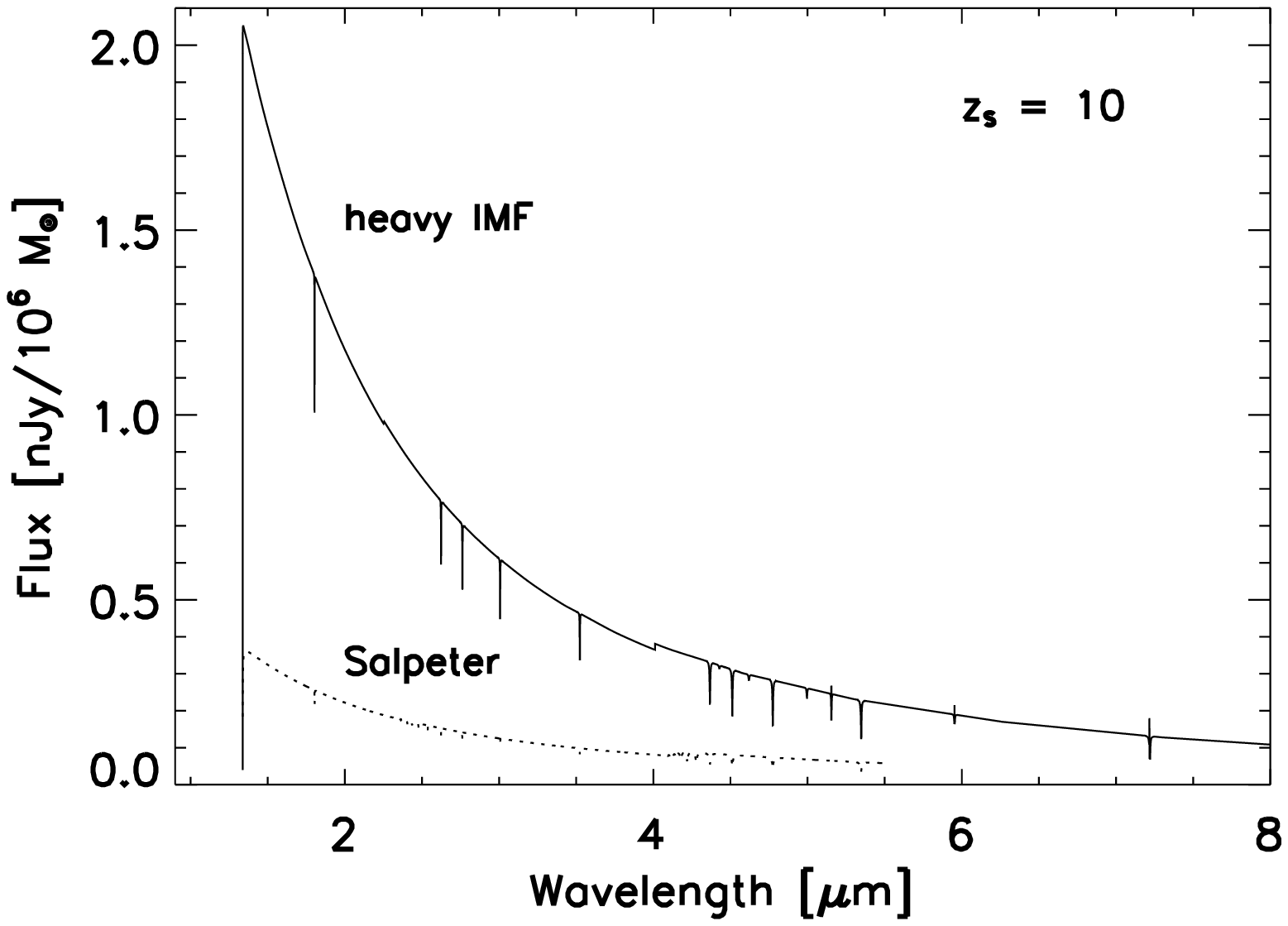}
\caption{ Comparison of the predicted flux from a Pop III star cluster
at $z_{s}=10$ for a Salpeter IMF (Tumlinson \& Shull 2000 \cite{TS00}) and a
massive IMF (Bromm et al.\ 2001 \cite{BKL2001}).  Plotted is the observed flux (in
$\mbox{nJy}$ per $10^{6}M_{\odot}$ of stars) vs.\ observed wavelength
(in $\mu$m) for a flat Universe with $\Omega_{\Lambda}=0.7$ and
$h=0.65$.  {\it Solid line:} The case of a heavy IMF.  {\it Dotted
line:} The fiducial case of a standard Salpeter IMF.  The cutoff below
$\lambda_{obs} = 1216\mbox{\,\AA \,} (1+z_{s})=1.34\mu$m is due to
complete Gunn-Peterson absorption (which is artificially assumed to be
sharp). Clearly, for the same total stellar mass, the observable flux
is larger by an order of magnitude for stars which are biased towards
having masses $\ga 100M_\odot\,$.  }
\label{fig4g}
\end{figure}

{\it Will it be possible to infer the initial mass function (IMF) of
the first stars from spectroscopic observations of the first
galaxies?}  Figure \ref{fig4g} compares the observed spectrum from a
Salpeter IMF ($dN_\star/dM\propto M^{-2.35}$) and a heavy IMF (with
all stars more massive than $100M_\odot$) for a galaxy at
$z_s=10$. The latter case follows from the assumption that each of the
dense clumps in the simulations described in the previous section ends
up as a single star with no significant fragmentation or mass
loss. The difference between the plotted spectra cannot be confused
with simple reddening due to normal dust. Another distinguishing
feature of the IMF is the expected flux in the hydrogen and helium
recombination lines, such as Ly$\alpha$ and He II 1640 \AA, from the
interstellar medium surrounding these stars. We discuss this next.

\noindent
  
\subsection{Emission of Recombination Lines from the First Galaxies}
\label{sec4.1.3}

The hard UV emission from a star cluster or a quasar at high redshift
is likely reprocessed by the surrounding interstellar medium,
producing very strong recombination lines of hydrogen and helium (Oh
1999 \cite{Oh99}; Tumlinson \& Shull 2000 \cite{TS00}; see also Baltz, Gnedin \& Silk
1998 \cite{Baltz98}). We define $\dot{N}_{\rm ion}$ to be the production rate of
ionizing photons by the source. The emitted luminosity $L_{\rm
line}^{\rm em}$ per unit stellar mass in a particular recombination
line is then estimated to be
\begin{equation}
L_{\rm line}^{\rm em} = p_{\rm line}^{\rm em} h\nu \dot{N}_{\rm ion}
(1 - p^{\rm esc}_{\rm cont}) p^{\rm esc}_{\rm line} \mbox{\ \ \ ,}
\end{equation}
where $p_{\rm line}^{\rm em}$ is the probability that a recombination leads
to the emission of a photon in the corresponding line, $\nu$ is the
frequency of the line and $p^{\rm esc}_{\rm cont}$ and $p^{\rm esc}_{\rm
line}$ are the escape probabilities for the ionizing photons and the line
photons, respectively. It is natural to assume that the stellar cluster is
surrounded by a finite H II region, and hence that $p^{\rm esc}_{\rm cont}$
is close to zero \cite{WL00,RS00}.  In addition, $p^{\rm esc}_{\rm line}$
is likely close to unity in the H II region, due to the lack of dust in the
ambient metal-free gas. Although the emitted line photons may be scattered
by neutral gas, they diffuse out to the observer and in the end survive if
the gas is dust free. Thus, for simplicity, we adopt a value of unity for
$p^{\rm esc}_{\rm line}$.

As a particular example we consider case B recombination which yields
$p_{\rm line}^{\rm em}$ of about 0.65 and 0.47 for the Ly${\alpha}$ and He
II 1640\,\AA \,lines, respectively. These numbers correspond to an electron
temperature of $\sim 3\times 10^4$K and an electron density of $\sim
10^{2}-10^{3}$ cm$^{-3}$ inside the H II region \cite{SH95}. For example,
we consider the extreme and most favorable case of metal-free stars all of
which are more massive than $\sim 100M_\odot$. In this case $L_{\rm
line}^{\rm em} = 1.7\times 10^{37}$ and $2.2\times 10^{36}$ erg
s$^{-1}M_{\odot}^{-1}$ for the recombination luminosities of Ly$\alpha$ and
He II 1640\,\AA\,per stellar mass \cite{BKL2001}. A cluster of $10^{6}
M_{\odot}$ in such stars would then produce 4.4 and 0.6 $\times
10^{9}L_{\odot}$ in the Ly$\alpha$ and He II
1640\,\AA\,lines. Comparably-high luminosities would be produced in other
recombination lines at longer wavelengths, such as He II 4686\,\AA\,and
H$\alpha$ \cite{Oh99,OHR00}.

The rest--frame equivalent width of the above emission lines measured
against the stellar continuum of the embedded star cluster at the line
wavelengths is given by
\begin{equation}
W_{\lambda} =\left(\frac{L_{\rm line}^{\rm em}}{L_{\lambda}}\right)
\mbox{\ \ \ ,}
\end{equation}
where $L_{\lambda}$ is the spectral luminosity per unit wavelength of
the stars at the line resonance.  The extreme case of metal-free stars
which are more massive than $100M_\odot$ yields a spectral luminosity
per unit frequency $L_{\nu} = 2.7\times 10^{21}$ and $1.8\times
10^{21}$ erg s$^{-1}$ Hz$^{-1}M_{\odot}^{-1}$ at the corresponding
wavelengths \cite{BKL2001}.  Converting to $L_{\lambda}$, this
yields rest-frame equivalent widths of $W_{\lambda}$ = 3100\,\AA\,and
1100\,\AA\,for Ly$\alpha$ and He II  1640\,\AA\,,
respectively. These extreme emission equivalent widths are more than
an order of magnitude larger than the expectation for a normal cluster
of hot metal-free stars with the same total mass and a Salpeter IMF
under the same assumptions concerning the escape probabilities and
recombination \cite{Kud00}. The equivalent widths are, of
course, larger by a factor of $(1+z_{s})$ in the observer frame.
Extremely strong recombination lines, such as Ly$\alpha$ and
He II  1640\,\AA, are therefore expected to be an additional
spectral signature that is unique to very massive stars in the early
Universe. The strong recombination lines from the first luminous
objects are potentially detectable with \NGST \cite{OHR00}.

\section{Supermassive Black holes}

\subsection{The Principle of Self-Regulation}

The fossil record in the present-day Universe indicates that every bulged
galaxy hosts a supermassive black hole (BH) at its center
\cite{Kor03}. This conclusion is derived from a variety of techniques which
probe the dynamics of stars and gas in galactic nuclei.  The inferred BHs
are dormant or faint most of the time, but ocassionally flash in a short
burst of radiation that lasts for a small fraction of the Hubble time. The
short duty cycle acounts for the fact that bright quasars are much less
abundant than their host galaxies, but it begs the more fundamental
question: {\it why is the quasar activity so brief?}  A natural explanation
is that quasars are suicidal, namely the energy output from the BHs
regulates their own growth.

Supermassive BHs make up a small fraction, $< 10^{-3}$, of the total mass
in their host galaxies, and so their direct dynamical impact is limited to
the central star distribution where their gravitational influence
dominates. Dynamical friction on the background stars keeps the BH close to
the center. Random fluctuations in the distribution of stars induces a
Brownian motion of the BH. This motion can be decribed by the same Langevin
equation that captures the motion of a massive dust particle as it responds
to random kicks from the much lighter molecules of air around it
\cite{Cha02}.  The characteristic speed by which the BH wanders around the
center is small, $\sim (m_\star/M_{\rm BH})^{1/2}\sigma_\star$, where
$m_\star$ and $M_{\rm BH}$ are the masses of a single star and the BH,
respectively, and $\sigma_\star$ is the stellar velocity dispersion. Since
the random force fluctuates on a dynamical time, the BH wanders across a
region that is smaller by a factor of $\sim (m_\star/M_{\rm BH})^{1/2}$
than the region traversed by the stars inducing the fluctuating force on
it.

The dynamical insignificance of the BH on the global galactic scale is
misleading. The gravitational binding energy per rest-mass energy of
galaxies is of order $\sim (\sigma_\star/c)^2< 10^{-6}$.  Since BH are
relativistic objects, the gravitational binding energy of material that
feeds them amounts to a substantial fraction its rest mass energy. Even if
the BH mass occupies a fraction as small as $\sim 10^{-4}$ of the baryonic
mass in a galaxy, and only a percent of the accreted rest-mass energy leaks
into the gaseous environment of the BH, this slight leakage can unbind the
entire gas reservoir of the host galaxy! This order-of-magnitude estimate
explains why quasars are short lived.  As soon as the central BH accretes
large quantities of gas so as to significantly increase its mass, it
releases large amounts of energy that would suppress further accretion onto
it. In short, the BH growth is {\it self-regulated}.

The principle of {\it self-regulation} naturally leads to a correlation
between the final BH mass, $M_{\rm bh}$, and the depth of the gravitational
potential well to which the surrounding gas is confined which can be
characterized by the velocity dispersion of the associated stars, $\sim
\sigma_\star^2$. Indeed such a correlation is observed in the present-day
Universe \cite{Tre02}. The observed power-law relation between $M_{\rm bh}$
and $\sigma_\star$ can be generalized to a correlation between the BH mass
and the circular velocity of the host halo, $v_c$ \cite{Fer02}, which in
turn can be related to the halo mass, $M_{\rm halo}$, and redshift, $z$
\cite{WL03}
\begin{eqnarray}
\label{eq:1}
\nonumber M_{\rm bh}(M_{\rm halo},z) &=&\mbox{const} \times v_c^5\\
&&\hspace{-25mm}= \epsilon_{\rm o} M_{\rm halo} \left(\frac{M_{\rm
halo}}{10^{12}M_{\odot}}\right)^{\frac{2}{3}}
[\zeta(z)]^\frac{5}{6}(1+z)^\frac{5}{2},
\end{eqnarray}
where $\epsilon_{\rm o}\approx 10^{-5.7}$ is a constant, and as before
$\zeta\equiv [(\Omega_m/\Omega_m^z)(\Delta_c/18\pi^2)]$, $\Omega_m^z \equiv
[1+(\Omega_\Lambda/\Omega_m)(1+z)^{-3}]^{-1}$,
$\Delta_c=18\pi^2+82d-39d^2$, and $d=\Omega_m^z-1$.  If quasars shine near
their Eddington limit as suggested by observations of low and high-redshift
quasars \cite{Flo03,Wil03}, then the above value of $\epsilon_{\rm o}$
implies that a fraction of $\sim 5$--$10\%$ of the energy released by the
quasar over a galactic dynamical time needs to be captured in the
surrounding galactic gas in order for the BH growth to be self-regulated
\cite{WL03}.

\begin{figure}
\centering
\includegraphics[height=6cm]{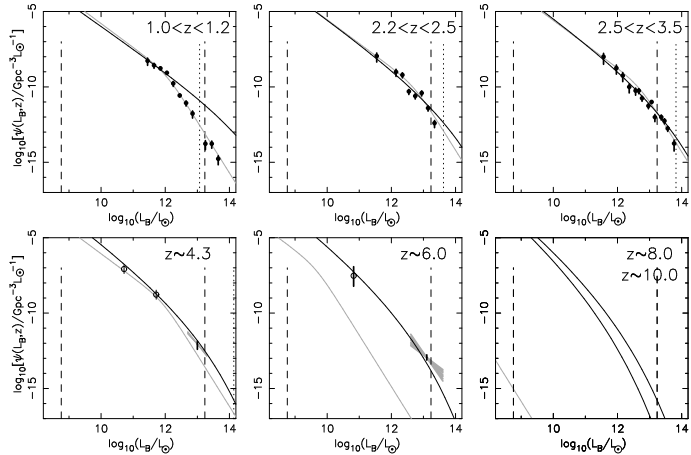}
\caption{Comparison of the observed and model luminosity functions (from
\cite{WL03}). The data points at $z<4$ are summarized in Ref. \cite{Pei95},
while the light lines show the double power-law fit to the {\it 2dF} quasar
luminosity function \cite{Boy00}.  At $z\sim4.3$ and $z\sim6.0$ the data is
from Refs. \cite{Fan01a}. The grey regions show the 1-$\sigma$ range of
logarithmic slope ($[-2.25,-3.75]$ at $z\sim4.3$ and $[-1.6,-3.1]$ at
$z\sim6$), and the vertical bars show the uncertainty in the
normalization. The open circles show data points converted from the X-ray
luminosity function \cite{Bar03} of low luminosity quasars using the median
quasar spectral energy distribution.  In each panel the vertical dashed
lines correspond to the Eddington luminosities of BHs bracketing the
observed range of the $M_{\rm bh}$--$v_{\rm c}$ relation, and the vertical
dotted line corresponds to a BH in a $10^{13.5}M_\odot$ galaxy.}
\label{fig1}
\end{figure}

With this interpretation, the $M_{\rm bh}$--$\sigma_\star$ relation
reflects the limit introduced to the BH mass by self-regulation; deviations
from this relation are inevitable during episodes of BH growth or as a
result of mergers of galaxies that have no cold gas in them.  A physical
scatter around this upper envelope could also result from variations in the
efficiency by which the released BH energy couples to the surrounding gas.

Various prescriptions for self-regulation were sketched by Silk \& Rees
\cite{Sil98}. These involve either energy or momentum-driven winds, where
the latter type is a factor of $\sim v_c/c$ ~less efficient
\cite{Beg04,Kin03,Mur04}. Wyithe \& Loeb \cite{WL03} demonstrated that a
particularly simple prescription for an energy-driven wind can reproduce
the luminosity function of quasars out to highest measured redshift, $z\sim
6$ (see Figs. \ref{fig1} and \ref{fig2}), as well as the observed
clustering properties of quasars at $z\sim 3$ \cite{WLcl} (see
Fig. \ref{fig3}). The prescription postulates that: {\it (i)}
self-regulation leads to the growth of $M_{\rm bh}$ up the
redshift-independent limit as a function of $v_c$ in Eq. (\ref{eq:1}), for
all galaxies throughout their evolution; and {\it (ii)} the growth of
$M_{\rm bh}$ to the limiting mass in Eq. (\ref{eq:1}) occurs through halo
merger episodes during which the BH shines at its Eddington luminosity
(with the median quasar spectrum) over the dynamical time of its host
galaxy, $t_{\rm dyn}$.  This model has only one adjustable parameter,
namely the fraction of the released quasar energy that couples to the
surrounding gas in the host galaxy. This parameter can be fixed based on
the $M_{\rm bh}$--$\sigma_\star$ relation in the local Universe
\cite{Fer02}.  It is remarkable that the combination of the above simple
prescription and the standard $\Lambda$CDM cosmology for the evolution and
merger rate of galaxy halos, lead to a satisfactory agreement with the rich
data set on quasar evolution over cosmic history. 

The cooling time of the heated gas is typically longer than its dynamical
time and so the gas should expand into the galactic halo and escape the
galaxy if its initial temperature exceeds the virial temperature of the
galaxy \cite{WL03}. The quasar remains active during the dynamical time of
the initial gas reservoir, $\sim 10^7$ years, and fades afterwards due to
the dilution of this reservoir.  Accretion is halted as soon as the quasar
supplies the galactic gas with more than its binding energy. The BH growth
may resume if the cold gas reservoir is replenished through a new merger.

Following the early analytic work, extensive numerical simulations by
Springel, Hernquist, \& Di Matteo (2005) \cite{SDH05} (see also Di Matteo et al. 2005 \cite{DSH05})
demonstrated that galaxy mergers do produce the observed correlations
between black hole mass and spheroid properties when a similar energy
feedback is incorporated. Because of the limited resolution near the galaxy
nucleus, these simulations adopt a simple prescription for the accretion
flow that feeds the black hole. The actual feedback in reality may depend
crucially on the geometry of this flow and the physical mechanism that
couples the energy or momentum output of the quasar to the surrounding gas.

\begin{figure}
\centering
\includegraphics[height=4.77cm]{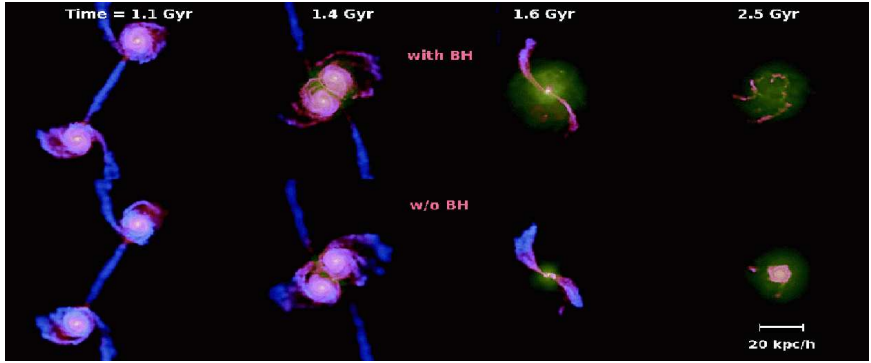}
\caption{Simulation images of a merger of galaxies resulting in quasar
activity that eventually shuts-off the accretion of gas onto the black hole
(from Di Matteo et al. 2005 \cite{DSH05}). The upper (lower) panels show a sequence of
snapshots of the gas distribution during a merger with (without) feedback
from a central black hole. The temperature of the gas is color coded.}
\label{merger}
\end{figure}

\begin{figure}
\centering
\includegraphics[height=6cm]{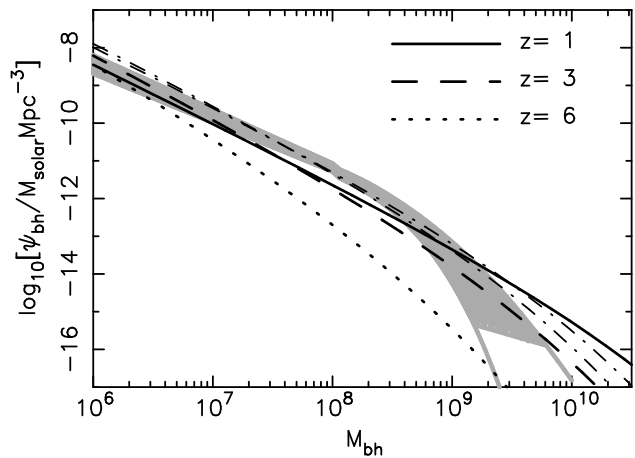}
\caption{The comoving density of supermassive BHs per unit BH mass (from
\cite{WL03}). The grey region shows the estimate based on the observed
velocity distribution function of galaxies in Ref. \cite{She03} and the
$M_{\rm bh}$--$v_c$ relation in Eq. (\ref{eq:1}). The lower bound
corresponds to the lower limit in density for the observed velocity
function while the grey lines show the extrapolation to lower densities. We
also show the mass function computed at $z=1$, 3 and 6 from the
Press-Schechter\cite{ps74} halo mass function and Eq.~(\ref{eq:1}), as well
as the mass function at $z\sim2.35$ and $z\sim3$ implied by the observed
density of quasars and a quasar lifetime of order the dynamical time of the
host galactic disk, $t_{\rm dyn}$ (dot-dashed lines).}
\label{fig2}
\end{figure}

Agreement between the predicted and observed correlation function of
quasars (Fig. \ref{fig3}) is obtained only if the BH mass scales with
redshift as in Eq. (\ref{eq:1}) and the quasar lifetime is of the
order of the dynamical time of the host galactic disk \cite{WLcl},
\begin{equation}
t_{\rm dyn}= 10^7~[\xi(z)]^{-1/2}\left({1+z\over 3}\right)^{-3/2}~{\rm yr}.
\label{eq:life}
\end{equation}
This characterizes the timescale it takes low angular momentum gas to
settle inwards and feed the black hole from across the galaxy before
feedback sets in and suppresses additional infall. It also characterizes
the timescale for establishing an outflow at the escape speed from the host
spheroid.

The inflow of cold gas towards galaxy centers during the growth phase of
the BH would naturally be accompanied by a burst of star formation.  The
fraction of gas that is not consumed by stars or ejected by supernovae,
will continue to feed the BH. It is therefore not surprising that quasar
and starburst activities co-exist in Ultra Luminous Infrared Galaxies
\cite{Gen02}, and that all quasars show broad metal lines indicating a
super-solar metallicity of the surrounding gas \cite{Ham03}. Applying a
similar self-regulation principle to the stars, leads to the expectation
\cite{WL03,Kau00} that the ratio between the mass of the BH and the mass in
stars is independent of halo mass (as observed locally \cite{Mag98}) but
increases with redshift as $\propto \xi(z)^{1/2}(1+z)^{3/2}$. A
consistent trend has indeed been inferred in an observed sample of
gravitationally-lensed quasars \cite{Rix99}.

\begin{figure}
\centering
\includegraphics[height=6cm]{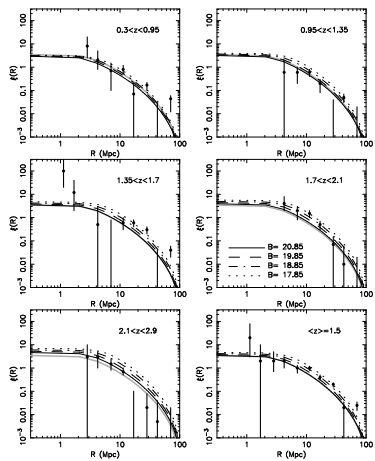}
\caption{Predicted correlation function of quasars at various redshifts
in comparison to the {\it 2dF} data \cite{Cro01} (from \cite{WLcl}). The
dark lines show the correlation function predictions for quasars of various
apparent B-band magnitudes. The {\it 2dF} limit is $B\sim20.85$. The lower
right panel shows data from entire {\it 2dF} sample in comparison to the
theoretical prediction at the mean quasar redshift of $\langle
z\rangle=1.5$. The $B=20.85$ prediction at this redshift is also shown by
thick gray lines in the other panels to guide the eye. The predictions are
based on the scaling $M_{\rm bh}\propto v_c^5$ in Eq. (\ref{eq:1}).  }
\label{fig3}
\end{figure}

The upper mass of galaxies may also be regulated by the energy output from
quasar activity. This would account for the fact that cooling flows are
suppressed in present-day X-ray clusters \cite{Fab04,Boe02,OhS04}, and that
massive BHs and stars in galactic bulges were already formed at $z\sim
2$. The quasars discovered by the {\it Sloan Digital Sky Survey} ({\it
SDSS}) at $z\sim 6$ mark the early growth of the most massive BHs and
galactic spheroids. The present-day abundance of galaxies capable of
hosting BHs of mass $\sim 10^9M_\odot$ (based on Eq. \ref{eq:1}) already
existed at $z\sim 6$ ~\cite{Loe03}. At some epoch, the quasar energy output
may have led to the extinction of cold gas in these galaxies and the
suppression of further star formation in them, leading to an apparent
``anti-hierarchical'' mode of galaxy formation where massive spheroids
formed early and did not make new stars at late times. In the course of
subsequent merger events, the cores of the most massive spheroids acquired
an envelope of collisionless matter in the form of already-formed stars or
dark matter \cite{Loe03}, without the proportional accretion of cold gas
into the central BH. The upper limit on the mass of the central BH and the
mass of the spheroid is caused by the lack of cold gas and cooling flows in
their X-ray halos. In the cores of cooling X-ray clusters, there is often
an active central BH that supplies sufficient energy to compensate for the
cooling of the gas \cite{Boe02,Fab04,Beg04}. The primary physical process
by which this energy couples to the gas is still unknown.

\begin{figure}
\begin{center}
\includegraphics[height=6cm]{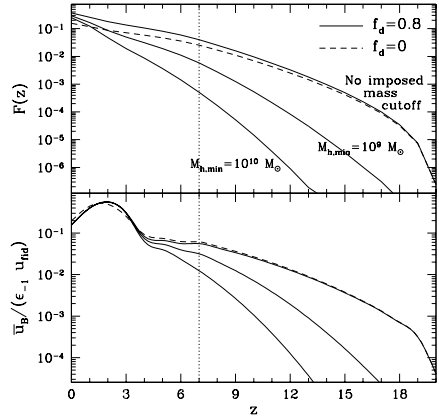}
\end{center}
\caption[]{ The global influence of magnetized quasar outflows on the
intergalactic medium (from \cite{Fur01}). {\it Upper Panel:} Predicted
volume filling fraction of magnetized quasar bubbles $F(z)$, as a function
of redshift.  \emph{Lower Panel:} Ratio of normalized magnetic energy
density, $\bar{u}_B/\epsilon_{-1}$, to the fiducial thermal energy density
of the intergalactic medium $u_{fid} = 3 n(z) k T_{IGM}$, where $T_{IGM} =
10^4 ~{\rm K}$, as a function of redshift (see \cite{Fur01} for more
details).  In each panel, the solid curves assume that the blast wave
created by quasar ouflows is nearly (80\%) adiabatic, and that the minimum
halo mass of galaxies, $M_{h,min}$, is determined by atomic cooling before
reionization and by suppression due to galactic infall afterwards (top
curve), $M_{h,min} = 10^9 M_\odot$ (middle curve), and $M_{h,min} = 10^{10}
M_\odot$ (bottom curve). The dashed curve assumes a fully-radiative blast
wave and fixes $M_{h,min}$ by the thresholds for atomic cooling and infall
suppression.  The vertical dotted line indicates the assumed redshift of
complete reionization, $z_r=7$.  }
\label{fig4}
\end{figure}

\subsection{Feedback on Large Intergalactic Scales}

Aside from affecting their host galaxy, quasars disturb their large-scale
cosmological environment. Powerful quasar outflows are observed in the form
of radio jets \cite{Beg84} or broad-absorption-line winds
\cite{Bran03}. The amount of energy carried by these outflows is largely
unknown, but could be comparable to the radiative output from the same
quasars. Furlanetto \& Loeb \cite{Fur01} have calculated the intergalactic
volume filled by such outflows as a function of cosmic time (see
Fig. \ref{fig4}). This volume is likely to contain magnetic fields and
metals, providing a natural source for the observed magnetization of the
metal-rich gas in X-ray clusters \cite{Kro01} and in galaxies \cite{Dal90}.
The injection of energy by quasar outflows may also explain the deficit of
Ly$\alpha$ absorption in the vicinity of Lyman-break galaxies
\cite{Ade03,Cro02} and the required pre-heating in X-ray clusters
\cite{Bor02,Boe02}.

\begin{figure}
\begin{center}
\includegraphics[height=6cm]{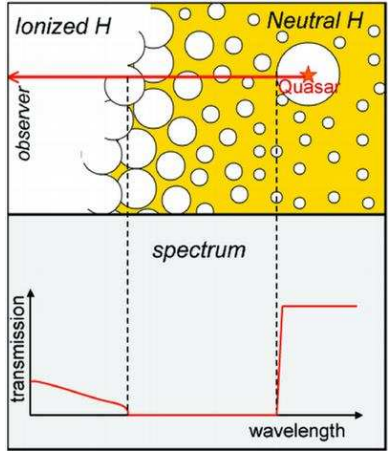}
\end{center}
\caption[]{Quasars serve as probes of the end of reionization. The measured
size of the HII regions around {\it SDSS} quasars can be used
\cite{WL04b,Mes04} to demonstrate that a significant fraction of the
intergalactic hydrogen was neutral at $z\sim 6.3$ or else the inferred size
of the quasar HII regions would have been much larger than observed
(assuming typical quasar lifetimes \cite{Mar03}).  Also, quasars can be
used to measure the redshift at which the intergalactic medium started to
transmit Ly$\alpha$ photons\cite{WB1,WL04c}. The upper panel illustrates how
the line-of-sight towards a quasar intersects this transition redshift. The
resulting Ly$\alpha$ transmission of the intrinsic quasar spectrum is shown
schematically in the lower panel.}
\label{fig5}
\end{figure}

Beyond the reach of their outflows, the brightest {\it SDSS} quasars at
$z>6$ are inferred to have ionized exceedingly large regions of gas (tens
of comoving Mpc) around them prior to global reionization (see
Fig. \ref{fig5} and Refs. \cite{WB1,WL04c}). Thus, quasars must have
suppressed the faint-end of the galaxy luminosity function in these regions
before the same occurred throughout the Universe.  The recombination time
is comparable to the Hubble time for the mean gas density at $z\sim 7$ and
so ionized regions persist \cite{Oh03} on these large scales where
inhomogeneities are small.  The minimum galaxy mass is increased by at
least an order of magnitude to a virial temperature of $\sim 10^5$K in
these ionized regions \cite{BL01}.  It would be particularly interesting
to examine whether the faint end ($\sigma_\star < 30{\rm km~s^{-1}}$) of
the luminosity function of dwarf galaxies shows any moduluation on
large-scales around rare massive BHs, such as M87.

To find the volume filling fraction of relic regions from $z\sim 6$, we
consider a BH of mass $M_{\rm bh}\sim3\times10^9M_\odot$. We can estimate
the comoving density of BHs directly from the observed quasar luminosity
function and our estimate of quasar lifetime.  At $z\sim 6$, quasars
powered by $M_{\rm bh}\sim3\times10^9M_\odot$ BHs had a comoving density of
$\sim 0.5\mbox{Gpc}^{-3}$\cite{WL03}.  However, the Hubble time exceeds
$t_{\rm dyn}$ by a factor of $\sim 2\times 10^2$ (reflecting the square
root of the density contrast of cores of galaxies relative to the mean
density of the Universe), so that the comoving density of the bubbles
created by the $z\sim 6$ BHs is $\sim10^2\mbox{Gpc}^{-3}$ (see
Fig. \ref{fig2}). The density implies that the volume filling fraction of
relic $z\sim 6$ regions is small, $<10\%$, and that the nearest BH that had
$M_{\rm bh}\sim3\times10^9M_\odot$ at $z\sim 6$ (and could have been
detected as an {\it SDSS} quasar then) should be at a distance $d_{\rm
bh}\sim \left(4\pi/3\times10^2\right)^{1/3}\mbox{Gpc} \sim140\mbox{Mpc}$
which is almost an order-of-magnitude larger than the distance of M87, a
galaxy known to possess a BH of this mass \cite{For94}.

{\it What is the most massive BH that can be detected dynamically in a
local galaxy redshift survey?} {\it SDSS} probes a volume of
$\sim1\mbox{Gpc}^3$ out to a distance $\sim30$ times that of M87.  At the
peak of quasar activity at $z\sim 3$, the density of the brightest quasars
implies that there should be $\sim100$ BHs with masses of
$3\times10^{10}M_\odot$ per $\mbox{Gpc}^{3}$, the nearest of which will be
at a distance $d_{\rm bh}\sim130\mbox{Mpc}$, or $\sim 7$ times the distance
to M87.  The radius of gravitational influence of the BH scales as $M_{\rm
bh}/v_{\rm c}^2\propto M_{\rm bh}^{3/5}$. We find that for the nearest
$3\times10^9M_\odot$ and $3\times10^{10}M_\odot$ BHs, the angular radius of
influence should be similar.  Thus, the dynamical signature of $\sim
3\times 10^{10}M_\odot$ BHs on their stellar host should be detectable.

\subsection{What seeded the growth of the supermassive black holes?}

The BHs powering the bright {\it SDSS} quasars possess a mass of a few
$\times 10^9 M_\odot$, and reside in galaxies with a velocity dispersion of
$\sim 500 {\rm km~s^{-1}}$\cite{nature}.  A quasar radiating at its
Eddington limiting luminosity, $L_E=1.4\times 10^{46}~{\rm
erg~s^{-1}}(M_{\rm bh}/10^8M_\odot)$, with a radiative efficiency,
$\epsilon_{\rm rad}=L_{E}/{\dot M}c^2$ would grow exponentially in mass as
a function of time $t$, $M_{\rm bh} =M_{\rm seed}\exp\{t/t_E\}$ on a time
scale, $t_E=4.1\times 10^7~{\rm yr} (\epsilon_{\rm rad}/0.1)$. Thus, the
required growth time in units of the Hubble time $t_{\rm hubble}= 9\times
10^8~{\rm yr}[(1+z)/7]^{-3/2}$ is
\begin{equation}
{t_{\rm growth}\over t_{\rm hubble}}=0.7 \left({\epsilon_{\rm rad} \over
10\%}\right) \left({1+z\over 7}\right)^{3/2}\ln \left({{M_{\rm
bh}/10^9M_\odot} \over M_{\rm seed}/100M_\odot}\right) ~.
\end{equation}
The age of the Universe at $z\sim 6$ provides just sufficient time to grow
an {\it SDSS} BH with $M_{\rm bh}\sim 10^9M_\odot$ out of a stellar mass
seed with $\epsilon_{\rm rad}=10 \%$ \cite{Hai01}. The growth time is
shorter for smaller radiative efficiencies, as expected if the seed
originates from the optically-thick collapse of a supermassive star (in
which case $M_{\rm seed}$ in the logarithmic factor is also larger).

\begin{figure}
\begin{center}
\includegraphics[height=6cm]{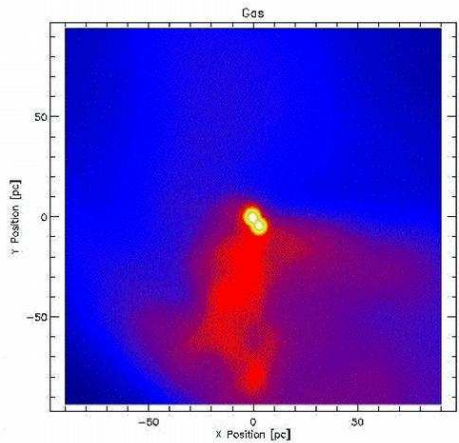}
\end{center}
\caption[]{SPH simulation of the collapse of an early dwarf galaxy with a
virial temperature just above the cooling threshold of atomic hydrogen and
no H$_2$ (from \cite{Bro03}).  The image shows a snapshot of the gas
density distribution at $z\approx 10$, indicating the formation of two
compact objects near the center of the galaxy with masses of $2.2\times
10^{6}M_{\odot}$ and $3.1\times 10^{6}M_{\odot}$, respectively, and radii
$<1$ pc. Sub-fragmentation into lower mass clumps is inhibited as long as
molecular hydrogen is dissociated by a background UV flux.  These
circumstances lead to the formation of supermassive stars
\cite{Loe94} that inevitably collapse and trigger the birth of supermassive
black holes \cite{Loe94,Bau02}.  The box size is 200 pc.  }
\label{fig6}
\end{figure}

{\it What was the mass of the initial BH seeds?  Were they planted in early
dwarf galaxies through the collapse of massive, metal free (Pop-III) stars
(leading to $M_{\rm seed}$ of hundreds of solar masses) or through the
collapse of even more massive, i.e. supermassive, stars \cite{Loe94} ?}
~Bromm \& Loeb \cite{Bro03} have shown through a hydrodynamical simulation
(see Fig. \ref{fig6}) that supermassive stars were likely to form in early
galaxies at $z\sim 10$ in which the virial temperature was close to the
cooling threshold of atomic hydrogen, $\sim 10^4$K. The gas in these
galaxies condensed into massive $\sim 10^6M_\odot$ clumps (the progenitors
of supermassive stars), rather than fragmenting into many small clumps (the
progenitors of stars), as it does in environments that are much hotter than
the cooling threshold. This formation channel requires that a galaxy be
close to its cooling threshold and immersed in a UV background that
dissociates molecular hydrogen in it. These requirements should make this
channel sufficiently rare, so as not to overproduce the cosmic mass density
of supermassive BH.

The minimum seed BH mass can be identified observationally through the
detection of gravitational waves from BH binaries with {\it Advanced LIGO}
\cite{WL04a} or with {\it LISA} \cite{WL03b}.  Most of the mHz binary
coalescence events originate at $z>7$ if the earliest galaxies included BHs
that obey the $M_{\rm bh}$--$v_c$ relation in Eq. (\ref{eq:1}). The number
of {\it LISA} sources per unit redshift per year should drop substantially
after reionization, when the minimum mass of galaxies increased due to
photo-ionization heating of the intergalactic medium.  Studies of the
highest redshift sources among the few hundred detectable events per year,
will provide unique information about the physics and history of BH growth
in galaxies \cite{Vol04}.

The early BH progenitors can also be detected as unresolved point sources,
using the future {\it James Webb Space Telescope} ({\it
JWST}). Unfortunately, the spectrum of metal-free massive and supermassive
stars is the same, since their surface temperature $\sim 10^5$K is
independent of mass \cite{BKL2001}. Hence, an unresolved cluster of massive
early stars would show the same spectrum as a supermassive star of the same
total mass.

It is difficult to ignore the possible environmental impact of
quasars on {\it anthropic} selection. One may wonder whether it is not a
coincidence that our Milky-Way Galaxy has a relatively modest BH mass of
only a few million solar masses in that the energy output from a much more
massive (e.g. $\sim 10^9M_\odot$) black hole would have disrupted the
evolution of life on our planet. A proper calculation remains to be done
(as in the context of nearby Gamma-Ray Bursts \cite{Sca02}) in order to
demonstrate any such link.


\section{\bf Radiative Feedback from the First Sources of Light}
\label{sec6}

\subsection{Escape of Ionizing Radiation from Galaxies}
\label{sec6.1}

The intergalactic ionizing radiation field, a key ingredient in the
development of reionization, is determined by the amount of ionizing
radiation escaping from the host galaxies of stars and quasars.  The value
of the escape fraction as a function of redshift and galaxy mass remains a
major uncertainty in all current studies, and could affect the cumulative
radiation intensity by orders of magnitude at any given redshift. Gas
within halos is far denser than the typical density of the IGM, and in
general each halo is itself embedded within an overdense region, so the
transfer of the ionizing radiation must be followed in the densest regions
in the Universe. Reionization simulations are limited in resolution and
often treat the sources of ionizing radiation and their immediate
surroundings as unresolved point sources within the large-scale
intergalactic medium (see, e.g., Gnedin 2000 \cite{g00}).  The
escape fraction is highly sensitive to the three-dimensional distribution
of the UV sources relative to the geometry of the absorbing gas within the
host galaxy (which may allow escape routes for photons along particular
directions but not others). 

The escape of ionizing radiation ($h\nu > 13.6$eV, $\lambda < 912$ {\AA})
from the disks of present-day galaxies has been studied in recent years in
the context of explaining the extensive diffuse ionized gas layers observed
above the disk in the Milky Way \cite{RTKMH95} and other galaxies
\cite{Rand96, HWR99}. Theoretical models predict that of order 3--14\% of
the ionizing luminosity from O and B stars escapes the Milky Way disk
\cite{DS94,DSF99}.  A similar escape fraction of $f_{\rm esc}=6$\% was
determined by Bland-Hawthorn \& Maloney (1999) \cite{BM99} based on
H$\alpha$ measurements of the Magellanic Stream.  From {\it Hopkins
Ultraviolet Telescope} observations of four nearby starburst galaxies
(Leitherer et al.\ 1995 \cite{LFHL95}; Hurwitz, Jelinsky, \& Dixon 1997
\cite{HJD97}), the escape fraction was estimated to be in the range
3\%$<f_{\rm esc} < 57$\%.  If similar escape fractions characterize high
redshift galaxies, then stars could have provided a major fraction of the
background radiation that reionized the IGM \cite{MS96,Madau99}.  However,
the escape fraction from high-redshift galaxies, which formed when the
Universe was much denser ($\rho\propto (1+z)^3$), may be significantly
lower than that predicted by models ment to describe present-day galaxies.
Current reionization calculations assume that galaxies are isotropic point
sources of ionizing radiation and adopt escape fractions in the range $5\%
< f_{\rm esc} < 60\%$ \cite{g00}.

Clumping is known to have a significant effect on the penetration and
escape of radiation from an inhomogeneous medium
\cite{Boi90,WG96,Neufeld91,HS99,BFDA00}.  The inclusion of clumpiness
introduces several unknown parameters into the calculation, such as the
number and overdensity of the clumps, and the spatial correlation between
the clumps and the ionizing sources.  An additional complication may arise
from hydrodynamic feedback, whereby part of the gas mass is expelled from
the disk by stellar winds and supernovae (\S \ref{sec7}).

\begin{figure} 
\centering
\includegraphics[height=6cm]{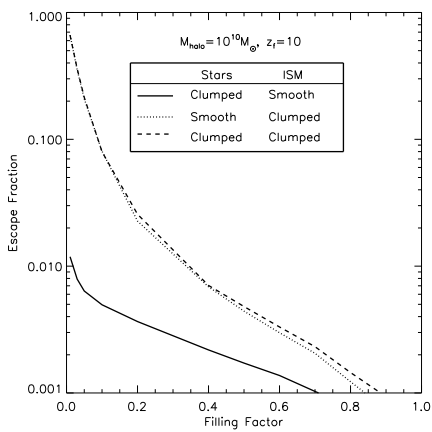}
\caption{Escape fractions of stellar ionizing photons from a gaseous
disk embedded within a $10^{10}M_\odot$ halo which have formed at
$z=10$ (from Wood \& Loeb 2000 \cite{WL00}). The curves show three different cases
of clumpiness within the disk. The volume filling factor refers to
either the ionizing emissivity, the gas clumps, or both, depending on
the case. The escape fraction is substantial ($\ga 1\%$) only if the
gas distribution is highly clumped.  }
\label{fig6a}
\end{figure}

Wood \& Loeb (2000) \cite{WL00} used a three-dimensional radiation transfer
code to calculate the steady-state escape fraction of ionizing photons from
disk galaxies as a function of redshift and galaxy mass. The gaseous disks
were assumed to be isothermal, with a sound speed $c_s\sim 10~{\rm
km~s^{-1}}$, and radially exponential, with a scale-length based on the
characteristic spin parameter and virial radius of their host halos. The
corresponding temperature of $\sim 10^4$ K is typical for a gas which is
continuousely heated by photo-ionization from stars.  The sources of
radiation were taken to be either stars embedded in the disk, or a central
quasar. For stellar sources, the predicted increase in the disk density
with redshift resulted in a strong decline of the escape fraction with
increasing redshift. The situation is different for a central quasar. Due
to its higher luminosity and central location, the quasar tends to produce
an ionization channel in the surrounding disk through which much of its
ionizing radiation escapes from the host. In a steady state, only
recombinations in this ionization channel must be balanced by ionizations,
while for stars there are many ionization channels produced by individual
star-forming regions and the total recombination rate in these channels is
very high. Escape fractions $\ga 10\%$ were achieved for stars at $z\sim
10$ only if $\sim 90\%$ of the gas was expelled from the disks or if dense
clumps removed the gas from the vast majority ($\ga 80\%$) of the disk
volume (see Fig. \ref{fig6a}). This analysis applies only to halos with
virial temperatures $\ga 10^4$ K. Ricotti \& Shull (2000) \cite{RS00}
reached similar conclusions but for a quasi-spherical configuration of
stars and gas.  They demonstrated that the escape fraction is substantially
higher in low-mass halos with a virial temperature $\la 10^4$ K.  However,
the formation of stars in such halos depends on their uncertain ability to
cool via the efficient production of molecular hydrogen.

\noindent
  
The main uncertainty in the above predictions involves the distribution of
the gas inside the host galaxy, as the gas is exposed to the radiation
released by stars and the mechanical energy deposited by supernovae.  Given
the fundamental role played by the escape fraction, it is desirable to
calibrate its value observationally.  Steidel, Pettini, \& Adelberger
\cite{Ste00} reported a  detection of significant Lyman
continuum flux in the composite spectrum of 29 Lyman break galaxies (LBG)
with redshifts in the range $z = 3.40\pm 0.09$. They co-added the spectra
of these galaxies in order to be able to measure the low flux. Another
difficulty in the measurement comes from the need to separate the
Lyman-limit break caused by the interstellar medium from that already
produced in the stellar atmospheres. After correcting for intergalactic
absorption, Steidel et al.\ \cite{Ste00} inferred a ratio between
the emergent flux density at 1500\AA~ and 900\AA~ (rest frame) of $4.6 \pm
1.0$. Taking into account the fact that the stellar spectrum should already
have an intrinsic Lyman discontinuity of a factor of $\sim 3$--5, but that
only $\sim 15$--$20\%$ of the 1500\AA~ photons escape from typical LBGs
without being absorbed by dust (Pettini et al.\ 1998 \cite{PKSDAG98a};
Adelberger et al.\ 2000 \cite{Ade00}), the inferred 900\AA~ escape fraction
is $f_{\rm esc} \sim 10$--$20\%$.  Although the galaxies in this sample
were drawn from the bluest quartile of the LBG spectral energy
distributions, the measurement implies that this quartile may itself
dominate the hydrogen-ionizing background relative to quasars at $z\sim 3$.

\subsection{Propagation of Ionization Fronts in the IGM}
\label{sec6.2}

The radiation output from the first stars ionizes hydrogen in a
growing volume, eventually encompassing almost the entire IGM within a
single H II  bubble. In the early stages of this process, each
galaxy produces a distinct H II  region, and only when the
overall H II  filling factor becomes significant do neighboring
bubbles begin to overlap in large numbers, ushering in the ``overlap
phase'' of reionization. Thus, the first goal of a model of
reionization is to describe the initial stage, when each source
produces an isolated expanding H II  region.

We assume a spherical ionized volume $V$, separated from the surrounding
neutral gas by a sharp ionization front. Indeed, in the case of a stellar
ionizing spectrum, most ionizing photons are just above the hydrogen
ionization threshold of 13.6 eV, where the absorption cross-section is high
and a very thin layer of neutral hydrogen is sufficient to absorb all the
ionizing photons. On the other hand, an ionizing source such as a quasar
produces significant numbers of higher energy photons and results in a
thicker transition region.

In the absence of recombinations, each hydrogen atom in the IGM would
only have to be ionized once, and the ionized proper volume $V_p$
would simply be determined by \beq \nb_H V_p=\Ng\ , \eeq where $\nb_H$
is the mean number density of hydrogen and $\Ng$ is the total number
of ionizing photons produced by the source. However, the increased
density of the IGM at high redshift implies that recombinations cannot
be neglected. Indeed, in the case of a steady ionizing source (and
neglecting the cosmological expansion), a steady-state volume would be
reached corresponding to the Str\"{o}mgren sphere, with recombinations
balancing ionizations: \beq \alpha_B \nb_H^2 V_p=\frac{d\, \Ng}{dt}\ ,
\eeq where the recombination rate depends on the square of the density
and on the case B recombination coefficient $\alpha_B=2.6\times 10^{-13}$
cm$^3$ s$^{-1}$ for hydrogen at $T=10^4$ K. The exact evolution for an
expanding H II  region, including a non-steady ionizing source,
recombinations, and cosmological expansion, is given by (Shapiro \&
Giroux 1987 \cite{SG87}) \beq \nb_H\left( \frac{dV_p}{dt}-3 H V_p\right)=
\frac{d\, \Ng}{dt} - \alpha_B \left<n_H^2\right> V_p\ . \label{front}
\eeq In this equation, the mean density $\nb_H$ varies with time as
$1/a^3(t)$. A critical physical ingredient is the dependence of
recombination on the square of the density. This means that if the IGM
is not uniform, but instead the gas which is being ionized is mostly
distributed in high-density clumps, then the recombination time is
very short. This is often dealt with by introducing a volume-averaged
clumping factor $C$ (in general time-dependent), defined
by\footnote{The recombination rate depends on the number density of
electrons, and in using equation~(\ref{clump}) we are neglecting the
small contribution caused by partially or fully ionized helium.} \beq
C=\left<n_H^2\right>/\nb_H^2 \label{clump}\ . \eeq

If the ionized volume is large compared to the typical scale of clumping,
so that many clumps are averaged over, then equation~(\ref{front}) can be
solved by supplementing it with equation~(\ref{clump}) and specifying
$C$. Switching to the comoving volume $V$, the resulting equation is \beq
\frac{dV}{dt}= \frac{1}{\nb_H^0} \frac{d\, \Ng}{dt}- \alpha_B \frac{C}{a^3}
\nb_H^0 V\ , \label{HIIreg} \eeq where the present number density of
hydrogen is \beq \nb_H^0=1.88\times 10^{-7} \left(\frac{\Omega_b
h^2}{0.022}\right)\ {\rm cm}^{-3}\ . \eeq This number density is lower than
the total number density of baryons $\nb_b^0$ by a factor of $\sim 0.76$,
corresponding to the primordial mass fraction of hydrogen. The solution for
$V(t)$ (generalized from Shapiro \& Giroux 1987 \cite{SG87}) around a source which
turns on at $t=t_i$ is \beq V(t)=\int_{t_i}^t \frac{1}{\nb_H^0} \frac{d\,
\Ng}{dt'} \,e^{F(t',t)} dt'\ ,\label{HIIsoln} \eeq where \beq
F(t',t)=-\alpha_B \nb_H^0 \int_{t'}^t \frac{C(t'')} {a^3(t'')}\, dt''\
\label{Fgen}. \eeq At high redshift (when $(1+z) \gg |\Omm^{-1}-1|$), the
scale factor varies as \beq a(t)\simeq \left(\frac{3}{2}\sqrt{\Omm} H_0
t\right)^{2/3}\ , \eeq and with the additional assumption of a constant $C$
the function $F$ simplifies as follows. Defining \beq f(t)=a(t)^{-3/2}
\label{foft}\ , \eeq we derive \beq F(t',t)=-\frac{2}{3}\frac{ \alpha_B
\nb_H^0} {\sqrt{\Omm} H_0}\,C \left[f(t')-f(t)\right]=-0.262 \left[f(t')
-f(t)\right] \ , \eeq where the last equality assumes $C=10$ and our
standard choice of cosmological parameters: $\Omm=0.3$, $\Oml=0.7$, and
$\Omega_b=0.045$. Although this expression for $F(t',t)$ is in general an
accurate approximation at high redshift, in the particular case of the
$\Lambda$CDM model (where $\Omm+\Oml=1$) we get the exact result by
replacing equation~(\ref{foft}) with \beq
f(t)=\sqrt{\frac{1}{a^3}+\frac{1-\Omm}{\Omm}}\ . \label{fLCDM} \eeq

\begin{figure}
\centering
\includegraphics[height=6cm]{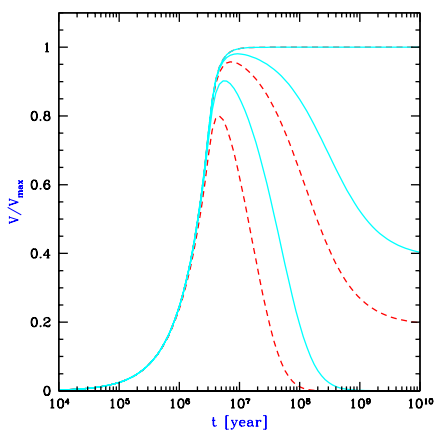}
\caption{Expanding H II  region around an isolated ionizing
source. The comoving ionized volume $V$ is expressed in units of the
maximum possible volume, $V_{\rm max}=4\pi r_{\rm max}^3/3$ [with
$r_{\rm max}$ given in equation~(\ref{rmax})], and the time is
measured after an instantaneous starburst which produces ionizing photons
according to equation~(\ref{zoltan}). We consider a source turning on at
$z=10$ (solid curves) or $z=15$ (dashed curves), with three cases for each:
no recombinations, $C=1$, and $C=10$, in order from top to bottom. The
no-recombination curve is identical for the different source redshifts.}
\label{fig6b}
\end{figure}

The size of the resulting H II  region depends on the halo which
produces it. Consider a halo of total mass $M$ and baryon fraction
$\Omega_b/\Omm$. To derive a rough estimate, we assume that baryons
are incorporated into stars with an efficiency of $f_{\rm star}=10\%$,
and that the escape fraction for the resulting ionizing radiation is
also $f_{\rm esc}=10\%$. If the stellar IMF is similar to the one
measured locally \cite{Sca98}, then $N_{\gamma}
\approx 4000$ ionizing photons are produced per baryon in stars (for a
metallicity equal to $1/20$ of the solar value). We define a parameter
which gives the overall number of ionizations per baryon, \beq \Ni
\equiv N_{\gamma} \, f_{\rm star}\, f_{\rm esc}\ . \eeq If we neglect
recombinations then we obtain the maximum comoving radius of the
region which the halo of mass $M$ can ionize, \beq r_{\rm max}=
\left(\frac{3}{4\pi}\, \frac{\Ng} {\nb_H^0} \right)^{1/3} =
\left(\frac{3}{4\pi}\, \frac{\Ni} {\nb_H^0}\, \frac{\Omega_b}{\Omm}\,
\frac{M}{m_p} \right)^{1/3}= 680\, {\rm kpc} \left( \frac{\Ni}{40}\,
\frac{M} {10^8 M_{\sun}}\right)^{1/3}\ , \label{rmax} \eeq for our
standard set of parameters. The actual radius never reaches this size
if the recombination time is shorter than the lifetime of the ionizing
source. For an instantaneous starburst with the Scalo (1998) 
\cite{Sca98} IMF, the production rate of ionizing photons can be
approximated as \beq \frac{d\,
\Ng}{dt}=\frac{\alpha-1}{\alpha} \frac{\Ng} {t_s}\times \left\{
\begin{array}{ll} \ \ \, 1 & \mbox{if
$t<t_s$,} \\ \left(\frac{t}{t_s}\right)^{-\alpha}\ & \mbox{otherwise,}
\end{array} \right. \label{zoltan} \eeq
where $\Ng=4000$, $\alpha=4.5$, and the most massive stars fade away
with the characteristic timescale $t_s=3\times 10^6$ yr. In figure
 \ref{fig6b} we show the time evolution of the volume ionized by such
a source, with the volume shown in units of the maximum volume $V_{\rm
max}$ which corresponds to $r_{\rm max}$ in equation~(\ref{rmax}). We
consider a source turning on at $z=10$ (solid curves) or $z=15$
(dashed curves), with three cases for each: no recombinations, $C=1$,
and $C=10$, in order from top to bottom (Note that the result is
independent of redshift in the case of no recombinations). When
recombinations are included, the volume rises and reaches close to
$V_{\rm max}$ before dropping after the source turns off. At large $t$
recombinations stop due to the dropping density, and the volume
approaches a constant value (although $V \ll V_{\rm max}$ at large $t$
if $C=10$).

We obtain a similar result for the size of the H II region around a galaxy
if we consider a mini-quasar rather than stars. For the typical quasar
spectrum (Elvis et al.\ 1994 \cite{ELV94}), roughly 11,000 ionizing photons
are produced per baryon incorporated into the black hole, assuming a
radiative efficiency of $\sim 6\%$. The efficiency of incorporating the
baryons in a galaxy into a central black hole is low ($\la 0.6\%$ in the
local Universe, e.g.\ Magorrian et al.\ 1998 \cite{Mag98}), but the escape
fraction for quasars is likely to be close to unity, i.e., an order of
magnitude higher than for stars (see previous sub-section). Thus, for every
baryon in galaxies, up to $\sim 65$ ionizing photons may be produced by a
central black hole and $\sim 40$ by stars, although both of these numbers
for $\Ni$ are highly uncertain. These numbers suggest that in either case
the typical size of H II regions before reionization may be $\la 1$ Mpc or
$\sim 10$ Mpc, depending on whether $10^8 M_{\sun}$ halos or $10^{12}
M_{\sun}$ halos dominate.

The ionization front around a bright transient source like a quasar expands
at early times at nearly the speed of light. This occurs when the HII
region is sufficiently small so that the production rate of ionizing
photons by the central source exceeds their consumption rate by hydrogen
atoms within this volume. It is straightforward to do the accounting for
these rates (including recombinations) taking the light propagation delay
into account.  This was done by Wyithe \& Loeb \cite{WL04b} [see also White
et al. (2003) \cite{WB1}] who derived the general equation for the
relativistic expansion of the {\em comoving} radius [$r=(1+z)r_{\rm p}$]
of the quasar \HII region in an IGM with a neutral filling fraction $x_{\rm
HI}$ (fixed by other ionizing sources) as,
\begin{equation}
\label{Vev2}
\frac{dr}{dt}=c(1+z)\left[\frac{\dot{N}_{\gamma} - \alpha_{\rm B}C x_{\rm
HI}\left(\bar{n}^0_{\rm H}\right)^2 \left(1+z\right)^3 \left({4\pi\over
3}r^3\right)} {\dot{N}_{\gamma} + 4\pi r^2 \left(1+z\right) c x_{\rm
HI}\bar{n}^0_{\rm H}}\right],
\end{equation}
where $c$ is the speed of light, $C$ is the clumping factor, $\alpha_{\rm
B}=2.6\times10^{-13}$cm$^3$s$^{-1}$ is the case-B recombination coefficient
at the characteristic temperature of $10^4$K, and $\dot{N}_{\gamma}$ is the
rate of ionizing photons crossing a shell at the radius of the HII region
at time $t$.  Indeed, for $\dot{N}_\gamma\rightarrow \infty$ the
propagation speed of the proper radius of the HII region $r_p=r/(1+z)$
approaches the speed of light in the above expression,
$(dr_p/dt)\rightarrow c$.  The actual size of the HII region along the
line-of-sight to a quasar can be inferred from the extent of the spectral
gap between the quasar's rest-frame Ly$\alpha$ wavelength and the start of
Ly$\alpha$ absorption by the IGM in the observed spectrum.  Existing data
from the SDSS quasars \cite{WL04b,Mes04, Wy05} provide typical values of
$r_p\sim 5$Mpc and indicate for plausible choices of the quasar lifetimes
that $x_{\rm HI}>0.1$ at $z>6$.  These ionized bubbles could be imaged
directly by future 21cm maps of the regions around the highest-redshift
quasars \cite{Tozzi,WyLo,WyBar}.

The profile of the Ly$\alpha$ emission line of galaxies has also been
suggested as a probe of the ionization state of the IGM
\cite{Loeb_Rybicki,Santos,Haiman_Cen,Haiman05,Madau_Rees,Loeb_Hernquist,Rhoads}.
If the IGM is neutral, then the damping wing of the Gunn-Peterson trough in
equation (\ref{eq:shift}) is modified since Ly$\alpha$ absorption starts
only from the near edge of the ionized region along the line-of-sight to
the source \cite{Haiman_Cen,Madau_Rees}. Rhoads \& Malhotra \cite{Rhoads}
showed that the observed abundance of galaxies with Ly$\alpha$ emission at
$z\sim 6.5$ indicates that a substantial fraction (tens of percent) of the
IGM must be ionized in order to allow transmission of the observed
Ly$\alpha$ photons.  However, if these galaxies reside in groups, then
galaxies with peculiar velocities away from the observer will
preferentially Doppler-shift the emitted Ly$\alpha$ photons to the red wing
of the Ly$\alpha$ resonance and reduce the depression of the line profile
\cite{Loeb_Hernquist,Haiman_Mesinger}. Additional uncertainties in the
intrinsic line profile based on the geometry and the stellar or gaseous
contents of the source galaxy \cite{Loeb_Hernquist,Santos}, as well as the
clustering of galaxies which ionize their immediate environment in groups
\cite{WL04c,Furlanetto_Hernquist}, limits this method from reaching
robust conclusions. Imaging of the expected halos of scattered Ly$\alpha$
radiation around galaxies embedded in a neutral IGM
\cite{Loeb_Rybicki,Rybicki_Loeb} provide a more definitive test of the
neutrality of the IGM, but is more challenging observationally.

\subsection{Reionization of Hydrogen}
\label{sec6.3}

In this section we summarize recent progress, both analytic and
numerical, made toward elucidating the basic physics of reionization
and the way in which the characteristics of reionization depend on the
nature of the ionizing sources and on other input parameters of
cosmological models.

The process of the reionization of hydrogen involves several distinct
stages. The initial, ``pre-overlap'' stage (using the terminology of Gnedin
\cite{g00}) consists of individual ionizing sources turning on and
ionizing their surroundings. The first galaxies form in the most massive
halos at high redshift, and these halos are biased and are preferentially
located in the highest-density regions. Thus the ionizing photons which
escape from the galaxy itself (see \S \ref{sec6.1}) must then make their
way through the surrounding high-density regions, which are characterized
by a high recombination rate. Once they emerge, the ionization fronts
propagate more easily into the low-density voids, leaving behind pockets of
neutral, high-density gas. During this period the IGM is a two-phase medium
characterized by highly ionized regions separated from neutral regions by
ionization fronts.  Furthermore, the ionizing intensity is very
inhomogeneous even within the ionized regions, with the intensity
determined by the distance from the nearest source and by the ionizing
luminosity of this source.

The central, relatively rapid ``overlap'' phase of reionization begins when
neighboring H II  regions begin to overlap. Whenever two ionized
bubbles are joined, each point inside their common boundary becomes exposed
to ionizing photons from both sources. Therefore, the ionizing intensity
inside H II  regions rises rapidly, allowing those regions to expand
into high-density gas which had previously recombined fast enough to remain
neutral when the ionizing intensity had been low. Since each bubble
coalescence accelerates the process of reionization, the overlap phase has
the character of a phase transition and is expected to occur rapidly, over
less than a Hubble time at the overlap redshift. By the end of this stage
most regions in the IGM are able to see several unobscured sources, and
therefore the ionizing intensity is much higher than before overlap and it
is also much more homogeneous. An additional ingredient in the rapid
overlap phase results from the fact that hierarchical structure formation
models predict a galaxy formation rate that rises rapidly with time at the
relevant redshift range. This process leads to a state in which the
low-density IGM has been highly ionized and ionizing radiation reaches
everywhere except for gas located inside self-shielded, high-density
clouds. This marks the end of the overlap phase, and this important
landmark is most often referred to as the 'moment of reionization'.

Some neutral gas does, however, remain in high-density structures
which correspond to Lyman Limit systems and damped Ly$\alpha$
systems seen in absorption at lower redshifts. The high-density
regions are gradually ionized as galaxy formation proceeds, and the
mean ionizing intensity also grows with time. The ionizing intensity
continues to grow and to become more uniform as an increasing number
of ionizing sources is visible to every point in the IGM. This
``post-overlap'' phase continues indefinitely, since collapsed objects
retain neutral gas even in the present Universe. The IGM does,
however, reach another milestone at $z \sim 1.6$, the breakthrough
redshift \cite{Mad99}. Below this redshift, all
ionizing sources are visible to each other, while above this redshift
absorption by the Ly$\alpha$ forest implies that only
sources in a small redshift range are visible to a typical point in
the IGM.

Semi-analytic models of the pre-overlap stage focus on the evolution
of the H II  filling factor, i.e., the fraction of the volume of
the Universe which is filled by H II  regions. We distinguish
between the naive filling factor $F_{\rm H\ II}$ and the actual
filling factor or porosity $Q_{\rm H\ II}$. The naive filling factor
equals the number density of bubbles times the average volume of each,
and it may exceed unity since when bubbles begin to overlap the
overlapping volume is counted multiple times. However, as explained
below, in the case of reionization the linearity of the physics means
that $F_{\rm H\ II}$ is a very good approximation to $Q_{\rm H\ II}$
up to the end of the overlap phase of reionization.

The model of individual H II  regions presented in the previous
section can be used to understand the development of the total filling
factor. Starting with equation~(\ref{HIIreg}), if we assume a common
clumping factor $C$ for all H II  regions then we can sum each
term of the equation over all bubbles in a given large volume of the
Universe, and then divide by this volume. Then $V$ is replaced by the
filling factor and $\Ng$ by the total number of ionizing photons
produced up to some time $t$, per unit volume. The latter quantity
equals the mean number of ionizing photons per baryon times the mean
density of baryons $\nb_b$. Following the arguments leading to
equation~(\ref{rmax}), we find that if we include only stars then \beq
\frac {\nb_\gamma} {\nb_b}= \Ni F_{\rm col}\ , \label{ngnb} \eeq where
the collapse fraction $F_{\rm col}$ is the fraction of all the baryons
in the Universe which are in galaxies, i.e., the fraction of gas which
settles into halos and cools efficiently inside them. In writing
equation~(\ref{ngnb}) we are assuming instantaneous production of
photons, i.e., that the timescale for the formation and evolution of
the massive stars in a galaxy is short compared to the Hubble time at
the formation redshift of the galaxy. In a model based on
equation~(\ref{HIIreg}), the near-equality between $F_{\rm H\ II}$ and
$Q_{\rm H\ II}$ results from the linearity of this equation. First,
the total number of ionizations equals the total number of ionizing
photons produced by stars, i.e., all ionizing photons contribute
regardless of the spatial distribution of sources; and second, the
total recombination rate is proportional to the total ionized volume,
regardless of its topology. Thus, even if two or more bubbles overlap
the model remains an accurate approximation for $Q_{\rm H\ II}$ (at
least until $Q_{\rm H\ II}$ becomes nearly equal to 1). Note, however,
that there still are a number of important simplifications in the
model, including the assumption of a homogeneous (though possibly
time-dependent) clumping factor, and the neglect of feedback whereby
the formation of one galaxy may suppress further galaxy formation in
neighboring regions. These complications are discussed in detail below
and in \S \ref{sec6.5} and \S \ref{sec7}.

Under these assumptions we convert equation~(\ref{HIIreg}), which describes
individual H II regions, to an equation which statistically describes the
transition from a neutral Universe to a fully ionized one (compare to Madau
et al.\ 1999 \cite{Mad99} and Haiman \& Loeb 1997 \cite{hl97}): \beq
\frac{dQ_{\rm H\ II}}{dt}=\frac{\Ni}{0.76} \frac{dF_{\rm col}}{dt}-
\alpha_B \frac{C}{a^3} \nb_H^0 Q_{\rm H\ II}
\label{QIIeqn}\ , \eeq where we assumed a primordial mass fraction of 
hydrogen of 0.76. The solution (in analogy with
equation~(\ref{HIIsoln})) is \beq Q_{\rm H\ II}(t) =\int_{0}^t
\frac{\Ni} {0.76} \frac{dF_{\rm col}}{dt'}\,e^{F(t',t)} dt'\ , \eeq
where $F(t',t)$ is determined by equations~(\ref{Fgen})--(
\ref{fLCDM}).
  
\begin{figure}
\centering
\includegraphics[height=6cm]{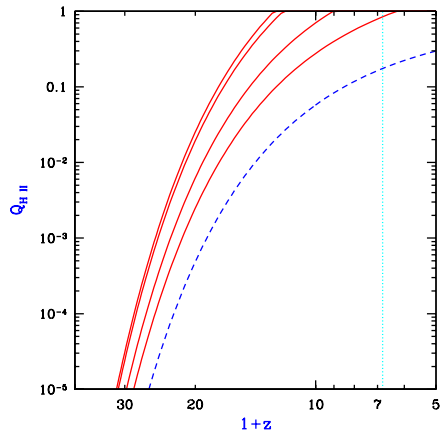}
\caption{Semi-analytic calculation of the reionization of the IGM (for
$\Ni=40$), showing the redshift evolution of the filling factor
$Q_{\rm H\ II}$. Solid curves show $Q_{\rm H\ II}$ for a clumping
factor $C=0$ (no recombinations), $C=1$, $C=10$, and $C=30$, in order
from left to right. The dashed curve shows the collapse fraction
$F_{\rm col}$, and the vertical dotted line shows the $z=5.8$
observational lower limit (Fan et al.\ 2000 \cite{f0}) on the reionization
redshift.}
\label{fig6c}
\end{figure}

A simple estimate of the collapse fraction at high redshift is the mass
fraction (given by equation~(\ref{PSerfc}) in the Press-Schechter model) in
halos above the cooling threshold, which is the minimum mass of halos in
which gas can cool efficiently. Assuming that only atomic cooling is
effective during the redshift range of reionization, the minimum mass
corresponds roughly to a halo of virial temperature $T_{\rm vir}=10^4$ K,
which can be converted to a mass using equation~(\ref{tvir}). With this
prescription we derive (for $\Ni=40$) the reionization history shown in
Fig. \ref{fig6c} for the case of a constant clumping factor $C$. The
solid curves show $Q_{\rm H\ II}$ as a function of redshift for a clumping
factor $C=0$ (no recombinations), $C=1$, $C=10$, and $C=30$, in order from
left to right. Note that if $C \sim 1$ then recombinations are unimportant,
but if $C \ga 10$ then recombinations significantly delay the reionization
redshift (for a fixed star-formation history). The dashed curve shows the
collapse fraction $F_{\rm col}$ in this model. For comparison, the vertical
dotted line shows the $z=5.8$ observational lower limit (Fan et al.\ 2000 \cite{f0})
on the reionization redshift.

Clearly, star-forming galaxies in CDM hierarchical models are capable of
ionizing the Universe at $z\sim 6$--15 with reasonable parameter
choices. This has been shown by a large number of theoretical,
semi-analytic calculations
\cite{fk94,sgb94,hl97,VS99,CO00,CFA00,Wyithe03,Cen03,Tumlinson04} as well
as numerical simulations
\cite{CO93,G097,g00,ANM99,RS99,Ciardi03,Sokasian03,Kohler05,Iliev05}. Similarly,
if a small fraction ($\la 1\%$) of the gas in each galaxy accretes onto a
central black hole, then the resulting mini-quasars are also able to
reionize the Universe, as has also been shown using semi-analytic models
\cite{fk94,HL98,VS99,Wyithe03}.

Although many models yield a reionization redshift around 7--12, the exact
value depends on a number of uncertain parameters affecting both the source
term and the recombination term in equation~(\ref{QIIeqn}). The source
parameters include the formation efficiency of stars and quasars and the
escape fraction of ionizing photons produced by these sources. The
formation efficiency of low mass galaxies may also be reduced by feedback
from galactic outflows. These parameters affecting the sources are
discussed elsewhere in this review (see \S \ref{sec6.1}, and
\ref{sec7}). Even when the clumping is inhomogeneous, the recombination
term in equation~(\ref{QIIeqn}) is generally valid if $C$ is defined as in
equation~(\ref{clump}), where we take a global volume average of the square
of the density inside ionized regions (since neutral regions do not
contribute to the recombination rate). The resulting mean clumping factor
depends on the density and clustering of sources, and on the distribution
and topology of density fluctuations in the IGM. Furthermore, the source
halos should tend to form in overdense regions, and the clumping factor is
affected by this cross-correlation between the sources and the IGM density.

Miralda-Escud\'e, Haehnelt, \& Rees (2000) \cite{MHR00} presented a simple model
for the distribution of density fluctuations, and more generally they
discussed the implications of inhomogeneous clumping during
reionization. They noted that as ionized regions grow, they more
easily extend into low-density regions, and they tend to leave behind
high-density concentrations, with these neutral islands being ionized
only at a later stage. They therefore argued that, since at
high-redshift the collapse fraction is low, most of the high-density
regions, which would dominate the clumping factor if they were
ionized, will in fact remain neutral and occupy only a tiny fraction
of the total volume. Thus, the development of reionization through the
end of the overlap phase should occur almost exclusively in the
low-density IGM, and the effective clumping factor during this time
should be $\sim 1$, making recombinations relatively unimportant (see
Fig. \ref{fig6c}). Only in the post-reionization phase,
Miralda-Escud\'e et al.\ (2000) \cite{MHR00} argued, do the high density clouds and
filaments become gradually ionized as the mean ionizing intensity
further increases.

\begin{figure}
\centering
\includegraphics[height=6cm]{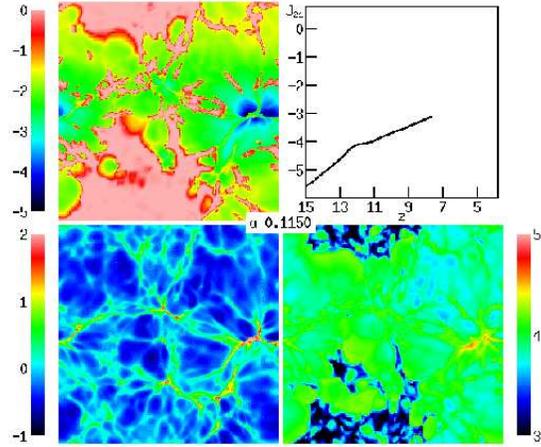}
\caption{Visualization at $z=7.7$ of a numerical simulation of
reionization, adopted from Figure 3c of \cite{g00}. The panels
display the logarithm of the neutral hydrogen fraction (upper left), the
gas density (lower left), and the gas temperature (lower right). Also shown
is the redshift evolution of the logarithm of the mean ionizing intensity
(upper right). Note the periodic boundary conditions.}
\label{fig6d}
\end{figure}

\begin{figure}
\centering
\includegraphics[height=6cm]{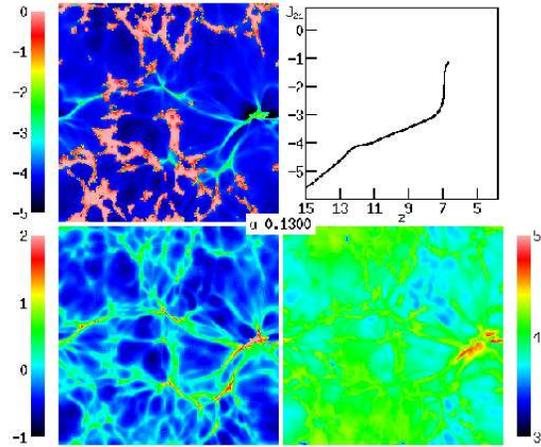}
\caption{Visualization at $z=6.7$ of a numerical simulation of
reionization, adopted from Figure 3e of \cite{g00}. The panels
display the logarithm of the neutral hydrogen fraction (upper left), the
gas density (lower left), and the gas temperature (lower right). Also shown
is the redshift evolution of the logarithm of the mean ionizing intensity
(upper right). Note the periodic boundary conditions.}
\label{fig6e}
\end{figure}

The complexity of the process of reionization is illustrated by the
numerical simulation of Gnedin \cite {g00} of stellar
reionization (in $\Lambda$CDM with $\Omm=0.3$). This simulation uses a
formulation of radiative transfer which relies on several rough
approximations; although it does not include the effect of shadowing behind
optically-thick clumps, it does include for each point in the IGM the
effects of an estimated local optical depth around that point, plus a local
optical depth around each ionizing source. This simulation helps to
understand the advantages of the various theoretical approaches, while
pointing to the complications which are not included in the simple
models. Figures \ref{fig6d} and \ref{fig6e}, taken from Figure 3 in 
\cite{g00}, show the state of the simulated Universe just
before and just after the overlap phase, respectively. They show a thin (15
$h^{-1}$ comoving kpc) slice through the box, which is 4 $h^{-1}$ Mpc on a
side, achieves a spatial resolution of $1 h^{-1}$ kpc, and uses $128^3$
each of dark matter particles and baryonic particles (with each baryonic
particle having a mass of $5\times 10^5 M_{\sun}$).  The figures show the
redshift evolution of the mean ionizing intensity $J_{21}$ (upper right
panel), and visually the logarithm of the neutral hydrogen fraction (upper
left panel), the gas density (lower left panel), and the gas temperature
(lower right panel). Note the obvious features resulting from the periodic
boundary conditions assumed in the simulation. Also note that the intensity
$J_{21}$ is defined as the intensity at the Lyman limit, expressed in units
of $10^{-21}\, \mbox{ erg cm}^{-2} \mbox{ s}^{-1} \mbox{ sr}
^{-1}\mbox{Hz}^{-1}$. For a given source emission, the intensity inside H
II regions depends on absorption and radiative transfer through the IGM
(e.g., Haardt \& Madau 1996 \cite{HM96}; Abel \& Haehnelt 1999 \cite{AH99})

Figure \ref{fig6d} shows the two-phase IGM at $z=7.7$, with ionized bubbles
emanating from one main concentration of sources (located at the right edge
of the image, vertically near the center; note the periodic boundary
conditions). The bubbles are shown expanding into low density regions and
beginning to overlap at the center of the image. The topology of ionized
regions is clearly complex: While the ionized regions are analogous to
islands in an ocean of neutral hydrogen, the islands themselves contain
small lakes of dense neutral gas. One aspect which has not been included in
theoretical models of clumping is clear from the figure. The sources
themselves are located in the highest density regions (these being the
sites where the earliest galaxies form) and must therefore ionize the gas
in their immediate vicinity before the radiation can escape into the low
density IGM. For this reason, the effective clumping factor is of order 100
in the simulation and also, by the overlap redshift, roughly ten ionizing
photons have been produced per baryon. Figure \ref{fig6e} shows that by
$z=6.7$ the low density regions have all become highly ionized along with a
rapid increase in the ionizing intensity. The only neutral islands left are
the highest density regions (compare the two panels on the left). However,
we emphasize that the quantitative results of this simulation must be
considered preliminary, since the effects of increased resolution and a
more accurate treatment of radiative transfer are yet to be
explored. Methods are being developed for incorporating a more complete
treatment of radiative transfer into three dimensional cosmological
simulations (e.g., \cite{ANM99,RS99,Ciardi03,Sokasian03,Kohler05,Iliev05}).

Gnedin, Ferrara, \& Zweibel (2000) \cite{GFZ00} investigated an additional effect
of reionization. They showed that the Biermann battery in cosmological
ionization fronts inevitably generates coherent magnetic fields of an
amplitude $\sim 10^{-19}$ Gauss. These fields form as a result of the
breakout of the ionization fronts from galaxies and their propagation
through the H I  filaments in the IGM. Although the fields are
too small to directly affect galaxy formation, they could be the seeds
for the magnetic fields observed in galaxies and X-ray clusters today.

If quasars contribute substantially to the ionizing intensity during
reionization then several aspects of reionization are modified compared to
the case of pure stellar reionization. First, the ionizing radiation
emanates from a single, bright point-source inside each host galaxy, and
can establish an escape route (H II  funnel) more easily than in the case
of stars which are smoothly distributed throughout the galaxy (\S
\ref{sec6.1}). Second, the hard photons produced by a quasar penetrate
deeper into the surrounding neutral gas, yielding a thicker ionization
front.  Finally, the quasar X-rays catalyze the formation of $H_2$
molecules and allow stars to keep forming in very small halos.

Oh (1999) \cite{Oh99} showed that star-forming regions may also produce
significant X-rays at high redshift. The emission is due to inverse Compton
scattering of CMB photons off relativistic electrons in the ejecta, as well
as thermal emission by the hot supernova remnant. The spectrum expected
from this process is even harder than for typical quasars, and the hard
photons photoionize the IGM efficiently by repeated secondary
ionizations. The radiation, characterized by roughly equal energy per
logarithmic frequency interval, would produce a uniform ionizing intensity
and lead to gradual ionization and heating of the entire IGM. Thus, if this
source of emission is indeed effective at high redshift, it may have a
crucial impact in changing the topology of reionization. Even if stars
dominate the emission, the hardness of the ionizing spectrum depends on the
initial mass function. At high redshift it may be biased toward massive,
efficiently ionizing stars, but this remains very much uncertain.

Semi-analytic as well as numerical models of reionization depend on an
extrapolation of hierarchical models to higher redshifts and lower-mass
halos than the regime where the models have been compared to observations
(see e.g. \cite{Wyithe03,Cen03,Tumlinson04}). These models have the
advantage that they are based on the current CDM paradigm which is
supported by a variety of observations of large-scale structure, galaxy
clustering, and the CMB. The disadvantage is that the properties of
high-redshift galaxies are derived from those of their host halos by
prescriptions which are based on low redshift observations, and these
prescriptions will only be tested once abundant data is available on
galaxies which formed during the reionization era (see \cite{Wyithe03} for
the sensitivity of the results to model parameters). An alternative
approach to analyzing the possible ionizing sources which brought about
reionization is to extrapolate from the observed populations of galaxies
and quasars at currently accessible redshifts. This has been attempted,
e.g., by Madau et al.\ (1999) \cite{Mad99} and Miralda-Escud\'e et al.\ (2000) \cite{MHR00}. The
general conclusion is that a high-redshift source population similar to the
one observed at $z=3$--4 would produce roughly the needed ionizing
intensity for reionization.  However, Dijkstra, Haiman, \& Loeb (2004)
\cite{Dijkstra04} constrained the role of quasars in reionizing the
Universe based on the unresolved flux of the X-ray background. At any
event, a precise conclusion remains elusive because of the same kinds of
uncertainties as those found in the models based on CDM: The typical escape
fraction, and the faint end of the luminosity function, are both not well
determined even at $z=3$--4, and in addition the clumping factor at high
redshift must be known in order to determine the importance of
recombinations. Future direct observations of the source population at
redshifts approaching reionization may help resolve some of these
questions.

\subsection{Photo-evaporation of Gaseous Halos After Reionization}
\label{sec6.4}

The end of the reionization phase transition resulted in the emergence
of an intense UV background that filled the Universe and heated the
IGM to temperatures of $\sim 1$--$2\times 10^4$K (see the previous
section). After ionizing the rarefied IGM in the voids and filaments
on large scales, the cosmic UV background penetrated the denser
regions associated with the virialized gaseous halos of the first
generation of objects. A major fraction of the collapsed gas had been
incorporated by that time into halos with a virial temperature $\la
10^4$K, where the lack of atomic cooling prevented the formation of
galactic disks and stars or quasars. Photoionization heating by the
cosmic UV background could then evaporate much of this gas back into
the IGM. The photo-evaporating halos, as well as those halos which did
retain their gas, may have had a number of important consequences just
after reionization as well as at lower redshifts.

In this section we focus on the process by which gas that had already
settled into virialized halos by the time of reionization was
evaporated back into the IGM due to the cosmic UV background. This
process was investigated by Barkana \& Loeb (1999) \cite{BL99} using 
semi-analytic methods and idealized numerical calculations. They first considered an 
isolated spherical, centrally-concentrated dark matter halo containing
gas. Since most of the photo-evaporation occurs at the end of overlap,
when the ionizing intensity builds up almost instantaneously, a sudden
illumination by an external ionizing background may be assumed.
Self-shielding of the gas implies that the halo interior sees a
reduced intensity and a harder spectrum, since the outer gas layers
preferentially block photons with energies just above the Lyman limit.
It is useful to parameterize the external radiation field by a
specific intensity per unit frequency, $\nu$, \beq J_{\nu}=10^{-21}\,
J_{21}\, \left(\frac{\nu}{\nu_L}\right)^ {-\alpha}\mbox{ erg
cm}^{-2}\mbox{ s}^{-1}\mbox{ sr} ^{-1}\mbox{Hz}^{-1}\ , \label{Inu}
\eeq where $\nu_L$ is the Lyman limit frequency, and $J_{21}$ is the
intensity at $\nu_L$ expressed in units of $10^{-21}\, \mbox{ erg
cm}^{-2} \mbox{ s}^{-1} \mbox{ sr} ^{-1}\mbox{Hz}^{-1}$. The intensity
is normalized to an expected post--reionization value of around unity
for the ratio of ionizing photon density to the baryon density.
Different power laws can be used to represent either quasar spectra
($\alpha \sim 1.8$) or stellar spectra ($\alpha \sim 5$).

Once the gas is heated throughout the halo, some fraction of it
acquires a sufficiently high temperature that it becomes unbound. This
gas expands due to the resulting pressure gradient and eventually
evaporates back into the IGM. The pressure gradient force (per unit
volume) 
$\kB \nabla (T \rho/\mu m_p)$ competes with the gravitational force of
$\rho\, G M/r^2$. Due to the density gradient, the ratio between the
pressure force and the gravitational force is roughly equal to the ratio
between the thermal energy $\sim \kB T$ and the gravitational binding
energy $\sim \mu m_p G M/r$ (which is $\sim \kB T_{\rm vir}$ at the virial
radius $r_{\rm vir}$) per particle. Thus, if the kinetic energy exceeds the
potential energy (or roughly if $T>T_{\rm vir}$), the repulsive pressure
gradient force exceeds the attractive gravitational force and expels the
gas on a dynamical time (or faster for halos with $T\gg T_{\rm vir}$).

\begin{figure}
\centering
\includegraphics[height=6cm]{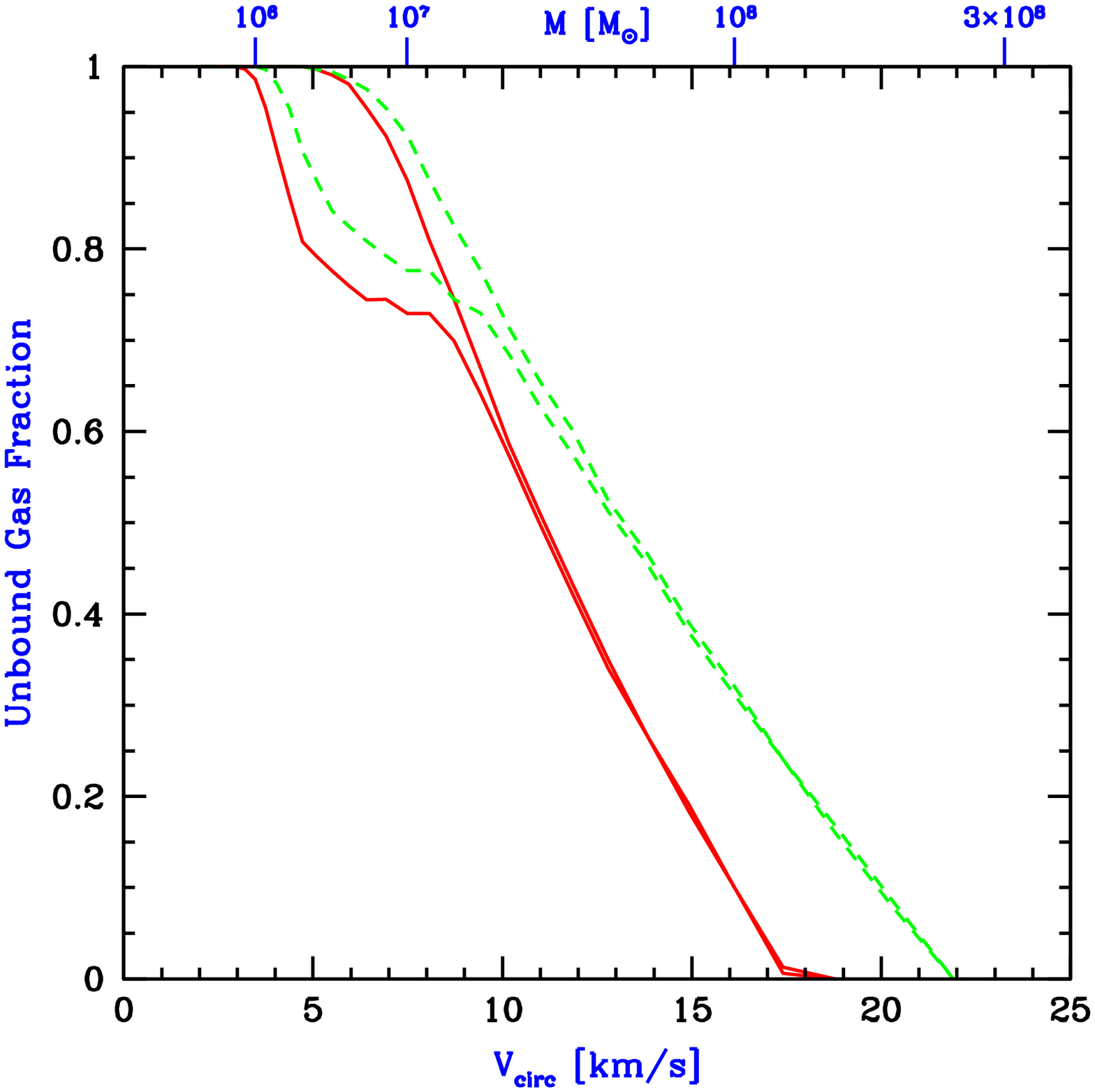}
\includegraphics[height=6cm]{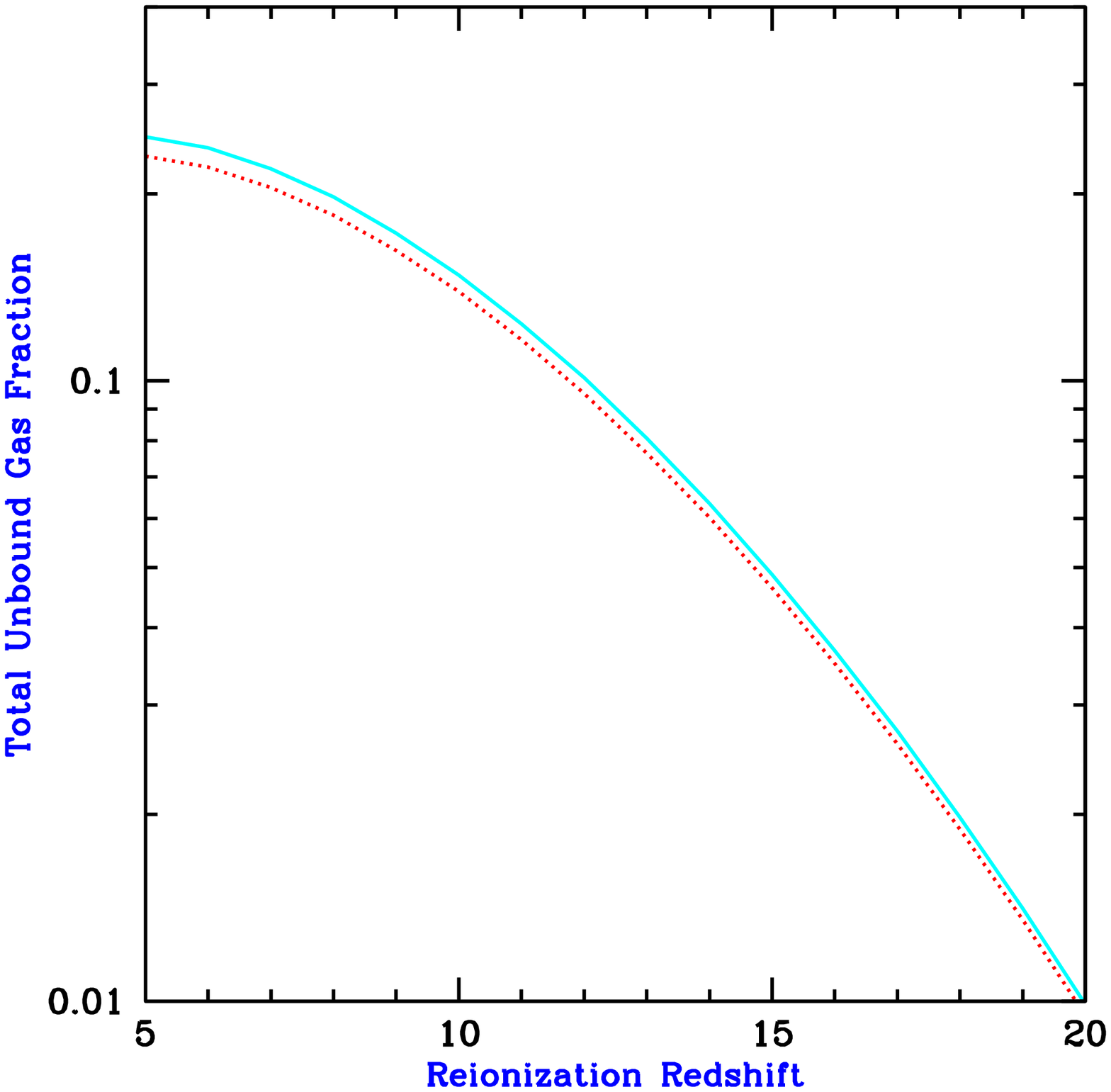}
\caption{Effect of photo-evaporation on individual halos and on the
overall halo population. The left panel shows the unbound gas fraction
(within the virial radius) versus total halo velocity dispersion or
mass, adopted from Figure 3 of Barkana \& Loeb (1999) \cite{BL99}. The two pairs
of curves correspond to spectral index $\alpha=5$ (solid) or
$\alpha=1.8$ (dashed), in each case at $z=8$. In each pair, assuming
an optically-thin halo leads to the upper curve, while the lower curve
shows the result of including radiative transfer and self
shielding. The right panel shows the total fraction of gas in the
Universe which evaporates from halos at reionization, versus the
reionization redshift, adopted from Figure 7 of Barkana \& Loeb
(1999) \cite{BL99}. The solid line assumes a spectral index $\alpha=1.8$, and the dotted line assumes 
$\alpha=5$.}
\label{fig6.7}
\end{figure}

The left panel of Figure \ref{fig6.7} (adopted from Fig. 3 of
Barkana \& Loeb 1999 \cite{BL99}) shows the fraction of gas within the virial
radius which becomes unbound after reionization, as a function of the
total halo circular velocity, with halo masses at $z=8$ indicated at
the top. The two pairs of curves correspond to spectral index
$\alpha=5$ (solid) or $\alpha=1.8$ (dashed). In each pair, a
calculation which assumes an optically-thin halo leads to the upper
curve, but including radiative transfer and self-shielding modifies
the result to the one shown by the lower curve. In each case
self-shielding lowers the unbound fraction, but it mostly affects only
a neutral core containing $\sim 30\%$ of the gas. Since high energy
photons above the Lyman limit penetrate deep into the halo and heat
the gas efficiently, a flattening of the spectral slope from
$\alpha=5$ to $\alpha=1.8$ raises the unbound gas fraction. This
figure is essentially independent of redshift if plotted in terms of
circular velocity, but the conversion to a corresponding mass does
vary with redshift. The characteristic circular velocity where most of
the gas is lost is $\sim 10$--$15~{\rm km~s^{-1}}$, but clearly the
effect of photo-evaporation is gradual, going from total gas removal
down to no effect over a range of a factor of $\sim 100$ in halo mass.

Given the values of the unbound gas fraction in halos of different masses,
the Press-Schechter mass function (\S \ref{sec2.4}) can be used to
calculate the total fraction of the IGM which goes through the process of
accreting onto a halo and then being recycled into the IGM at
reionization. The low-mass cutoff in this sum over halos is given by the
lowest mass halo in which gas has assembled by the reionization
redshift. This mass can be estimated by the linear Jeans mass $\mjeans$ in
equation (\ref{eq:m_j}). The Jeans mass does not in general precisely equal
the limiting mass for accretion (see the discussion in the next
section). Indeed, at a given redshift some gas can continue to fall into
halos of lower mass than the Jeans mass at that redshift. On the other
hand, the larger Jeans mass at higher redshifts means that a time-averaged
Jeans mass may be more appropriate, as indicated by the filtering mass. In
practice, the Jeans mass is sufficiently accurate since at $z\sim 10$--20
it agrees well with the values found in the numerical spherical collapse
calculations of Haiman, Thoul, \& Loeb (1996) \cite{Haiman}.

The right panel of Figure  \ref{fig6.7} (adopted from Fig. 7 of
Barkana \& Loeb 1999 \cite{BL99}) shows the total fraction of gas in the 
Universe which evaporates from halos at reionization, versus the reionization
redshift. The solid line assumes a spectral index $\alpha=1.8$, and
the dotted line assumes $\alpha=5$, showing that the result is
insensitive to the spectrum. Even at high redshift, the amount of gas
which participates in photo-evaporation is significant, which suggests
a number of possible implications as discussed below. The gas fraction
shown in the figure represents most ($\sim 60$--$80\%$ depending on
the redshift) of the collapsed fraction before reionization, although
some gas does remain in more massive halos.

The photo-evaporation of gas out of large numbers of halos may have
interesting implications. First, gas which falls into halos and is expelled
at reionization attains a different entropy than if it had stayed in the
low-density IGM. The resulting overall reduction in the entropy is expected
to be small -- the same as would be produced by reducing the temperature of
the entire IGM by a factor of $\sim 1.5$ -- but localized effects near
photo-evaporating halos may be more significant. Furthermore, the resulting
$\sim 20~{\rm km~s^{-1}}$ outflows induce small-scale fluctuations in
peculiar velocity and temperature. These outflows are usually well below
the resolution limit of most numerical simulations, but some outflows were
resolved in the simulation of Bryan et al.\ (1998) \cite{BMAN98}. The
evaporating halos may consume a significant number of ionizing photons in
the post-overlap stage of reionization \cite{HAM00,Iliev05}, but a
definitive determination requires detailed simulations which include the
three-dimensional geometry of source halos and sink halos.

Although gas is quickly expelled out of the smallest halos,
photo-evaporation occurs more gradually in larger halos which retain some
of their gas. These surviving halos initially expand but they continue to
accrete dark matter and to merge with other halos. These evaporating gas
halos could contribute to the high column density end of the Ly$\alpha$
forest \cite{BSS88}. Abel \& Mo (1998) \cite{AM98} suggested that, based on
the expected number of surviving halos, a large fraction of the Lyman limit
systems at $z\sim 3$ may correspond to mini-halos that survived
reionization. Surviving halos may even have identifiable remnants in the
present Universe. These ideas thus offer the possibility that a population
of halos which originally formed prior to reionization may correspond
almost directly to several populations that are observed much later in the
history of the Universe. However, the detailed dynamics of
photo-evaporating halos are complex, and detailed simulations are required
to confirm these ideas.  Photo-evaporation of a gas cloud has been followed
in a two dimensional simulation with radiative transfer, by Shapiro \& Raga
(2000) \cite{SR00}. They found that an evaporating halo would indeed appear
in absorption as a damped Ly$\alpha$ system initially, and as a weaker
absorption system subsequently. Future simulations \cite{Iliev05} will
clarify the contribution to quasar absorption lines of the entire
population of photo-evaporating halos.

\subsection{Suppression of the Formation of Low Mass Galaxies}
\label{sec6.5}

At the end of overlap, the cosmic ionizing background increased sharply,
and the IGM was heated by the ionizing radiation to a temperature $\ga
10^4$ K. Due to the substantial increase in the IGM temperature, the
intergalactic Jeans mass increased dramatically, changing the minimum mass
of forming galaxies \cite{MJR86,EF92,G097,MR98}.

Gas infall depends sensitively on the Jeans mass. When a halo more massive
than the Jeans mass begins to form, the gravity of its dark matter
overcomes the gas pressure. Even in halos below the Jeans mass, although
the gas is initially held up by pressure, once the dark matter collapses
its increased gravity pulls in some gas \cite{Haiman}. Thus, the Jeans mass
is generally higher than the actual limiting mass for accretion. Before
reionization, the IGM is cold and neutral, and the Jeans mass plays a
secondary role in limiting galaxy formation compared to cooling. After
reionization, the Jeans mass is increased by several orders of magnitude
due to the photoionization heating of the IGM, and hence begins to play a
dominant role in limiting the formation of stars. Gas infall in a reionized
and heated Universe has been investigated in a number of numerical
simulations. Thoul \& Weinberg (1996) \cite{Th96} inferred, based on a
spherically-symmetric collapse simulation, a reduction of $\sim 50\%$ in
the collapsed gas mass due to heating, for a halo of circular velocity
$V_c\sim 50\ {\rm km\ s}^{-1}$ at $z=2$, and a complete suppression of
infall below $V_c \sim 30\ {\rm km\ s}^{-1}$. Kitayama \& Ikeuchi (2000)
\cite{KI00} also performed spherically-symmetric simulations but included
self-shielding of the gas, and found that it lowers the circular velocity
thresholds by $\sim 5\ {\rm km\ s}^{-1}$. Three dimensional numerical
simulations \cite{QKE96,WHK97,NS97} found a significant suppression of gas
infall in even larger halos ($V_c \sim 75\ {\rm km\ s}^{-1}$), but this was
mostly due to a suppression of late infall at $z\la 2$.

When a volume of the IGM is ionized by stars, the gas is heated to a
temperature $T_{\rm IGM}\sim 10^4$ K. If quasars dominate the UV background
at reionization, their harder photon spectrum leads to $T_{\rm IGM}>
2\times 10^4$ K. Including the effects of dark matter, a given temperature
results in a linear Jeans mass corresponding to a halo circular velocity of
\beq V_J=81 \left(\frac{T_{\rm IGM}}{1.5\times 10^4 {\rm K}}\right)^{1/2}\
\left[\frac{1}{\Ommz}\ \frac{\Delta_c}{18 \pi^2}\right]^{1/6}\ {\rm km\
s}^{-1}, \eeq where we used equation~(\ref{Vceqn}) and assumed
$\mu=0.6$. In halos with $V_c>V_J$, the gas fraction in infalling gas
equals the universal mean of $\Omega_b/\Omega_m$, but gas infall is
suppressed in smaller halos. Even for a small dark matter halo, once it
collapses to a virial overdensity of $\Delta_c/\Ommz$ relative to the mean,
it can pull in additional gas. A simple estimate of the limiting circular
velocity, below which halos have essentially no gas infall, is obtained by
substituting the virial overdensity for the mean density in the definition
of the Jeans mass. The resulting estimate is \beq V_{\rm lim}=34
\left(\frac{T_{\rm IGM}}{1.5\times 10^4 {\rm K}}\right)^{1/2}\ {\rm km\
s}^{-1}. \eeq This value is in rough agreement with the numerical
simulations mentioned before.  A more recent study by Dijkstra et
al. (2004) \cite{Dijkstra04} indicates that at the high redshifts of $z>10$ gas could
nevertheless assemble into halos with circular velocities as low as
$v_c\sim 10~{\rm km~s^{-1}}$, even in the presence of a UV background.

Although the Jeans mass is closely related to the rate of gas infall at a
given time, it does not directly yield the total gas residing in halos at a
given time. The latter quantity depends on the entire history of gas
accretion onto halos, as well as on the merger histories of halos, and an
accurate description must involve a time-averaged Jeans mass. Gnedin
\cite{Gnedin2000b} showed that the gas content of halos in simulations is
well fit by an expression which depends on the filtering mass, a particular
time-averaged Jeans mass (Gnedin \& Hui 1998 \cite{Gnedin98}). Gnedin
\cite{Gnedin2000b} calculated the Jeans and filtering masses using the mean
temperature in the simulation to define the sound speed, and found the
following fit to the simulation results: \beq \bar{M_g}=\frac{f_b M}{\left
[1+ \left(2^{1/3}-1\right) M_C/M \right]^3}\ , \eeq where $\bar{M_g}$ is
the average gas mass of all objects with a total mass $M$,
$f_b=\Omega_b/\Omm$ is the universal baryon fraction, and the
characteristic mass $M_C$ is the total mass of objects which on average
retain $50\%$ of their gas mass. The characteristic mass was well fit by
the filtering mass at a range of redshifts from $z=4$ up to $z\sim 15$.

The reionization process was not perfectly synchronized throughout the
Universe. Large-scale regions with a higher density than the mean tend to
form galaxies first and reionize earlier than underdense regions (see
detailed discussion in \S \ref{scatter}). The suppression of low-mass
galaxies by reionization will therefore be modulated by the fluctuations in
the timing of reionization.  Babich \& Loeb (2005) \cite{Bab} considered
the effect of inhomogeneous reionization on the power-spectrum of low-mass
galaxies.  They showed that the shape of the high redshift galaxy power
spectrum on small scales in a manner which depends on the details of epoch
of reionization. This effect is significantly larger than changes in the
galaxy power spectrum due to the current uncertainty in the inflationary
parameters, such as the tilt of the scalar power spectrum $n$ and the
running of the tilt $\alpha$.  Therefore, future high redshift galaxies
surveys hoping to constrain inflationary parameters must properly model the
effects of reionization, but conversely they will also be sensitive to the
thermal history of the high redshift intergalactic medium.


\section{\bf Feedback from Galactic Outflows}
\label{sec7}

\subsection{Propagation of Supernova Outflows in the IGM}
\label{sec7.1}

Star formation is accompanied by the violent death of massive stars in
supernova explosions. In general, if each halo has a fixed baryon fraction
and a fixed fraction of the baryons turns into massive stars, then the
total energy in supernovae outflows is proportional to the halo mass. The
binding energy of the gas in the halo is proportional to the halo mass
squared. Thus, outflows are expected to escape more easily out of low-mass
galaxies, and to expel a greater fraction of the gas from dwarf
galaxies. At high redshifts, most galaxies form in relatively low-mass
halos, and the high halo merger rate leads to vigorous star
formation. Thus, outflows may have had a great impact on the earliest
generations of galaxies, with consequences that may include metal
enrichment of the IGM and the disruption of dwarf galaxies. In this
subsection we present a simple model for the propagation of individual
supernova shock fronts in the IGM. We discuss some implications of this
model, but we defer to the following subsection the brunt of the discussion
of the cosmological consequences of outflows.

For a galaxy forming in a given halo, the supernova rate is related to the
star formation rate. In particular, for a Scalo (1998) \cite{Sca98} initial
stellar mass function, if we assume that a supernova is produced by each
$M>8 M_{\odot}$ star, then on average one supernova explodes for every 126
$M_{\odot}$ of star formation, expelling an ejecta mass of $\sim 3\,
M_{\odot}$ including $\sim 1\, M_{\odot}$ of heavy elements. We assume that
the individual supernovae produce expanding hot bubbles which merge into a
single overall region delineated by an outwardly moving shock front. We
assume that most of the baryons in the outflow lie in a thin shell, while
most of the thermal energy is carried by the hot interior. The total
ejected mass equals a fraction $\fg$ of the total halo gas which is lifted
out of the halo by the outflow. This gas mass includes a fraction $\fe$ of
the mass of the supernova ejecta itself (with $\fe \le 1$ since some metals
may be deposited in the disk and not ejected). Since at high redshift most
of the halo gas is likely to have cooled onto a disk, we assume that the
mass carried by the outflow remains constant until the shock front reaches
the halo virial radius. We assume an average supernova energy of
$10^{51}E_{51}$ erg, a fraction $\fw$ of which remains in the outflow after
it escapes from the disk. The outflow must overcome the gravitational
potential of the halo, which we assume to have a Navarro, Frenk, \& White
(1997) \cite{Na97} density profile [NFW; see equation (\ref{NFW})]. Since the entire
shell mass must be lifted out of the halo, we include the total shell mass
as well as the total injected energy at the outset. This assumption is
consistent with the fact that the burst of star formation in a halo is
typically short compared to the total time for which the corresponding
outflow expands.

The escape of an outflow from an NFW halo depends on the concentration
parameter $\cN$ of the halo. Simulations by Bullock et al.\ (2000) \cite{Bu00}
indicate that the concentration parameter decreases with redshift, and
their results may be extrapolated to our regime of interest (i.e., to
smaller halo masses and higher redshifts) by assuming that \beq
\cN=\left(\frac{M}{10^9 M_{\sun}}\right) ^{-0.1}\, \frac{25}{(1+z)}\
. \eeq Although we calculate below the dynamics of each outflow in
detail, it is also useful to estimate which halos can generate
large-scale outflows by comparing the kinetic energy of the outflow to
the potential energy needed to completely escape (i.e., to infinite
distance) from an NFW halo. We thus find that the outflow can escape
from its originating halo if the circular velocity is below a critical
value given by \beq V_{\rm crit} = 200 \sqrt{\frac{E_{51}\fw
(\eta/0.1)} {\fg\ g(\cN)}}\ {\rm km\ s}^{-1} \ , \label{Vcrit} \eeq
where the efficiency $\eta$ is the fraction of baryons incorporated in
stars, and \beq g(x)=\frac{x^2}{(1+x)\ln(1+x)-x} \ .  \eeq Note that
the contribution to $\fg$ of the supernova ejecta itself is $0.024
\eta \fe$, so the ejecta mass is usually negligible unless $\fg \la
1\%$. Equation (\ref{Vcrit}) can also be used to yield the maximum gas
fraction $\fg$ which can be ejected from halos, as a function of their
circular velocity. Although this equation is most general, if we
assume that the parameters $\fg$ and $\fw$ are independent of $M$ and
$z$ then we can normalize them based on low-redshift observations. If
we specify $\cN \sim 10$ (with $g(10)=6.1$) at $z=0$, then setting
$E_{51}=1$ and $\eta=10\%$ yields the required energy efficiency as a
function of the ejected halo gas fraction: \beq \fw = 1.5 \fg
\left[\frac{V_{\rm crit}}{100\ {\rm km\ s}^{-1}} \right]^2\ . \eeq A
value of $V_{\rm crit} \sim 100\ {\rm km\ s}^{-1}$ is suggested by
several theoretical and observational arguments which are discussed in
the next subsection. However, these arguments are not conclusive, and
$V_{\rm crit}$ may differ from this value by a large factor,
especially at high redshift (where outflows are observationally
unconstrained at present). Note the degeneracy between $\fg$ and $\fw$
which remains even if $V_{\rm crit}$ is specified. Thus, if $V_{\rm
crit} \sim 100\ {\rm km\ s}^{-1}$ then a high efficiency $\fw \sim 1$
is required to eject most of the gas from all halos with $V_c < V_{\rm
crit}$, but only $\fw \sim 10\%$ is required to eject 5--10$\%$ of the
gas. The evolution of the outflow does depend on the value of $\fw$
and not just the ratio $\fw/\fg$, since the shell accumulates material
from the IGM which eventually dominates over the initial mass carried
by the outflow.

We solve numerically for the spherical expansion of a galactic outflow,
elaborating on the basic approach of Tegmark, Silk, \& Evrard (1993)
\cite{TSE93}. We assume that most of the mass $m$ carried along by the
outflow lies in a thin, dense, relatively cool shell of proper radius
$R$. The interior volume, while containing only a fraction $\fin \ll 1$ of
the mass $m$, carries most of the thermal energy in a hot, isothermal
plasma of pressure $p_{\rm int}$ and temperature $T$. We assume a uniform
exterior gas, at the mean density of the Universe (at each redshift), which
may be neutral or ionized, and may exert a pressure $p_{\rm ext}$ as
indicated below. We also assume that the dark matter distribution follows
the NFW profile out to the virial radius, and is at the mean density of the
Universe outside the halo virial radius. Note that in reality an overdense
distribution of gas as well as dark matter may surround each halo due to
secondary infall.

The shell radius $R$ in general evolves as follows: \beq m \frac{d^2R}
{dt^2}= 4 \pi R^2 \delta p-\left (\frac{dR}{dt} - H R\right) \frac{dm}
{dt}- \frac{G m} {R^2} \left(M(R)+ \frac{1} {2}m \right) + \frac{8}{3}
\pi G R m \rho_{\Lambda} \ , \eeq where the right-hand-side includes
forces due to pressure, sweeping up of additional mass, gravity, and a
cosmological constant, respectively.  The shell is accelerated by
internal pressure and decelerated by external pressure, i.e., $\delta
p=p_{\rm int}-p_{\rm ext}$. In the gravitational force, $M(R)$ is the
total enclosed mass, not including matter in the shell, and $\frac{1}
{2}m$ is the effective contribution of the shell mass in the
thin-shell approximation \cite {OM88}. The interior
pressure is determined by energy conservation, and evolves according
to \cite{TSE93}: \beq \frac{d p_{\rm int}} {dt}=\frac{L}{2
\pi R^3}-5\, \frac{p_{\rm int}} {R}\, \frac{d R}{dt} \ , \eeq where
the luminosity $L$ incorporates heating and cooling terms. We include
in $L$ the supernova luminosity $L_{\rm sn}$ (during a brief initial
period of energy injection), cooling terms $L_{\rm cool}$, ionization
$L_{\rm ion}$, and dissipation $L_{\rm diss}$. For simplicity, we
assume ionization equilibrium for the interior plasma, and a
primordial abundance of hydrogen and helium. We include in $L_{\rm
cool}$ all relevant atomic cooling processes in hydrogen and helium,
i.e., collisional processes, Bremsstrahlung emission, and Compton
cooling off the CMB. Compton scattering is the dominant cooling
process for high-redshift outflows. We include in $L_{\rm ion}$ only
the power required to ionize the incoming hydrogen upstream, at the
energy cost of 13.6 eV per hydrogen atom. The interaction between the
expanding shell and the swept-up mass dissipates kinetic energy. The
fraction $f_d$ of this energy which is re-injected into the interior
depends on complex processes occurring near the shock front, including
turbulence, non-equilibrium ionization and cooling, and so (following
Tegmark et al.\ 1993 \cite{TSE93}) we let \beq L_{\rm diss}=\frac{1}{2} f_d
\frac{dm}{dt} \left( \frac{dR}{dt} - H R \right)^2\ , \eeq where we
set $f_d=1$ and compare below to the other extreme of $f_d=0$.

In an expanding Universe, it is preferable to describe the propagation of
outflows in terms of comoving coordinates since, e.g., the critical result
is the maximum {\it comoving}\, size of each outflow, since this size
yields directly the total IGM mass which is displaced by the outflow and
injected with metals. Specifically, we apply the following transformation
\cite{Shan80,Voit96}: \beq d\hat{t}=a^{-2} dt,\ \ \hat{R}=a^{-1}R,\ \
\hat{p}=a^5 p,\ \hat{\rho}= a^3 \rho\ . \eeq For $\Oml=0$, Voit (1996)
\cite{Voit96} obtained (with the time origin $\hat{t}=0$ at redshift
$z_1$): \beq \hat{t}=\frac{2}{\Omm H_0} \left[ \sqrt{1+\Omm
z_1}-\sqrt{1+\Omm z}\, \right]\ , \eeq while for $\Omm+\Oml=1$ there is no
simple analytic expression. We set $\beta=\hat{R}/\hat{r}_{\rm vir}$, in
terms of the virial radius $r_{\rm vir}$ [equation (\ref{rvir})] of the
source halo. We define $\alpha_S^1$ as the ratio of the shell mass $m$ to
$\frac{4}{3} \pi \hat{\rho}_b\, \hat{r}_{\rm vir}^3$, where
$\hat{\rho}_b=\rho_b(z=0)$ is the mean baryon density of the Universe at
$z=0$. More generally, we define \beq \alpha_S(\beta) \equiv
\frac{m}{\frac{4}{3} \pi \hat{\rho}_b \, \hat{\rho}^3} = \left\{
\begin{array}{ll} \alpha_S^1 / \beta^3 & \mbox{if $\beta<1$} \\
1+\left(\alpha_S^1-1 \right)/\beta^3 & \mbox{otherwise.}
\end{array} \right. \eeq Here we assumed, as noted above, that the shell
mass is constant until the halo virial radius is reached, at which point
the outflow begins to sweep up material from the IGM. We thus derive the
following equations: \beq \frac{d^2 \hat{R}} {d \hat{t}^2}= \left\{
\begin{array}{ll} \frac{3}{\alpha_S(\beta)} \frac{\hat{p}}{\hat{\rho}_b\,
\hat{R} }-\frac{a}{2} \hat{R} H_0^2 \Omm \bar{\delta} (\beta) & \mbox{if
$\beta<1$} \vspace{.1in} \\ \frac{3}{\alpha_S(\beta) \hat{R}}
\left[\frac{\hat{p}}{\hat{\rho}_b}-\left( \frac{d\hat{R}} {d\hat{t}}
\right)^2 \right]-\frac{a}{2} \hat{R} H_0^2 \Omm \bar{\delta} (\beta) +
\frac{a}{4} \hat{R} H_0^2 \Omega_b \alpha_S(\beta) & \mbox{otherwise,}
\end{array} \right. \eeq along with \beq \frac{d}{d\hat{t}} \left(
\hat{R}^5 \hat{p}_{\rm int} \right)= \frac{a^4}{2 \pi} L \hat{R}^2\ . \eeq
In the evolution equation for $\hat{R}$, for $\beta < 1$ we assume for
simplicity that the baryons are distributed in the same way as the dark
matter, since in any case the dark matter halo dominates the gravitational
force. For $\beta>1$, however, we correct (via the last term on the
right-hand side) for the presence of mass in the shell, since at $\beta \gg
1$ this term may become important. The $\beta > 1$ equation also includes
the braking force due to the swept-up IGM mass. The enclosed mean
overdensity for the NFW profile [Eq. (\ref{NFW})] surrounded by matter at
the mean density is \beq \bar\delta(\beta)= \left\{
\begin{array}{ll} \frac{ \Delta_c}{\Ommz \beta^3} \frac{\ln (1+\cN
\beta) - \cN \beta/(1+ \cN \beta)} {\ln (1+c )-c/ (1+c)}& \mbox{if
$\beta<1$} \vspace{.1in} \\ \left( \frac{ \Delta_c}{\Ommz}-1
\right)\frac{1} {\beta^3} & \mbox{otherwise.}
\end{array} \right. \eeq

The physics of supernova shells is discussed in Ostriker \& McKee (1988)
\cite{OM88} along with a number of analytical solutions.  The propagation of
cosmological blast waves has also been computed by Ostriker \& Cowie (1981)
\cite{OC81}, Bertschinger (1985) \cite{Be85} and Carr \& Ikeuchi (1985)
\cite{CI85}.  Voit (1996) \cite{Voit96} derived an exact analytic solution to the fluid equations
which, although of limited validity, is nonetheless useful for understanding
roughly how the outflow size depends on several of the parameters.  The solution
requires an idealized case of an outflow which at all times expands into a
homogeneous IGM.  Peculiar gravitational forces, and the energy lost in escaping
from the host halo, are neglected, cooling and ionization losses are also assumed
to be negligible, and the external pressure is not included.  The dissipated energy
is assumed to be retained, i.e., $f_d$ is set equal to unity.  Under these
conditions, the standard Sedov self-similar solution 
\cite{Sed59,Sed93} generalizes to the cosmological case as follows 
\cite{Voit96}:
\beq \hat{R}=\left( \frac{\xi \hat{E}_0}{ \hat{\rho}_b} \right) ^{1/5}
\hat{t}^{\,2/5}\ , \label{Voit} \eeq where $\xi=2.026$ and $\hat{E}_0=
E_0/(1+z_1)^2$ in terms of the initial (i.e., at $t=\hat{t}=0$ and $z=z_1$) energy
$E_0$.  Numerically, the comoving radius is \beq \hat{R}= 280 \left(
\frac{0.022}{\Omega_b h^2}\, \frac{E_0}{10^{56}{\rm erg}} \right) ^{1/5}
\left(\frac{10}{1+z_1} \, \frac{\hat{t}}{10^{10}{\rm yr}} \right) ^{2/5}\ {\rm kpc}\ .  \eeq

In solving the equations described above, we assume that the shock
front expands into a pre-ionized region which then recombines after a
time determined by the recombination rate. Thus, the external pressure
is included initially, it is turned off after the pre-ionized region
recombines, and it is then switched back on at a lower redshift when
the Universe is reionized. When the ambient IGM is neutral and the
pressure is off, the shock loses energy to ionization. In practice we
find that the external pressure is unimportant during the initial
expansion, although it {\it is}\, generally important after
reionization. Also, at high redshift ionization losses are much
smaller than losses due to Compton cooling. In the results shown
below, we assume an instantaneous reionization at $z=9$.

\begin{figure}
\centering
\includegraphics[height=6cm]{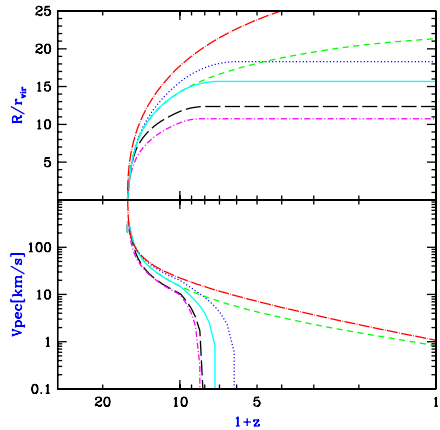}
\caption{Evolution of a supernova outflow from a $z=15$ halo of circular
velocity $V_c=20$ km/s. Plotted are the shell comoving radius in units of
the virial radius of the source halo (top panel), and the physical peculiar
velocity of the shock front (bottom panel). Results are shown for the
standard parameters $\fin=0.1$, $f_d=1$, $\fw=75\%$, and $\fg=50\%$ (solid
curve). Also shown for comparison are the cases of no cooling (dotted
curve), no reionization (short-dashed curve), $f_d=0$ (long-dashed curve),
or $\fw=15\%$ and $\fg=10\%$ (dot-short dashed curve), as well as the
simple Voit (1996) \cite{Voit96} solution of equation (\ref{Voit}) for the
standard parameter set (dot-long dashed curve). In cases where the outflow
halts, we freeze the radius at the point of maximum expansion.}
\label{figVII1}
\end{figure}

Figure \ref{figVII1} shows the results for a starting redshift $z=15$, for
a halo of mass $5.4 \times 10^7 M_{\sun}$, stellar mass $8.0 \times 10^5
M_{\sun}$, comoving $\hat{r}_{\rm vir}=12$ kpc, and circular velocity
$V_c=20$ km/s. We show the shell comoving radius in units of the virial
radius of the source halo (top panel), and the physical peculiar velocity
of the shock front (bottom panel). Results are shown (solid curve) for the
standard set of parameters $\fin=0.1$, $f_d=1$, $\fw=75\%$, and
$\fg=50\%$. For comparison, we show several cases which adopt the standard
parameters except for no cooling (dotted curve), no reionization
(short-dashed curve), $f_d=0$ (long-dashed curve), or $\fw=15\%$ and
$\fg=10\%$ (dot-short dashed curve). When reionization is included, the
external pressure halts the expanding bubble. We freeze the radius at the
point of maximum expansion (where $d \hat{R}/d\hat{t}=0$), since in reality
the shell will at that point begin to spread and fill out the interior
volume due to small-scale velocities in the IGM. For the chosen parameters,
the bubble easily escapes from the halo, but when $\fw$ and $\fg$ are
decreased the accumulated IGM mass slows down the outflow more
effectively. In all cases the outflow reaches a size of 10--20 times
$\hat{r}_{\rm vir}$, i.e., 100--200 comoving kpc. If all the metals are
ejected (i.e., $\fe=1$), then this translates to an average metallicity in
the shell of $\sim 1$--5$\times 10^{-3}$ in units of the solar metallicity
(which is $2\%$ by mass). The asymptotic size of the outflow varies roughly
as $\fw^{1/5}$, as predicted by the simple solution in equation
(\ref{Voit}), but the asymptotic size is rather insensitive to $\fg$ (at a
fixed $\fw$) since the outflow mass becomes dominated by the swept-up IGM
mass once $\hat{R} \ga 4 \hat{r}_{\rm vir}$. With the standard parameter
values (i.e., those corresponding to the solid curve), Figure \ref{figVII1}
also shows (dot-long dashed curve) the Voit (1996) \cite{Voit96} solution
of equation (\ref{Voit}). The Voit solution behaves similarly to the
no-reionization curve at low redshift, although it overestimates the shock
radius by $\sim 30\%$, and the overestimate is greater compared to the more
realistic case which does include reionization.

\begin{figure}
\centering
\includegraphics[height=6cm]{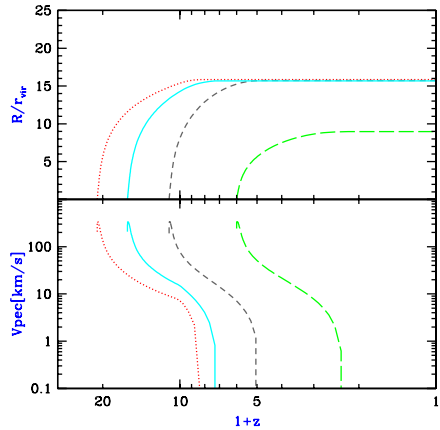}
\caption{Evolution of supernova outflows at different redshifts. The
top and bottom panels are arranged similarly to Figure \ref{figVII1}.
The $z=15$ outflow (solid curve) is repeated from Figure
\ref{figVII1}, and it is compared here to outflows with the same
parameters but starting at $z=20$ (dotted curve), $z=10$ (short-dashed
curve), and $z=5$ (long-dashed curve). A $V_c=20$ km/s halo is assumed
except for $z=5$, in which case a $V_c=42$ km/s halo is assumed to
produce the outflow (see text).}
\label{figVII2}
\end{figure}

Figure \ref{figVII2} shows different curves than Figure \ref{figVII1} but
on an identical layout. A single curve starting at $z=15$ (solid curve) is
repeated from Figure \ref{figVII1}, and it is compared here to outflows
with the same parameters but starting at $z=20$ (dotted curve), $z=10$
(short-dashed curve), and $z=5$ (long-dashed curve). A $V_c=20$ km/s halo,
with a stellar mass equal to $1.5\%$ of the total halo mass, is chosen at
the three higher redshifts, but at $z=5$ a $V_c=42$ km/s halo is
assumed. Because of the suppression of gas infall after reionization, we
assume that the $z=5$ outflow is produced by supernovae from a stellar mass
equal to only $0.3\%$ of the total halo mass (with a similarly reduced
initial shell mass), thus leading to a relatively small final shell
radius. The main conclusion from both figures is the following: In all
cases, the outflow undergoes a rapid initial expansion over a fractional
redshift interval $\delta z/z \sim 0.2$, at which point the shell has
slowed down to $\sim 10$ km/s from an initial 300 km/s. The rapid
deceleration is due to the accumulating IGM mass. External pressure from
the reionized IGM completely halts all high-redshift outflows, and even
without this effect most outflows would only move at $\sim 10$ km/s after
the brief initial expansion. Thus, it may be possible for high-redshift
outflows to pollute the Lyman alpha forest with metals without affecting
the forest hydrodynamically at $z \la 4$. While the bulk velocities of
these outflows may dissipate quickly, the outflows do sweep away the IGM
and create empty bubbles. The resulting effects on observations of the
Lyman alpha forest should be studied in detail (some observational
signatures of feedback have been suggested recently by Theuns, Mo, \&
Schaye 2000 \cite{TMS00}).

Furlanetto \& Loeb (2003) \cite{FL03} derived the evolution of the
characteristic scale and filling fraction of supernova-driven bubbles based
on a refinement of this formalism (see also their 2001 paper for
quasar-driven outflows).  The role of metal-rich outflows in smearing the
transition epoch between Pop-III (metal-free) and Pop II (metal-enriched)
stars, was also analysed by Furlanetto \& Loeb (2005) \cite{FL05}, who
concluded that a double-reionization history in which the ionization
fraction goes through two (or more) peaks is unlikely.


\subsection{Effect of Outflows on Dwarf Galaxies and on the IGM}
\label{sec7.2}

Galactic outflows represent a complex feedback process which affects
the evolution of cosmic gas through a variety of phenomena. Outflows
inject hydrodynamic energy into the interstellar medium of their host
galaxy. As shown in the previous subsection, even a small fraction of
this energy suffices to eject most of the gas from a dwarf galaxy,
perhaps quenching further star formation after the initial burst. At
the same time, the enriched gas in outflows can mix with the
interstellar medium and with the surrounding IGM, allowing later
generations of stars to form more easily because of metal-enhanced
cooling. On the other hand, the expanding shock waves may also strip
gas in surrounding galaxies and suppress star formation.

Dekel \& Silk (1986) \cite{DS86} attempted to explain the different
properties of diffuse dwarf galaxies in terms of the effect of galactic
outflows. They noted the observed trends whereby lower-mass dwarf galaxies
have a lower surface brightness and metallicity, but a higher mass-to-light
ratio, than higher mass galaxies. They argued that these trends are most
naturally explained by substantial gas removal from an underlying dark
matter potential. Galaxies lying in small halos can eject their remaining
gas after only a tiny fraction of the gas has turned into stars, while
larger galaxies require more substantial star formation before the
resulting outflows can expel the rest of the gas. Assuming a wind
efficiency $\fw \sim 100\%$, Dekel \& Silk showed that outflows in halos
below a circular velocity threshold of $V_{\rm crit} \sim 100$ km/s have
sufficient energy to expel most of the halo gas. Furthermore, cooling is
very efficient for the characteristic gas temperatures associated with
$V_{\rm crit} \la 100$ km/s halos, but it becomes less efficient in more
massive halos. As a result, this critical velocity is expected to signify a
dividing line between bright galaxies and diffuse dwarf galaxies. Although
these simple considerations may explain a number of observed trends, many
details are still not conclusively determined. For instance, even in
galaxies with sufficient energy to expel the gas, it is possible that this
energy gets deposited in only a small fraction of the gas, leaving the rest
almost unaffected.

Since supernova explosions in an inhomogeneous interstellar medium lead to
complicated hydrodynamics, in principle the best way to determine the basic
parameters discussed in the previous subsection ($\fw$, $\fg$, and $\fe$)
is through detailed numerical simulations of individual galaxies. Mac Low
\& Ferrara (1999) \cite{Mac99} simulated a gas disk within a $z=0$ dark
matter halo. The disk was assumed to be azimuthally symmetric and initially
smooth. They represented supernovae by a central source of energy and mass,
assuming a constant luminosity which is maintained for 50 million
years. They found that the hot, metal-enriched ejecta can in general escape
from the halo much more easily than the colder gas within the disk, since
the hot gas is ejected in a tube perpendicular to the disk without
displacing most of the gas in the disk. In particular, most of the metals
were expelled except for the case with the most massive halo considered
(with $10^9 M_{\sun}$ in gas) and the lowest luminosity ($10^{37}$ erg/s,
or a total injection of $2 \times 10^{52}$ erg). On the other hand, only a
small fraction of the total gas mass was ejected except for the least
massive halo (with $10^6 M_{\sun}$ in gas), where a luminosity of $10^{38}$
erg/s or more expelled most of the gas. We note that beyond the standard
issues of numerical resolution and convergence, there are several
difficulties in applying these results to high-redshift dwarf
galaxies. Clumping within the expanding shells or the ambient interstellar
medium may strongly affect both the cooling and the hydrodynamics. Also,
the effect of distributing the star formation throughout the disk is
unclear since in that case several characteristics of the problem will
change; many small explosions will distribute the same energy over a larger
gas volume than a single large explosion [as in the Sedov (1959)
\cite{Sed59} solution; see, e.g., equation (\ref{Voit})], and the geometry
will be different as each bubble tries to dig its own escape route through
the disk. Also, high-redshift disks should be denser by orders of magnitude
than $z=0$ disks, due to the higher mean density of the Universe at early
times.  Thus, further numerical simulations of this process are required in
order to assess its significance during the reionization epoch.

Some input on these issues also comes from observations. Martin (1999)
\cite{Mart99} showed that the hottest extended X-ray emission in galaxies
is characterized by a temperature of $\sim 10^{6.7}$ K. This hot gas, which
is lifted out of the disk at a rate comparable to the rate at which gas
goes into new stars, could escape from galaxies with rotation speeds of
$\la 130$ km/s. However, these results are based on a small sample which
includes only the most vigorous star-forming local galaxies, and the
mass-loss rate depends on assumptions about the poorly understood transfer
of mass and energy among the various phases of the interstellar medium.

Many authors have attempted to estimate the overall cosmological effects of
outflows by combining simple models of individual outflows with the
formation rate of galaxies, obtained via semi-analytic methods
\cite{CR86,TSE93,Voit96,NT97,FPS00,SB00} or numerical simulations
\cite{G097,Gne98,CO99,AHWKG2000a}. The main goal of these calculations is
to explain the characteristic metallicities of different environments as a
function of redshift. For example, the IGM is observed to be enriched with
metals at redshifts $z\la 5$. Identification of C IV, Si IV an O VI
absorption lines which correspond to Ly$\alpha$ absorption lines in the
spectra of high-redshift quasars has revealed that the low-density IGM has
been enriched to a metal abundance (by mass) of $Z_{\rm IGM}\sim 10^{-2.5
(\pm 0.5)}Z_\odot$, where $Z_\odot=0.019$ is the solar metallicity
\cite{MY87,Tyt95,SC96,LSBR98,CS98,Song97,ESSP00}. The metal enrichment has
been clearly identified down to H I column densities of $\sim
10^{14.5}~{\rm cm^{-2}}$. The detailed comparison of cosmological
hydrodynamic simulations with quasar absorption spectra has established
that the forest of Ly$\alpha$ absorption lines is caused by the
smoothly-fluctuating density of the neutral component of the IGM
\cite{CMOR94,ZAN95,HKWM96}. The simulations show a strong correlation
between the H I column density and the gas overdensity $\delta_{\rm gas}$
\cite{DHKW99}, implying that metals were dispersed into regions with an
overdensity as low as $\delta_{\rm gas}\sim 3$ or possibly even lower.

In general, dwarf galaxies are expected to dominate metal enrichment at
high-redshift for several reasons. As noted above and in the previous
subsection, outflows can escape more easily out of the potential wells of
dwarfs. Also, at high redshift, massive halos are rare and dwarf halos are
much more common. Finally, as already noted, the Sedov (1959) \cite{Sed59}
solution [or equation (\ref{Voit})] implies that for a given total energy
and expansion time, multiple small outflows fill large volumes more
effectively than would a smaller number of large outflows. Note, however,
that the strong effect of feedback in dwarf galaxies may also quench star
formation rapidly and reduce the efficiency of star formation in dwarfs
below that found in more massive galaxies.

Cen \& Ostriker (1999) \cite{CO99} showed via numerical simulation that
metals produced by supernovae do not mix uniformly over cosmological
volumes.  Instead, at each epoch the highest density regions have much
higher metallicity than the low-density IGM.  They noted that early star
formation occurs in the most overdense regions, which therefore reach a
high metallicity (of order a tenth of the solar value) by $z \sim 3$, when
the IGM metallicity is lower by 1--2 orders of magnitude.  At later times,
the formation of high-temperature clusters in the highest-density regions
suppresses star formation there, while lower-density regions continue to
increase their metallicity.  Note, however, that the spatial resolution of
the hydrodynamic code of Cen \& Ostriker is a few hundred kpc, and anything
occurring on smaller scales is inserted directly via simple parametrized
models.  Scannapieco \& Broadhurst (2000) \cite{SB00} implemented expanding
outflows within a numerical scheme which, while not a full gravitational
simulation, did include spatial correlations among halos.  They showed that
winds from low-mass galaxies may also strip gas from nearby galaxies (see
also Scannapieco, Ferrara, \& Broadhurst 2000 \cite{SFB00}), thus
suppressing star formation in a local neighborhood and substantially
reducing the overall abundance of galaxies in halos below a mass of $\sim
10^{10} M_{\sun}$.  Although quasars do not produce metals, they may also
affect galaxy formation in their vicinity via energetic outflows
\cite{ER88,BW91,Sil98,NSS98}.

Gnedin \& Ostriker (1997) \cite{G097} and Gnedin (1998) \cite{Gne98}
identified another mixing mechanism which, they argued, may be dominant at
high redshift ($z \ga 4$).  In a collision between two protogalaxies, the
gas components collide in a shock and the resulting pressure force can
eject a few percent of the gas out of the merger remnant.  This is the
merger mechanism, which is based on gravity and hydrodynamics rather than
direct stellar feedback.  Even if supernovae inject most of their metals in
a local region, larger-scale mixing can occur via mergers.  Note, however,
that Gnedin's (1998) \cite{Gne98} simulation assumed a comoving star
formation rate at $z \ga 5$ of $\sim 1 M_{\sun}$ per year per comoving
Mpc$^3$, which is 5--10 times larger than the observed rate at redshift
3--4.  Aguirre et al.\ \cite{AHWKG2000a} used outflows implemented in
simulations to conclude that winds of $\sim 300$ km/s at $z \la 6$ can
produce the mean metallicity observed at $z \sim 3$ in the Ly$\alpha$
forest.  In a separate paper Aguirre et al. \cite{AHKGW2000b} explored
another process, where metals in the form of dust grains are driven to
large distances by radiation pressure, thus producing large-scale mixing
without displacing or heating large volumes of IGM gas.  The success of
this mechanism depends on detailed microphysics such as dust grain
destruction and the effect of magnetic fields.  The scenario, though, may
be directly testable because it leads to significant ejection only of
elements which solidify as grains.

Feedback from galactic outflows encompasses a large variety of processes
and influences. The large range of scales involved, from stars or quasars
embedded in the interstellar medium up to the enriched IGM on cosmological
scales, make possible a multitude of different, complementary approaches,
promising to keep galactic feedback an active field of research.

\section{The Frontier of 21cm Cosmology}

\subsection{Mapping Hydrogen Before Reionization}

The small residual fraction of free electrons after cosmological
recombination coupled the temperature of the cosmic gas to that of the
cosmic microwave background (CMB) down to a redshift, $z\sim 200$
\cite{Peebles}. Subsequently, the gas temperature dropped adiabatically as
$T_{\rm gas}\propto (1+z)^2$ below the CMB temperature $T_{\gamma}\propto
(1+z)$.  The gas heated up again after being exposed to the photo-ionizing
ultraviolet light emitted by the first stars during the {\it reionization
epoch} at $z\lesssim 20$. Prior to the formation of the first stars, the
cosmic neutral hydrogen must have resonantly absorbed the CMB flux through
its spin-flip 21cm transition \cite{Field,Scott,Tozzi,Zalda04}.  The linear
density fluctuations at that time should have imprinted anisotropies on the
CMB sky at an observed wavelength of $\lambda=21.12[(1+z)/100]$ meters. We
discuss these early 21cm fluctuations mainly for pedagogical purposes.
Detection of the earliest 21cm signal will be particularly challenging
because the foreground sky brightness rises as $\lambda^{2.5}$ at long
wavelengths in addition to the standard $\sqrt{\lambda}$ scaling of the
detector noise temperature for a given integration time and fractional
bandwidth.  The discussion in this section follows Loeb \& Zaldarriaga
(2004) \cite{Loeb04}.

\begin{figure} 
\centering
\includegraphics[width=0.7\columnwidth,angle=-90]{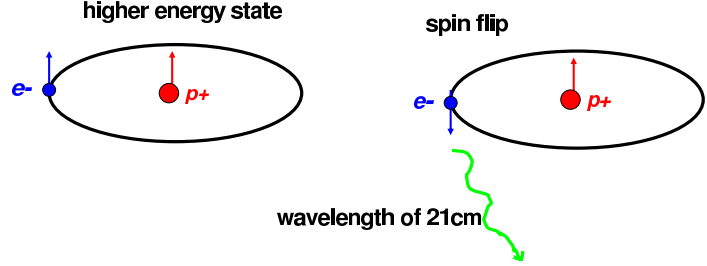} 
\caption{The 21cm transition of hydrogen. The higher energy level the spin
of the electron (e-) is aligned with that of the proton (p+).  A spin flip
results in the emission of a photon with a wavelength of 21cm (or a
frequency of 1420MHz).}
\label{21cm}
\end{figure}

We start by calculating the history of the spin temperature, $T_s$, defined
through the ratio between the number densities of hydrogen atoms in the
excited and ground state levels, ${n_1/ n_0}=(g_1/
g_0)\exp\left\{-{T_\star/ T_s}\right\},$ 
\begin{equation} 
{n_1\over
n_0}={g_1\over g_0}\exp\left\{-{T_\star\over T_s}\right\}, 
\label{eq:spin}
\end{equation} 
where subscripts $1$ and $0$ correspond to the excited and
ground state levels of the 21cm transition, $(g_1/g_0)=3$ is the ratio of
the spin degeneracy factors of the levels, $n_{\rm H}=(n_0+n_1)\propto
(1+z)^3$ is the total hydrogen density, and $T_\star=0.068$K is the
temperature corresponding to the energy difference between the levels.  The
time evolution of the density of atoms in the ground state is given by,
\begin{eqnarray} \left( {\partial\over \partial t} + 
3{{\dot a}\over a} \right) & n_0 & =-n_0\left(C_{01}+B_{01}I_\nu\right)
\nonumber \\ + & n_1 & \left(C_{10}+A_{10}+B_{10}I_\nu\right),
\label{eq:evolution} \end{eqnarray} where $a(t)=(1+z)^{-1}$ is the cosmic
scale factor, $A$'s and $B$'s are the Einstein rate coefficients, $C$'s are
the collisional rate coefficients, and $I_\nu$ is the blackbody intensity
in the Rayleigh-Jeans tail of the CMB, namely
$I_\nu=2kT_{\gamma}/\lambda^2$ with $\lambda=21$ cm \cite{RL}.  Here a dot
denotes a time-derivative.  The $0\rightarrow 1$ transition rates can be
related to the $1\rightarrow 0$ transition rates by the requirement that in
thermal equilibrium with $T_s=T_\gamma=T_{\rm gas}$, the right-hand-side of
Eq. (\ref{eq:evolution}) should vanish with the collisional terms balancing
each other separately from the radiative terms. The Einstein coefficients
are $A_{10}=2.85\times 10^{-15}~{\rm s^{-1}}$, $B_{10}=(\lambda^3/2hc)
A_{10}$ and $B_{01}=(g_1/g_0)B_{10}$ \cite{Field,RL}.  The collisional
de-excitation rates can be written as $C_{10}={4\over 3} \kappa(1-0) n_{\rm
H}$, where $\kappa(1-0)$ is tabulated as a function of $T_{\rm gas}$
\cite{AD,Zyg}.

\begin{figure}[th]
\centering
\includegraphics[height=6cm]{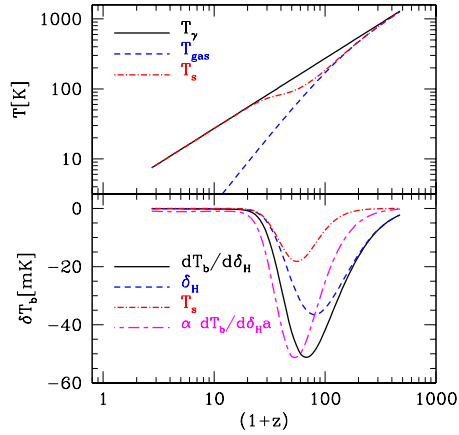}
\caption{{\it Upper panel:} Evolution of the gas, CMB and spin temperatures
with redshift [4].  {\it Lower panel:} ${dT_b/d\dh}$ as function of
redshift.  The separate contributions from fluctuations in the density and
the spin temperature are depicted. We also show ${dT_b/d\dh} a \propto
{dT_b/d\dh} \times \dh$, with an arbitrary normalization.}
\label{dtb}
\end{figure}

Equation (\ref{eq:evolution}) can
be simplified to the form,
\begin{eqnarray}
{d\Upsilon \over dz} & = & -\left[H(1+z)\right]^{-1}
\left[-\Upsilon(C_{01}+B_{01}I_\nu) \right. \nonumber \\ 
&& \left. +
(1-\Upsilon)(C_{10}+A_{10}  + B_{10}I_\nu)\right],
\label{eq:upsilon}
\end{eqnarray}
where $\Upsilon\equiv n_0/n_{\rm H}$, $H\approx
H_0\sqrt{\Omega_m}(1+z)^{3/2}$ is the Hubble parameter at high redshifts
(with a present-day value of $H_0$), and $\Omega_m$ is the density
parameter of matter.  The upper panel of Fig. \ref{dtb} shows the results
of integrating Eq.~(\ref{eq:upsilon}). Both the spin temperature and the
kinetic temperature of the gas track the CMB temperature down to $z\sim
200$. Collisions are efficient at coupling $T_s$ and $T_{gas}$ down to
$z\sim 70$ and so the spin temperature follows the kinetic temperature
around that redshift. At much lower redshifts, the Hubble expansion makes
the collision rate subdominant relative the radiative coupling rate to the
CMB, and so $T_s$ tracks $T_{\gamma}$ again. Consequently, there is a
redshift window between $30\lesssim z \lesssim 200$, during which the
cosmic hydrogen absorbs the CMB flux at its resonant 21cm
transition. Coincidentally, this redshift interval precedes the appearance
of collapsed objects \cite{BL01} and so its signatures are not contaminated by
nonlinear density structures or by radiative or hydrodynamic feedback
effects from stars and quasars, as is the case at lower redshifts
\cite{Zalda04}.

During the period when the spin temperature is smaller than the CMB
temperature, neutral hydrogen atoms absorb CMB photons. The resonant 21cm
absorption reduces the brightness temperature of the CMB by,
\begin{equation}
T_b =\tau \left( T_s-T_{\gamma}\right)/(1+z) ,
\end{equation}
where the optical depth for resonant 21cm absorption is,
\begin{equation}
\tau= {3c\lambda^2hA_{10}n_{\rm H}\over 32 \pi k T_s H(z)} .
\label{eq:tau} 
\end{equation}

Small inhomogeneities in the hydrogen density $\dh\equiv (n_{\rm H}-{\bar
n_{\rm H}})/{\bar n}_{\rm H}$ result in fluctuations of the 21cm absorption
through two separate effects. An excess of neutral hydrogen directly
increases the optical depth and also alters the evolution of the spin
temperature. For now, we ignore the additional effects of peculiar
velocities (Bharadwaj \& Ali 2004 \cite{Indian}; Barkana \& Loeb 2004 \cite{BL04a}) as well as
fluctuations in the gas kinetic temperature due to the adiabatic
compression (rarefaction) in overdense (underdense) regions \cite{BLinf}.
Under these approximations, we can write an equation for the resulting
evolution of $\Upsilon$ fluctuations,
\begin{eqnarray}
{d\delta \Upsilon \over dz} & = & \left[H(1+z)\right]^{-1}
\left\{[C_{10}+C_{01}+(B_{01}+B_{10})I_\nu]\delta\Upsilon \right. \nonumber \\
 && \left. + \left[ C_{01} \Upsilon
-C_{10}(1-\Upsilon)\right]\dh\right\},
\label{eq:dup}
\end{eqnarray}
leading to spin temperature fluctuations,
\begin{equation}
{\delta T_s\over {\bar T}_s}= -{1\over
\ln[3\Upsilon/(1-\Upsilon)]}{\delta\Upsilon\over \Upsilon (1-\Upsilon)}.
\label{eq:dts}
\end{equation}
The resulting brightness temperature fluctuations can be related to the derivative,
\begin{equation}
{\delta T_b\over {\bar T}_b}=\dh +  {T_{\gamma}\over ({\bar T}_s-T_{\gamma})}
{\delta T_s\over {\bar T}_s}.
\label{eq:dtb}
\end{equation}
The spin temperature fluctuations ${\delta T_s/ {T}_s}$ are proportional to
the density fluctuations and so we define, 
\beq
\label{weight} 
{d T_b \over d \dh} \equiv {\bar T}_b + {T_{\gamma} {\bar T}_b
\over ({\bar T}_s-T_{\gamma})} {\delta T_s\over {\bar T}_s \dh}, 
\eeq
through ${\delta T_b}=({d T_b /d \dh}) \dh$.  We ignore fluctuations
in $C_{ij}$ due to
fluctuations in $T_{\rm gas}$ which are very small \cite{AD}.
Figure \ref{dtb} shows ${dT_b/d\dh}$ as a function of redshift, including
the two contributions to ${dT_b/d\dh}$, one originating directly from
density fluctuations and the second from the associated changes in the spin
temperature \cite{Scott}.  Both contributions have the same sign, because
an increase in density raises the collision rate and lowers the spin
temperature and so it allows $T_s$ to better track $T_{\rm gas}$.  Since
$\dh$ grows with time as $\dh \propto a$, the signal peaks at $z\sim 50$, a
slightly lower redshift than the peak of ${dT_b/d\dh}$.

Next we calculate the angular power spectrum of the brightness temperature
on the sky, resulting from density perturbations with a power spectrum
$P_{\delta}(k)$, 
\beq
\label{pdel} \langle \dh(\k_1) \dh(\k_2) \rangle = (2\pi)^3
\delta^D(\k_1+\k_2) P_{\delta}(k_1).  \eeq where $\dh(\k)$ is the Fourier
tansform of the hydrogen density field, $\k$ is the comoving wavevector,
and $\langle \cdots \rangle$ denotes an ensemble average (following the
formalism described in \cite{Zalda04}).  The 21cm brightness temperature
observed at a frequency $\nu$ corresponding to a distance $r$ along the
line of sight, is given by \beq
\label{los} \delta T_b(\n,\nu)=\int dr W_{\nu}(r)
\ {dT_b \over d\dh} \dh(\n,r), \eeq where $\n$ denotes the direction of
observation, $W_{\nu}(r)$ is a narrow function of $r$ that peaks at the
distance corresponding to $\nu$. The details of this function depend on the
characteristics of the experiment.  The brightness fluctuations in Eq.
\ref{los}
can be expanded in spherical harmonics with expansion coefficients
$a_{lm}(\nu)$. The angular power spectrum of map $C_{l}(\nu)= \langle
|a_{lm}(\nu)|^2 \rangle$ can be expressed in terms of the 3D power spectrum
of fluctuations in the density $P_{\delta}(k)$, \beqa
\label{fcldef} C_{l}(\nu)&=&4
\pi \int {d^3k \over (2\pi)^3} P_{\delta}(k) \alpha_l^2(k,\nu) \nonumber \\
\alpha_l(k,\nu)&=& \int dr W_{r_0}(r) {dT_b\over d\dh}(r) j_l(kr).  \eeqa
Our calculation ignores inhomogeneities in the hydrogen ionization
fraction, since they freeze at the earlier recombination epoch ($z\sim
10^3$) and so their amplitude is more than an order of magnitude smaller
than $\dh$ at $z\lesssim 100$.  The gravitational
potential perturbations induce a redshift distortion effect that is of
order $\sim (H/ck)^2$ smaller than $\dh$ for the
high--$l$ modes of interest here.  

\begin{figure}
\centering
\includegraphics[height=6cm]{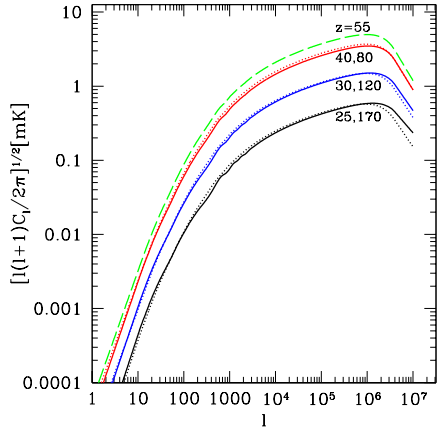}
\caption{Angular power spectrum of 21cm anisotropies on the sky at
various redshifts.  From
top to bottom, $z=55,40,80,30,120,25,170$.}
\label{clfig}
\end{figure}

Figure \ref{clfig} shows the angular power spectrum at various
redshifts. 
The signal peaks around $z\sim 50$ but maintains a substantial
amplitude over the full range of $30\lesssim z\lesssim 100$.  
The ability to probe the small scale power of density fluctuations is only
limited by the Jeans scale, below which the dark matter inhomogeneities are
washed out by the finite pressure of the gas. Interestingly, the
cosmological Jeans mass reaches its minimum value, $\sim 3\times 10^4
M_\odot$, within the redshift interval of interest here which 
corresponds to modes of angular scale $\sim$ arcsecond on the sky. During
the epoch of reionization, photoionization heating raises the Jeans mass by
several orders of magnitude and broadens spectral features, thus limiting
the ability of other probes of the intergalactic medium, such as the
Ly$\alpha$ forest, from accessing the same very low mass scales. The 21cm
tomography has the additional advantage of probing the majority of the
cosmic gas, instead of the trace amount ($\sim 10^{-5}$) of neutral
hydrogen probed by the Ly$\alpha$ forest after reionization.  Similarly to
the primary CMB anisotropies, the 21cm signal is simply shaped by gravity,
adiabatic cosmic expansion, and well-known atomic physics, and is not
contaminated by complex astrophysical processes that affect the
intergalactic medium at $z\lesssim 30$.

Characterizing the initial fluctuations is one of the
primary goals of observational cosmology, as it offers a window into the
physics of the very early Universe, namely the epoch of inflation during
which the fluctuations are believed to have been produced. 
In most models of inflation, the evolution of the Hubble parameter during
inflation leads to departures from a scale-invariant spectrum that are of
order $1/N_{\rm efold}$ with $N_{\rm efold}\sim 60$ being the number of
$e$--folds between the time when the scale of our horizon was of order the
horizon during inflation and the end of inflation \cite{lidlith}. 
Hints that the standard
$\Lambda$CDM model may have too much power on galactic scales have inspired
several proposals for suppressing the power on small scales. Examples
include the possibility that the dark matter is warm and it decoupled while
being relativistic so that its free streaming erased small-scale power
\cite{Ba01}, or direct modifications of inflation that produce a cut-off in
the power on small scales \cite{kamlidd}. An unavoidable collisionless
component of the cosmic mass budget beyond CDM, is provided by massive
neutrinos (see \cite{neut} for a review). Particle physics experiments
established the mass splittings among different species which translate
into a lower limit on the fraction of the dark matter accounted for by
neutrinos of $f_\nu > 0.3 \%$, while current constraints based on galaxies
as tracers of the small scale power imply $f_\nu < 12 \%$ \cite{tegsdss}.

\begin{figure}
\centering
\includegraphics[height=6cm]{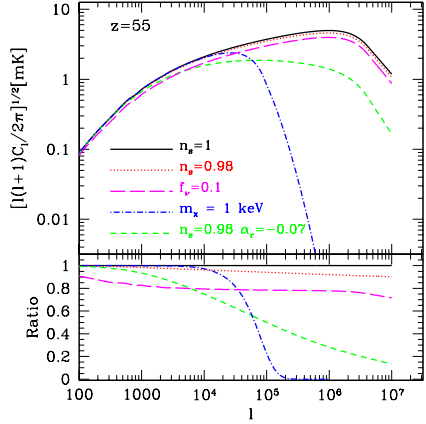}
\caption{{\it Upper panel:} Power spectrum of 21cm anisotropies at $z=55$
for a $\Lambda$CDM scale-invariant power spectrum, a model with $n=0.98$, a
model with $n=0.98$ and $\alpha_r\equiv {1\over 2} (d^2\ln P/d\ln
k^2)=-0.07$, a model of warm dark matter particles with a mass of 1 keV,
and a model in which $f_\nu=10\% $ of the matter density is in three
species of massive neutrinos with a mass of $0.4~{\rm eV}$ each. {\it Lower
panel:} Ratios between the different power spectra and the scale-invariant
spectrum. }
\label{clcomp}
\end{figure}

Figure \ref{clcomp} shows the 21cm power spectrum for various
models that differ in their level of small scale power. It is clear that a
precise measurement of the 21cm power spectrum will dramatically improve
current constraints on alternatives to the standard $\Lambda$CDM spectrum.

The 21cm signal contains a wealth of information about the 
initial fluctuations. A full sky map at a single photon 
frequency measured up to $l_{\rm max}$, can 
probe the power spectrum up to $k_{\rm max}\sim (l_{\rm max}/10^4) {\rm
Mpc}^{-1}$. Such a map contains $l_{\rm max}^2$ independent samples. By
shifting the photon frequency, one may obtain many independent measurements
of the power. When measuring a mode $l$, which corresponds to a wavenumber
$k\sim l/r$, two maps at different photon frequencies will be independent
if they are separated in radial distance by $1/k$. Thus, an experiment that
covers a spatial range $\Delta r$ can probe a total of $k\Delta r\sim l
\Delta r/r$ independent maps. An experiment that detects the 21cm signal
over a range $\Delta\nu$ centered on a frequency $\nu$, is sensitive to
$\Delta r/r\sim 0.5 (\Delta\nu/\nu)(1+z)^{-1/2}$, and so it 
measures a total of $N_{\rm 21cm}\sim 3 \times 10^{16} (l_{\rm max}/10^6)^3
(\Delta\nu/\nu) (z/100)^{-1/2}$ independent samples.

This detection capability cannot be reproduced even remotely by other
techniques. For example, the primary CMB anisotropies are damped on small
scales (through the so-called Silk damping), and probe only modes with $l
\leq 3000$ ($k\leq 0.2 \ {\rm Mpc}^{-1}$). The total number of modes
available in the full sky is $N_{\rm cmb} = 2 l_{\rm max}^2\sim 2\times
10^7 (l_{\rm max}/3000)^2$, including both temperature and polarization
information.

The sensitivity of an experiment depends strongly on its particular design,
involving the number and distribution of the antennae for an
interferometer. Crudely speaking, the uncertainty in the measurement of
$[{l(l+1)C_l/2\pi}]^{1/2}$ is dominated by noise, $N_\nu$, which is
controlled by 
the sky brightness $I_\nu$ at the observed frequency $\nu$ \cite{Zalda04}, 
\beqa
\label{err} 
N_\nu && \sim 0.4 {\rm mK }\left({I_\nu [50{\rm MHz}]\over 5\times 10^{5} {
\rm Jy\ sr^{-1}}}\right) \left({l_{\rm min}\over 35}\right) \left( {5000
\over l_{\rm max}}\right) \left( {0.016 \over f_{\rm cover}}\right)
\nonumber \\&& \times \left({1 \ {\rm year} \over t_0}\right)^{1/2}
\left({\Delta \nu \over 50{\rm MHz}
}\right)^{-1/2} \ \left({50\ {\rm MHz }\over \nu}\right)^{2.5} , \eeqa
where $l_{\rm min}$ is the minimum observable $l$ as determined by the
field of view of the instruments, $l_{\rm max}$ is the maximum observable
$l$ as determined by the maximum separation of the antennae, $f_{\rm
cover}$ is the fraction of the array area thats is covered by telescopes,
$t_0$ is the observation time and $\Delta \nu$ is the frequency range over
which the signal can be detected.  Note that the assumed sky temperature of
$0.7\times 10^4$K at $\nu=50$MHz (corresponding to $z\sim 30$) is more than
six orders of magnitude larger than the signal.  We have already included
the fact that several independent maps can be produced by varying the
observed frequency.  The numbers adopted above are appropriate for the
inner core of the {\it LOFAR} array ({\it http://www.lofar.org}), planned
for initial operation in 2006. The predicted signal is $\sim 1{\rm mK}$,
and so a year of integration or an increase in the covering fraction are
required to observe it with {\it LOFAR}.  Other experiments whose goal is
to detect 21cm fluctuations from the subsequent epoch of reionization at
$z\sim 6-12$ (when ionized bubbles exist and the fluctuations are larger)
include the Mileura Wide-Field Array (MWA;
http://web.haystack.mit.edu/arrays/MWA/), the Primeval Structure Telescope
({\it PAST}; {\it http://arxiv.org/abs/astro-ph/0502029}), and in the more
distant future the Square Kilometer Array ({\it SKA}; {\it
http://www.skatelescope.org}).  The main challenge in detecting the
predicted signal from higher redshifts involves its appearance at low
frequencies where the sky noise is high.  Proposed space-based instruments
\cite{RadioAstr} avoid the terrestrial radio noise and the increasing
atmospheric opacity at $\nu< 20 \ {\rm MHz}$ (corresponding to $z> 70$).

\begin{figure}
\centering
\includegraphics[height=6cm]{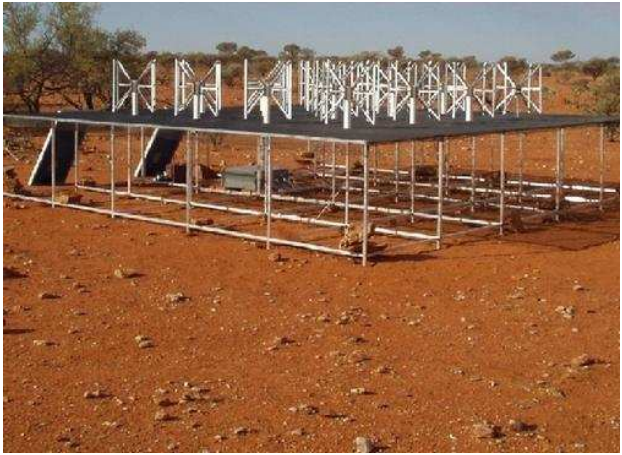}
\caption{Prototype of the tile design for the {\it Mileura Wide-Field
Array} (MWA) in western Australia, aimed at detecting redshifted 21cm from
the epoch of reionization. Each 4m$\times$4m tile contains 16 dipole
antennas operating in the frequency range of 80--300MHz. Altogether the
initial phase of MWA (the so-called ``Low-Frequency Demostrator'') will
include 500 antenna tiles with a total collecting area of 8000 m$^2$ at
150MHz, scattered across a 1.5 km region and providing an angular
resolution of a few arcminutes.}
\end{figure}

The 21cm absorption is replaced by 21cm emission from neutral hydrogen as
soon as the intergalactic medium is heated above the CMB temperature by
X-ray sources during the epoch of reionization \cite{ChenMi}. This occurs
long before reionization since the required heating requires only a modest
amount of energy, $\sim 10^{-2}~{\rm eV}[(1+z)/30]$, which is three orders
of magnitude smaller than the amount necessary to ionize the Universe.  As
demonstrated by Chen \& Miralda-Escude (2004) \cite{ChenMi}, heating due
the recoil of atoms as they absorb Ly$\alpha$ photons \cite{Madau} is not
effective; the Ly$\alpha$ color temperature reaches equilibrium with the
gas kinetic temperature and suppresses subsequent heating before the level
of heating becomes substantial. Once most of the cosmic hydrogen is
reionized at $z_{\rm reion}$, the 21cm signal is diminished. The optical
depth for free-free absorption after reionization, $\sim 0.1 [(1+z_{\rm
reion})/20]^{5/2}$, modifies only slightly the expected 21cm anisotropies.
Gravitational lensing should modify the power spectrum \cite{Pen} at high
$l$, but can be separated as in standard CMB studies (see \cite{lensing}
and references therein).  The 21cm signal should be simpler to clean as it
includes the same lensing foreground in independent maps obtained at
different frequencies.

\begin{figure}
\centering
\includegraphics[height=10cm,angle=-90]{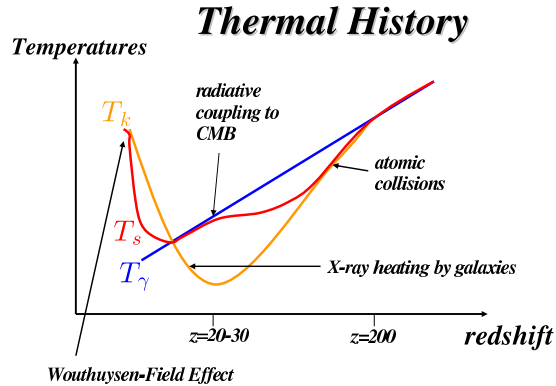}
\caption{Schematic sketch of the evolution of the kinetic temperature
($T_k$) and spin temperature ($T_s$) of cosmic hydrogen. Following
cosmological recombination at $z\sim 10^3$, the gas temperature (orange
curve) tracks the CMB temperature (blue line; $T_\gamma\propto (1+z)$) down
to $z\sim 200$ and then declines below it ($T_k\propto (1+z)^2$) until the
first X-ray sources (accreting black holes or exploding supernovae) heat it
up well above the CMB temperature. The spin temperature of the 21cm
transition (red curve) interpolates between the gas and CMB
temperatures. Initially it tracks the gas temperature through collisional
coupling; then it tracks the CMB through radiative coupling; and eventually
it tracks the gas temperature once again after the production of a cosmic
background of UV photons between the Ly$\alpha$ and the Lyman-limit
frequencies that redshift or cascade into the Ly$\alpha$ resonance (through
the Wouthuysen-Field effect [Wouthuysen 1952 \cite{Wout}; Field 1959
\cite{Field}]). Parts of the curve are exaggerated for pedagogic purposes.
The exact shape depends on astrophysical details about the first galaxies,
such as their production of X-ray binaries, supernovae, nuclear accreting
black holes, and their generation of relativistic electrons in
collisionless shocks which produce UV and X-ray photons through
inverse-Compton scattering of CMB photons. }
\label{spin}
\end{figure}

The large number of independent modes probed by the 21cm signal would
provide a measure of non-Gaussian deviations to a level of $\sim N_{\rm 21
cm}^{-1/2}$, constituting a test of the inflationary origin of the
primordial inhomogeneities which are expected to possess deviations
$\gtrsim 10^{-6}$ \cite{malda}.
 
\subsection{The Characteristic Observed Size of Ionized Bubbles at the
End of Reionization} 
\label{size}

\begin{figure}
\centering
\includegraphics[height=6cm]{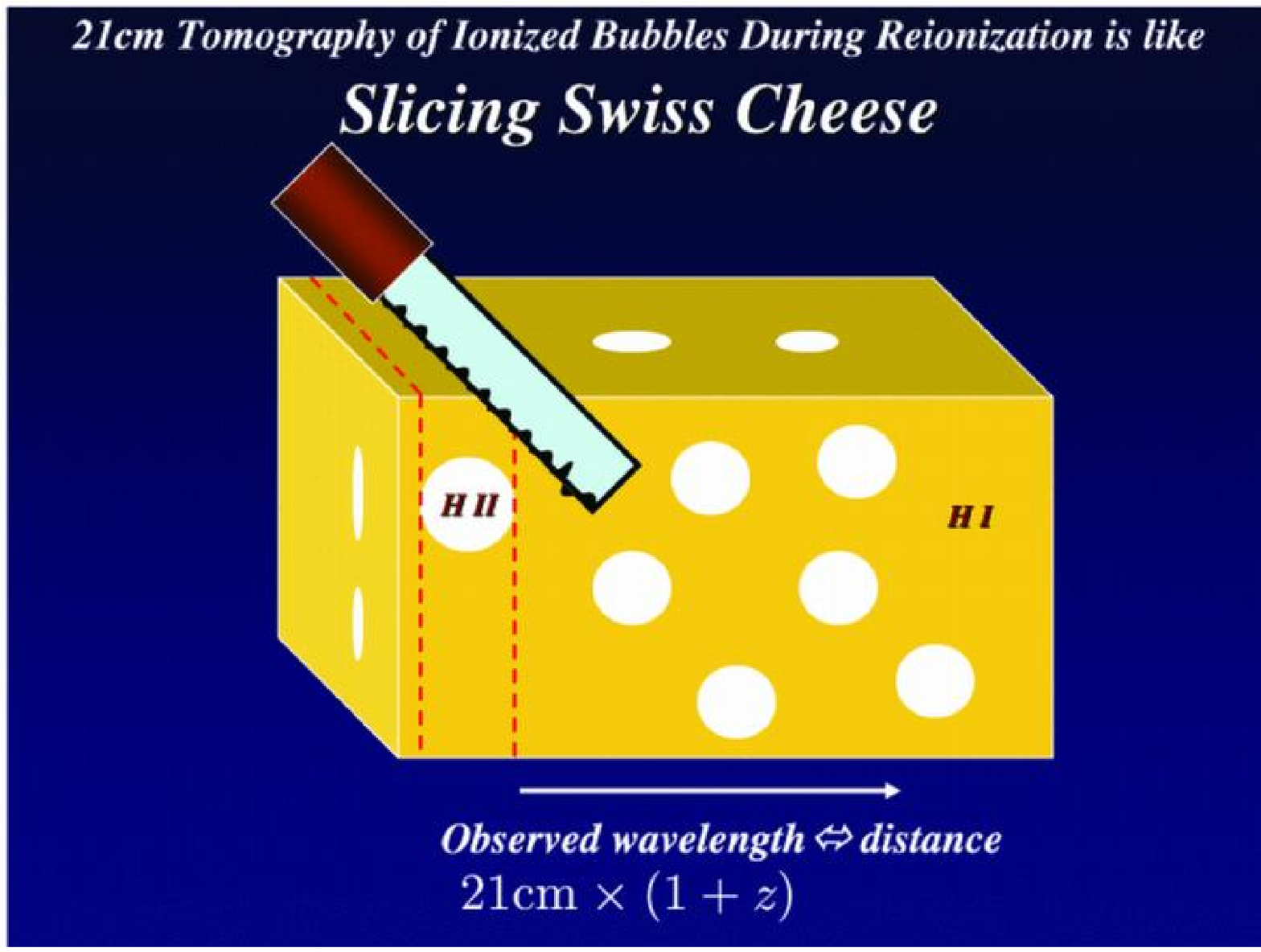}
\caption{21cm imaging of ionized bubbles during the epoch of reionization
is analogous to slicing swiss cheese. The technique of slicing at intervals
separated by the typical dimension of a bubble is optimal for revealing
different pattens in each slice.}  \label{swiss}
\end{figure}

The first galaxies to appear in the Universe at redshifts $z\ga 20$ created
ionized bubbles in the intergalactic medium (IGM) of neutral hydrogen (\HI)
left over from the Big-Bang. It is thought that the ionized bubbles grew
with time, surrounded clusters of dwarf galaxies\cite{BL04,FZH} and
eventually overlapped quickly throughout the Universe over a narrow
redshift interval near $z\sim 6$. This event signaled the end of the
reionization epoch when the Universe was a billion years old.  Measuring
the unknown size distribution of the bubbles at their final overlap phase
is a focus of forthcoming observational programs aimed at highly redshifted
21cm emission from atomic hydrogen.  In this sub-section we follow Wyithe
\& Loeb (2004) \cite{Bubble} and show that the combined constraints of cosmic variance
and causality imply an observed bubble size at the end of the overlap epoch
of $\sim 10$ physical Mpc, and a scatter in the observed redshift of
overlap along different lines-of-sight of $\sim 0.15$. This scatter is
consistent with observational constraints from recent spectroscopic data on
the farthest known quasars. This result implies that future radio
experiments should be tuned to a characteristic angular scale of $\sim
0.5^\circ$ and have a minimum frequency band-width of $\sim 8$ MHz for an
optimal detection of 21cm flux fluctuations near the end of reionization.

During the reionization epoch, the characteristic bubble size (defined here
as the spherically averaged mean radius of the \HII regions that contain
most of the ionized volume\cite{FZH}) increased with time as smaller
bubbles combined until their overlap completed and the diffuse IGM was
reionized.  However the largest size of isolated bubbles (fully surrounded
by \HI boundaries) that can be {\em observed} is finite, because of the
combined phenomena of cosmic variance and causality. Figure \ref{ffig1}
presents a schematic illustration of the geometry. There is a surface on
the sky corresponding to the time along different lines-of-sight when the
diffuse (uncollapsed) IGM was {\em most recently neutral}. We refer to it
as the Surface of Bubble Overlap (SBO). There are two competing sources for
fluctuations in the SBO, each of which is dependent on the characteristic
size, $R_{\rm SBO}$, of the ionized regions just before the final
overlap. First, the finite speed of light implies that 21cm photons
observed from different points along the curved boundary of an \HII region
must have been emitted at different times during the history of the
Universe.  Second, bubbles on a comoving scale $R$ achieve reionization
over a spread of redshifts due to cosmic variance in the initial conditions
of the density field smoothed on that scale.  The characteristic scale of
\HII bubbles grows with time, leading to a decline in the spread of their
formation redshifts\cite{BL04} as the cosmic variance is averaged over an
increasing spatial volume.  However the 21cm light-travel time across a
bubble rises concurrently. Suppose a signal 21cm photon which encodes the
presence of neutral gas, is emitted from the far edge of the ionizing
bubble. If the adjacent region along the line-of-sight has not become
ionized by the time this photon reaches the near side of the bubble, then
the photon will encounter diffuse neutral gas. Other photons emitted at
this lower redshift will therefore also encode the presence of diffuse
neutral gas, implying that the first photon was emitted prior to overlap,
and not from the SBO. Hence the largest observable scale of \HII regions
when their overlap completes, corresponds to the first epoch at which the
light crossing time becomes larger than the spread in formation times of
ionized regions.  Only then will the signal photon leaving the far side of
the HII region have the lowest redshift of any signal photon along that
line-of-sight.

\begin{figure}
\centering
\includegraphics[height=6cm]{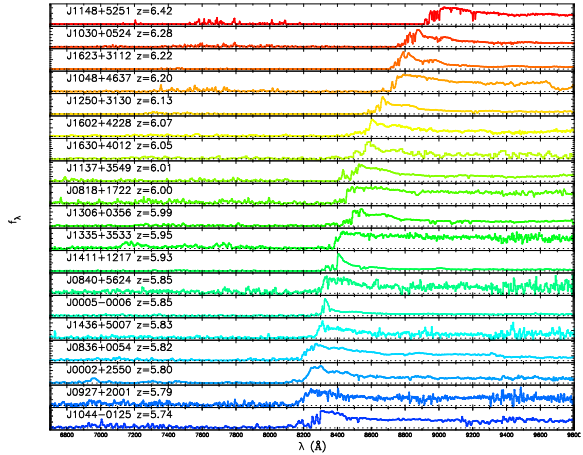}
\caption{Spectra of 19 quasars with redshifts $5.74<z<6.42$ from the {\it
Sloan Digital Sky Survey} \cite{Fan05}. For some of the highest-redshift
quasars, the spectrum shows no transmitted flux shortward of the Ly$\alpha$
wavelength at the quasar redshift (the so-called ``Gunn-Peterson trough''),
indicating a non-negligible neutral fraction in the IGM (see the analysis
of Fan et al. \cite{Fan05} for details).  }
\end{figure}

The observed spectra of some quasars beyond $z\sim6.1$ show a Gunn-Peterson
trough\cite{GP,f2} (Fan et al. 2005 \cite{Fan05}), a blank spectral region at
wavelengths shorter than \lya at the quasar redshift, implying the presence
of \HI in the diffuse IGM.  The detection of Gunn-Peterson troughs
indicates a rapid change\cite{f1,Pe,WB1} in the neutral content of the IGM
at $z\sim6$, and hence a rapid change in the intensity of the background
ionizing flux. This rapid change implies that overlap, and hence the
reionization epoch, concluded near $z\sim6$. The most promising
observational probe\cite{Zalda04,Miguel1} of the reionization epoch is redshifted
21cm emission from intergalactic \HI.  Future observations using low
frequency radio arrays (e.g. LOFAR, MWA, and PAST) will allow a direct
determination of the topology and duration of the phase of bubble
overlap. In this section we determine the expected angular scale and
redshift width of the 21cm fluctuations at the SBO theoretically, and show
that this determination is consistent with current observational
constraints.

We start by quantifying the constraints of causality and cosmic
variance. First suppose we have an \HII region with a physical radius
$R/(1+\langle z\rangle)$. For a 21cm photon, the light crossing time of
this radius is
\begin{equation}
\label{causality}
\langle\Delta z^2\rangle^{1/2} =
\left|\frac{dz}{dt}\right|_{\langle z\rangle}
\frac{R}{c(1+\langle z\rangle)},
\end{equation}
where at the high-redshifts of interest
$(dz/dt)=-(H_0\sqrt{\Omega_m})(1+z)^{5/2}$.  Here, $c$ is the speed of
light, $H_0$ is the present-day Hubble constant, $\Omega_m$ is the present
day matter density parameter, and $\langle z\rangle$ is the mean redshift
of the SBO.  Note that when discussing this crossing time, we are referring
to photons used to probe the ionized bubble (e.g. at 21cm), rather than
photons involved in the dynamics of the bubble evolution.

Second, overlap would have occurred at different times in different regions
of the IGM due to the cosmic scatter in the process of structure formation
within finite spatial volumes\cite{BL04}. Reionization should be completed
within a region of comoving radius $R$ when the fraction of mass
incorporated into collapsed objects in this region attains a certain
critical value, corresponding to a threshold number of ionizing photons
emitted per baryon. The ionization state of a region is governed by the
enclosed ionizing luminosity, by its over-density, and by dense pockets of
neutral gas that are self shielding to ionizing radiation.  There is an
offset \cite{BL04} $\delta z$ between the redshift when a region of mean
over-density $\bar{\delta}_{\rm R}$ achieves this critical collapsed
fraction, and the redshift ${\bar z}$ when the Universe achieves the same
collapsed fraction on average.  This offset may be computed\cite{BL04} from
the expression for the collapsed fraction\cite{bond91} $F_{\rm col}$ within a
region of over-density $\bar{\delta}_{\rm R}$ on a comoving scale $R$,
\begin{equation}
\label{scatter}
F_{\rm col}(M_{\rm min})=\mbox{erfc}\left[\frac{\delta_{\rm
c}-\bar{\delta}_{\rm R}}{\sqrt{2[\sigma_{\rm R_{\rm min}}^2-\sigma_{\rm
R}^2]}}\right]
\rightarrow 
\frac{\delta
z}{(1+\bar{z})}=\frac{\bar{\delta}_{\rm R}}{\delta_{\rm
c}(\bar{z})}-\left[1-\sqrt{1-\frac{\sigma_{\rm R}^2}{\sigma_{\rm R_{\rm
min}}^2}}\right],
\end{equation}
where $\delta_{\rm c}(\bar{z})\propto (1+\bar{z})$ is the collapse
threshold for an over-density at a redshift $\bar{z}$; $\sigma_{\rm R}$ and
$\sigma_{R_{\rm min}}$ are the variances in the power-spectrum linearly
extrapolated to $z=0$ on comoving scales corresponding to the region of
interest and to the minimum galaxy mass $M_{\rm min}$, respectively.  The
offset in the ionization redshift of a region depends on its linear
over-density, $\bar{\delta}_{\rm R}$. As a result, the distribution of
offsets, and therefore the scatter in the SBO may be obtained directly from
the power spectrum of primordial inhomogeneities. As can be seen from
equation~(\ref{scatter}), larger regions have a smaller scatter due to
their smaller cosmic variance.

Note that equation~(\ref{scatter}) is independent of the critical
value of the collapsed fraction required for reionization. Moreover,
our numerical constraints are very weakly dependent on the minimum
galaxy mass, which we choose to have a virial temperature of $10^4$K
corresponding to the cooling threshold of primordial atomic gas.  The
growth of an \HII bubble around a cluster of sources requires that the
mean-free-path of ionizing photons be of order the bubble radius or
larger. Since ionizing photons can be absorbed by dense pockets of
neutral gas inside the \HII region, the necessary increase in the
mean-free-path with time implies that the critical collapsed fraction
required to ionize a region of size $R$ increases as well. This larger
collapsed fraction affects the redshift at which the region becomes
ionized, but not the scatter in redshifts from place to place which is
the focus of this sub-section. Our results are therefore independent
of assumptions about unknown quantities such as the star formation
efficiency and the escape fraction of ionizing photons from galaxies,
as well as unknown processes of feedback in galaxies and clumping of
the IGM.

Figure~\ref{ffig2} displays the above two fundamental constraints.  The
causality constraint (Eq.~\ref{causality}) is shown as the blue line,
giving a longer crossing time for a larger bubble size. This contrasts with
the constraint of cosmic variance (Eq.~\ref{scatter}), indicated by the red
line, which shows how the scatter in formation times decreases with
increasing bubble size. The scatter in the SBO redshift and the
corresponding fluctuation scale of the SBO are given by the intersection of
these curves. We find that the thickness of the SBO is $\langle\Delta
z^2\rangle^{1/2}\sim0.13$, and that the bubbles which form the SBO have a
characteristic comoving size of $\sim60$Mpc (equivalent to 8.6 physical
Mpc). At $z\sim6$ this size corresponds to angular scales of $\theta_{\rm
SBO}\sim0.4$ degrees on the sky.

A scatter of $\sim0.15$ in the SBO is somewhat larger than the value
extracted from existing numerical
simulations\cite{g00,yoshida}. The difference is most likely due
to the limited size of the simulated volumes; while the simulations
appropriately describe the reionization process within limited regions
of the Universe, they are not sufficiently large to describe the
global properties of the overlap phase\cite{BL04}. The scales over
which cosmological radiative transfer has been simulated are smaller
than the characteristic extent of the SBO, which we find to be $R_{\rm
SBO}\sim70$ comoving Mpc.

We can constrain the scatter in the SBO redshift observationally using the
spectra of the highest redshift quasars. Since only a trace amount of
neutral hydrogen is needed to absorb Ly$\alpha$ photons, the time where the
IGM becomes \lya transparent need not coincide with bubble
overlap. Following overlap the IGM was exposed to ionizing sources in all
directions and the ionizing intensity rose rapidly. After some time the
ionizing background flux was sufficiently high that the \HI fraction fell
to a level at which the IGM allowed transmission of resonant \lya
photons. This is shown schematically in Figure~\ref{ffig1}.  The lower
wavelength limit of the Gunn-Peterson trough corresponds to the \lya
wavelength at the redshift when the IGM started to allow transmission of
\lya photons {\em along that particular line-of-sight}.  In addition to the
SBO we therefore also define the Surface of \lya Transmission (hereafter
SLT) as the redshift along different lines-of-sight when the diffuse IGM
became transparent to Ly$\alpha$ photons.

The scatter in the SLT redshift is an observable which we would like to
compare with the scatter in the SBO redshift.  The variance of the density
field on large scales results in the biased clustering of
sources\cite{BL04}.  \HII regions grow in size around these clusters of
sources. In order for the ionizing photons produced by a cluster to advance
the walls of the ionized bubble around it, the mean-free-path of these
photons must be of order the bubble size or larger.  After bubble overlap,
the ionizing intensity at any point grows until the ionizing photons have
time to travel across the scale of the new mean-free-path, which represents
the horizon out to which ionizing sources are visible.  Since the
mean-free-path is larger than $R_{\rm SBO}$, the ionizing intensity at the
SLT averages the cosmic scatter over a larger volume than at the SBO.  This
constraint implies that the cosmic variance in the SLT redshift must be
smaller than the scatter in the SBO redshift. However, it is possible that
opacity from small-scale structure contributes additional scatter to the
SLT redshift.

If cosmic variance dominates the observed scatter in the SLT redshift, then
based on the spectra of the three $z>6.1$ quasars\cite{f2,WB1} we
would expect the scatter in the SBO redshift to satisfy $\langle\Delta
z^2\rangle^{1/2}_{\rm obs}\ga0.05$. In addition, analysis of the {\it
proximity effect} for the size of the \HII regions around the two highest
redshift quasars\cite{WL04b,Mes04} implies a neutral fraction that is
of order unity (i.e. pre-overlap) at $z\sim6.2-6.3$, while the transmission
of Ly$\alpha$ photons at $z\la6$ implies that overlap must have completed
by that time. This restricts the scatter in the SBO to be $\langle\Delta
z^2\rangle^{1/2}_{\rm obs}\la0.25$. The constraints on values for the
scatter in the SBO redshift are shaded gray in Figure~\ref{ffig2}.  It is
reassuring that the theoretical prediction for the SBO scatter of
$\langle\Delta z^2\rangle^{1/2}_{\rm obs}\sim0.15$, with a characteristic
scale of $\sim70$ comoving Mpc, is bounded by these constraints.

The possible presence of a significantly neutral IGM just beyond the
redshift of overlap\cite{WL04b,Mes04} is encouraging for upcoming
21cm studies of the reionization epoch as it results in emission near an
observed frequency of 200 MHz where the signal is most readily
detectable. Future observations of redshifted 21cm line emission at $6\la
z\la 6.5$ with instruments such as LOFAR, MWA, and PAST, will be
able to map the three-dimensional distribution of HI at the end of
reionization. The intergalactic \HII regions will imprint a 'knee' in the
power-spectrum of the 21cm anisotropies on a characteristic angular scale
corresponding to a typical isolated \HII region\cite{Zalda04}. Our results
suggest that this characteristic angular scale is large at the end of
reionization, $\theta_{\rm SBO}\sim 0.5$ degrees, motivating the
construction of compact low frequency arrays. An SBO thickness of $\langle
\Delta z^2\rangle^{1/2}\sim0.15$ suggests a minimum frequency band-width of
$\sim8$ MHz for experiments aiming to detect anisotropies in 21cm emission
just prior to overlap. These results will help guide the design of the next
generation of low-frequency radio observatories in the search for 21cm
emission at the end of the reionization epoch.
\begin{figure}
\centering
\includegraphics[height=8cm]{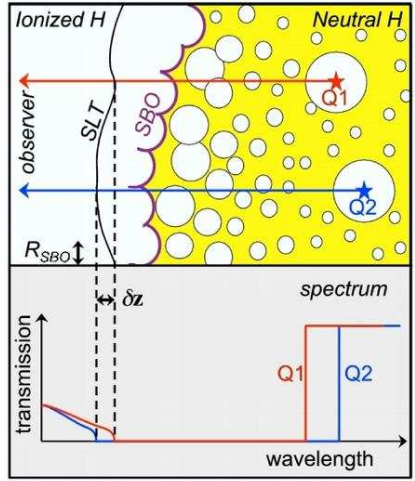}
\caption{The distances to the observed Surface of Bubble
Overlap (SBO) and Surface of Ly$\alpha$ Transmission (SLT) fluctuate
on the sky. The SBO corresponds to the first region of diffuse neutral
IGM {\em observed} along a random line-of-sight. It fluctuates across
a shell with a minimum width dictated by the condition that the light
crossing time across the characteristic radius $R_{\rm SBO}$ of
ionized bubbles equals the cosmic scatter in their formation
times. Thus, {\it causality} and {\it cosmic variance} determine the
characteristic scale of bubbles at the completion of bubble overlap.
After some time delay the IGM becomes transparent to Ly$\alpha$
photons, resulting in a second surface, the SLT.  The upper panel
illustrates how the lines-of-sight towards two quasars (Q1 in red and
Q2 in blue) intersect the SLT with a redshift difference $\delta
z$. The resulting variation in the observed spectrum of the two
quasars is shown in the lower panel.  Observationally, the ensemble of
redshifts down to which the Gunn-Peterson troughs are seen in the
spectra of $z>6.1$ quasars is drawn from the probability distribution
$dP/dz_{\rm SLT}$ for the redshift at which the IGM started to allow
\lya transmission along random lines-of-sight. The observed values of
$z_{\rm SLT}$ show a small scatter\cite{f2} in the SLT redshift around
an average value of $\langle z_{\rm SLT}\rangle\approx 5.95$. Some
regions of the IGM may have also become transparent to Ly$\alpha$
photons prior to overlap, resulting in windows of transmission inside
the Gunn-Peterson trough (one such region may have been seen\cite{WB1}
in SDSS J1148+5251).  In the existing examples, the portions of the
Universe probed by the lower end of the Gunn-Peterson trough are
located several hundred comoving Mpc away from the background quasar,
and are therefore not correlated with the quasar host galaxy. The
distribution $dP/dz_{\rm SLT}$ is also independent of the redshift
distribution of the quasars. Moreover, lines-of-sight to these quasars
are not causally connected at $z\sim 6$ and may be considered
independent. }
\label{ffig1}
\end{figure}

\begin{figure}
\centering
\includegraphics[height=8cm]{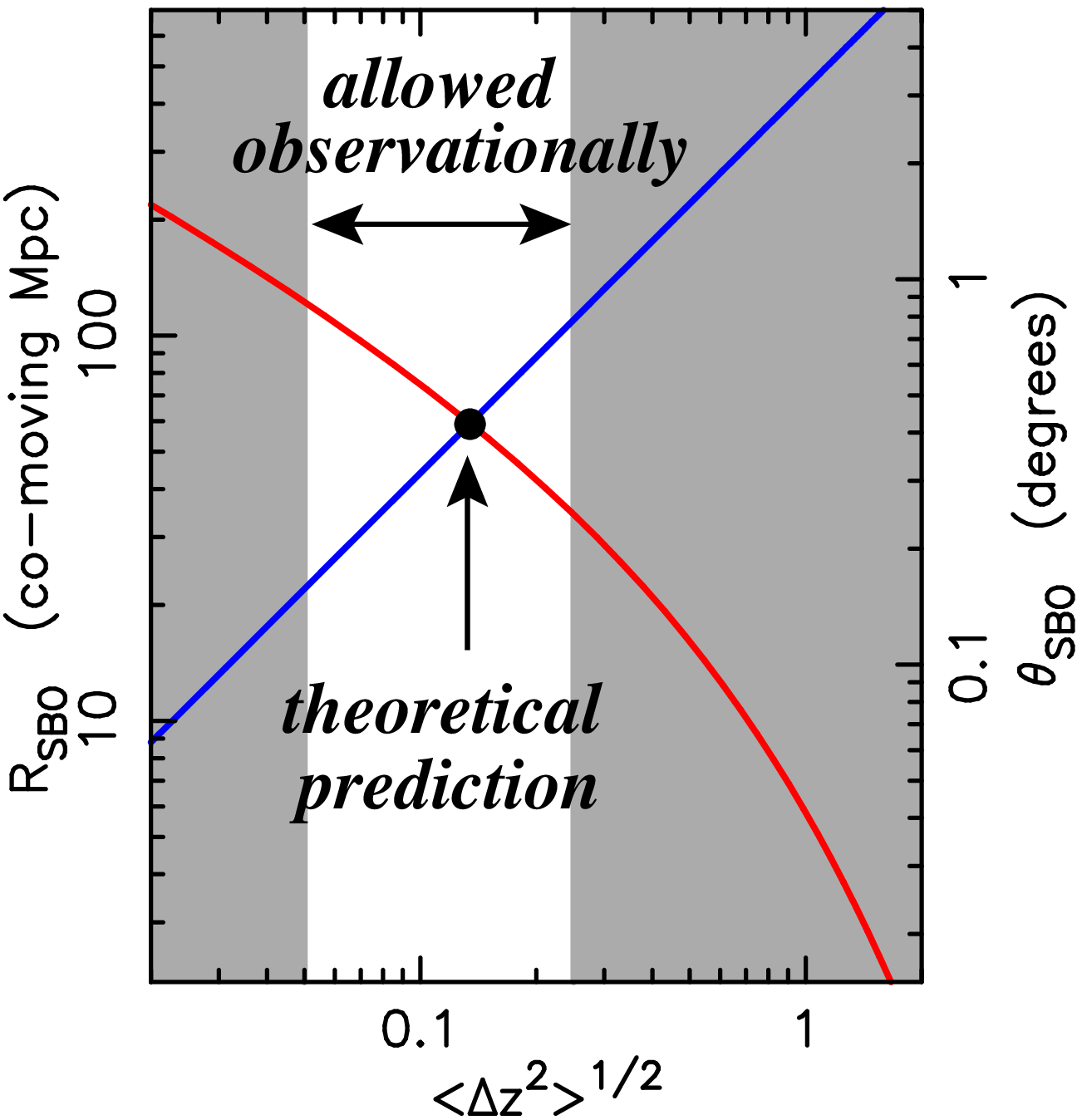}
\caption{Constraints on the scatter in the SBO redshift and
the characteristic size of isolated bubbles at the final overlap stage,
$R_{\rm SBO}$ (see Fig. 1). The characteristic size of \HII regions grows
with time. The SBO is observed for the bubble scale at which the light
crossing time (blue line) first becomes smaller than the cosmic scatter in
bubble formation times (red line).  At $z\sim6$, the implied scale $R_{\rm
SBO}\sim60$ comoving Mpc (or $\sim 8.6$ physical Mpc), corresponds to a
characteristic angular radius of $\theta_{\rm SLT}\sim 0.4$ degrees on the
sky.  After bubble overlap, the ionizing intensity grows to a level at
which the IGM becomes transparent to Ly$\alpha$ photons. The collapsed
fraction required for \lya transmission within a region of a certain size
will be larger than required for its ionization. However, the scatter in
equation~(\ref{scatter}) is not sensitive to the collapsed fraction, and so
may be used for both the SBO and SLT.  The scatter in the SLT is smaller
than the cosmic scatter in the structure formation time on the scale of the
mean-free-path for ionizing photons.  This mean-free-path must be longer
than $R_{\rm SBO}\sim60$Mpc, an inference which is supported by analysis of
the Ly$\alpha$ forest at $z\sim4$ where the mean-free-path is
estimated\cite{ME1} to be $\sim 120$ comoving Mpc at the Lyman limit (and
longer at higher frequencies). If it is dominated by cosmic variance, then
the scatter in the SLT redshift provides a lower limit to the SBO
scatter. The three known quasars at $z>6.1$ have \lya transmission
redshifts of\cite{WB1,f2} $z_{\rm SLT}=5.9$, 5.95 and 5.98,
implying that the scatter in the SBO must be $\ga0.05$ (this scatter may
become better known from follow-up spectroscopy of Gamma Ray Burst
afterglows at $z>6$ that might be discovered by the {\it SWIFT}
satellite\cite{GRBquasar,BL02}).  
The observed scatter in the SLT redshift is somewhat smaller than the
predicted SBO scatter, confirming the expectation that cosmic variance is
smaller at the SLT. The scatter in the SBO redshift must also be $\la0.25$
because the lines-of-sight to the two highest redshift quasars have a
redshift of \lya transparency at $z\sim6$, but a neutral fraction that is
known from the {\it proximity effect}\cite{WL04b} to be substantial at
$z\ga6.2-6.3$. The excluded regions of scatter for the SBO are shown in
gray.  }
\label{ffig2}
\end{figure}

The full size distribution of ionized bubbles has to be calculated from a
numerical cosmological simulation that includes gas dynamics and radiative
transfer. The simulation box needs to be sufficiently large for it to
sample an unbiased volume of the Universe with little cosmic variance, but
at the same time one must resolve the scale of individual dwarf galaxies
which provide (as well as consume) ionizing photons (see discussion at the
last section of this review). Until a reliable simulation of this magnitude
exists, one must adopt an approximate analytic approach to estimate the
bubble size distribution. Below we describe an example for such a method,
developed by Furlanetto, Zaldarriaga, \& Hernquist (2004) \cite{FZH}.

The criterion for a region to be ionized is that galaxies inside of it
produce a sufficient number of ionizing photons per baryon.  This condition
can be translated to the requirement that the collapsed fraction of mass in
halos above some threshold mass $M_{\rm min}$ will exceed some threshold,
namely $F_{\rm col}>\zeta^{-1}$. The minimum halo mass most likely
corresponds to a virial temperature of $10^4$K relating to the threshold
for atomic cooling (assuming that molecular hydrogen cooling is suppressed
by the UV background in the Lyman-Werner band).  We would like to find the
largest region around every point that satisfies the above condition on the
collapse fraction and then calculate the abundance of ionized regions of
this size.  Different regions have different values of $F_{\rm col}$
because their mean density is different. In the extended Press-Schechter
model (Bond et al. 1991 \cite{bond91}; Lacey \& Cole 1993 \cite{LC93}), the collapse fraction in a
region of mean overdensity $\delta_M$ is
\begin{equation}
F_{\rm col}={\rm erfc}\left({\delta_c-\delta_M\over \sqrt{2[\sigma_{\rm
min}^2-\sigma^2(M,z)]}}\right). 
\end{equation}
where $\sigma^2(M,z)$ is the variance of density fluctuations on 
mass scale $M$, $\sigma^2_{\rm min}\equiv\sigma^2(M_{\rm min},z)$,
and $\delta_c$ is the collapse threshold. This equation can be used
to derive the condition on the mean overdensity within
a region of mass $M$ in order for it to be ionized,
\begin{equation}
\delta_M>\delta_B(M,z)\equiv \delta_c-{\sqrt{2}}K(\zeta)[\sigma_{\rm
min}^2-\sigma^2(M,z)]^{1/2},
\label{barrier}
\end{equation}
where $K(\zeta)={\rm erfc}^{-1}(1-\zeta^{-1})$. Furlanetto et al.
\cite{FZH} showed how to construct the mass function of ionized regions
from $\delta_B$ in analogy with the halo mass function (Press \& Schechter
1974 \cite{Press}; Bond et al. 1991 \cite{bond91}). The barrier in equation (\ref{barrier}) is well
approximated by a linear dependence on $\sigma^2$,
\begin{equation}
\delta_B\approx B(M)=B_0+B_1\sigma^2(M),
\end{equation}
in which case the mass function has an analytic solution (Sheth 1998 \cite{sheth98}),
\begin{equation}
n(M)={\sqrt{2\over \pi}}{{\bar{\rho}}\over M^2}\left | {d\ln \sigma
\over d\ln M}\right | {B_0\over \sigma(M)}\exp\left[-{B^2(M)\over
2\sigma^2(M)}\right],
\end{equation}
where ${\bar{\rho}}$ is the mean mass density. This solution provides the
comoving number density of ionized bubbles with mass in the range of
$(M,M+dM)$. The main difference of this result from the Press-Schechter
mass function is that the barrier in this case becomes more difficult to
cross on smaller scales because $\delta_B$ is a decreasing function of mass
$M$.  This gives bubbles a characteristic size. The size evolves with
redshift in a way that depends only on $\zeta$ and $M_{\rm min}$.

One limitation of the above analytic model is that it ignores the non-local
influence of sources on distant regions (such as voids) as well as the
possible shadowing effect of intervening gas.  Radiative transfer effects
in the real Universe are inherently three-dimensional and cannot be fully
captured by spherical averages as done in this model. Moreover, the value
of $M_{\rm min}$ is expected to increase in regions that were already
ionized, complicating the expectation of whether they will remain ionized
later.  The history of reionization could be complicated and non monotonic
in individual regions, as described by Furlanetto \& Loeb (2005) \cite{FL05}. Finally,
the above analytic formalism does not take the light propagation delay into
account as we have done above in estimating the characteristic bubble size
at the end of reionization. Hence this formalism describes the observed
bubbles only as long as the characteristic bubble size is sufficiently
small, so that the light propagation delay can be neglected compared to
cosmic variance.  The general effect of the light propagation delay on the
power-spectrum of 21cm fluctuations was quantified by Barkana \& Loeb
(2005) \cite{BLinf}.

\subsection{Separating the ``Physics'' from the ``Astrophysics'' of the 
Reionization Epoch with 21cm Fluctuations}

The 21cm signal can be seen from epochs during which the cosmic gas was
largely neutral and deviated from thermal equilibrium with the cosmic
microwave background (CMB). The signal vanished at redshifts $z\ga
200$, when the residual fraction of free electrons after cosmological
recombination kept the gas kinetic temperature, $T_{k}$, close to the
CMB temperature, $T_\gamma$. But during $200\ga z\ga 30$ the gas
cooled adiabatically and atomic collisions kept the spin temperature
of the hyperfine level population below $T_\gamma$, so that the gas
appeared in absorption \cite{Scott,Loeb04}. As the Hubble expansion
continued to rarefy the gas, radiative coupling of $T_s$ to $T_\gamma$
began to dominate and the 21cm signal faded. When the first galaxies
formed, the UV photons they produced between the Ly$\alpha$ and Lyman
limit wavelengths propagated freely through the Universe, redshifted
into the Ly$\alpha$ resonance, and coupled $T_s$ and $T_{k}$ once
again through the Wouthuysen-Field \cite{Wout,Field} effect by which
the two hyperfine states are mixed through the absorption and
re-emission of a Ly$\alpha$ photon \citep{Madau, Ciardi}. Emission
above the Lyman limit by the same galaxies initiated the
process of reionization by creating ionized bubbles in the neutral
cosmic gas, while X-ray photons propagated farther and heated $T_{k}$
above $T_\gamma$ throughout the Universe. Once $T_s$ grew larger than
$T_\gamma$, the gas appeared in 21cm emission. The ionized bubbles
imprinted a knee in the power spectrum of 21cm fluctuations
\citep{Zalda04}, which traced the H I  topology until the
process of reionization was completed \citep{FZH}.

The various effects that determine the 21cm fluctuations can be separated
into two classes. The density power spectrum probes basic cosmological
parameters and inflationary initial conditions, and can be calculated
exactly in linear theory. However, the radiation from galaxies, both
Ly$\alpha$ radiation and ionizing photons, involves the complex, non-linear
physics of galaxy formation and star formation. If only the sum of all
fluctuations could be measured, then it would be difficult to extract the
separate sources, and in particular, the extraction of the power spectrum
would be subject to systematic errors involving the properties of
galaxies. Barkana \& Loeb (2005) \cite{BL05a} showed that the unique
three-dimensional properties of 21cm measurements permit a separation of
these distinct effects. Thus, 21cm fluctuations can probe astrophysical
(radiative) sources associated with the first galaxies, while at the same
time separately probing the physical (inflationary) initial conditions of
the Universe. In order to affect this separation most easily, it is
necessary to measure the three-dimensional power spectrum of 21cm
fluctuations. The discussion in this section follows Barkana \& Loeb
(2005) \cite{BL05a}.

\noindent{\bf Spin temperature history}


As long as the spin-temperature $T_s$ is smaller than the CMB temperature
$T_{\gamma} = 2.725 (1+z)$ K, hydrogen atoms absorb the CMB, whereas if
$T_s > T_{\gamma}$ they emit excess flux. In general, the resonant 21cm
interaction changes the brightness temperature of the CMB by
\citep{Scott,Madau} $T_b =\tau \left( T_s-T_{\gamma}\right)/(1+z)$, where
the optical depth at a wavelength $\lambda=21$cm is \beq \label{tau} \tau=
\frac {3c\lambda^2h A_{10}n_{\rm H}} {32 \pi k T_s\, (1+z)\, (dv_r/dr)}
x_{\rm HI}\ , \eeq where $n_H$ is the number density of hydrogen,
$A_{10}=2.85\times 10^{-15}~{\rm s^{-1}}$ is the spontaneous emission
coefficient, $x_{\rm HI}$ is the neutral hydrogen fraction, and $dv_r/dr$
is the gradient of the radial velocity along the line of sight with $v_r$
being the physical radial velocity and $r$ the comoving distance; on
average $dv_r/dr = H(z)/ (1+z)$ where $H$ is the Hubble parameter.  The
velocity gradient term arises because it dictates the path length over
which a 21cm photon resonates with atoms before it is shifted out of
resonance by the Doppler effect \citep{Sobolev}.

For the concordance set of cosmological parameters \citep{WMAP}, the
mean brightness temperature on the sky at redshift $z$ is 
\begin{equation}
T_b = 28\,
{\rm mK}\,
\left( \frac{\Omega_b h}{.033} \right) \left(
\frac{\Omega_m}{.27} \right)^{-\frac{1}{2}} 
\left[{({1+z})\over{10}}\right]^{1/2} \left[{({T_s - T_{\gamma}})\over
{T_s}}\right]
\bar{x}_{\rm HI}, 
\end{equation}
where $\bar{x}_{\rm HI}$ is the mean neutral
fraction of hydrogen.  The spin temperature itself is coupled to $T_k$
through the spin-flip transition, which can be excited by collisions
or by the absorption of \Lya photons.  As a result, the combination
that appears in $T_b$ becomes \citep{Field} $(T_s - T_{\gamma})/T_s =
[x_{\rm tot}/(1+ x_{\rm tot})] \left(1 - T_{\gamma}/T_k \right)$,
where $x_{\rm tot} = x_{\alpha} + x_c$ is the sum of the radiative and
collisional threshold parameters. These parameters are $x_{\alpha} =
{4 P_{\alpha} T_\star}/{27 A_{10} T_{\gamma}}$ and $x_c = {4
\kappa_{1-0}(T_k)\, n_H T_\star}/{3 A_{10} T_{\gamma}}$, where
$P_{\alpha}$ is the \Lya scattering rate which is proportional to the
\Lya intensity, and $\kappa_{1-0}$ is tabulated as a function of $T_k$
\citep{AD, Zyg}. The coupling of the spin temperature
to the gas temperature becomes substantial when $x_{\rm tot} \ga 1$.

\noindent{\bf Brightness temperature fluctuations}

Although the mean 21cm emission or absorption is difficult to measure due
to bright foregrounds, the unique character of the fluctuations in $T_b$
allows for a much easier extraction of the signal \citep{Shaver, Zalda04,
Miguel1, Miguel2, Santos}. We adopt the notation $\delta_A$ for the
fractional fluctuation in quantity $A$ (with a lone $\delta$ denoting
density perturbations). In general, the fluctuations in $T_b$ can be
sourced by fluctuations in gas density ($\delta$), \Lya flux (through
$\delta_{x_{\alpha}}$) neutral fraction ($\delta_{x_{\rm HI}}$), radial
velocity gradient ($\delta_{d_rv_r}$), and temperature, so we find \beqa
\label{dsignal} \delta_{T_b} & = & \left( 1 + \frac{x_c} {\tilde{x}_{\rm
tot}} \right) \delta + \frac{x_{\alpha}} {\tilde{x}_{\rm tot}}
\delta_{x_{\alpha}} + \delta_{x_{\rm HI}} - \delta_{d_rv_r} \nonumber \\ &
& + (\gamma_a - 1) \left[ \frac{T_{\gamma}} {T_k - T_{\gamma}} + \frac{x_c}
{\tilde{x}_{ \rm tot}}\, \frac{d \log(\kappa_{1-0})} {d \log(T_k)}
\right]\, \delta \ , \eeqa where the adiabatic index is $\gamma_a = 1 +
(\delta_{T_k} / \delta)$, and we define $\tilde{x}_{\rm tot} \equiv (1 +
x_{\rm tot}) x_{\rm tot}$. Taking the Fourier transform, we obtain the
power spectrum of each quantity; e.g., the total power spectrum $P_{T_b}$
is defined by \beq \label{pTb} \langle \td_{T_b} (\bk_1) \td_{T_b} (\bk_2)
\rangle = (2\pi)^3 \delta^D(\bk_1+\bk_2) P_{T_b}(\bk_1)\ , \eeq where
$\td_{T_b} (\bk)$ is the Fourier transform of $\delta_{T_b}$, $\bk$ is the
comoving wavevector, $\delta^D$ is the Dirac delta function, and $\langle
\cdots \rangle$ denotes an ensemble average.  In this analysis, we consider
scales much bigger than the characteristic bubble size and the early phase
of reionization (when ${\bar{\delta_{x_{\rm HI}}}}<<1$), so that the
fluctuations $\delta_{x_{\rm HI}}$ are also much smaller than unity. For a
more general treatment, see McQuinn et al. (2005) \cite{McQuinn}.

\bigskip
\medskip
\noindent{\bf The separation of powers}

The fluctuation $\delta_{T_b}$ consists of a number of isotropic
sources of fluctuations plus the peculiar velocity term
$-\delta_{d_rv_r}$. Its Fourier transform is simply proportional to
that of the density field \citep{kaiser, Indian}, \beq \label{vgrad}
\td_{d_rv_r} = -\mu^2 \td , \eeq where $\mu = \cos\theta_k$ in terms 
of the angle $\theta_k$ of $\bk$ with respect to the line of
sight. The $\mu^2$ dependence in this equation results from taking the
radial (i.e., line-of-sight) component ($\propto \mu$) of the peculiar
velocity, and then the radial component ($\propto \mu$) of its
gradient. Intuitively, a high-density region possesses a velocity
infall towards the density peak, implying that a photon must travel
further from the peak in order to reach a fixed relative redshift,
compared with the case of pure Hubble expansion. Thus the optical
depth is always increased by this effect in regions with $\delta >
0$. This phenomenon is most properly termed {\it velocity
compression}.

We therefore write the fluctuation in Fourier space as 
\beq \label{Tbk} \td_{T_b} (\bk) = \mu^2 \td(\bk) + \beta \td(\bk) + 
\td_{\rm rad}(\bk)\ , \eeq
where we have defined a coefficient $\beta$ by collecting all terms
$\propto \delta$ in Eq.~(\ref{dsignal}), and have also combined
the terms that depend on the radiation fields of \Lya photons and
ionizing photons, respectively. We assume that these radiation fields
produce isotropic power spectra, since the physical processes that
determine them have no preferred direction in space. The total power
spectrum is
\beqa \label{powTb}
P_{T_b}(\bk) & = & \mu^4 P_{\delta}(k) + 2 \mu^2 [\beta P_{\delta}(k)
+ P_{\delta\cdot{\rm rad}}(k)]+ \nonumber \\ & & [\beta^2
P_{\delta}(k) + P_{\rm rad}(k) + 2 \beta P_{\delta\cdot{\rm rad}}(k)]\
,
\eeqa 
where we have defined the power spectrum $P_{\delta\cdot{\rm
rad}}$ as the Fourier transform of the cross-correlation 
function,
\beq \label{xi} 
\xi_{\delta\cdot{\rm rad}} (r) =
\langle \delta(\br_1)\, \delta_{\rm rad} (\br_1 + \br) \rangle\ .  
\eeq 

We note that a similar anisotropy in the power spectrum has been
previously derived in a different context, i.e., where the use of
galaxy redshifts to estimate distances changes the apparent
line-of-sight density of galaxies in redshift surveys
\citep{kaiser,lilje,hamilton,fisher}. However, galaxies are
intrinsically complex tracers of the underlying density field, and in
that case there is no analog to the method that we demonstrate below
for separating in 21cm fluctuations the effect of initial conditions
from that of later astrophysical processes.

The velocity gradient term has also been examined for its global effect on
the sky-averaged power and on radio visibilities \citep{t00, Indian}.
The other sources of 21cm perturbations are isotropic and would
produce a power spectrum $P_{T_b}(k)$ that could be measured by averaging
the power over spherical shells in $\bk$ space. In the simple case where
$\beta = 1$ and only the density and velocity terms contribute, the
velocity term increases the total power by a factor of $\langle (1+\mu^2)^2
\rangle = 1.87$ in the spherical average. However, instead of averaging the
signal, we can use the angular structure of the power spectrum to greatly
increase the discriminatory power of 21cm observations. We may break up
each spherical shell in $\bk$ space into rings of constant $\mu$ and
construct the observed $P_{T_b}(k,\mu)$. Considering Eq.~(\ref{powTb}) as a
polynomial in $\mu$, i.e., $\mu^4 P_{\mu^4} + \mu^2 P_{\mu^2} + P_{\mu^0}$,
we see that the power at just three values of $\mu$ is required in order to
separate out the coefficients of 1, $\mu^2$, and $\mu^4$ for each $k$.

If the velocity compression were not present, then only the
$\mu$-independent term (times $T_b^2$) would have been observed, and its
separation into the five components ($T_b$, $\beta$, and three power
spectra) would have been difficult and subject to degeneracies. Once the
power has been separated into three parts, however, the $\mu^4$ coefficient
can be used to measure the density power spectrum directly, with no
interference from any other source of fluctuations. Since the overall
amplitude of the power spectrum, and its scaling with redshift, are well
determined from the combination of the CMB temperature fluctuations and
galaxy surveys, the amplitude of $P_{\mu^4}$ directly determines the mean
brightness temperature $T_b$ on the sky, which measures a combination of
$T_s$ and $\bar{x}_{\rm HI}$ at the observed redshift. McQuinn et
al. (2005) \cite{McQuinn} analysed in detail the parameters that can be
constrained by upcoming 21cm experiments in concert with future CMB
experiments such as Planck
(http://www.rssd.esa.int/index.php?project=PLANCK).  Once $P_{\delta}(k)$
has been determined, the coefficients of the $\mu^2$ term and the
$\mu$-independent term must be used to determine the remaining unknowns,
$\beta$, $P_{\delta\cdot{\rm rad}}(k)$, and $P_{\rm rad}(k)$. Since the
coefficient $\beta$ is independent of $k$, determining it and thus breaking
the last remaining degeneracy requires only a weak additional assumption on
the behavior of the power spectra, such as their asymptotic behavior at
large or small scales. If the measurements cover $N_k$ values of wavenumber
$k$, then one wishes to determine $2 N_k + 1$ quantities based on $2 N_k$
measurements, which should not cause significant degeneracies when $N_k \gg
1$. Even without knowing $\beta$, one can probe whether some sources of
$P_{\rm rad}(k)$ are uncorrelated with $\delta$; the quantity $P_{\rm
un-\delta}(k) \equiv P_{\mu^0}- P_{\mu^2}^2/(4 P_{\mu^4})$ equals $P_{\rm
rad} - P_{\delta\cdot{\rm rad}}^2 / P_{\delta}$, which receives no
contribution from any source that is a linear functional of the density
distribution (see the next subsection for an example).

\noindent{\bf Specific epochs}

At $z \sim 35$, collisions are effective due to the high gas density,
so one can measure the density power spectrum \citep{Loeb04} and the
redshift evolution of $n_{\rm HI}$, $T_{\gamma}$, and $T_k$. At $z\la
35$, collisions become ineffective but the first stars produce a
cosmic background of \Lya photons (i.e. photons that redshift into
the \Lya resonance) that couples $T_s$ to $T_k$. During
the period of initial \Lya coupling, fluctuations in the
\Lya flux translate into fluctuations in the 21cm brightness
\citep{BarkL05}. This signal can be observed from $z \sim 25$ until the
\Lya coupling is completed (i.e., $x_{\rm tot} \gg 1$) at $z \sim
15$. At a given redshift, each atom sees \Lya photons that were
originally emitted at earlier times at rest-frame wavelengths between
\Lya and the Lyman limit. Distant sources are time retarded, and since
there are fewer galaxies in the distant, earlier Universe, each atom
sees sources only out to an apparent source horizon of $\sim 100$
comoving Mpc at $z \sim 20$. A significant portion of the flux comes
from nearby sources, because of the $1/r^2$ decline of flux with
distance, and since higher Lyman series photons, which are degraded to
\Lya photons through scattering, can only be seen from a small
redshift interval that corresponds to the wavelength interval between
two consecutive atomic levels.

\begin{figure}
\centering
\includegraphics[height=6cm]{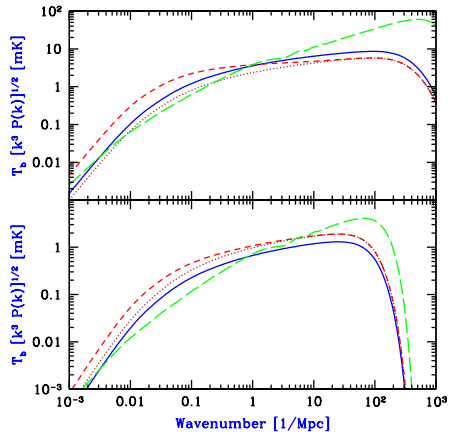}
\caption{Observable power spectra during the period of initial \Lya 
coupling. {\it Upper panel:} Assumes adiabatic cooling. {\it Lower
panel:} Assumes pre-heating to 500 K by X-ray sources. Shown are
$P_{\mu^4}=P_{\delta}$ (solid curves), $P_{\mu^2}$ (short-dashed
curves), and $P_{\rm un-\delta}$ (long-dashed curves), as well as
for comparison $2 \beta P_{\delta}$ (dotted curves).}
\label{Pkfig} 
\end{figure}

There are two separate sources of fluctuations in the \Lya flux
\citep{BarkL05}. The first is density inhomogeneities.  Since gravitational
instability proceeds faster in overdense regions, the biased
distribution of rare galactic halos fluctuates much more than the
global dark matter density.  When the number of sources seen by each
atom is relatively small, Poisson fluctuations provide a second source
of fluctuations. Unlike typical Poisson noise, these fluctuations are
correlated between gas elements at different places, since two nearby
elements see many of the same sources. Assuming a scale-invariant
spectrum of primordial density fluctuations, and that $x_{\alpha}=1$
is produced at $z=20$ by galaxies in dark matter halos where the gas
cools efficiently via atomic cooling, Figure \ref{Pkfig} shows the
predicted observable power spectra. The figure suggests that $\beta$ can
be measured from the ratio $P_{\mu^2} / P_{\mu^4}$ at $k \ga 1$
Mpc$^{-1}$, allowing the density-induced fluctuations in flux to be
extracted from $P_{\mu^2}$, while only the Poisson fluctuations
contribute to $P_{\rm un-\delta}$. Each of these components probes the
number density of galaxies through its magnitude, and the distribution
of source distances through its shape. Measurements at $k \ga 100$
Mpc$^{-1}$ can independently probe $T_k$ because of the smoothing
effects of the gas pressure and the thermal width of the 21cm line.

After \Lya coupling and X-ray heating are both completed, reionization
continues. Since $\beta = 1$ and $T_k \gg T_{\gamma}$, the
normalization of $P_{\mu^4}$ directly measures the mean neutral
hydrogen fraction, and one can separately probe the density
fluctuations, the neutral hydrogen fluctuations, and their
cross-correlation.

\noindent{\bf Fluctuations on large angular scales}

Full-sky observations must normally be analyzed with an angular and
radial transform \citep{FZH, Santos, Indian}, rather than a
Fourier transform which is simpler and yields more directly the
underlying 3D power spectrum \citep{Miguel1, Miguel2}. The 21cm
brightness fluctuations at a given redshift -- corresponding to a
comoving distance $r_0$ from the observer -- can be expanded in
spherical harmonics with expansion coefficients $a_{lm}(\nu)$, where
the angular power spectrum is
\beqa \label{cldef} C_{l}(r_0) & = & \langle |a_{lm}(\nu)|^2 \rangle 
=4 \pi \int \frac {k^2 dk} {2 \pi^2} \biggl[ G_l^2(k r_0)
P_{\delta}(k) + \nonumber \\ & & 2 P_{\delta\cdot{\rm rad}}(k) G_l(k
r_0) j_l(k r_0) + P_{\rm rad}(k) j_l^2(k r_0) \biggl] \ , \eeqa with
$G_l(x) \equiv J_l(x) + (\beta - 1) j_l(x)$ and $J_l(x)$ being a
linear combination of spherical Bessel functions \citep{Indian}.

In an angular transform on the sky, an angle of $\theta$ radians
translates to a spherical multipole $l \sim 3.5/ \theta$. For
measurements on a screen at a comoving distance $r_0$, a multipole $l$
normally measures 3D power on a scale of $k^{-1} \sim \theta r_0 \sim
35/l$ Gpc for $l\gg 1$, since $r_0 \sim 10$ Gpc at $z \ga 10$. This
estimate fails at $l \la 100$, however, when we consider the sources
of 21cm fluctuations. The angular projection implied in $C_l$ involves
a weighted average (Eq.~\ref{cldef}) that favors large scales
when $l$ is small, but density fluctuations possess little large-scale
power, and the $C_l$ are dominated by power around the peak of $k
P_\delta(k)$, at a few tens of comoving Mpc.

\begin{figure}
\centering
\includegraphics[height=6cm]{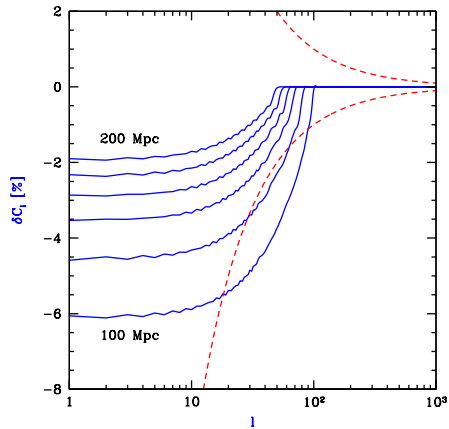}
\caption{Effect of large-scale power on the angular 
power spectrum of 21cm anisotropies on the sky. This example shows the
power from density fluctuations and velocity compression, assuming a
warm IGM at $z=12$ with $T_s=T_k \gg T_{\gamma}$. Shown is the $\%$
change in $C_l$ if we were to cut off the power spectrum above $1/k$
of 200, 180, 160, 140, 120, and 100 Mpc (top to bottom). Also shown
for comparison is the cosmic variance for averaging in bands of
$\Delta l \sim l$ (dashed lines).}
\label{fclfig} 
\end{figure}

Figure \ref{fclfig} shows that for density and velocity fluctuations, even
the $l=1$ multipole is affected by power at $k^{-1} > 200$ Mpc only at the
$2\%$ level. Due to the small number of large angular modes available on
the sky, the expectation value of $C_l$ cannot be measured precisely at
small $l$. Figure \ref{fclfig} shows that this precludes new information
from being obtained on scales $k^{-1} \ga 130$ Mpc using angular structure
at any given redshift. Fluctuations on such scales may be measurable using
a range of redshifts, but the required $\Delta z \ga 1$ at $z \sim 10$
implies significant difficulties with foreground subtraction and with the
need to account for time evolution.

\section{Major Challenge for Future Theoretical Research: 
{\it radiative transfer during reionization requires a large dynamic range,
challenging the capabilities of existing simulation codes}}

Observations of the cosmic microwave background \citep{WMAP}
have confirmed the notion that the present large-scale structure in
the Universe originated from small-amplitude density fluctuations at
early cosmic times. Due to the natural instability of gravity, regions
that were denser than average collapsed and formed bound halos, first
on small spatial scales and later on larger and larger scales. At each
snapshot of this cosmic evolution, the abundance of collapsed halos,
whose masses are dominated by cold dark matter, can be computed from
the initial conditions using numerical simulations and can be
understood using approximate analytic models \citep{ps74, bond91}. The
common understanding of galaxy formation is based on the notion that
the constituent stars formed out of the gas that cooled and
subsequently condensed to high densities in the cores of some of these
halos \citep{wr78}.

The standard analytic model for the abundance of halos \citep{ps74,
bond91} considers the small density fluctuations at some early,
initial time, and attempts to predict the number of halos that will
form at some later time corresponding to a redshift $z$. First, the
fluctuations are extrapolated to the present time using the growth
rate of linear fluctuations, and then the average density is computed
in spheres of various sizes. Whenever the overdensity (i.e., the
density perturbation in units of the cosmic mean density) in a sphere
rises above a critical threshold $\delta_c(z)$, the corresponding
region is assumed to have collapsed by redshift $z$, forming a halo
out of all the mass that had been included in the initial spherical
region. In analyzing the statistics of such regions, the model
separates the contribution of large-scale modes from that of
small-scale density fluctuations. It predicts that galactic halos will
form earlier in regions that are overdense on large scales \citep{k84,
b86, ck89, mw96}, since these regions already start out from an
enhanced level of density, and small-scale modes need only supply the
remaining perturbation necessary to reach $\delta_c(z)$. On the other
hand, large-scale voids should contain a reduced number of halos at
high redshift. In this way, the analytic model describes the
clustering of massive halos.

As gas falls into a dark matter halo, it can fragment into stars only if
its virial temperature is above $10^4$K for cooling mediated by atomic
transitions [or $\sim 500$ K for molecular ${\rm H}_2$ cooling; see
Fig. \ref{cooling}]. The abundance of dark matter halos with a virial
temperature above this cooling threshold declines sharply with increasing
redshift due to the exponential cutoff in the abundance of massive halos at
early cosmic times. Consequently, a small change in the collapse threshold
of these rare halos, due to mild inhomogeneities on much larger spatial
scales, can change the abundance of such halos dramatically. Barkana \&
Loeb (2004) \cite{BL04a} have shown that the modulation of galaxy formation
by long wavelength modes of density fluctuations is therefore amplified
considerably at high redshift; the discussion in this section follows their
analysis.

\noindent{\bf Amplification of Density Fluctuations}

Galaxies at high redshift are believed to form in all halos above some
minimum mass $M_{\rm min}$ that depends on the efficiency of atomic and
molecular transitions that cool the gas within each halo. This makes useful
the standard quantity of the collapse fraction $F_{\rm col}(M_{\rm min})$,
which is the fraction of mass in a given volume that is contained in halos
of individual mass $M_{\rm min}$ or greater (see Fig. \ref{collapsed}). If
we set $M_{\rm min}$ to be the minimum halo mass in which efficient cooling
processes are triggered, then $F_{\rm col}(M_{\rm min})$ is the fraction of
all baryons that reside in galaxies. In a large-scale region of comoving
radius $R$ with a mean overdensity $\bar{\delta}_R$, the standard result is
\citep{bond91}
\begin{equation} F_{\rm col}(M_{\rm min})={\rm erfc}\left[
\frac{\delta_c(z)- \bar{\delta}_R} {\sqrt{2 \left[S(R_{\rm min}) - S(R)
\right]}} \right]\ , \label{eq:Fcol} \end{equation} where
$S(R)=\sigma^2(R)$ is the variance of fluctuations in spheres of radius
$R$, and $S(R_{\rm min})$ is the variance in spheres of radius $R_{\rm
min}$ corresponding to the region at the initial time that contained a mass
$M_{\rm min}$. In particular, the cosmic mean value of the collapse
fraction is obtained in the limit of $R \rightarrow \infty$ by setting
$\bar{\delta}_R$ and $S(R)$ to zero in this expression. Throughout this
section we shall adopt this standard model, known as the extended
Press-Schechter model. Whenever we consider a cubic region, we will
estimate its halo abundance by applying the model to a spherical region of
equal volume. Note also that we will consistently quote values of comoving
distance, which equals physical distance times a factor of $(1+z)$.

At high redshift, galactic halos are rare and correspond to high peaks in
the Gaussian probability distribution of initial fluctuations. A modest
change in the overall density of a large region modulates the threshold for
high peaks in the Gaussian density field, so that the number of galaxies is
exponentially sensitive to this modulation. This amplification of
large-scale modes is responsible for the large statistical fluctuations
that we find.

\begin{figure} 
\centering
\includegraphics[height=6cm]{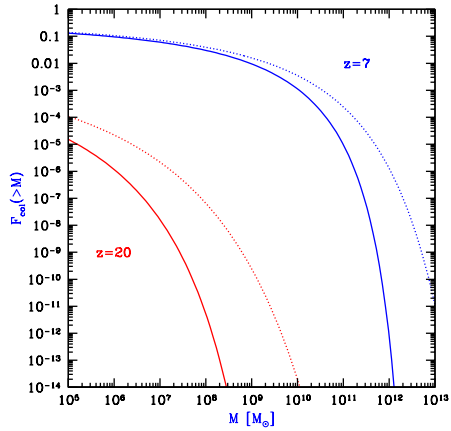} 
\caption{Bias in the halo mass distribution in simulations. Shown is the
amount of mass contained in all halos of individual mass $M_{\rm min}$ or
greater, expressed as a fraction of the total mass in a given volume. This
cumulative fraction $F_{\rm col}(M_{\rm min})$ is illustrated as a function
of the minimum halo mass $M_{\rm min}$. We consider two cases of redshift
and simulation box size, namely $z=7$, $l_{\rm box}=6$ Mpc (upper curves),
and $z=20$, $l_{\rm box}=1$ Mpc (lower curves). At each redshift, we
compare the true average distribution in the Universe (dotted curve) to the
biased distribution (solid curve) that would be measured in a simulation
box with periodic boundary conditions (for which $\bar{\delta}_R$ is
artificially set to zero).}
\label{bias}
\end{figure}

In numerical simulations, periodic boundary conditions are usually assumed,
and this forces the mean density of the box to equal the cosmic mean
density. The abundance of halos as a function of mass is then biased in
such a box (see Fig.~\ref{bias}), since a similar region in the real
Universe will have a distribution of different overdensities
$\bar{\delta}_R$.  At high redshift, when galaxies correspond to high
peaks, they are mostly found in regions with an enhanced large-scale
density. In a periodic box, therefore, the total number of galaxies is
artificially reduced, and the relative abundance of galactic halos with
different masses is artificially tilted in favor of lower-mass halos. Let
us illustrate these results for two sets of parameters, one corresponding
to the first galaxies and early reionization ($z=20$) and the other to the
current horizon in observations of galaxies and late reionization
($z=7$). Let us consider a resolution equal to that of state-of-the-art
cosmological simulations that include gravity and gas
hydrodynamics. Specifically, let us assume that the total number of dark
matter particles in the simulation is $N = 324^3$, and that the smallest
halo that can form a galaxy must be resolved into 500 particles;
\citet{converge} showed that this resolution is necessary in order to
determine the star formation rate in an individual halo reliably to within
a factor of two. Therefore, if we assume that halos that cool via molecular
hydrogen must be resolved at $z=20$ (so that $M_{\rm min}=7 \times 10^5
M_{\odot}$), and only those that cool via atomic transitions must be
resolved at $z=7$ (so that $M_{\rm min}=10^8 M_{\odot}$), then the maximum
box sizes that can currently be simulated in hydrodynamic comological
simulations are $l_{\rm box}=1$ Mpc and $l_{\rm box}=6$ Mpc at these two
redshifts, respectively.

\begin{figure}
\centering
\includegraphics[height=8cm]{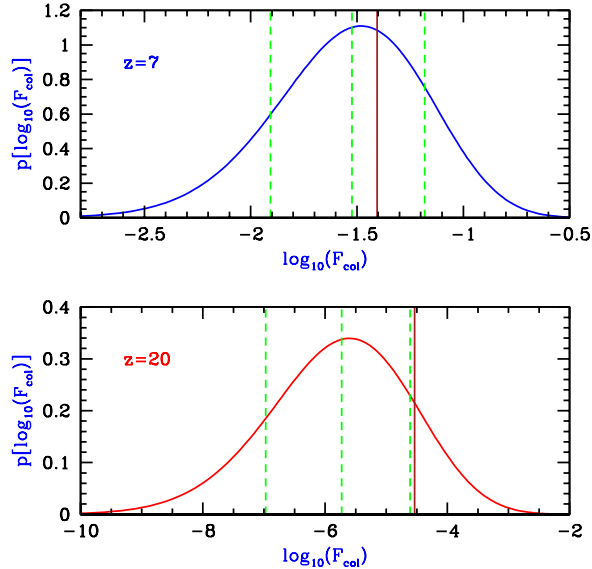}
\caption{Probability distribution within a small volume of the total
mass fraction in galactic halos. The normalized distribution of the
logarithm of this fraction $F_{\rm col}(M_{\rm min})$ is shown for two
cases: $z=7$, $l_{\rm box}=6$ Mpc, $M_{\rm min}=10^8 M_{\odot}$ (upper
panel), and $z=20$, $l_{\rm box}=1$ Mpc, $M_{\rm min}=7 \times 10^5
M_{\odot}$ (bottom panel). In each case, the value in a periodic box
($\bar{\delta}_R=0$) is shown along with the value that would be
expected given a plus or minus $1-\sigma$ fluctuation in the mean
density of the box (dashed vertical lines). Also shown in each case is
the mean value of $F_{\rm col}(M_{\rm min})$ averaged over large
cosmological volumes (solid vertical line).}  
\label{prob}
\end{figure}

At each redshift we only consider cubic boxes large enough so that the
probability of forming a halo on the scale of the entire box is
negligible. In this case, $\bar{\delta}_R$ is Gaussian distributed with
zero mean and variance $S(R)$, since the no-halo condition $\sqrt{S(R)} \ll
\del_c(z)$ implies that at redshift $z$ the perturbation on the scale $R$
is still in the linear regime. We can then calculate the probability
distribution of collapse fractions in a box of a given size (see
Fig.\ref{prob}). This distribution corresponds to a real variation in the
fraction of gas in galaxies within different regions of the Universe at a
given time. In a numerical simulation, the assumption of periodic boundary
conditions eliminates the large-scale modes that cause this cosmic
scatter. Note that Poisson fluctuations in the number of halos within the
box would only add to the scatter, although the variations we have
calculated are typically the dominant factor. For instance, in our two
standard examples, the mean expected number of halos in the box is 3 at
$z=20$ and 900 at $z=7$, resulting in Poisson fluctuations of a factor of
about 2 and 1.03, respectively, compared to the clustering-induced scatter
of a factor of about 16 and 2 in these two cases.

\begin{figure}
\centering
\includegraphics[height=8cm]{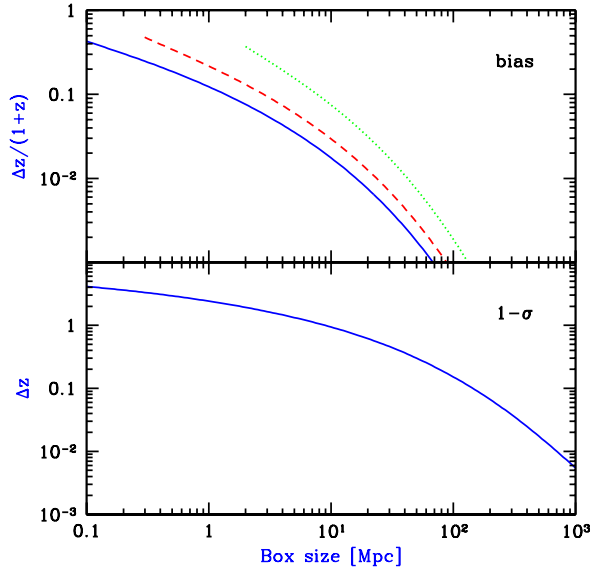}
\caption{Cosmic scatter and numerical bias, expressed as the change in
redshift needed to get the correct cosmic mean of the collapse
fraction. The plot shows the $1-\sigma$ scatter (about the biased value) in
the redshift of reionization, or any other phenomenon that depends on the
mass fraction in galaxies (bottom panel), as well as the redshift bias
[expressed as a fraction of $(1+z)$] in periodic simulation boxes (upper
panel). The bias is shown for $M_{\rm min}=7 \times 10^5 M_{\odot}$ (solid
curve), $M_{\rm min}=10^8 M_{\odot}$ (dashed curve), and $M_{\rm min}=3
\times 10^{10} M_{\odot}$ (dotted curve). The bias is always negative, and
the plot gives its absolute value. When expressed as a shift in redshift, the
scatter is independent of $M_{\rm min}$.}
\label{shifti}
\end{figure}

Within the extended Press-Schechter model, both the numerical bias and the
cosmic scatter can be simply described in terms of a shift in the redshift
(see Fig. \ref{shifti}). In general, a region of radius $R$ with a mean
overdensity $\bar{\delta}_R$ will contain a different collapse fraction
than the cosmic mean value at a given redshift $z$. However, at some wrong
redshift $z + \Delta z$ this small region will contain the cosmic mean
collapse fraction at $z$. At high redshifts ($z > 3$), this shift in
redshift was derived by Barkana \& Loeb \cite{BL04a}
from equation~(\ref{eq:Fcol})
[and was already mentioned in Eq. (\ref{scatter})]
\begin{equation} \Delta z = \frac{\bar{\delta}_R}{\delta_0} - (1+z)
\times \left[ 1 - \sqrt{1 - \frac{S(R)} {S(R_{\rm min})}}\ \right]\ ,
\end{equation} where $\delta_0 \equiv \delta_c(z)/(1+z)$ is
approximately constant at high redshifts \citep{p80}, and equals 1.28 for
the standard cosmological parameters (with its deviation from the
Einstein-de Sitter value of 1.69 resulting from the existence of a
cosmological constant). Thus, in our two examples, the bias is -2.6 at
$z=20$ and -0.4 at $z=7$, and the one-sided $1-\sigma$ scatter is 2.4 at
$z=20$ and 1.2 at $z=7$.

\noindent{\bf Matching Numerical Simulations}

\label{massfn}

Next we may develop an improved model that fits the results
of numerical simulations more accurately. The model constructs the
halo mass distribution (or mass function); cumulative quantities such
as the collapse fraction or the total number of galaxies can then be
determined from it via integration. We first define $f(\del_c(z),S)\,
dS$ to be the mass fraction contained at $z$ within halos with mass in
the range corresponding to $S$ to $S+d S$. 
As derived earlier, the Press-Schechter halo abundance is 
\beq \frac{dn}{dM} = \frac{\bar{\rho}_0}{M} \left|\frac{d S}{d M}
\right| f(\del_c(z),S)\ , \label{eq:abundance} \eeq where $dn$ is the
comoving number density of halos with masses in the range $M$ to
$M+dM$, and
\beq f_{\rm PS}(\del_c(z),S) =
\frac{1} {\sqrt{2 \pi}} \frac{\nu }{S} \exp\left[-\frac{\nu^2}{2}
\right]\ , \label{eq:PS} \eeq where $\nu=\del_c(z)/\sqrt{S}$ is the
number of standard deviations that the critical collapse overdensity
represents on the mass scale $M$ corresponding to the variance $S$.

However, the Press-Schechter mass function fits numerical simulations only
roughly, and in particular it substantially underestimates the abundance of
the rare halos that host galaxies at high redshift. The halo mass function
of \cite{shetht99} [see also \cite{shethmot}] adds two free parameters that
allow it to fit numerical simulations much more accurately
\citep{jenkins}. These N-body simulations followed very large volumes
at low redshift, so that cosmic scatter did not compromise their
accuracy. The matching mass function is given by \beq f_{\rm
ST}(\del_c(z),S) = A' \frac{\nu }{S} \sqrt{\frac{a'} {2 \pi}} \left[
1+\frac{1}{(a' \nu^2)^{q'}} \right] \exp\left[-\frac{a' \nu^2}{2} \right]\
, \label{eq:ST} \eeq with best-fit parameters \citep{shetht02} $a'=0.75$
and $q'=0.3$, and where normalization to unity is ensured by taking
$A'=0.322$.

In order to calculate cosmic scatter we must determine the biased halo
mass function in a given volume at a given mean density. Within the
extended Press-Schechter model \citep{bond91}, the halo mass
distribution in a region of comoving radius $R$ with a mean
overdensity $\bar{\delta}_R$ is given by \beq f_{\rm
bias-PS}(\del_c(z), \bar{\delta}_R,R,S)=f_{\rm PS}(\del_c(z)-
\bar{\delta}_R,S-S(R))\ \label{eq:ePS}. \eeq The corresponding collapse
fraction in this case is given simply by eq.~(\ref{eq:Fcol}). Despite
the relatively low accuracy of the Press-Schechter mass function, the
{\it relative change} is predicted rather accurately by the extended
Press-Schechter model. In other words, the prediction for the halo
mass function in a given volume compared to the cosmic mean mass
function provides a good fit to numerical simulations over a wide
range of parameters \citep{mw96,casas02}.

\begin{figure}
\centering
\includegraphics[height=8cm]{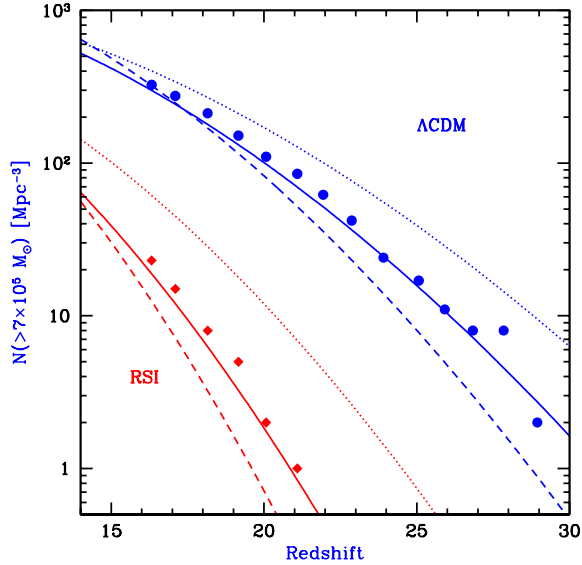} 
\caption{Halo mass function at high redshift in a 1 Mpc box at the
cosmic mean density. The prediction  (solid lines) of
the hybrid model of Barkana \& Loeb (2004) \cite{BL04a} is
compared with the number of halos above mass $7 \times 10^5 M_{\odot}$
measured in the simulations of \citet{yoshida} [data points are taken from
their Figure~5]. The cosmic mean of the halo mass function
(dotted lines) deviates significantly from the simulated values, since
the periodic boundary conditions within the finite simulation box
artificially set the amplitude of large-scale modes to zero. The
hybrid model starts with the Sheth-Tormen mass function and applies a
correction based on the extended Press-Schechter model; in doing so,
it provides a better fit to numerical simulations than the pure
extended Press-Schechter model (dashed lines) used in the previous
figures. We consider two sets of cosmological parameters, the
scale-invariant $\Lambda$CDM model of \citet{yoshida} (upper curves),
and their running scalar index (RSI) model (lower curves).}
\label{fig-Lars} 
\end{figure}

For the improved model (derived in \cite{BL04a}), we adopt a hybrid
approach that combines various previous models with each applied where it
has been found to closely match numerical simulations. We obtain the halo
mass function within a restricted volume by starting with the Sheth-Torme
formula for the cosmic mean mass function, and then adjusting it with a
relative correction based on the extended Press-Schechter model. In other
words, we set \ba & & f_{\rm bias}(\del_c(z),\bar{\delta}_R,R,S) =
\nonumber \\ & & f_{\rm ST}(\del_c(z),S)\ \times \left[ \frac{f_{\rm PS}
(\del_c(z)-\bar{\delta}_R,S-S(R))} {f_{\rm PS}(\del_c(z),S)} \right]\
. \label{eq:bias} \end{eqnarray} As noted, this model is based on fits to
simulations at low redshifts, but we can check it at high redshifts as
well. Figure~\ref{fig-Lars} shows the number of galactic halos at $z \sim
15-30$ in two numerical simulations run by \citet{yoshida}, and our
predictions given the cosmological input parameters assumed by each
simulation. The close fit to the simulated data (with no additional free
parameters) suggests that our hybrid model (solid lines) improves on the
extended Press-Schechter model (dashed lines), and can be used to calculate
accurately the cosmic scatter in the number of galaxies at both high and
low redshifts. The simulated data significantly deviate from the expected
cosmic mean [eq.~(\ref{eq:ST}), shown by the dotted line], due to the
artificial suppression of large-scale modes outside the simulated box.

As an additional example, we consider the highest-resolution first
star simulation \citep{ABN02}, which used $l_{\rm box}=128$ kpc and
$M_{\rm min}=7 \times 10^5 M_{\odot}$.  The first star forms within
the simulated volume when the first halo of mass $M_{\rm min}$ or
larger collapses within the box. To compare with the simulation, we
predict the redshift at which the probability of finding at least one
halo within the box equals $50\%$, accounting for Poisson
fluctuations. We find that if the simulation formed a population of
halos corresponding to the correct cosmic average [as given by
eq.~(\ref{eq:ST})], then the first star should have formed already at
$z=24.0$. The first star actually formed in the simulation box only at
$z=18.2$ \citep{ABN02}. Using eq.~(\ref{eq:bias}) we can account for
the loss of large-scale modes beyond the periodic box, and predict a
first star at $z=17.8$, a close match given the large Poisson
fluctuations introduced by considering a single galaxy within the box.

The artificial bias in periodic simulation boxes can also be seen in
the results of extensive numerical convergence tests carried out by
\citet{converge}. They presented a large array of numerical
simulations of galaxy formation run in periodic boxes over a wide
range of box size, mass resolution, and redshift. In particular, we
can identify several pairs of simulations where the simulations in
each pair have the same mass resolution but different box sizes; this
allows us to separate the effect of large-scale numerical bias from
the effect of having poorly-resolved individual halos. 

\bigskip
\noindent{\bf Implications}

\noindent
{\it (i) The nature of reionization} 

A variety of papers in the literature \citep{aw72, fk94, sgb94,
hl97, g00, BL01, BL01a} maintain that reionization ended with a fast,
simultaneous, overlap stage throughout the Universe. This view has been
based on simple arguments and has been supported by numerical simulations
with small box sizes. The underlying idea was that the ionized hydrogen (H
II ) regions of individual sources began to overlap when the typical size
of each H II bubble became comparable to the distance between nearby
sources. Since these two length scales were comparable at the critical
moment, there is only a single timescale in the problem -- given by the
growth rate of each bubble -- and it determines the transition time between
the initial overlap of two or three nearby bubbles, to the final stage
where dozens or hundreds of individual sources overlap and produce large
ionized regions. Whenever two ionized bubbles were joined, each point
inside their common boundary became exposed to ionizing photons from both
sources, reducing the neutral hydrogen fraction and allowing ionizing
photons to travel farther before being absorbed. Thus, the ionizing
intensity inside H II regions rose rapidly, allowing those regions to
expand into high-density gas that had previously recombined fast enough to
remain neutral when the ionizing intensity had been low. Since each bubble
coalescence accelerates the process, it has been thought that the overlap
phase has the character of a phase transition and occurs rapidly. Indeed,
the simulations of reionization \citep{g00} found that the
average mean free path of ionizing photons in the simulated volume rises by
an order of magnitude over a redshift interval $\Delta z = 0.05$ at $z=7$.

These results imply that overlap is still expected to occur rapidly, but
only in localized high-density regions, where the ionizing intensity and
the mean free path rise rapidly even while other distant regions are still
mostly neutral. In other words, the size of the bubble of an individual
source is about the same in different regions (since most halos have masses
just above $M_{\rm min}$), but the typical distance between nearby sources
varies widely across the Universe. The strong clustering of ionizing
sources on length scales as large as 30--100 Mpc introduces long timescales
into the reionization phase transition. The sharpness of overlap is
determined not by the growth rate of bubbles around individual sources, but
by the ability of large groups of sources within overdense regions to
deliver ionizing photons into large underdense regions. 

Note that the recombination rate is higher in overdense regions
because of their higher gas density. These regions still reionize
first, though, despite the need to overcome the higher recombination
rate, since the number of ionizing sources in these regions is
increased even more strongly as a result of the dramatic amplification
of large-scale modes discussed earlier.

\noindent
{\it (ii) Limitations of current simulations} 

The shortcomings of current simulations do not amount simply to a shift of
$\sim 10\%$ in redshift and the elimination of scatter. The effect
mentioned above can be expressed in terms of a shift in redshift only
within the context of the extended Press-Schechter model, and only if the
total mass fraction in galaxies is considered and not its distribution as a
function of galaxy mass. The halo mass distribution should still have the
wrong shape, resulting from the fact that $\Delta z$ depends on $M_{\rm
min}$. A self-contained numerical simulation must directly evolve a very
large volume.

Another reason that current simulations are limited is that at high
redshift, when galaxies are still rare, the abundance of galaxies grows
rapidly towards lower redshift. Therefore, a $\sim 10\%$ relative error in
redshift implies that at any given redshift around $z \sim 10$--20, the
simulation predicts a halo mass function that can be off by an order of
magnitude for halos that host galaxies (see Fig. \ref{fig-Lars}). This
large underestimate suggests that the first generation of galaxies formed
significantly earlier than indicated by recent simulations.  Another
element missed by simulations is the large cosmic scatter. This scatter can
fundamentally change the character of any observable process or feedback
mechanism that depends on a radiation background. Simulations in periodic
boxes eliminate any large-scale scatter by assuming that the simulated
volume is surrounded by identical periodic copies of itself. In the case of
reionization, for instance, current simulations neglect the collective
effects described above, whereby groups of sources in overdense regions may
influence large surrounding underdense regions. In the case of the
formation of the first stars due to molecular hydrogen cooling, the effect
of the soft ultraviolet radiation from these stars, which tends to
dissociate the molecular hydrogen around them \citep{hrl97, rgs02, Oh03},
must be reassessed with cosmic scatter included.

\noindent{\it (iii) Observational consequences} 

The spatial fluctuations that we have calculated also affect current and
future observations that probe reionization or the galaxy population at
high redshift. For example, there are a large number of programs searching
for galaxies at the highest accessible redshifts (6.5 and beyond) using
their strong Ly$\alpha$ emission \citep{h02, r03, m03, k03}. These programs
have previously been justified as a search for the reionization redshift,
since the intrinsic emission should be absorbed more strongly by the
surrounding IGM if this medium is neutral. For any particular source, it
will be hard to clearly recognize this enhanced absorption because of
uncertainties regarding the properties of the source and its radiative and
gravitational effects on its surroundings
\citep{nature,GRBquasar,s03}. However, if the luminosity function of
galaxies that emit Ly$\alpha$ can be observed, then faint sources, which do
not significantly affect their environment, should be very strongly
absorbed in the era before reionization. Reionization can then be detected
statistically through the sudden jump in the number of faint
sources\cite{Rhoads}. The above results alter the expectation for such
observations. Indeed, no sharp ``reionization redshift'' is
expected. Instead, a Ly$\alpha$ luminosity function assembled from a large
area of the sky will average over the cosmic scatter of $\Delta z \sim
1$--2 between different regions, resulting in a smooth evolution of the
luminosity function over this redshift range. In addition, such a survey
may be biased to give a relatively high redshift, since only the most
massive galaxies can be detected, and as we have shown, these galaxies will
be concentrated in overdense regions that will also get reionized
relatively early.

The distribution of ionized patches during reionization will likely be
probed by future observations, including small-scale anisotropies of the
cosmic microwave background photons that are rescattered by the ionized
patches \citep{a96, gh98, san03}, and observations of 21 cm emission by the
spin-flip transition of the hydrogen in neutral regions \citep{t00, cgo02,
fsh03}. Previous analytical and numerical estimates of these signals have
not included the collective effects discussed above, in which rare groups
of massive galaxies may reionize large surrounding areas. The transfer of
photons across large scales will likely smooth out the signal even on
scales significantly larger than the typical size of an H II bubble due to
an individual galaxy. Therefore, even the characteristic angular scales
that are expected to show correlations in such observations must be
reassessed.

The cosmic scatter also affects observations in the present-day Universe
that depend on the history of reionization. For instance, photoionization
heating suppresses the formation of dwarf galaxies after reionization,
suggesting that the smallest galaxies seen today may have formed prior to
reionization \citep{bkw01, s02, b02}. Under the popular view that assumed a
sharp end to reionization, it was expected that denser regions would have
formed more galaxies by the time of reionization, possibly explaining the
larger relative abundance of dwarf galaxies observed in galaxy clusters
compared to lower-density regions such as the Local Group of galaxies
\citep{t02, b03}. The above results undercut the basic assumption of this
argument and suggest a different explanation. Reionization
occurs roughly when the number of ionizing photons produced starts to
exceed the number of hydrogen atoms in the surrounding IGM. If the
processes of star formation and the production of ionizing photons are
equally efficient within galaxies that lie in different regions, then
reionization in each region will occur when the collapse fraction reaches
the same critical value, even though this will occur at different times in
different regions. Since the galaxies responsible for reionization have the
same masses as present-day dwarf galaxies, this estimate argues for a
roughly equal abundance of dwarf galaxies in all environments today. This
simple picture is, however, modified by several additional effects. First,
the recombination rate is higher in overdense regions at any given time, as
discussed above. Furthermore, reionization in such regions is accomplished
at an earlier time when the recombination rate was higher even at the mean
cosmic density; therefore, more ionizing photons must be produced in order
to compensate for the enhanced recombination rate. These two effects
combine to make overdense regions reionize at a higher value of $F_{\rm
col}$ than underdense regions. In addition, the overdense regions, which
reionize first, subsequently send their extra ionizing photons into the
surrounding underdense regions, causing the latter to reionize at an even
lower $F_{\rm col}$. Thus, a higher abundance of dwarf galaxies today is
indeed expected in the overdense regions.

The same basic effect may be even more critical for understanding the
properties of large-scale voids, 10--30 Mpc regions in the present-day
Universe with an average mass density that is well below the cosmic
mean. In order to predict their properties, the first step is to
consider the abundance of dark matter halos within them. Numerical
simulations show that voids contain a lower relative abundance of rare
halos \citep{mw02,co00,bh03}, as expected from the raising of the
collapse threshold for halos within a void. On the other hand,
simulations show that voids actually place a larger fraction of their
dark matter content in dwarf halos of mass below $10^{10} M_{\odot}$
\citep{gottl03}. This can be understood within the extended
Press-Schechter model. At the present time, a typical region in the
Universe fills halos of mass $10^{12} M_{\odot}$ and higher with most
of the dark matter, and very little is left over for isolated dwarf
halos. Although a large number of dwarf halos may have formed at early
times in such a region, the vast majority later merged with other
halos, and by the present time they survive only as substructure
inside much larger halos. In a void, on the other hand, large halos
are rare even today, implying that most of the dwarf halos that formed
early within a void can remain as isolated dwarf halos till the
present. Thus, most isolated dwarf dark matter halos in the present
Universe should be found within large-scale voids \citep{infall}.

However, voids are observed to be rather deficient in dwarf galaxies as
well as in larger galaxies on the scale of the Milky Way mass of $\sim
10^{12}M_\odot$ \cite{BigVoid,el-ad,pVoids}. A deficit of large galaxies
is naturally expected, since the total mass density in the void is
unusually low, and the fraction of this already low density that assembles
in large halos is further reduced relative to higher-density regions. The
absence of dwarf galaxies is harder to understand, given the higher
relative abundance expected for their host dark matter halos. The standard
model for galaxy formation may be consistent with the observations if some
of the dwarf halos are dark and do not host stars. Large numbers of dark
dwarf halos may be produced by the effect of reionization in suppressing
the infall of gas into these halos. Indeed, exactly the same factors
considered above, in the discussion of dwarf galaxies in clusters compared
to those in small groups, apply also to voids. Thus, the voids should
reionize last, but since they are most strongly affected by ionizing
photons from their surroundings (which have a higher density than the voids
themselves), the voids should reionize when the abundance of galaxies
within them is relatively low.

\bigskip
\bigskip
\bigskip
\noindent
{\bf Acknowledgements}

I thank my young collaborators with whom my own research in this field was
accomplished: Dan Babich, Rennan Barkana, Volker Bromm, Benedetta Ciardi,
Daniel Eisenstein, Steve Furlanetto, Zoltan Haiman, Rosalba Perna, Stuart
Wyithe, and Matias Zaldarriaga. I thank Donna Adams for her highly
professional assistance with the latex file, and Dan Babich \& 
Matt McQuinn for their helpful comments on the manuscript.

%
%
%


\printindex
\end{document}